\documentclass[twocolumn,a4paper,accepted=2025-06-16]{quantumarticle}
\pdfoutput=1
\usepackage[numbers,sort&compress]{natbib}
\usepackage[utf8]{inputenc}
\usepackage[english]{babel}
\usepackage[T1]{fontenc}
\usepackage{amsmath}
\usepackage{hyperref}
\usepackage{multirow}
\usepackage{amsmath}
\usepackage{amssymb}
\usepackage{physics}
\usepackage{graphicx}
\usepackage{enumitem}
\usepackage{float}
\usepackage{booktabs,multirow}
\usepackage{makecell}
\hypersetup{
    colorlinks=true,
    linkcolor=blue,
    filecolor=blue,      
    urlcolor=blue,
    citecolor=blue,
}

\begin{document}
\title{Subsystem Information Capacity in Random Circuits and Hamiltonian Dynamics}

\author{Yu-Qin Chen}
\affiliation{Graduate School of China Academy of Engineering Physics, Beijing 100193, China}
\author{Shuo Liu}
\affiliation{Institute for Advanced Study, Tsinghua University, Beijing 100084, China}
\author{Shi-Xin Zhang}
\email{shixinzhang@iphy.ac.cn}
\affiliation{Institute of Physics, Chinese Academy of Sciences, Beijing 100190, China}

\begin{abstract}
In this study, we explore the information capacity of open quantum systems, focusing on the effective channels formed by the subsystem of random quantum circuits and quantum Hamiltonian evolution. By analyzing the subsystem information capacity, which is closely linked to quantum coherent information of these effective quantum channels, we uncover a diverse range of dynamical and steady behaviors depending on the types of evolution. Therefore, the subsystem information capacity serves as a valuable tool for studying the intrinsic nature of various dynamical phases, such as integrable, localized, thermalized, and topological systems. We also reveal the impact of different initial information encoding schemes on information dynamics including one-to-one, one-to-many, and many-to-many. To support our findings, we provide representative examples for numerical simulations, including random quantum circuits with or without mid-circuit measurements, random Clifford Floquet circuits, free and interacting Aubry-André models, and Su-Schrieffer-Heeger models. These numerical results are further quantitatively explained using the effective statistical model mapping and the quasiparticle picture in the cases of random circuits and non-interacting Hamiltonian dynamics, respectively. 

\end{abstract}

\maketitle
\section{Introduction} 
Quantum information can present rich and intriguing dynamical behaviors in different non-equilibrium phases. The principles that govern the propagation of information often defy intuition, contrasting with the more familiar laws of matter and energy transfer. Despite their counterintuitive nature, these principles are of vital importance for advancing our comprehension of quantum thermalization \cite{Deutsch1991, Srednicki1994, Rigol2008}, quantum chaos \cite{Maldacena2016, Xu2020, Dowling2023},  quantum gravity \cite{Hayden2007, Brown2016}, quantum computation and error correction \cite{Lloyd2000, PhysRevLett.125.030505, PhysRevB.103.174309, PhysRevB.103.104306}.

Entanglement \cite{Amico2008} and its growth \cite{Calabrese2005, Kim2013, Liu2014a, Alba2017a, Ho2017, Mezei2017, PhysRevX.7.031016, Bertini2022} constitute fundamental building blocks for modern condensed matter physics and usually offer valuable insights from information perspectives for the analysis of quantum many-body phases.
Out-of-time correlators (OTOC) \cite{Maldacena2016, Lewis-Swan2019, Xu2024}, as another commonly used information probe, also attract intensive academic interest owing to its multifaceted phenomena, experimental relevance and relations with information scrambling \cite{Hayden2007, Sekino2008, Lashkari2013a, Hosur2016} and holographic duality. 

In this work, we introduce the subsystem information capacity (SIC) to investigate information dynamics in generic non-equilibrium systems. SIC is defined on effective channels formed by the subsystem under specific dynamical evolution.  In essence, we are investigating the proportion of the initial information that can be faithfully transmitted and preserved across varying temporal scales and subsystem sizes after Hamiltonian quench or quantum circuit dynamics. SIC is closely related to the intrinsic nature of corresponding dynamical phases and reflects the information scrambling and information protection capabilities from quantum information theory perspectives. 

We demonstrate the equivalence between the subsystem information capacity investigated here and the concept of quantum coherent information \cite{Schumacher1996} for the effective subsystem channels in some cases. This equivalence further connects the subsystem information capacity to the single-shot quantum channel capacity \cite{Lloyd1997, DiVincenzo1998, Barnum1998, Shor2003, Devetak2003, Gyongyosi2018} in quantum communication. The monotonically decreasing nature of quantum coherent information with the application of any quantum channels indicates that only the effective channels that maintain the initial value of quantum coherent information can admit perfect quantum error correction \cite{Colmenarez2023}. We also discuss the similarities and differences between subsystem information capacity and Holevo information, serving as bounds for quantum and classical information transmission capacities, respectively. The relation between OTOC and SIC is also explored.

In this work, we examine the information aspects of effective subsystem channels in a variety of systems that hold significant experimental relevance or offer clear analytical insights within certain limits. Our investigation encompasses quantum circuits with brickwall two-qubit gates, incorporating randomness in spatial or temporal dimensions, with or without mid-circuit measurements. Additionally, we study information dynamics in time evolution controlled by both non-interacting and interacting Aubry-André (AA) models and Su-Schrieffer-Heeger (SSH) models.

This paper is organized as follows. In Sec. \ref{sec:setup} we introduce the setups in this work including the models, the encoding schemes, and the information metrics of interest. In Sec. \ref{sec:ruc}, we present the results from random quantum circuits and apply the entanglement membrane picture to quantitatively understand these results. In Sec. \ref{sec:fermion}, we present numerical results for the time evolution of several one-dimensional representative fermionic Hamiltonians and understand the results with the quasiparticle picture in non-interacting limits. In Sec. \ref{sec:discuss}, we discuss possible future directions and conclude. Several technical details and supplemental results are presented in the Appendices.

\section{General setups}\label{sec:setup}

\begin{figure}[!htb]
\includegraphics[width=0.49\textwidth]{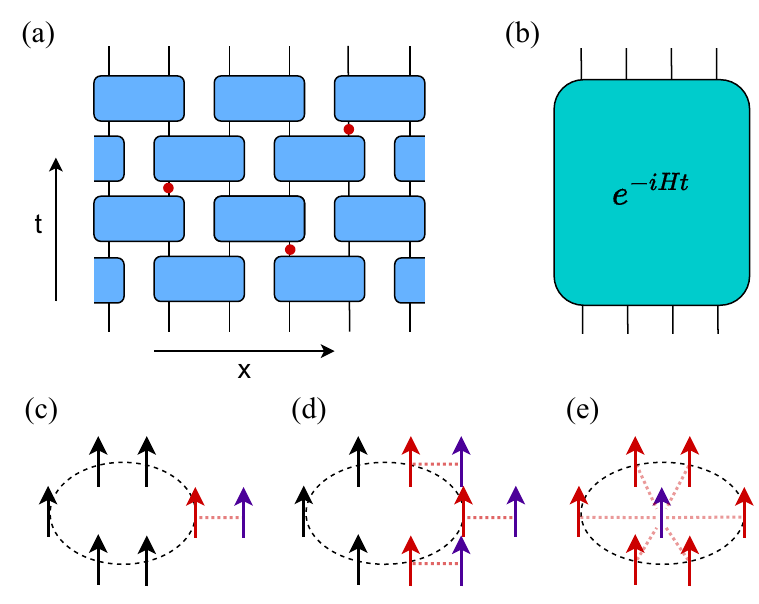}
\caption{Schematic representation of the information dynamics settings for (a) random quantum circuits and (b) time-independent Hamiltonian evolution, each accompanied by distinct information encoding schemes for the initial state. These schemes include (c) one-to-one encoding (d) finite rate (many-to-many) encoding (e) one-to-all encoding.
} \label{fig:setup}   
\end{figure}

In this section, we introduce the setups utilized for the definition of SIC in terms of models that describe the dynamics including quantum circuits and time-independent Hamiltonian evolution, schemes for information encoding at the initial state, and the probes employed to characterize information dynamics. An overview of our general setups and encoding schemes is provided in Fig. \ref{fig:setup}. We also discuss the relations between SIC and other information metrics and the unique advantages for SIC.

SIC provides a direct measure of how effectively a specific spatial subsystem $A$ can preserve or recover quantum information initially encoded in the region $E$ after undergoing time evolution $U$. Its core strengths include:
\begin{itemize}
    \item \textbf{Direct Information-Theoretic Meaning:} SIC is fundamentally linked to the quantum coherent information \cite{Schumacher1996, Lloyd1997}, quantifying the capacity of the effective subsystem channel to transmit quantum information. This provides a clear operational interpretation relevant to quantum communication, memory, and error correction.
    \item \textbf{Rich Spatio-Temporal Information:} SIC, defined as a function of subsystem size $x=|A|$ and time $t$, provides a detailed map $MI(x,t)$ revealing not only the speed of information propagation but also its spatial distribution and structure, particularly in the long-time limit.
    \item \textbf{Broad Applicability and Sensitivity:} SIC effectively distinguishes diverse dynamical phases, including thermalizing, integrable, Anderson localized, many-body localized (MBL), and topological phases, revealing characteristic ``fingerprints'' in its late-time spatial profile (see Fig. \ref{fig:resultsummary}).
    \item \textbf{Experimental Advantages:} Compared to standard protocols for OTOCs or Tripartite Mutual Information (TMI), measuring SIC often requires simpler state preparation (fewer entangled qubits) and measurement protocols (only one forward time evolution $U$), rendering it potentially more feasible on near-term quantum hardware.
\end{itemize}

\subsection{Models}
We explore two categories of dynamical models: (1) quantum circuit models including random circuits and random Clifford Floquet circuits, and (2) time-independent Hamiltonian evolution driven by Aubry-André models without or with many-body interactions as well as SSH model of topological characters.

In our analysis of circuit models, we focus on one-dimensional random quantum circuits \cite{Fisher2023}, featuring a structured arrangement of even-odd brickwall layers with random two-qudit gates drawn from Haar or Clifford ensembles. We also consider the cases with mid-circuit measurements where the properties of corresponding dynamics are averaged over different measurement history trajectories \cite{PhysRevB.98.205136, PhysRevB.100.134306, PhysRevX.9.031009, PhysRevB.99.224307}. Such mid-circuit measurements can significantly influence the global information structure and dynamics, offering a contrast to the local unitary dynamics. 

We also consider the random Clifford circuit setup where the layers of two-qudit gates are time-periodic, leading to what we term random Clifford Floquet dynamics. Previous studies reported that 1D random Clifford Floquet systems show Anderson localization behaviors \cite{Farshi2022, Farshi2023}, while 1D random Haar Floquet circuits tend to thermalize \cite{Sunderhauf2018}. 

In terms of Hamiltonian dynamics, we consider one-dimensional spinless fermionic Hamiltonians. These Hamiltonians $H$ can exhibit thermal, localization, topology features and the details of these Hamiltonians are reported in Sec. \ref{sec:fermion}. Consequently, the quench dynamics under investigation is described by the Hamiltonian evolution unitary $e^{-iHt}$. 

\subsection{Encoding schemes}\label{subsec:encoding}
We adopt the initial state for the dynamics as simple short-range entangled states, e.g. product states. In the random circuit case, we simply use $\vert 0^L\rangle$ as initial states as different product states lead to the same results due to the randomness in the gates. Since the Hamiltonian dynamics has an extra $U(1)$ symmetry associated with charge conservation, the initial state is better chosen as N\'eel states $\vert 0101\cdots\rangle$ due to the large Hilbert subspace that can be explored in the dynamics. For both circuit models and Hamiltonian models, we consider periodic boundary conditions (PBC) unless otherwise specified.

To encode quantum information in the initial state and investigate its dynamics, we employ three different schemes, and they can result in very different dynamical behaviors.

The most common encoding scheme, dubbed one-to-one encoding in our work, is to entangle a single ancilla qudit (purple arrow in Fig. \ref{fig:setup}(c)) with one of the systems qudits (red arrow in Fig. \ref{fig:setup}(c)). We label the reference qudits as $R$ and the system qudits entangled with reference qudits as $E$ or simply entangled qudits. Specifically, in our random circuit case, $R$ and $E$ forming an EPR pair at the beginning as $\frac{1}{\sqrt{d}}\sum_{i=1}^d \vert i\rangle_R\otimes\vert i\rangle_E$.
In the context of fermionic Hamiltonian cases, we consider the Bell pair $\frac{1}{\sqrt{2}}(\vert 01\rangle -i\vert 10\rangle)$ to keep the initial state in a fixed charge sector. This Bell pair can be achieved through the action of a quadratic Hamiltonian evolution, as exemplified by the following time evolution:
\begin{align}
    e^{-it (c_1^\dagger c_2 + c_2^\dagger c_1)}\vert 01\rangle = \cos t\,\vert 01\rangle-i\sin t \,\vert 10\rangle.
\end{align}
Setting the free Hamiltonian evolution time to $t=\pi/4$ leads to the desired Bell pair encoding.

In the meantime, other system qudits $\bar{E}$ remain in the same product state as described above for initial states.
The one-to-one encoding scheme is different from the typical information scrambling settings in defining tripartite mutual information (TMI) or connecting OTOC with R\'enyi entropy, where other system qubits $\bar{E}$ are also initialized with fully mixed states \cite{Hosur2016}. Therefore, our encoding schemes are different from conventional information scrambling settings and more qubit-efficient for real quantum hardware experiments (see Appendix \ref{sec:exp} for detailed discussions).

If the encoded quantum information is proportional to the size of the system, we call the encoding scheme finite-rate encoding or many-to-many encoding. In this work, we focus on the case where $L/2$ reference qudits $R$ are entangled with $L/2$ system qudits $E$ in a one-to-one fashion. We can thus study whether the amount of information encoded can drastically reshape the information dynamics and information protection capacity.

Besides, we also investigate the one-to-many encoding scheme. Specifically, we focus on the one-to-all cases, where one qudit of information is encoded in the global system at the beginning stage. The information dynamics in this case is of particular interest for its comparison with one-to-one local encoding. Practically, we encode one ancilla qudit $R$ with all system qudits into a GHZ state as $\frac{1}{\sqrt{d}}\sum_{i=1}^d\vert i\rangle_R \otimes \vert i^L\rangle_E  $ and evolve such a state under time evolution.

\subsection{SIC Probes}\label{subsec:probes}
We explore the information dynamics of a quantum evolution process by focusing on the subsystem information capacity. In this subsection, we introduce the probe of subsystem information capacity and present the differences and similarities between SIC and other commonly used information indicators.

First of all, we establish the definitions and conventions for entropy-based information measures used throughout this work. We adopt the practice of defining entropies with a logarithmic base of 
$\log_d$ for $d$-qudits. The von Neumann entropy and R\'enyi entropy for a quantum state $\rho$ are defined as:
\begin{align}
    S &= -\text{Tr}(\rho \log_d \rho), \\
    S^{(n)} &= \frac{1}{1-n}\log_d \text{Tr}(\rho^n),
\end{align}
respectively. 
The entanglement entropy of subsystem A is defined via the reduced density matrix $\rho_A$, which is obtained from the full system state $\rho$ by partial trace:
\begin{align}
    \rho_A = \text{Tr}_{\bar{A}}(\rho),
\end{align}
where $\bar{A}$ is the complementary subsystem to $A$.
According to our convention, the entanglement entropy of one qudit in a $d$-qudit EPR pair is $S_A=1$. 

The mutual information between two systems A and B is defined as:
\begin{align}
    I(A:B)=S_A+S_B-S_{AB}
\end{align}
The R\'enyi mutual information can be defined similarly using R\'enyi entropies.

In our settings, as discussed in Sec. \ref{subsec:encoding}, the system qubits are divided into $E$ and $\bar{E}$ for the initial states. We further categorize the system qubits into intervals $A$ and $\bar{A}$ for the output states at time $t$. Generally, the number of freedoms of the region $E$ and the region $A$ are not necessarily the same.  We identify the effective channel $\mathcal{E}_{sub}$ formed by the subsystem evolution mapping $E$ to $A$ as 
\begin{align}
    \mathcal{E}_{sub}(\rho_E)=\text{Tr}_{\bar{A}}\left (U(\rho_E\otimes \rho_{\bar{E}})U^\dagger\right),
    \label{eq:subsystem}
\end{align}
where $U$ describes the dynamics given by the quantum circuit or Hamiltonian evolution, $\rho_E=\text{Tr}_R(\rho_{ER})$ for initial pure state $\rho_{ER}$ under given encoding scheme and $\rho_{\bar{E}}$ is the initial state for subsystem $\bar{E}$. 
Such an effective channel is formed by the subsystem of some dynamics after tracing out the complementary system as given in Eq. \eqref{eq:subsystem}, we call it the ``subsystem channel'' that can be regarded as an evolution with some form of dissipation induced by the complementary subsystem. Such subsystem channels are interesting and explicitly useful for exploring the intrinsic properties of a system as they are formed by tracing out homogeneous physical freedoms instead of prototypical models where the freedoms for the system and environment bath are heterogeneous.  

With the lens of the subsystem channel, we introduce SIC to describe the system's information dynamics
\begin{align} \label{eq:SIC}
MI(x, t)= I(A:R)(t)
\end{align}
where $x = \vert A\vert$, with $A$ being a continuous region in real space, and the entangled qubits $E$ being located in the middle of this region. It is obvious that the definition depends on the encoding scheme as well as the initial states $\rho_{\bar{E}}$. Unless explicitly specified, we focus on the case where $\rho_{\bar{E}}$ is a pure state. $I(A:R)(t)$ is the mutual information between subsystem $A$ and reference qubits $R$ at time $t$. This probe quantitatively characterizes how much quantum information is left within region $A$ compared to the initial information encoded under the subsystem channel.  Specifically, this quantity captures the spatial-temporal profile of the information distribution via the two variables $x$ and $t$, and shows rich dynamical behaviors with varying $t$ as well as steady behaviors with $t\rightarrow \infty$ and varying subsystem size $x$.

\subsection{Late-time SIC as ``fingerprints'' for dynamical phases}
Following the presentation of our general frameworks, we now highlight one of its most powerful applications: using the late-time spatial profile of SIC as a characteristic ``fingerprint'' to identify and distinguish different quantum dynamical phases. While the temporal evolution $MI(x, t)$ reveals the dynamics of information propagation, the focus here is on the steady-state or long-time limit, $MI(x, t \to \infty)$. This quantity, when plotted as a function of the subsystem size fraction $x/L = L_A/L$ (typically for the one-to-one encoding scheme with $E$ initially at the center or boundary), reflects the stable spatial distribution achieved by the initially localized quantum information at long time limits.

\begin{figure}[!htb]
\includegraphics[width=0.48\textwidth]{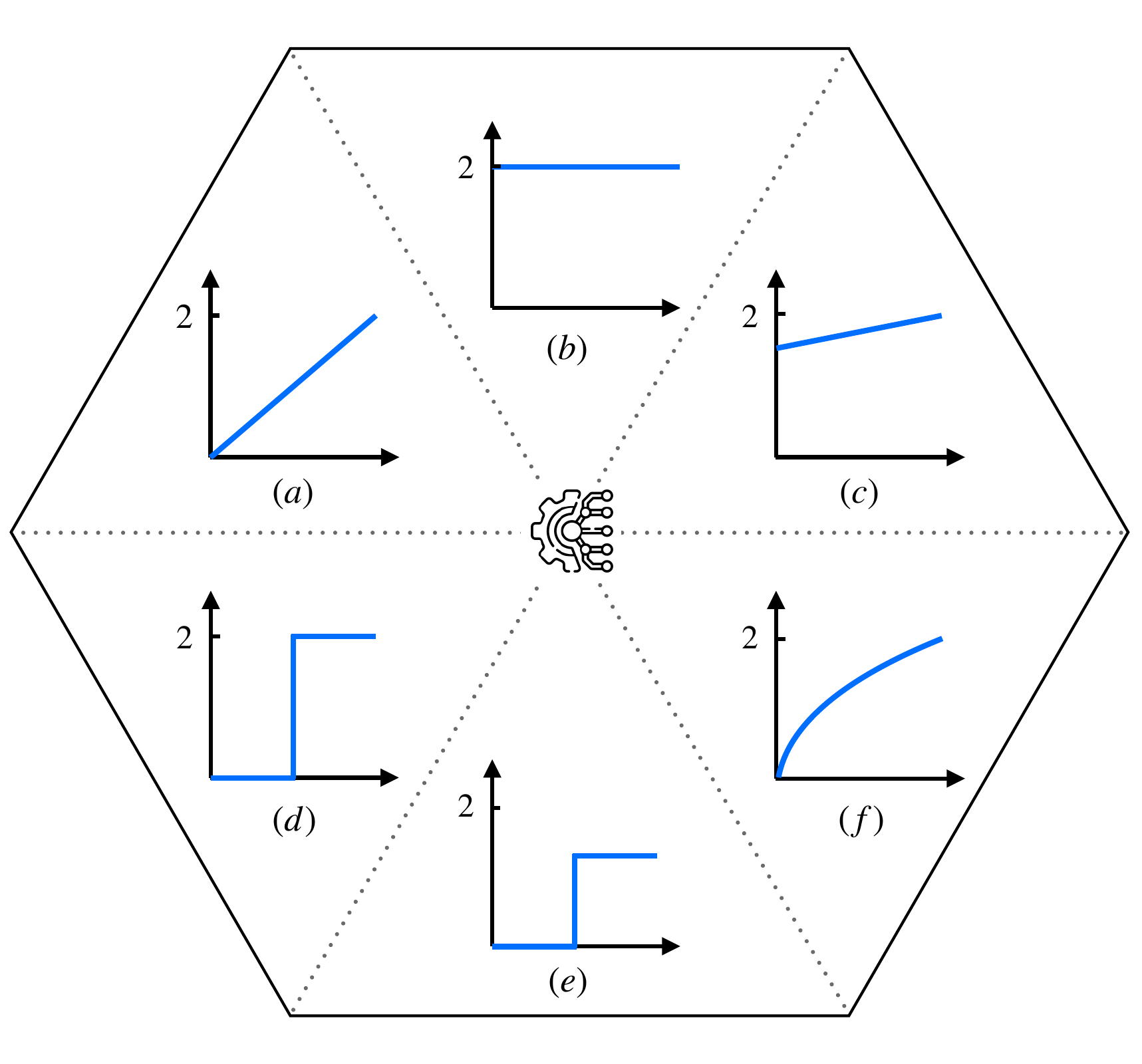}
\caption{Late-time subsystem information capacity with one-to-one encoding scheme as a function of the subsystem ratio $L_A/L$, highlighting the qualitative differences in behavior across various dynamical phases. Each panel presents the x-axis as 
$L_A/L$ and the y-axis as the average late-time SIC, with the initial value 2. (a) Extended phase of non-interacting fermion, with some integrable models following similar trends.  (b) Anderson localization phase of non-interacting fermion. Anderson localization phase induced by random Clifford Floquet circuit also follows similar curves. (c) Topological phase with edge states with the entangled qubit at the spatial boundary. (d) Chaotic phases including random circuits and thermal phases of non-integrable interacting systems. Notably, some non-thermal phases of integrable interacting systems also display comparable behavior. (e) Noise-resilient phase in monitored random circuits. (f) Many-body localization phases.
} \label{fig:resultsummary}   
\end{figure}

Our central finding, concisely illustrated in Fig.~\ref{fig:resultsummary}, is that this late-time SIC curve exhibits qualitatively distinct and characteristic shapes depending fundamentally on the nature of the underlying dynamics (e.g., chaotic, integrable, localized, topological). This allows SIC to serve as an insightful diagnostic tool. By treating the evolution $U$ as a ``black box'', measuring the $MI(x, t \to \infty)$ profile provides a remarkably clear means to discern intrinsic properties of the system's dynamical phase. The shape of this curve encapsulates how effectively information initially encoded in $E$ can be recovered from subsystems $A$ of varying size in the long run, revealing crucial details about information transport and memory within the system and is potentially helpful in quantum error correction context.

This ``fingerprint'' capability offers potentially richer information than simpler metrics. For instance, while steady-state entanglement entropy often follows volume or area laws for broad classes of phases, it might not fully capture the structural differences within those classes. As summarized in Table~\ref{tab:summary}, different phases that might all exhibit volume-law entanglement (e.g., thermalizing chaotic systems vs. certain interacting integrable systems) can display markedly different late-time SIC profiles. SIC probes the structure and accessibility of specific initially encoded quantum information distributed across spatial partitions.

The origin of these distinct fingerprints lies in the fundamental mechanisms governing information transport in different phases.
\begin{itemize}
    \item In chaotic/thermalizing systems, information scrambles rapidly throughout the system, leading to a near-complete loss of recoverable information from any small local subsystem $|A|<L/2$, resulting in a profile (Fig.~\ref{fig:resultsummary}(d)) until $A$ encompasses a significant fraction of the system.
    \item Anderson localized systems completely halt information transport. Information initially in $E$ remains localized, yielding an SIC profile as in Fig.~\ref{fig:resultsummary}(b).
    \item Many-body localized systems exhibit an intermediate behavior with slow (logarithmic) information spreading but incomplete delocalization. This results in a unique SIC profile (Fig.~\ref{fig:resultsummary}(f)) distinct from both thermal and Anderson phases, reflecting partial information kept across larger distances.
    \item Systems with localized modes, such as topological edge states, can trap information near the corresponding modes. If $E$ is placed near such a mode, the SIC profile can exhibit plateaus reflecting this trapping (Fig.~\ref{fig:resultsummary}(c)).
    \item Integrable systems can show varied behaviors depending on the nature of their quasiparticle excitations (e.g., compare Fig.~\ref{fig:resultsummary}(a) for free fermions vs. (d) for certain interacting cases).
\end{itemize}
Therefore, the late-time SIC profile $MI(x, t \to \infty)$ provides a nuanced and powerful lens through which to classify and understand the fundamental nature of quantum dynamics, complementing and extending insights gained from traditional entanglement and scrambling measures. It paves the way for a deeper comprehension of how information behaves in complex quantum systems, which is essential for both fundamental physics and applications in quantum information processing.

\begin{table*}[!htb]
\caption{Summary of main results: steady and dynamical entanglement behaviors with initial product state as well as steady-state information distribution $MI(x, t\rightarrow\infty)$ under one-to-one encoding scheme in corresponding quantum many-body phases.\\}\label{tab:summary}
\begin{tabular}{@{} cccc @{}}
\toprule
                       Phase&Entanglement growth&Steady-state entanglement&Steady-state SIC \\ 
\midrule
\makecell[c]{~\\Chaotic phase\\(Sec.~\ref{sec:ruc_1to1}, \ref{subsec:mbl})\\~} &Linear growth&Volume law (Haar value)&(d) in Fig.~\ref{fig:resultsummary}\\
\makecell[c]{~\\Many-body localization\\(Sec. \ref{subsec:mbl})\\~} & Logarithmic growth & \makecell[c]{Volume law \\(smaller than Haar value)} & (f) in Fig.~\ref{fig:resultsummary}\\
\makecell[c]{~\\Anderson localization\\(Sec. \ref{sec:1to1floquet}, \ref{sec:freefermion})\\~} & No growth & Area law & (b) in Fig.~\ref{fig:resultsummary}\\ \makecell[c]{~\\Noise-resilient phase\\(Sec. \ref{sec:ruc_1to1})\\~} & Linear growth & \makecell[c]{Volume law \\(smaller than Haar value)} & (e) in Fig.~\ref{fig:resultsummary}
\\
\makecell[c]{~\\Free fermion extended phase\\(Sec. \ref{sec:freefermion})\\~} & Linear growth & Volume law & (a) in Fig.~\ref{fig:resultsummary}\\
\makecell[c]{~\\Interacting integrable systems\\(Sec. \ref{subsec:integrable})\\~} & Linear growth & Volume law & (d) in Fig.~\ref{fig:resultsummary}\\
\makecell[c]{~\\Topological edge states\\(Sec. \ref{sec:topological_edge})\\~} & Linear growth & \makecell[c]{Volume law \\(smaller than Haar value)} & (c) in Fig.~\ref{fig:resultsummary}\\
\makecell[c]{~\\SWAP circuits\\(Sec. \ref{sec:freefermion})\\~}& No growth & Area law & (a) in Fig.~\ref{fig:resultsummary}\\
\makecell[c]{~\\CNOT circuits\\(Sec. \ref{subsec:ruc-one-to-all})\\~} & No growth & \makecell[c]{Area law \\(possible long-range entanglement)} & Fig. \ref{fig:floquet_1toall}\\
\bottomrule
\end{tabular}
\end{table*}

\subsection{Relation with other information metrics}

There are a variety of indicators for evaluating information propagation and protection, such as quantum coherent information, OTOC, tripartite mutual information, and Holevo information. In the following discussions, we clarify the differences and similarities between SIC and these quantities and compares the scope and advantages of SIC relative to those metrics.

\subsubsection{Relation with quantum coherent information}
For an arbitrary quantum channel $\mathcal{E}$ that maps the state from $\rho$ defined on system $Q$ to $\rho'=\mathcal{E}(\rho)$ defined on system $Q'$ (the number of freedom can be different for $\rho$ and $\rho'$, i.e. $\vert Q\vert \neq \vert Q'\vert$), we introduce the ancilla system $R$ and the pure state $\vert \Psi_{RQ} \rangle$ such that it reproduces the input state $\rho$ on $Q$ as $\rho = \text{Tr}_R \vert \Psi_{RQ}\rangle\langle \Psi_{RQ}\vert$. The quantum coherent information is defined in terms of the output state $\rho'_{RQ'} = (\mathcal{I}\otimes\mathcal{E})(\vert \Psi_{RQ}\rangle\langle \Psi_{RQ}\vert) $:
\begin{align}
I_\mathcal{E} = S_{Q'}-S_{Q'\cup R}.
\end{align}
Given any unitary evolution $U$ in our settings, $S_R=1$ remains unchanged, and thus the mutual information $I(A:R)$ defined in Eq. \eqref{eq:SIC} reduces to the quantum coherent information for the subsystem channel $\mathcal{E}_{sub}$,
\begin{align}
I(A:R) = S_A+1-S_{AR} = 1 + I_{\mathcal{E}_{sub}},
\end{align}
where $A$ and $E$ in our settings correspond to $Q$ and $Q'$ in the definition of quantum coherent information.

The advantage of quantum coherent information lies in that it has a data processing inequality \cite{Schumacher1996}, which states that the quantity cannot increase with the application of quantum channels. The initial coherent quantum information is the entropy of the input state $S(\rho_{E})=S(\rho_{A})$ which sets the upper limit of the coherent quantum information of any output state. As quantum error correction operations are also quantum channels, the monotonicity of this quantity suggests that perfect information protection and decoding are only feasible with channels that preserve the initial value of coherent information ($I(A:R)=2$ for unitary evolution $U$). Consequently, this measure serves as a faithful reflection of the channel's information capacity. Indeed, this quantity is used to lower bound the single-shot quantum channel capacity \cite{Lloyd1997, DiVincenzo1998, Barnum1998, Shor2003, Devetak2003, Gyongyosi2018}, which gives the maximal amount of quantum information that can be reliably transferred via the quantum channel.

\subsubsection{Relation with OTOC}\label{subsec:otoc}

OTOC is an important quantity that can diagnose information scrambling. It is defined as $F=\text{Tr}(W_i(t)V_jW_i(t)V_j)$ for infinite temperature systems. If the two local operators $W$ and $V$ average over all Pauli operators on two regions $\bar{A}$ and $E$, the corresponding averaged OTOC is related to the R\'enyi-2 entropy $S^{(2)}_{AR}$ \cite{Hosur2016}. However, $S^{(2)}_{AR}$ here is defined in a different setting from our case, where the input state in $\bar{E}$ is a full mixed state $\rho_{\bar{E}}\propto I$. Specifically, all system qubits are entangled with reference qubits individually, leaving the input to the unitary $U$ a fully mixed state on $E\cup\bar{E}$. On the contrary, we employ the input state on $\bar{E}$ as a pure state $\rho_{\bar{E}}=\vert \psi_{\bar{E}}\rangle \langle  \psi_{\bar{E}}\vert$. In sum, our settings differ from conventional settings for mutual information and are highly qubit-efficient for numerical simulations and experiments.

We establish a similar equivalence between mutual information $I(A:R)$ and OTOC for our setup with pure initial states on $\bar{E}$, see Appendix \ref{sec:tnproof} for the proof. The equivalent OTOC for SIC in our case is defined over some projector operators specified by the pure initial states in $\bar{E}$ and thus lacks the desired locality. In other words, SIC is much easier to measure from entropy-based quantities than from average OTOCs for long-range operators.

The SIC proposed in this work has several advantages over the commonly used probe OTOC for diagnosing information scrambling. From experimental perspectives, SIC only requires one call on the evolution operator $U$, thereby eliminating the need for an extra invocation on reversed time evolution $U^\dagger$, which is typically required for measuring OTOCs \cite{Mi2021}. (See Appendix \ref{sec:exp} for detailed comparisons of experimental protocols).
While OTOCs of local operators are frequently utilized to investigate when and how local information could reach distant locations, they in general may not resolve the detailed spatial distribution of conserved or slowly scrambling information within the system at late times. Consequently, they are not straightforward to differentiate phases without information scrambling. On the contrary, SIC provides a more comprehensive insight into the global structure of information propagation, evolution, and distribution and gives a finer classification on non-chaotic phases. Notably, this research is the first to unveil the steady-state information structures across a variety of dynamical phases with the lens of SIC. 

\subsubsection{Relation with tripartite mutual information}
Tripartite mutual information has been proposed to explore the intrinsic information scrambling capacity of unitary evolutions \cite{Hosur2016}.
In settings for TMI, all system qubits are entangled with external ancilla qubits, as already discussed in Sec. \ref{subsec:otoc}.

Indeed, if $\rho_{\bar{E}} $ defined in Eq. \eqref{eq:SIC} is $I$ instead of a pure state $\vert \psi_{\bar{E}}\rangle \langle  \psi_{\bar{E}}\vert$, our setup can be reduced to the case in \cite{Hosur2016}, where TMI is given as 
\begin{align}
I(R:A:\bar{A}) = I(R: A) + I(R:\bar{A}) - I(R:A\cup \bar{A}).
\end{align}
On the contrary, for our one-to-one encoding case with pure initial states in $\bar{E}$, the TMI defined above is trivially zero, for any partition $A$ and $\bar{A}$ of the system i.e.
$I(R:A:\bar{A}) = 0.$
This is guaranteed by the presence of an effective information conservation law as long as the state for the whole system $A\cup\bar{A}\cup R$ is pure:
\begin{align}
I(A:R)+I(\bar{A}:R) = 2. \label{eq:conservation}
\end{align}
Since the input state is pure for all encoding schemes in the joint system $ E\cup\bar{E}\cup R$, the information conservation law holds as long as the system evolution is unitary. 

From the experimental perspective, the fully mixed initial state preparation for standard TMI protocols is resource-intensive, typically demanding $2L$
 qubits ($L$ system qubits entangled with $L$ reference qubits). Conversely, the SIC with one-to-one encoding scheme is considerably more efficient, requiring only a single reference qubit and a total of $L+1$ qubits.

\subsubsection{Relation with Holevo information}
Another closely related information metric is Holevo information \cite{Yuan2022}. The quantity provides a lower bound on the classical information transmission capacity for a quantum channel, analogous to how quantum coherent information bounds the quantum information transmission capacity. The definition of Holevo information, in terms of the effective subsystem channel introduced in this work, is 
\begin{align}
    \chi_A = S\left(\sum_j^m\rho_A^{(j)}/m\right )-\frac{1}{m}\sum_j^m S(\rho_A^{(j)}),\label{eq:holevo}
\end{align}
where $\rho_A^{(j)} = \text{Tr}_{\bar{A}}\left (U\vert\psi_0^{(j)}\rangle \langle \psi_0^{(j)}\vert U^\dagger\right)$ is the reduced density matrix after evolution on subsystem $A$, starting from some pure initial state in the system $\vert \psi_0^{(j)}\rangle$. Unlike quantum coherent information, the definition above doesn't explicitly introduce ancilla or reference systems. In terms of the one-to-one encoding scheme, we have $m=2$ and $\vert \psi_0^{(2)}\rangle= X_E\vert \psi_0^{(1)}\rangle= X_E\vert \psi_{0E}\rangle \otimes \vert \psi_{0\bar{E}}\rangle$, where the two initial states only differ by a local perturbation on qubit $E$ initially. The first term in Holevo information then is the entropy of the reduced density matrix on $A$, with initial reference qubit $R$ entangled with the qubit $E$ as a Bell pair. Specifically, we employ the following relation:
\begin{equation} 
\resizebox{0.97\hsize}{!}{
$\begin{aligned} 
&\frac{\vert \psi_{0E}\rangle \langle \psi_{0E}\vert + X_E \vert \psi_{0E}\rangle \langle \psi_{0E}\vert X_E}{2} =  \\
&\text{Tr}_R \left(\frac{(\vert 0\rangle_R\vert \psi_{0E}\rangle+\vert 1\rangle_R X_E\vert \psi_{0E}\rangle)(\langle 0\vert_R\langle \psi_{0E}\vert+\langle 1\vert_R \langle \psi_{0E}\vert X_E)}{2}\right).
\end{aligned}$
} 
\end{equation}
Accordingly, under the one-to-one encoding scheme, the first term of Holevo information is reduced to $S_A$ in the language of our main settings with reference qubits $R$. The second term of Holevo information is straightforward as the steady-state entanglement for subsystem $A$ averaged over different initial product states. According to the statistical model mapping introduced in Sec. \ref{subsec:mapping}, the second term differs from $S_A$ term in our main settings by the free bottom boundary condition for $E$. For comparison, the bottom boundary condition for $E$ for $S_A$ and $S_{AR}$ are $\mathbb{I}$ and $\mathbb{C}$, respectively (Notations on the effective statistical model are introduced in Sec. \ref{subsec:mapping}).  Holevo information ranges from 0 to 1, with the two limits implying complete information loss and complete information protection, respectively.

Unlike SIC, Holevo information has no effective conservation laws similar to Eq. \eqref{eq:conservation}. Specifically, $\chi_A + \chi_{\bar{A}}\neq \chi_{A\cup\bar{A}}$. For example, under one-to-all encoding, where the two initial states in Eq. \eqref{eq:holevo} are $\vert \psi_0^{1}\rangle = \vert 0\rangle^{\otimes L}$ and $\vert \psi_0^{2}\rangle = \vert 1\rangle^{\otimes L}$, we have $\chi_A = \chi_{\bar{A}}=\chi_{A\cup\bar{A}}=1$. The presence and absence of information conservation laws for quantum coherent information and classical Holevo information imply the subtle differences in characterizing information from classical and quantum perspectives.

We remark that the SIC $I(A:R)$ and Holevo information $\chi_A$ can be equal to each other under a specific condition: $\text{Tr}_{\bar{A}}(\vert \psi_0^{(1)}\rangle\langle \psi_0^{(2)}\vert)=0$, see Appendix \ref{sec:holevo} for the proof. 

~\newline

In the subsequent sections, we present our findings for the random circuit systems and Hamiltonian quench dynamics. The numerical data were obtained using state-of-the-art computational tools, including the Julia package {\sf QuantumClifford.jl} for large-scale Clifford circuit simulations, the Python package {\sf TensorCircuit-NG} \cite{Zhang2022} for scalable free fermion simulations, and the Python package {\sf QuSpin} \cite{Weinberg2017, Weinberg2019} for simulations involving interacting fermions. The code implementation can be found at \cite{code}.

Supplementing these numerical results, we provide an in-depth theoretical analysis aimed at offering deeper insights into our observations. In the context of random circuit dynamics, we employ a mapping to an effective statistical model that allows us to quantitatively reproduce the numerical findings. 
For the Hamiltonian dynamics, we leverage the quasiparticle picture to gain an intuitive grasp of the system's behavior, especially in the non-interacting limit. By combining these numerical and theoretical tools, we aim to provide a comprehensive understanding of the quantum information dynamics in various physical regimes.

\section{Results for random circuit dynamics} \label{sec:ruc}

Random quantum circuits \cite{Fisher2023} emerge as a promising and powerful platform for exploring a variety of non-equilibrium phenomena, particularly from information perspectives including entanglement growth \cite{PhysRevX.7.031016, Nahum2018c} and information scrambling characterized by OTOC and other metrics \cite{Nahum2018b, VonKeyserlingk2018b, Rakovszky2018, Khemani2018, Zhuang2023}. Recently, the noise-resilience phase in monitored random circuits that can protect information from mid-circuit measurements also attracted a lot of interest \cite{PhysRevB.98.205136, PhysRevB.100.134306, PhysRevX.9.031009, PhysRevLett.126.060501, PRXQuantum.2.040319, PhysRevX.12.011045, PhysRevX.10.041020, PhysRevLett.125.030505, PhysRevB.99.224307, PhysRevB.100.064204,  PhysRevB.103.174309, PhysRevB.103.104306, hokeMeasurementinducedEntanglementTeleportation2023,  PhysRevB.102.014315, PhysRevB.106.214316, Gullans2020}. We remark that mid-circuit measurement is the only ingredient we consider in this work that leads to open system evolution and information leakage, i.e. $I(\vert A\vert=L: R)(t)<2$. It is an interesting future direction to explore possibilities and properties of other gadgets leading to information leakage in information dynamics, which include quantum noises \cite{PhysRevLett.129.080501, PhysRevB.107.L201113, Liu2024, liu2024noise, Coding_Vijay, Turkeshi2024} or construction with explicit ancillary environments \cite{Weinstein2023}. 

To investigate the dynamical behavior in random circuits with the lens of SIC, we consider systems with qubits $d=2$ numerically, and in the theoretical analysis, we consider systems composed of $L$ $d$-qudits with both large $d$ limit and finite $d$ corrections ensuring a comprehensive understanding of the dynamics. 

\subsection{Analytical framework}\label{subsec:mapping}

In this subsection, we provide a concise review of the analytical method that establishes the correspondence between random quantum circuits and effective classical spin statistical models, with further details elaborated in Appendix \ref{sec:mapping}. 
Within this framework, the entanglement entropy is interpreted as a so-called entanglement membrane in space-time \cite{Jonay2018, Zhou2020c, Sierant2023}, i.e. minimal free energy domain walls between different permutation valued spin in one dimension case. For an in-depth discussion of the mapping and the field theory, we direct the reader to \cite{Skinner2019a,Zhou2019, Bao2020, Jian2020a, PhysRevLett.129.080501, Liu2024}.

The quantum circuit under investigation is composed of random two-qudit unitary gates arranged in a brickwall layout as shown in Fig \ref{fig:setup}(a). In this work, we adopt the convention by taking the unit time $\Delta t$ = 1 for an even-odd layer of two-qudit gates. Note that the convention gives the lightcone speed $v=2$, which may be different from some literature where $\Delta t=1$ refers to one even or one odd layer. After $T$ layers of gates, the output density matrix $\rho$ is 
\begin{eqnarray}
    \rho = \prod_{t=1}^{T} \tilde{U}_{t} \rho_{0} \tilde{U}_{t}^{\dagger},
\end{eqnarray}
where $\rho_{0}$ is the density matrix of the initial state and $\tilde{U}_t$ is one even-odd layer of Haar random two-qudit gates $U$ at time step $t$.
To obtain the von Neumann entropy of the subsystem for the output state, the density matrix can be rewritten in an $r$-fold replicated Hilbert space
\begin{eqnarray}
    \vert \rho \rangle^{\otimes r} &=& \prod_{t=1}^{T} \left[ \tilde{U}_{t} \otimes \tilde{U}_{t}^{*}  \right]^{\otimes r} \vert \rho_{0} \rangle^{\otimes r}.
\end{eqnarray}

The mapping to the effective statistical model is achieved through the average over Haar random two-qudit unitary gates $U$:
\begin{eqnarray}
    \mathbb{E}_{\mathcal{U}}(U \otimes U^{*})^{\otimes r} = \sum_{\sigma, \tau \in S_{r}} \text{Wg}_{d^2}^{(r)}(\sigma \tau^{-1}) \vert \tau \tau \rangle \langle \sigma \sigma \vert,
\end{eqnarray}
where $S_{r}$ is the permutation group of degree $r$, $d$ is the local Hilbert space dimension of qudit, and $\text{Wg}_{d^{2}}^{(r)}$ is the Weingarten function~\cite{Zhou2019, collinsIntegrationRespectHaar2006}. By independently averaging over all two-qudit gates, the quantum circuit has been transformed into a classical statistical model, where the degrees of freedom are formed by permutation-valued spins $\sigma$, $\tau$ on a honeycomb lattice. The partition function $Z$ of this statistical model is obtained by summing the total weights of different spin configurations. We further trace over all $\tau$ spins to make each configuration weight positive definite.

 The total weight of a specific spin configuration after tracing $\tau$ freedom is the product of the weights of all downward triangles $W(\sigma_1, \sigma_2; \sigma_3)$.
In the large $d$ limit, we focus on the most dominant spin configuration that has the largest total weight, i.e., the partition function $Z$ is determined by the weight of the dominant spin configuration.

The statistical model is ferromagnetic by considering the weights of each downward triangle with specific spin configurations as
$W^{0}(\sigma,\sigma; \sigma) 
    \approx d^{0}
    $ and  
    $W^{0}(\sigma^{\prime},\sigma; \sigma) 
    \approx d^{-\vert (\sigma^{\prime})^{-1}) \sigma \vert}<d^0$ in the large $d$ limit, where $\vert \sigma \vert$ is the number of transpositions required to arrive $\sigma$ from the identity permutation.
Therefore, all the spins tend to align in the same direction to achieve the largest total weight. However, as discussed below, due to the particular boundary conditions, domain walls may appear with free energy of $\log (W^{0}(\sigma^{\prime},\sigma; \sigma))$ for unit length. It is also worth noting that the weight $W^{0}(\sigma, \sigma;\sigma^{\prime})=0$ due to unitary constraint~\cite{Uaverage_Qi, PRXQuantum.4.010331}, which further restricts the possible configurations of horizontal domain walls.

We now interpolate the von Neumann entropy $S_{\alpha}$ as $n=1$ limit for R\'enyi-$n$ entropy
\begin{eqnarray}
    S_{\alpha} = \underset{n \rightarrow 1}{\lim} S^{(n)}_{\alpha} = \underset{n \rightarrow 1}{\lim} \frac{1}{1-n} \mathbb{E}_{\mathcal{U}} \log\frac{\tr \rho_{\alpha}^{n}}{(\tr \rho)^{n}},
\end{eqnarray}
where $\rho_{\alpha}$ is the reduced density matrix of region $\alpha$ and $S^{(n)}_{\alpha}$ is the $n$-th order R\'enyi entropy. We can represent $S^{(n)}_{\alpha}$ in $n$-fold replicated Hilbert space as
\begin{align}
    S_{\alpha}^{(n)} &= \frac{1}{1-n} \mathbb{E}_{\mathcal{U}} \log \frac{\tr \rho_{\alpha}^{n}}{(\tr \rho)^{n}} \nonumber\\&= \frac{1}{1-n} \mathbb{E}_{\mathcal{U}} \log \frac{\Tr((C_{\alpha} \otimes I_{\bar{\alpha}}) \rho^{\otimes n})}{\Tr ( (I_{\alpha} \otimes I_{\bar{\alpha}}) \rho^{\otimes n})} \nonumber\\&= \frac{1}{1-n} \mathbb{E}_{\mathcal{U}} \log \frac{Z^{(n)}_{\alpha}}{Z^{(n)}_{0}},
\end{align}
where $C = \begin{pmatrix} 1 & 2 & ... & n  \\ 2 & 3 & ... & 1\end{pmatrix} $ and $I = \begin{pmatrix} 1 & 2 & ... & n  \\ 1 & 2 & ... & n\end{pmatrix} $ are the cyclic and identity permutations in $S_n$ group applied on each qudit, respectively. With the help of the  replica trick~\cite{nishimori2001statistical, kardar2007statistical}, we can overcome the difficulty of averaging outside the logarithmic function
\begin{eqnarray}
    \mathbb{E}_{\mathcal{U}} \log Z_{\alpha}^{(n)} &=&   \underset{k \rightarrow 0}{\lim} \frac{1}{k} \log Z_{\alpha}^{(n,k)},
    \nonumber  \\
    \mathbb{E}_{\mathcal{U}} \log Z_{0}^{(n)} &=&  \underset{k \rightarrow 0 }{\lim} \frac{1}{k} \log Z_{0}^{(n,k)},
\end{eqnarray}
where 
\begin{eqnarray}
    \label{eq:topboundary}
    Z_{\alpha}^{(n,k)} &=& \Tr  \left\{ \mathbb{C}_{\alpha} \otimes \mathbb{I}_{\bar{\alpha}}  \left[ \mathbb{E}_{\mathcal{U}} \rho^{\otimes nk} \right] \right\}, \nonumber\\ 
    Z_{0}^{(n,k)} &=& \Tr \left\{ \mathbb{I}_{\alpha} \otimes \mathbb{I}_{\bar{\alpha}}  \left[ \mathbb{E}_{\mathcal{U}} \rho^{\otimes nk} \right] \right\},
\end{eqnarray}
with $\mathbb{C} = \begin{pmatrix} 1 & 2 & ... & n  \\ 2 & 3 & ... & 1\end{pmatrix}^{\otimes k} $ and $\mathbb{I} = \begin{pmatrix} 1 & 2 & ... & n  \\ 1 & 2 & ... & n\end{pmatrix}^{\otimes k} $ are permutations in the $r$-fold replicated Hilbert space with $r=nk$. Therefore,
\begin{eqnarray}
    S_{\alpha} = \underset{n \rightarrow 1}{\underset{k \rightarrow 0 }{\lim}} \frac{1}{k(1-n)} \log \left\{ \frac{ Z_{\alpha}^{(n,k)}}{ Z_{0}^{(n,k)}} \right\},
    \label{eq:smzz}
\end{eqnarray}
where $Z$ is the partition function for the classical spin model via the mapping. In the large $d$ limit, the partition function can be reduced to the weight of the dominant spin configuration with the largest weight with particular top boundary conditions:
$\mathbb{C}_{\alpha} \otimes \mathbb{I}_{\bar{\alpha}}$ for $Z_{\alpha}$ and $\mathbb{I}_{\alpha} \otimes \mathbb{I}_{\bar{\alpha}}$ for $Z_{0}$. Therefore, $S_{\alpha}$ can be represented as the free energy difference:
\begin{eqnarray}
    S_{\alpha}^{(n,k)} = \frac{1}{k(n-1)} \left[ F_{\alpha}^{(n,k)} - F_{0}^{(n,k)}\right].
\end{eqnarray}
We note that the free energy $F^{(n,k)}$ is proportional to the length of the domain wall with unit energy $k(n-1)$, and thus $\frac{1}{k(n-1)} F^{(n,k)}$ is independent of the index $(n,k)$ in the large $d$ limit.  Moreover, the above discussion has assumed that the initial state is a product state with a free bottom boundary condition. In the case with other input states, we have to figure out the bottom boundary conditions case by case, i.e. to compute the overlap between different spin configurations and the input state $\rho_0$, such as computing the boundary weight contribution $\text{Tr}(\mathbb{C}_\alpha^{\otimes L}\rho_0)$.

\begin{figure}[!htb]
\includegraphics[width=0.45\textwidth]{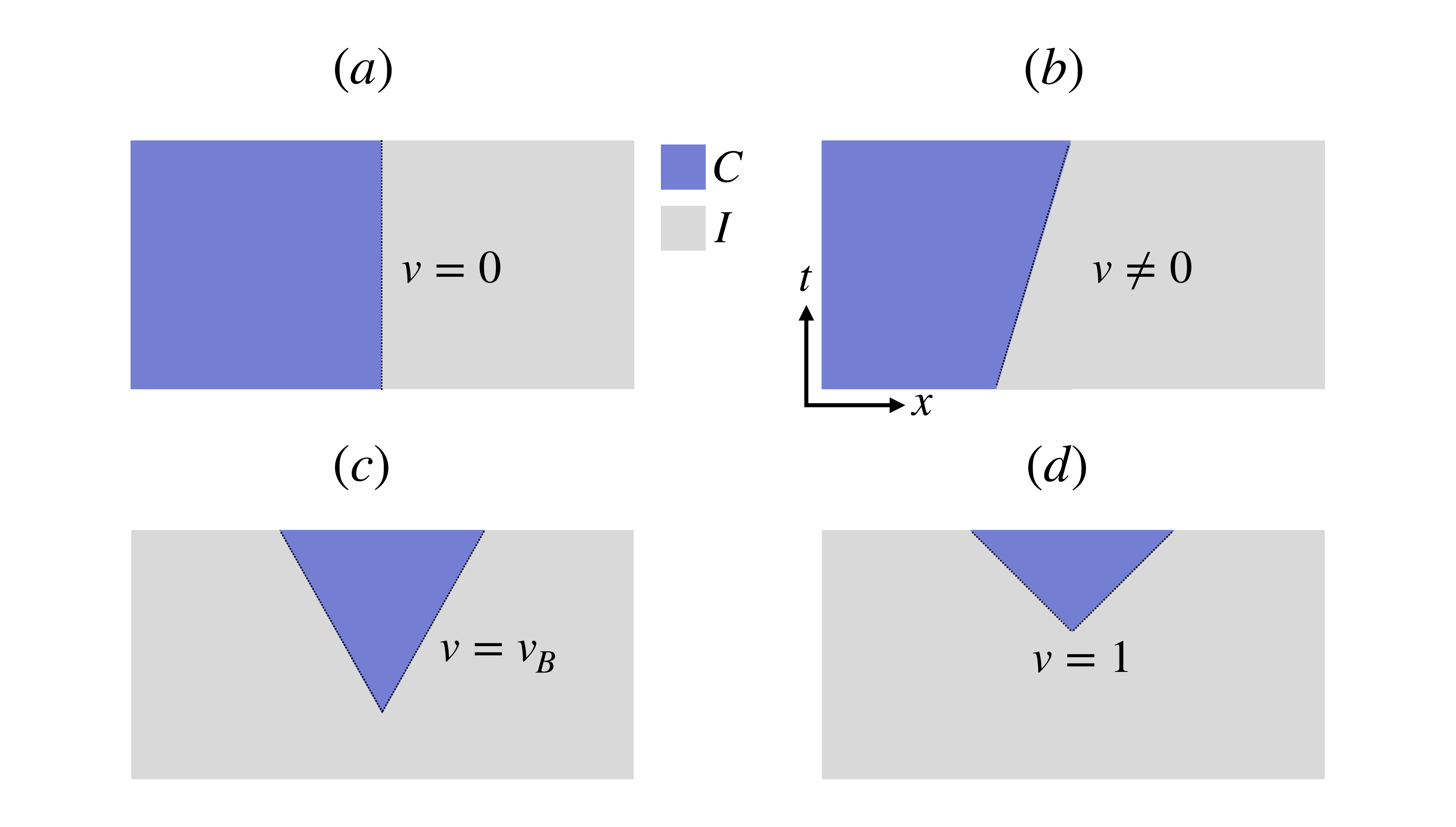}
\caption{Entanglement membrane picture for random quantum circuits. The time direction is bottom up as shown by the axis. The top boundary conditions in (a)(b) and (c)(d) correspond to the half-chain entanglement entropy in an infinite system and finite interval $L_A$ entanglement entropy in late times, respectively. Due to the free bottom boundary conditions fixed by the product initial states, the entanglement entropy is solely determined by the free energy contribution from the entanglement membrane as $\int_t\mathbb{E}(v)dt$. For the former quantity shown in (a)(b), due to the fact that $\text{argmin}_v\mathbb{E}(v)=0$, we conclude that (a) is the dominant configuration, with the free energy contribution as $\mathbb{E}(0)T =v_E T$. In the latter case shown in (c)(d), the minimal free energy configuration gives $v=v_B$ as explained in the main text. And (d) corresponds to the large $d$ limit. Interestingly, both (c) in the finite $d$ case and (d) in the large $d$ limit happen to give the identical free energy contribution of $L_A$. 
}  \label{fig:dmintro}
\end{figure}

Deviating from the large $d$ limit to the qubit case ($d=2$), there are several new contributing factors beyond the picture of dominant configuration with unit domain wall free energy $1$ (in the unit $k(n-1)$). 

\begin{enumerate}[label=(\alph*)]
    \item The least impacted factor is the separation of the annealed and quenched averaged results for finite $d$, i.e. the difference between $-\ln \mathbb{E}_{\mathcal{U}} Z$ and $-\mathbb{E}_{\mathcal{U}} \ln Z$. This difference can be properly addressed by considering the effective interaction between replicas and the term is in the order of $1/(d^8\ln d)$ as revealed in \cite{Zhou2019}. Therefore, this difference can be safely ignored for most needs even with $d=2$. In the following, we only consider $k=1$ replica as we are satisfied with annealed average results. 

    \item The domain wall unit energy approximation is also affected by the finite $d$ correction. Recall that in the large $d$ limit, the triangular weight with a domain wall separating $\mathbb{C}$ and $\mathbb{I}$ crossing is $d^{-k(n-1)}$, namely for R\'enyi-2 entropy with $k=1$ replica, the weight is $d^{-1}$ ($1$ in the context of free energy). However, in the finite $d$ case for R\'enyi-2 entropy $n=2, k=1$, the domain wall separating $\mathbb{C}$ and $\mathbb{I}$ gives an exact unit weight $d/(d^2+1)$ which is $2/5$ for $d=2$ qubit, rendering a large deviation from $1/d$ from large $d$ approximation. This significant deviation from the large $d$ approximation must be taken into account.

    \item Lastly, with finite $d$, other configurations that are less favorable in energy can give significant contributions to the final partition function, as the weights of configurations, each with terms that are small by several orders in $d$, can be compensated by the extensive freedom to arrange the domain wall. Consequently, the entropy (number of configurations) contribution to the final partition function or free energy must be carefully considered.
\end{enumerate}

The last two factors mentioned above (b), (c) for finite $d$ cases can be unified as the entanglement membrane picture \cite{Jonay2018}. In the coarse-grained space-time geometry, the domain wall/entanglement membrane can still take some fixed configuration with the speed $v=\frac{dx}{dt}$ at each location. The total free energy contribution of the domain wall is given by $\int_t \mathbb{E}(v)dt$ plus the possible bottom boundary overlap weights. To consider both the entropy term and the energy term, we have the following line tension for R\'enyi-2 entropy membrane in the random quantum circuit:
\begin{align}
    \mathbb{E}_2(v) = \log_d \frac{d^2+1}{d} + \frac{1+v}{2}\log_d \frac{1+v}{2}\nonumber\\+ \frac{1-v}{2}\log_d \frac{1-v}{2}. \label{eq:tension}
\end{align}
The first term in Eq. \eqref{eq:tension} is from the domain wall energy as indicated by (b), which is $\log_d\frac{d^2+1}{d}$ instead of $1$ for large $d$ limit. The remaining terms are from the configuration entropy, with the Stirling approximation applied to the binomial coefficients for the total count of possible domain wall configurations:
\begin{align}
    &\log_d\binom{{T(1+v)}/{2}}{T}^{1/T}\approx \nonumber\\ &\frac{1+v}{2}\log_d \frac{1+v}{2} +\frac{1-v}{2}\log_d \frac{1-v}{2}.
\end{align}
Essentially, in the large $d$ limit, the effective temperature for the classical statistical model is zero as other energy unfavorable configurations are frozen with the weights smaller than the dominant configuration in the order of $d$, which is the largest scale in the large $d$ limit. In some cases, zero temperature entropy also exists when many different domain wall configurations give the largest weight of the same values, such as in the case of a vertical domain wall with free bottom boundary conditions. 

On the contrary, finite $d$ is too small to separate different weights and the configuration with less weights can significantly contribute to the free energy. 
Calculating the partition function by summing the exponentially many weight terms of different configurations remains challenging. Instead, we seek an alternative base to $d$ that can differentiate the weight scales, necessitating the consideration of only dominant configuration collections in the finite $d$ case.
In the coarse-grained space-time lattice, we categorize these configurations based on their effective speed in each time slice, meaning a group of configurations shares the same 
 $v(t)$  for the domain walls. 
The sum of weights of such a group of configurations is $d^{-\int_t\mathbb{E}_2(v)dt}$. The weight sum ratio between different groups of configurations is thus $d^{\int_t (\mathbb{E}_2(v_1)-\mathbb{E}_2(v_2)) dt}$, which can be very large for favored $v_1(t)$ with sufficiently long domain wall (large $T=\int_t dt$). Namely, although $d$ is too small to suppress other weights, $d^{T}$ is still large enough to suppress other configuration groups in favor of the dominant group characterized by optimal $v(t)$. The optimal $v(t)$ is consequently taken to minimize the tension energy of the entanglement membrane $v(t)=\text{argmin}_v \int_t \mathbb{E}_2(v) dt$. In this work, we only focus on cases where optimal $v$ is constant over time, i.e. the preferred representative domain wall is a straight line in the space-time lattice.

We now focus on some special limits of the entanglement membrane picture for a better understanding. Firstly, we consider the problem of half-chain entanglement growth in an infinitely long random quantum circuit. This quantity can map to the ferromagnetic statistical model with a top boundary condition of half chain in $\mathbb{C}$ and half chain in $\mathbb{I}$. Since the space extent is infinite, the dominant domain wall must be vertical. Noticing the bottom boundary is free due to the product state input, the vertical domain wall with velocity $v(t)=v$ gives the free energy as $E_2(v)T$. According to Eq. \eqref{eq:tension}, the minimum value of $\mathbb{E}$ is achieved when $v=0$, we have $v_E = \mathbb{E}(0) = \log_d\frac{d^2+1}{2d}\approx 0.322$.  Note that the definition of time unit in our work is twice as \cite{Zhou2019}, i.e. our $v_E=0.644$.  Further analysis including the difference between annealed and quenched averages gives a more accurate result
as in Eq. (6) in \cite{Zhou2019}, the speed of R\'enyi-2 entropy in random Haar circuit is given by the field theory up to $O(\frac{1}{d^{10}\ln d})$ as   
$v_E\approx 0.643$. We can confirm that the difference between annealed and quenched average is small as discussed in factor (a) above. The entanglement membrane picture is summarized in Fig. \ref{fig:dmintro}(a)(b).

We now consider another limit, focusing on the entanglement entropy of the finite interval $L_A$. In late times, the entanglement is saturated, the statistical model has a top boundary condition with $L_A$ sites having $\mathbb{C}$ configuration and others having $\mathbb{I}$ configurations. Since now the time direction is sufficiently long, the vertical domain wall is not favorable anymore. The favored domain wall would be a downward triangle with velocity $v$ on both sides near the top boundary as shown in Fig. \ref{fig:dmintro}(c)(d). The free energy is then $\mathbb{E}_2(v)\frac{L_A}{v}$. To minimize this objective, we set the derivative zero: $v\mathbb{E}_2'(v)=\mathbb{E}_2(v)$, this gives a velocity $v_B$, which corresponds to the operator spreading speed, satisfying the assumption that $\mathbb{E}_2(v_B)=v_B$ and $\mathbb{E}_2'(v_B)=1$ \cite{Jonay2018}. In random quantum circuit case, $v_B=\frac{d^2-1}{d^2+1}$. Considering the time unit definition is our work, we have $v_B = 1.2$ in $d=2$ qubit system. Plugging the speed $v_B$ into the free energy, we have $S_2=L_A$. This result is particularly intriguing, as it contrasts with the vertical domain wall; the free energy of the triangular domain wall near the top boundary remains identical for the large 
$d$ limit and finite $d$ case. Therefore, in the following domain wall pictures, we simply use $v=1$ domain wall in the visualization, which represents the free energy contribution in terms of the spatial length $L_A$ both in the large $d$ and finite $d$ case.

\subsection{Numerical results}

We now present the numerical results for different circuit models and different encoding schemes using random Clifford circuits \cite{Aaronson2004a}, along with quantitative predictions based on the analytical framework presented above. We demonstrate that the numerical simulations align well with the analytical predictions.

\subsubsection{Random circuit with one-to-one encoding}\label{sec:ruc_1to1}

\begin{figure}[!htb]
\includegraphics[width=0.45\textwidth]{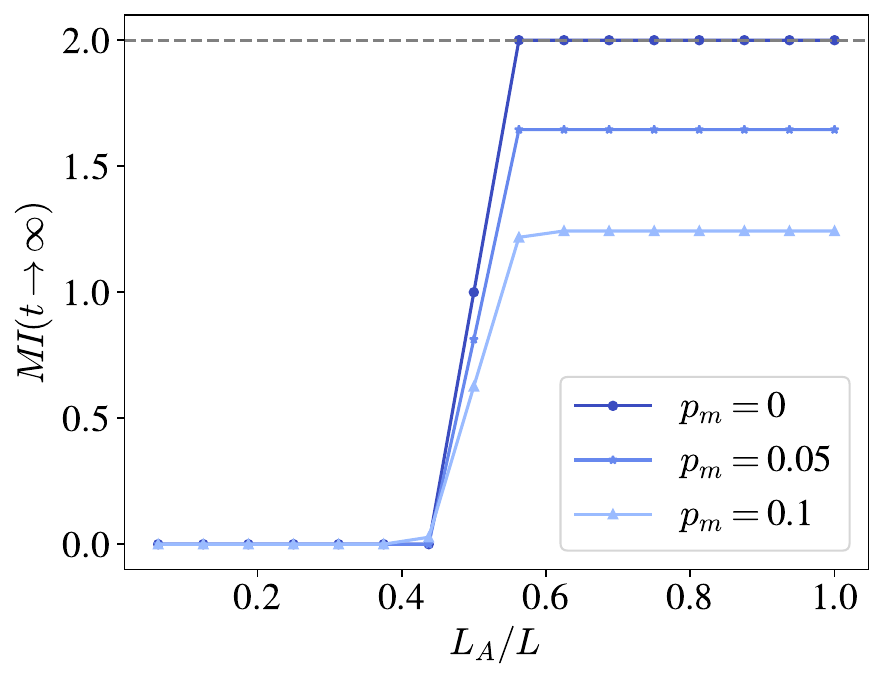}
\caption{Steady-state SIC in random quantum circuit with different subsystem sizes $L_A$ using one-to-one encoding scheme. The system size $L=512$, the result is averaged over 1000 random configurations. $p_m$ represents the mid-circuit measurement probability. 
} \label{fig:ruc_1to1}   
\end{figure}

We first investigate the results for one-to-one encoding as shown in Fig.~\ref{fig:setup}(c). For random circuits with and without mid-circuit measurements, the late-time steady-state mutual information between the reference qubit and the subregion $A$ in the system centered around the entangled qubit  $MI(L_A, t\rightarrow \infty)$ is shown in Fig. \ref{fig:ruc_1to1}. In terms of mid-circuit measurements, we only focus on the noise-resilient phase, where the steady-state entanglement is volume law and the measurement probability $p^m<p^m_c\approx 0.16$. 

We find that the steady mutual information for the random unitary circuit without mid-circuit measurements shows a step function behavior: for $L_A<L/2$, $MI=0$ while for $L_A >L/2$, $MI=2$ (pattern (d) in Fig. \ref{fig:resultsummary}). In other words, the information can be perfectly decoded from a subsystem with more than half the system while the information is totally lost for a subsystem of less than half of the system. This result indicates full information scrambling and is consistent with the philosophy of Hayden-Preskill thought experiment \cite{Hayden2007}. Considering the information conservation law $I(A:R)+I(\bar{A}:R)=2$, the steady-state mutual information applies to any subregion of a given size no matter whether $A$ has included the initial $E$ qubit. This is understandable since the information is fully scrambled in structureless random quantum circuits rendering the initial position of the encoding irrelevant. It is worth noting that this typical step function behavior in the steady-state mutual information is a universal feature of thermalized systems, which we will revisit in the context of fermionic Hamiltonian dynamics.

In the presence of mid-circuit measurements with  $p_m>0$, the mutual information is lower than the saturating value $2$ even with $L_A=L$ indicating information leakage. This is because a portion of the information has been extracted through measurements and transferred to the environment. The steady-state curve is described by pattern (e) in Fig. \ref{fig:resultsummary}.


\begin{figure}[!htb]
\includegraphics[width=0.45\textwidth]{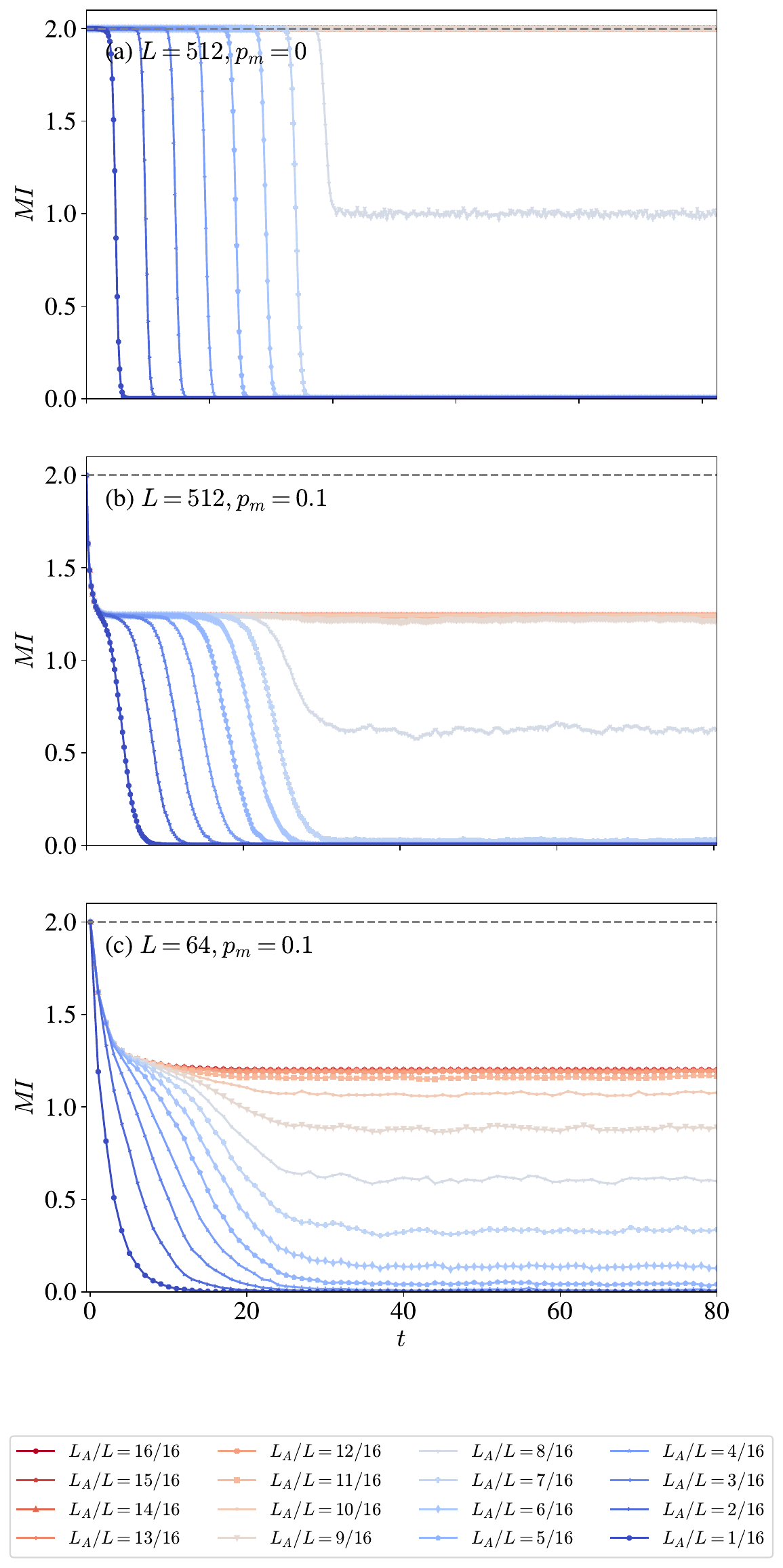}
\caption{ SIC dynamics in random quantum circuit with different subsystem sizes $L_A$ using one-to-one encoding scheme with or without mid-circuit measurements. (a) System size $L=512$, mid-circuit measurements $p_m=0$, (b) $L=512$, $p_m=0.1$, (c) $L=64$, $p_m=0.1$,  average over 1000 random circuit configurations.
} \label{fig:ruc_1to1_pm_dynamics}   
\end{figure}

\begin{figure}[!htb]
\includegraphics[width=0.5\textwidth]{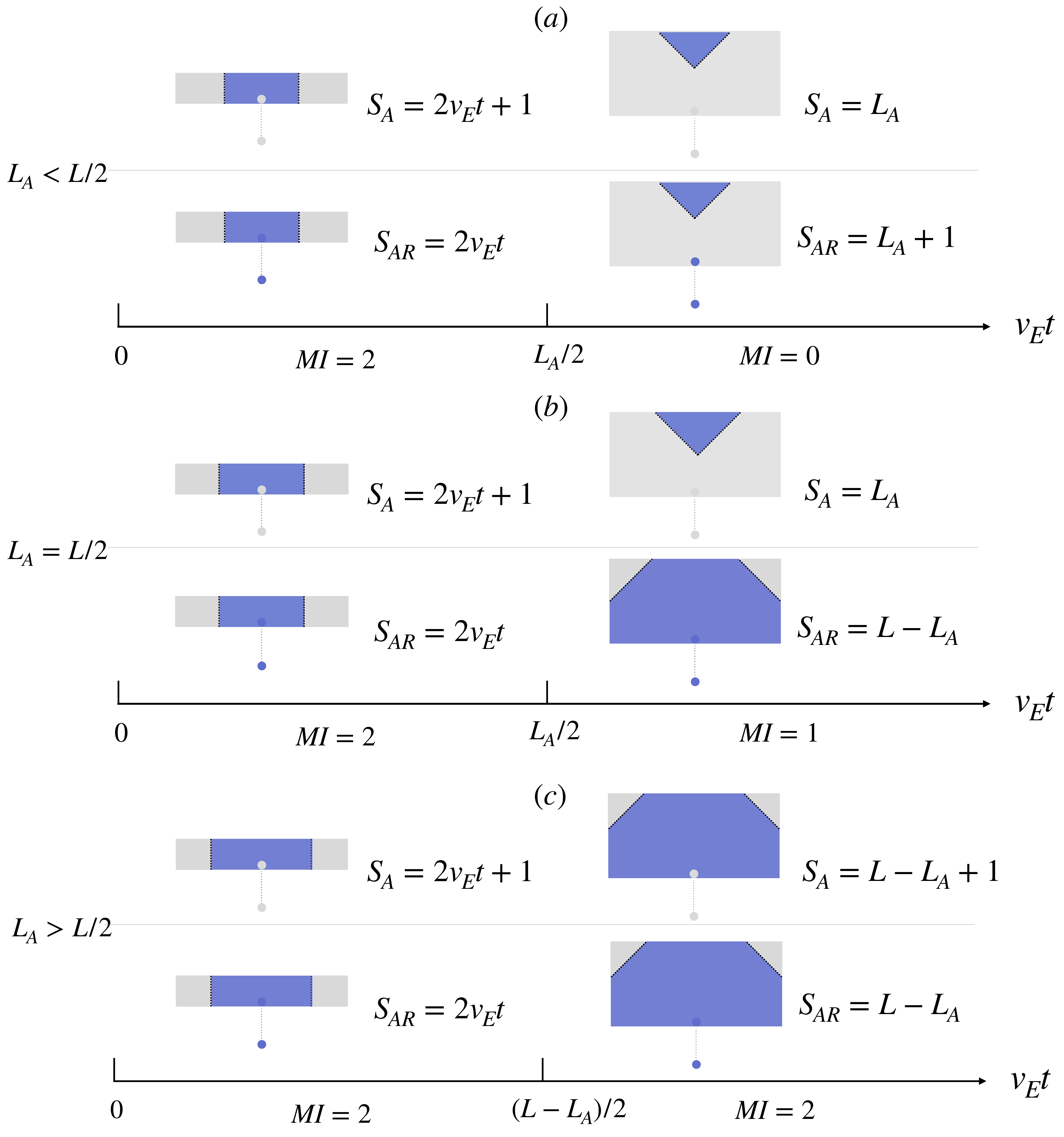}
\caption{Entanglement membrane explanation for the information dynamics in random quantum circuit with one-to-one encoding. $v_E\approx 0.643$. Each domain wall configuration gives a free energy contribution in terms of the effective statistical model and thus an entropy value in terms of the original circuit model. $S_R=1$ is constant over time and is ignored in the figure. The target quantity $MI=S_R+S_A-S_{AR}$ can be constructed from the above dominant configurations at different times and different subsystem sizes.
} \label{fig:dm_1to1}   
\end{figure}

Next, we investigate the dynamics of $MI(x, t)$ with varying $t$ before reaching steady states. Fig. \ref{fig:ruc_1to1_pm_dynamics}(a) illustrates the information dynamics for random quantum circuits without mid-circuit measurements. In this figure, we also present the predictions from the analytical framework using the entanglement membrane picture and find that the results match remarkably well with the theoretical predictions. To gain a deeper understanding of the dynamics, we show the dominant entanglement membrane picture for different times in Fig. \ref{fig:dm_1to1}. In each spin configuration plot, the time direction is oriented upwards. The upper panel in each case shows the dominant configuration for $S_A$ with top boundary conditions $\mathbb{C}$ (purple color) in the region $A$ and $\mathbb{I}$ on the region $\bar{A}$ as well as the reference qubit $R$. The lower panel in each case shows the dominant configuration for $S_{AR}$, and the only difference compared to the effective model for $S_A$ is that the reference qubit is applied with the boundary condition $\mathbb{C}$. To compute the free energy contribution of each given configuration, we compute the contribution from the tension energy of the domain walls and the overlaps from the bottom boundary. 

For example, for $L_A<L/2$, the top-boundary triangular domain wall is only geometrically possible for $v_Et>L_A/2$ and contributes to the free energy as $L_A$ as we explained in Sec. \ref{subsec:mapping}. Similarly, the vertical domain wall contributes free energy as $v_E t$, where $t$ is the domain wall length. 
As the vertical domain wall and the triangular domain wall near the top boundaries compete, the phase transition occurs for the time $t=L_A/(2v_E)$. To compute the entropy, we have to also consider the contribution from the bottom boundary overlap. Since the input state beyond the entangled qubit is in the form of a product state, the bottom boundary beyond the entangled qubit is all free, in the sense that for each qubit $i$ we have $\text{Tr}(\mathbb{C}_i\rho_i^{\otimes r})=\text{Tr}(\mathbb{I}_i\rho_i^{\otimes r})=1$. For the entangled qubit with reference qubit $E\cup R$, we have $\text{Tr}(\mathbb{C}_E\mathbb{C}_R \rho_{B}^{\otimes r})=\text{Tr}(\mathbb{I}_E\mathbb{I}_R \rho_{B})=1$ and $\text{Tr}(\mathbb{C}_E\mathbb{I}_R \rho_{B}^{\otimes r})=\text{Tr}(\mathbb{I}_E\mathbb{C}_R \rho_{B}^{\otimes r})=d^{-r}$, where $\rho_B = \frac{1}{2}(\vert 0_E0_R\rangle +\vert 1_E1_R\rangle)(\langle 0_E0_E\vert +\langle 1_E1_R\vert)$ is the initial Bell pair. In sum, for $L_A<L/2$ and $v_Et<L_A/2$, we have $S_A  = 2v_Et+1$. Similarly, we have $S_{AR} = 2v_Et$ which lead to the mutual information $I(A:R) = S_A + 1 - S_{AR} =2$. The physical meaning is that the information is fully kept in the subsystem at short times $t<L_A/(2v_E)$. On the other hand, the top-boundary triangular weight will dominate, and gives $S_A=L_A$ and $S_{AR}=L_A+1$ at late times, leaving the mutual information $I(A:R)=0$. The prediction is consistent with numerical results in Fig. \ref{fig:ruc_1to1_pm_dynamics}(a). We find that for subsystems smaller than half the system, the mutual information first is 2 and at some later time $t^*$ suddenly drops to zero. The timescale $t^*$ for the mutual information drop is exactly given by $L_A/(2v_E)$, and this setting explains why $v_E$ can be regarded as a velocity.

A similar analysis can be applied for $L_A=L/2$ and $L_A>L/2$. For $L_{A}=L/2$, we have $S_A= 2v_Et+1$ and $S_{AR}=2v_Et$ for $t<L_A/(2v_E)$, and $S_A = L_A$ and $S_{AR}= L-L_A = L_A$ for $t>L_A/(2v_E t)$. This gives us a mutual information drop from $2$ to a plateau of value $1$. For $L_{A}>L/2$, the analysis is the same for the short time stage of $L_{A}=L/2$ . For $t>L_A/(2v_E)$, now the dominant domain wall for both $S_{AR}$ and $S_A$ is of length $L-L_A$, leaving $MI = (L-L_A+1)+1-(L-L_A)=2$, which explains why the information can always be fully held in any subsystem of more than half the size.

The analysis utilizing the effective statistical model can be extended to other information metrics, such as Holevo information mentioned in Sec. \ref{subsec:probes}. In this case, the upper panel for each time slice and $L_A$ are kept unchanged to account for the first term in Holevo information, $S_A$. The second term for Holevo information is characterized by the statistical model with free boundary conditions on the entangled qubit $E$. For small $t$, the entanglement membrane picture remains the same as $S_{AR}$, yielding the same plateau dynamics starting from $\chi_A=1$. For steady-state Holevo information, the second term is $L_A$ for $L_A\leq L/2$ and $L-L_A$ for $L_A>L/2$. The steady-state $\chi_A(t\rightarrow \infty)$ is thus $0$ and $1$ for $L_A\leq L/2$ and $L_A>L/2$, respectively. Remarkably, the results for mutual information and Holevo information show qualitative difference when $L_A=L/2$: while the steady-state mutual information is half as the initial value, the steady-state Holevo information is zero.

With the mid-circuit measurement $p_m>0$, the information dynamics is shown in Fig. \ref{fig:ruc_1to1_pm_dynamics}(b)(c). For large-size systems, the information decay is categorized into two stages. In the early stage, the information stored in the subsystem universally drops due to the measurements. Since we are in the noise-resilient phase of MIPT, the final steady information is non-zero. The time scale $t^{*}$ to approach the measurement-induced steady state is rather small in the order of $O(10)$ for $p_m=0.1$ and agnostic to the system size. For any subsystem size $L_A>2v_E t^{*}$, the corresponding mutual information with the reference qubit $R$ will experience two stages of dynamics separately. At the first stage, the information drops due to the monitored dynamics and stays at the plateau induced by measurements for $t^*<t<L_A/(2 v_E)$. At the second stage with $t>L_A/(2v_E)$, the information begins to drop again from the universal plateau due to the information scrambling to the complementary system $\bar{A}$. For small systems where $L_A\sim 2v_E t^{*}$, the two stages of information dynamics become mixed as shown in Fig. \ref{fig:ruc_1to1_pm_dynamics}(c), i.e. the information is spreading to the subsystem $\bar{A}$ and leaks to the environment at the same time.

\subsubsection{Random Floquet circuit with one-to-one encoding}\label{sec:1to1floquet}

In this subsection, we investigate the information dynamics in random Floquet circuits, i.e. random circuits with time periodicity. These circuits exhibit distinct dynamical phases depending on whether the local random gates are sampled from a random Haar ensemble or a random Clifford ensemble. The former ensemble leads to thermal phases \cite{Sunderhauf2018} while the latter ensemble leads to Anderson localization phases in one dimension \cite{Farshi2022} and thermal phases in higher dimensions \cite{Farshi2023}. To facilitate the large-scale numerical simulation, we focus on the 1D random Clifford Floquet circuits as the testbed. This model can reveal the information dynamics in Anderson localization systems while other chaotic Floquet systems are believed to show similar full information scrambling dynamics as discussed in Sec. \ref{sec:ruc_1to1}.

\begin{figure}[!htb]
\includegraphics[width=0.42\textwidth]{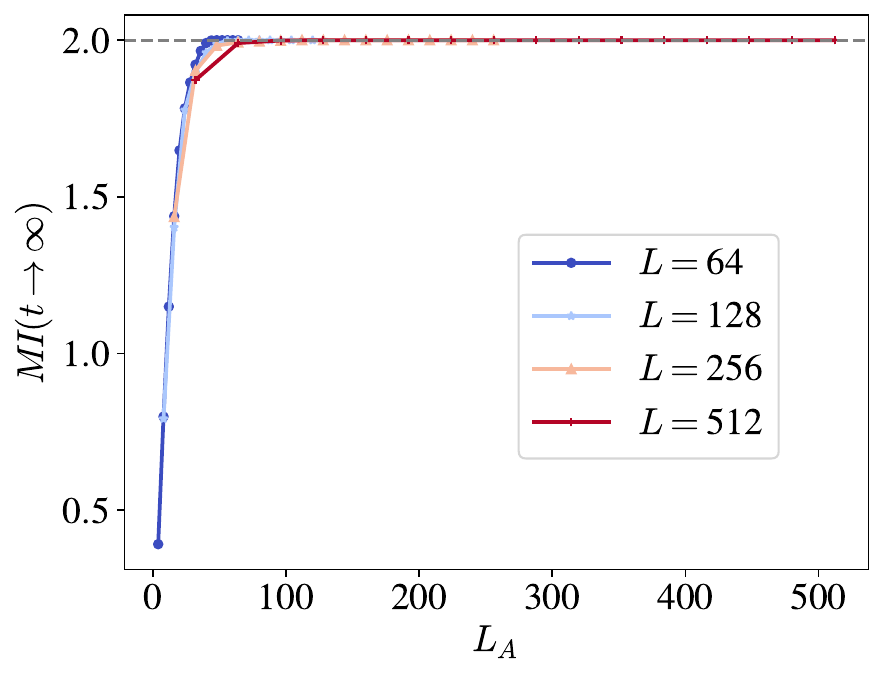}
\caption{Steady-state SIC in 1D random Clifford Floquet circuit with different subsystem sizes $L_A$ under one-to-one encoding scheme. $L=64, 128, 256, 512$, average over 1000 configurations. We can see that curves of different system sizes collapse implying Anderson localization behavior with the typical localization length around 50.
} \label{fig:floquet_1to1}   
\end{figure}

Fig. \ref{fig:floquet_1to1} shows the steady-state mutual information with respect to different subsystem sizes $L_A$. In Anderson localization phases, the mutual information curve converges in thermodynamic limit with $L_A$ as x-axis, indicating a system size irrelevant localization length. The curve will be always at value $2$ if the x-axis is $L_A/L$ in the thermodynamic limit. In other words, the localization length restricts the capacity of information scrambling, and no information can spread far away beyond the localization length. Therefore, the output subsystem with size on the order of localization length is sufficient to extract most of information stored at the beginning.

\subsubsection{Random circuit with one-to-all encoding}\label{subsec:ruc-one-to-all}

In this subsection, we focus on the one-to-all encoding where the reference qubit is entangled with all system qubits in a GHZ state. When mapping to the statistical model, one-to-all encoding introduces a nontrivial bottom boundary condition compared to free boundary conditions for product initial states. We can show that 
\begin{align}
&\text{Tr}(\prod_{i=0}^L\mathbb{C}_i (\vert GHZ\rangle\langle GHZ\vert)^{\otimes r}) \nonumber \\&= \text{Tr}(\prod_{i=0}^L\mathbb{I}_i (\vert GHZ\rangle\langle GHZ\vert)^{\otimes r}) \nonumber \\&= d^0=1,
\end{align}
where the reference qubit corresponds $i=0$. Besides, the overlap between GHZ state and all other permutations such as $\prod_{i=0}^k\mathbb{C}_i\prod_{i=k+1}^L\mathbb{I}_i$ gives $d^{-r}$. Notably, the bottom boundary overlap is agnostic to the system size $L$. In other words, as long as not all the spins at the bottom layer of the statistical model share the same permutation configuration as the reference qubit $R$, there is an extra contribution of $1$ to the free energy/entanglement entropy. This is often favored since the relaxation of the boundary condition can arrange a more compatible configuration pattern between the bottom layer and the bulk so that the energy gain is larger than $1$.

The initial mutual information behavior $MI(t=0)$ in the one-to-all encoding case is also distinct from one-to-one encoding. For one-to-one encoding, the initial mutual information $MI(t=0)=2$ for any subsystem that covers the entangled qubit $E$. However, for the GHZ state, it is straightforward to show that for any subsystem $0<L_A<L$, $MI(t=0)=1$. Such spatial distribution of the information can also be regarded as the steady state from so-called cnot circuits where each two-qubit gate is randomly selected from identity or cnot gate.

\begin{figure}[!htb]
\includegraphics[width=0.48\textwidth]{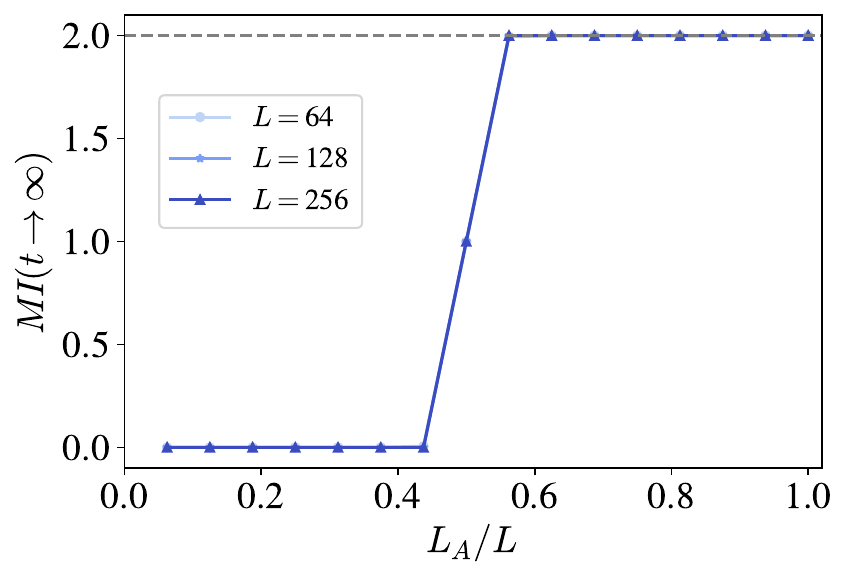}
\caption{Steady-state SIC in random quantum circuit with different subsystem sizes $L_A$ under one-to-all encoding scheme. The system sizes are $L=64, 128, 256$, the results are averaged over 1000 configurations. The result is consistent with one-to-one encoding for $p_m=0$. However, with the introduction of mid-circuit measurements $p_m>0$, MI immediately decays to zero as long as the first mid-circuit measurement occurs (not shown in the figure).
} \label{fig:random_1toall}   
\end{figure}

\begin{figure}[!htb]
\includegraphics[width=0.45\textwidth]{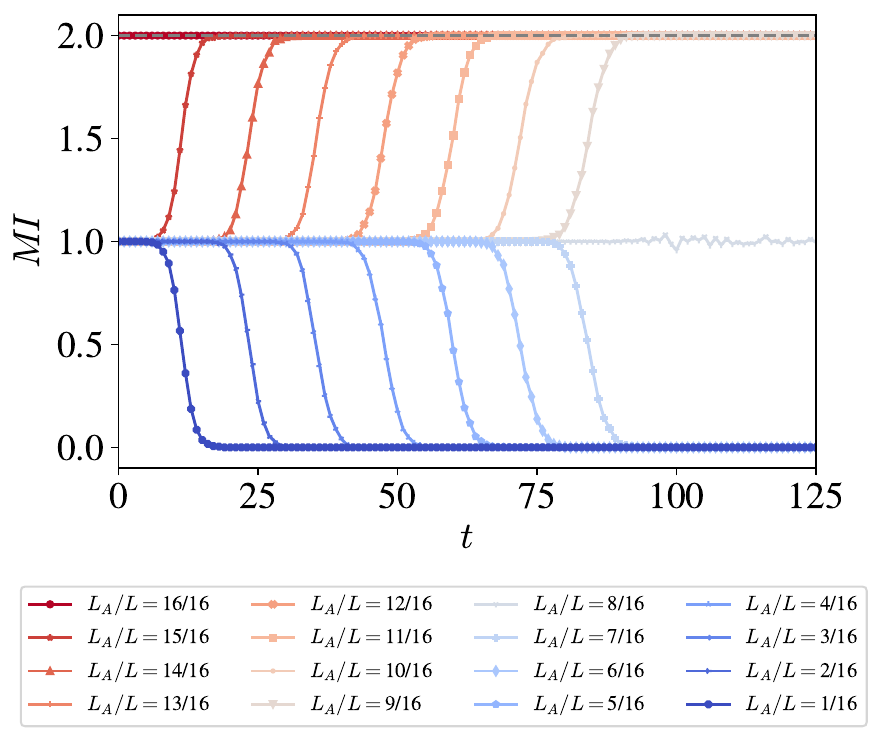}
\caption{SIC in random quantum circuit with different subsystem sizes $L_A$ under one-to-all encoding scheme at the beginning. $L=256$,  average over 1000 configurations. 
} \label{fig:random_1toall_dynamics}   
\end{figure}

\begin{figure}[!htb]
\includegraphics[width=0.5\textwidth]{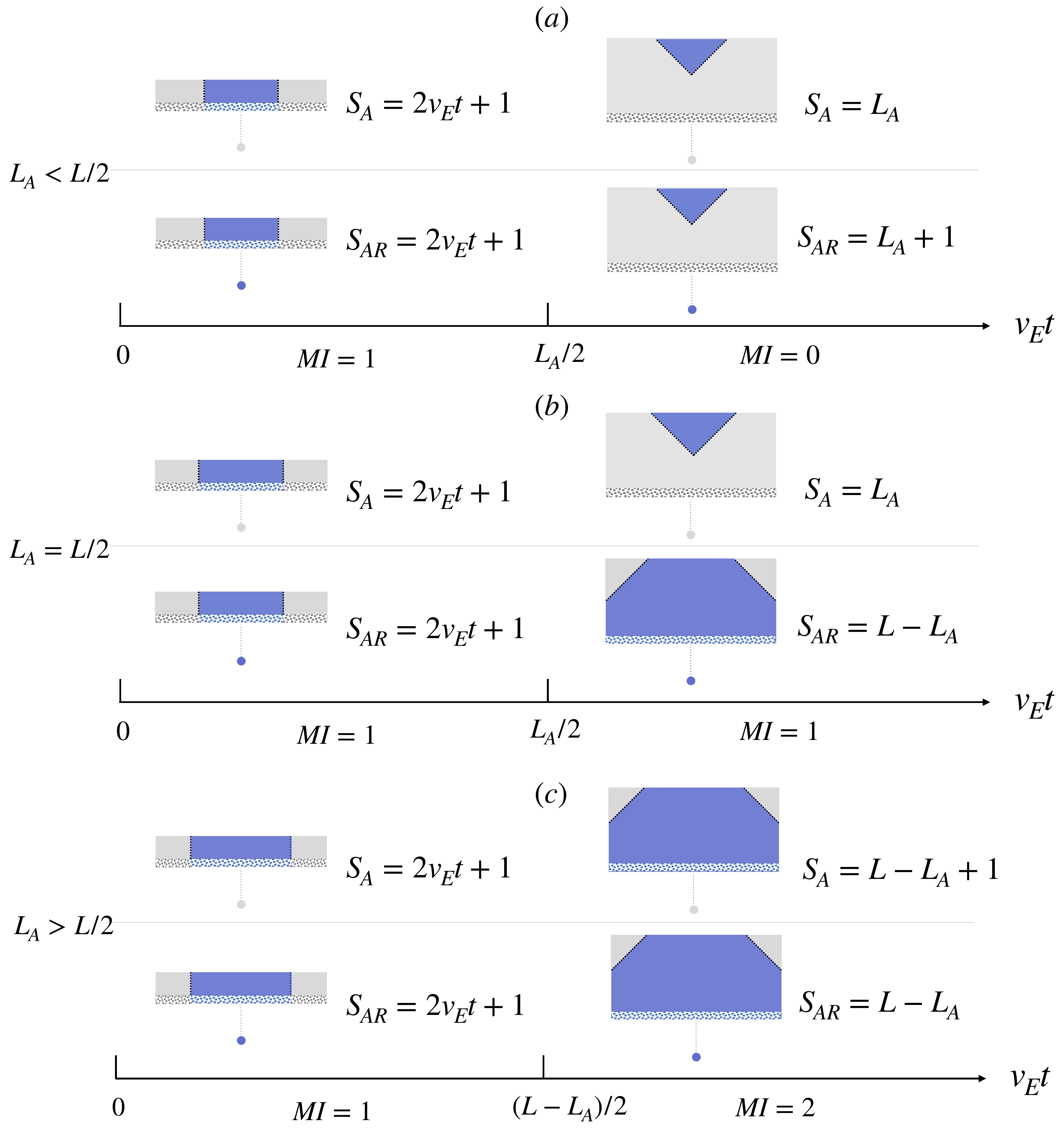}
\caption{Entanglement membrane explanation for the information dynamics in random quantum circuit with one-to-all encoding. $v_E\approx 0.643$. Each domain wall configuration gives a free energy contribution in terms of the effective statistical model and thus an entropy value in terms of the original circuit model. $S_R=1$ is constant over time and is ignored in the figure. The target quantity $MI=S_R+S_A-S_{AR}$ can be constructed from the above dominant configurations at different times and different subsystem sizes. 
} \label{fig:random_1toall_dw}   
\end{figure}

Fig. \ref{fig:random_1toall} presents steady-state mutual information with different subsystem sizes $L_A$. We find the result is consistent with one-to-one encoding and the curves of different sizes collapse with rescaled subsystem size $L_A/L$, contrary to Fig. \ref{fig:floquet_1to1}, where late-time curves could collapse with respect to subsystem sizes $L_A$. Again, the results only rely on the subsystem size and are independent of the subsystem's position which is unsurprising as there is no specific entangled qubit at all due to the global nature of the one-to-all encoding scheme. The corresponding information dynamics is shown in Fig.~\ref{fig:random_1toall_dynamics}. The initial value for $MI$ is $1$ for different system sizes as expected. For subsystems with size $L_A>L/2$ ($L_A<L/2$), the information capacity increases (decreases) with the evolved time from $1$ and vice versa.  The entanglement membrane picture, as shown in Fig. \ref{fig:random_1toall_dw}, offers a comprehensive explanation for the steady and dynamical behaviors of subsystem information capacity. It is worth noting that the nonuniform configuration at the bottom boundary gives $1$ extra unit of free energy due to the bottom boundary condition imposed by GHZ states. The timescale at which the mutual information begins to deviate from the initial value $1$ is also determined by  $L_A/(2v_E)$ for $L_A<L/2$, which is consistent with the numerical results.

We now discuss the dynamics of Holevo information $\chi_A$ with the help of the entanglement membrane picture. Under one-to-all encoding, the two initial states, as employed in the definition of Holevo information in Eq. \eqref{eq:holevo}, are $\vert \psi_0^{1}\rangle = \vert 0\rangle ^{\otimes L}$ and $\vert \psi_0^{2}\rangle = \prod_{i}^L X_i\vert \psi_0^{1}\rangle=\vert 1\rangle^{\otimes L} $. The two terms of Holevo information are thus determined by the upper panel for $S_A$ in Fig. \ref{fig:random_1toall_dw} and the same statistical model with free bottom boundary conditions, respectively. The initial value of Holevo information $\chi_A(t=0)$ for one-to-all encoding is $1$, the saturating value for Holevo information, contrasting with half of the saturating value for initial mutual information. In terms of the steady-state value, $\chi_A(t\rightarrow \infty)=1$ or $0$ depending on whether $L_A>L/2$ or $L_A\leq L/2$. Consequently, the dynamics for mutual information and Holevo information is distinct with $L_A>L/2$: while mutual information starts from $1$ and increases to $2$ at the time scale $L_A/(2v_E)$, Holevo information stays at the initial value $1$ for the whole evolution.
The qualitative differences between mutual information and Holevo information enrich our understanding of quantum information theory by revealing the nuanced interplay between various information-theoretic quantities. 

We further discuss the implications of monitoring circuits with one-to-all encoding. Since the information is encoded into the GHZ state globally at the initial stage, a single mid-circuit measurement is sufficient to extract the one-qubit information from the system. Therefore, with any $p_m>0$, SIC drops to zero as long as $t>0$, rendering the information dynamics trivial. Therefore, although full scrambling by local unitaries after one-to-one local encoding and initial states by direct one-to-all encoding are both of global nature, the former is resilient to mid-circuit measurements in terms of information protection while the latter is fragile. This difference can be revealed by different spatial information distributions: the former case has a distribution like the curve in Fig. \ref{fig:random_1toall}, while the latter case gives a horizontal line of value $1$ for any $L_A$.

\subsubsection{Random Floquet circuit with one-to-all encoding}

\begin{figure}[!htb]
\includegraphics[width=0.45\textwidth]{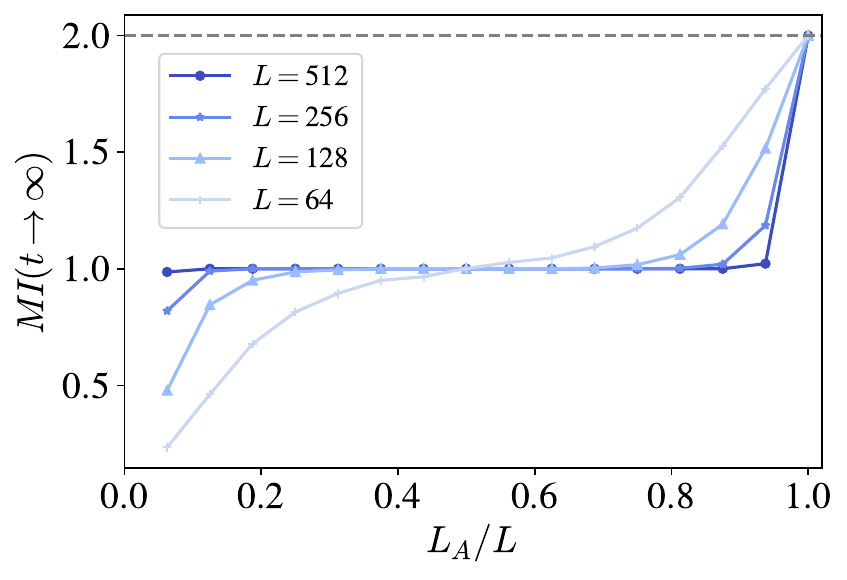}
\caption{Steady-state SIC in 1D random Clifford Floquet circuits with different subsystem sizes $L_A$ under one-to-all encoding scheme at the beginning. System sizes $L=64, 128, 256$, $t>L$, and the results are averaged over 1000 configurations. The results imply $MI(t\rightarrow\infty)=1$ for any $L_A<L$ in the thermodynamic limit. 
} \label{fig:floquet_1toall}   
\end{figure}

\begin{figure}[!htb]
\includegraphics[width=0.45\textwidth]{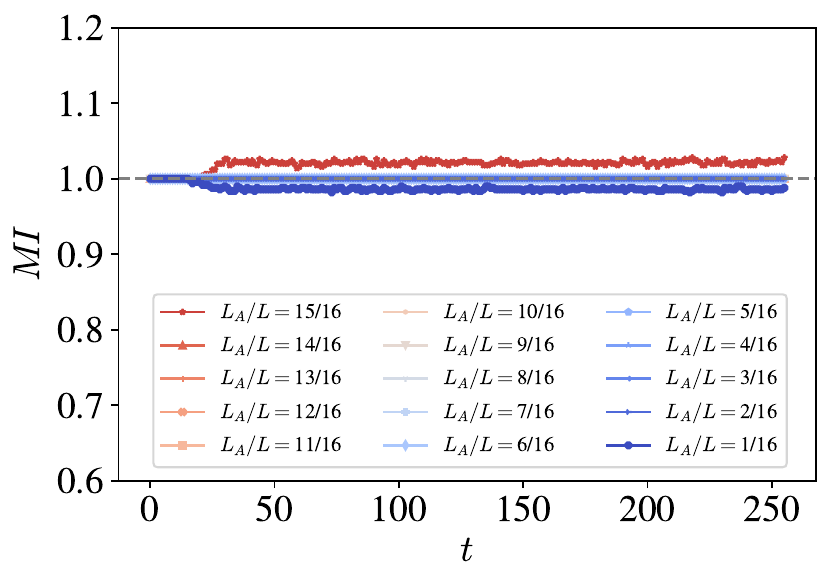}
\caption{SIC dynamics in 1D random Clifford Floquet circuits with different subsystem sizes $L_A$ under one-to-all encoding scheme. System size $L=512$, no mid-circuit measurement presents $p_m=0$, and the results are averaged over 1000 configurations. The temporal fluctuations can be attributed to finite size effects. 
} \label{fig:floquet_1toall_dynamics}   
\end{figure}

Since the initial values of $MI$ are $1$ for all subsystems with one-to-all encoding scheme, an interesting question arises on the fate of such information dynamics in Anderson localization systems. We here report the information dynamics results with one-to-all encoding in random Clifford Floquet circuits. The late-time steady-state mutual information curve is depicted in Fig.~\ref{fig:floquet_1toall}. With increasing system sizes, the steady-state mutual information also converges to $1$ irrespective of the subsystem size $L_A<L$. Thus we conclude in the thermodynamic limit, the steady-state information capacity is always $1$ for any $L_A<L$ ($MI=2$ for $L_A=L$). This fact suggests that Anderson localization can fully stop the information spreading, a phenomenon that becomes more pronounced as shown in Fig.~\ref{fig:floquet_1toall_dynamics}. The mutual information for any subsystem size remains essentially unchanged, with minor fluctuations attributed to finite-size effects. In the thermodynamic limit, mutual information is expected to stabilize at $1$ for all times and across all subsystem sizes $L_A<L$.  The global encoding strategy thus offers a clearer insight into the intrinsic characteristics of Anderson localization, where information spreading is effectively halted.

\subsubsection{Random circuit with many-to-many encoding}

\begin{figure}[!htb]
\includegraphics[width=0.4\textwidth]{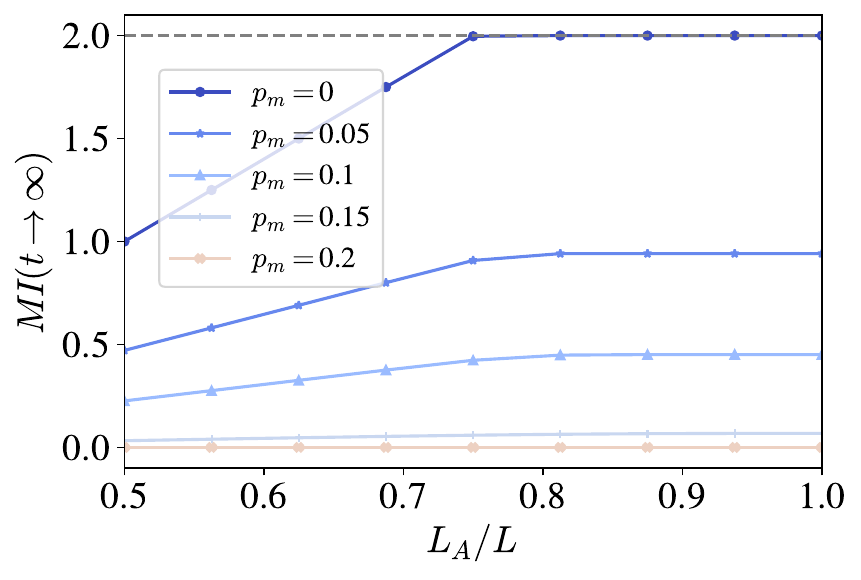}
\caption{Steady-state SIC in 1D random quantum circuit with different subsystem sizes $L_A$ under finite rate encoding scheme which encodes $L/2$ Bell pairs with $L/2$ ancilla qubits at the beginning. System size is $L=512$. Results are presented with different $p_m$ and are averaged over 200 configurations. 
} \label{fig:ruc_finiterate}   
\end{figure}


\begin{figure}[!htb]
\includegraphics[width=0.5\textwidth]{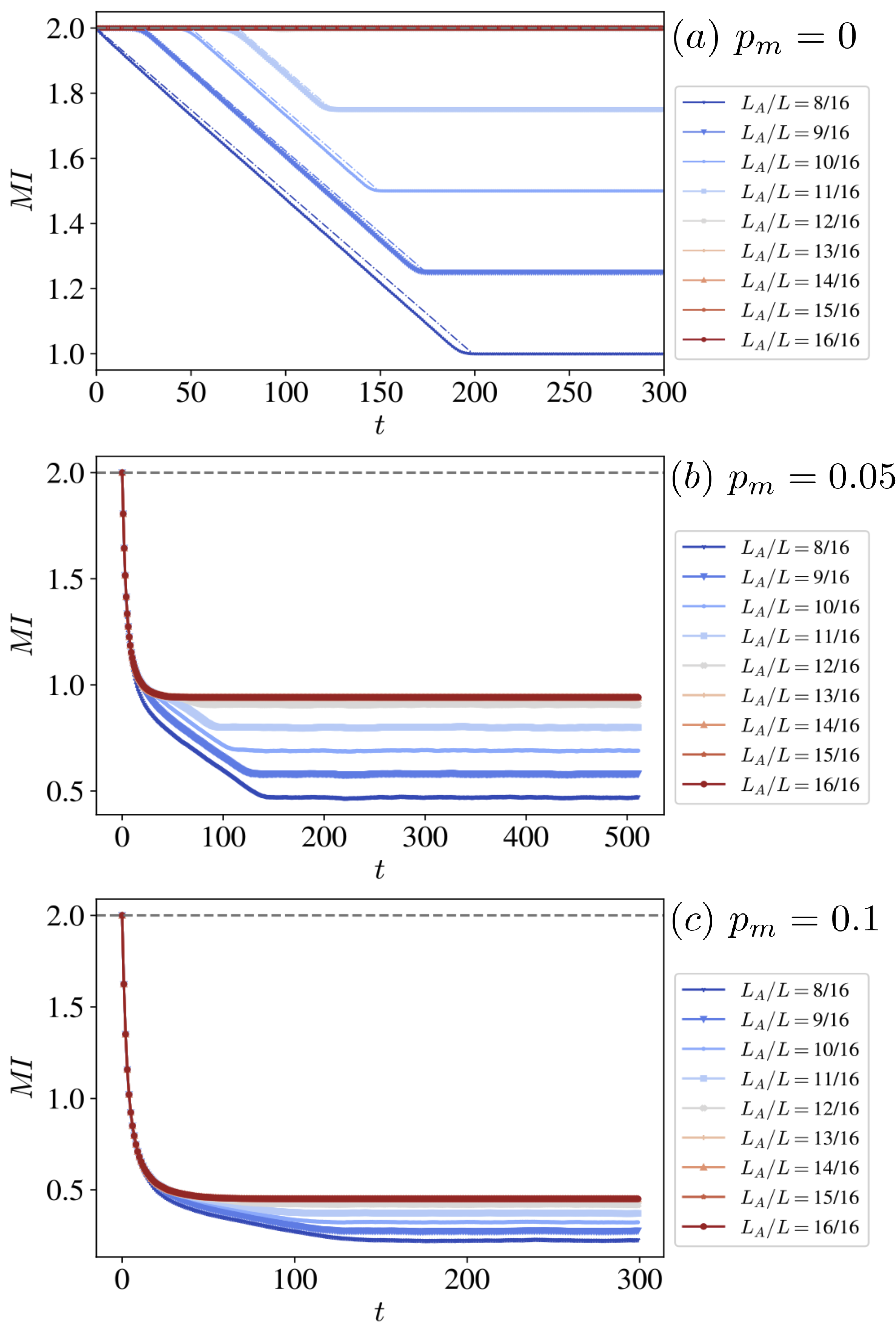}
\caption{SIC dynamics in 1D random quantum circuit with different subsystem sizes $L_A$ under finite rate encoding scheme which encodes $L/2$ Bell pairs with $L/2$ ancilla qubits at the beginning. System size is $L=512$ with $p_m=0$ (a) $p_m=0.05$ (b) and $p_m=0.1$ (c). The results are averaged over 200 circuit configurations. The information decay is very slow compared to dynamics in the one-to-one encoding case. The dashed line in (a) is the analytical result predicted by the entanglement membrane picture.
} \label{fig:ruc_finiterate_dynamics}   
\end{figure}

\begin{figure*}[!htb]
\includegraphics[width=0.9\textwidth]{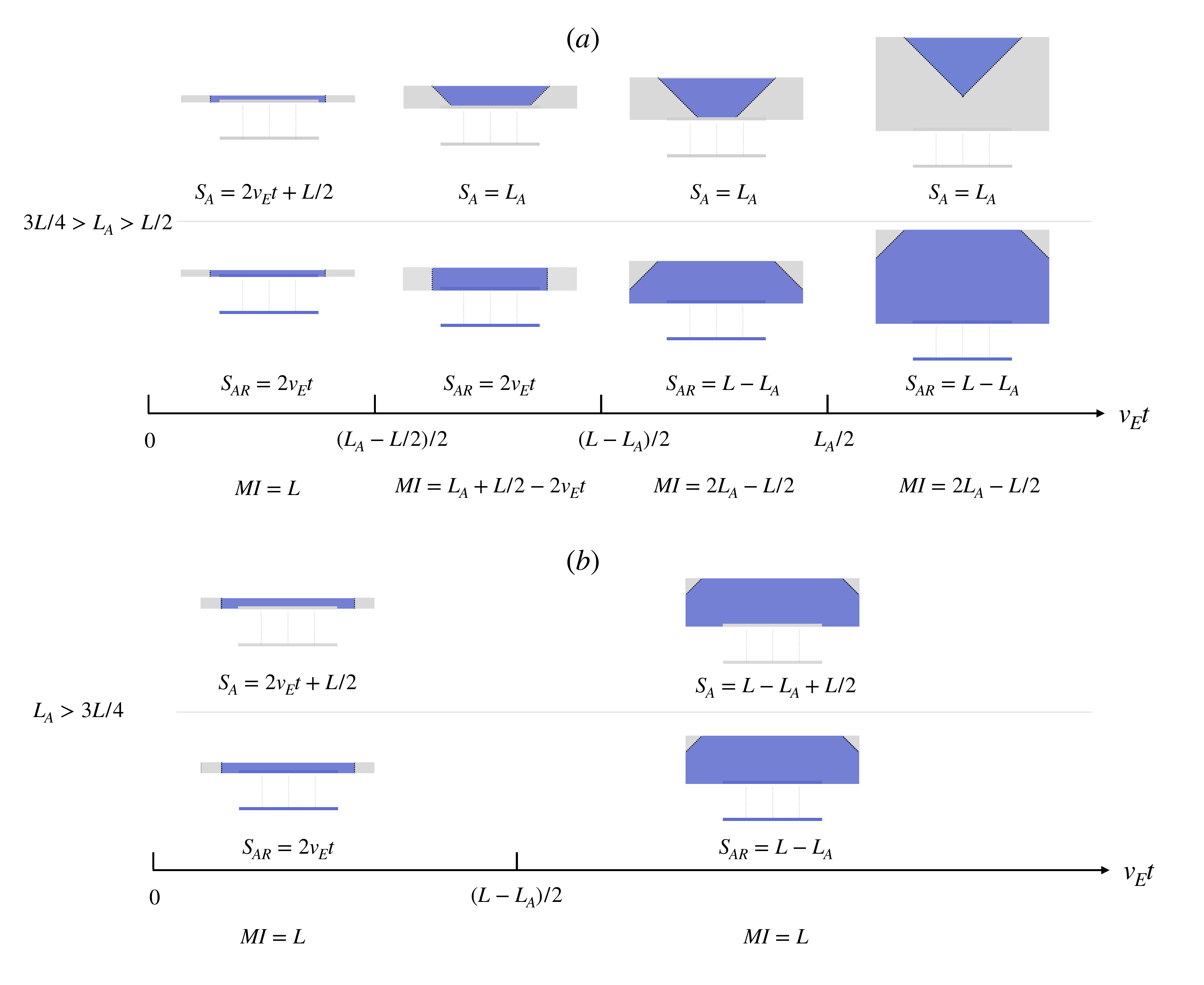}
\caption{Entanglement membrane explanation for the information dynamics in random quantum circuit with many-to-many (finite rate) encoding. Each domain wall configuration gives a free energy contribution in terms of the effective statistical model and thus an entropy value in terms of the original circuit model. $S_R=L/2$ is constant over time and is ignored in the figure. The target quantity $MI=S_R+S_A-S_{AR}$ can be constructed from the above favorable configurations at different times and different subsystem sizes. (a) $L/2<L_A<3L/4$ and (b) $L_A>3L/4$.
} \label{fig:ruc_finiterate_dw}  
\end{figure*}

We now consider the finite rate encoding scheme, specifically focusing on the case where $L/2$ reference qubits are entangled in Bell pairs with $L/2$ system qubits $\abs{E}=\abs{R}=L/2$. The final mutual information is normalized by $L/2$ in this subsection so that the value of $MI$ is comparable with previous cases with initial value $MI=2$ for subsystem size $L_A>L/2$. In this setting, we always focus on the subsystems that include all the entangled system qubits $\abs{A}\geq \abs{E}$. The steady-state mutual information behaviors are shown in Fig. \ref{fig:ruc_finiterate} and the dynamics without and with mid-circuit measurements are shown in Fig. \ref{fig:ruc_finiterate_dynamics}.

The steady-state results demonstrate that when the mid-circuit measurement probability $p$ exceeds the threshold $p_m$, information is completely lost as expected. Furthermore, the mutual information attains its maximal value $2$ when the subsystem size $L_A>3L/4$, reminiscent of the one-to-one encoding where $L_A>L/2$ is sufficient for full information restoration. In terms of the dynamics, we observe a stage where the mutual information decays linearly, contrasting with the one-to-one encoding case where the mutual information suddenly drops to the steady value at $t=L_A/(2v_E)$. 

The three stages of the information dynamics, predicted by the dashed line in Fig. \ref{fig:ruc_finiterate_dynamics}(a), can also be perfectly explained via the entanglement membrane picture as shown in Fig. \ref{fig:ruc_finiterate_dw}. Unlike one-to-one and one-to-all encoding cases, now the bottom boundary condition can have extended overhead. If the $L/2$ ancilla qubits are imposed with boundary condition $\mathbb{C}$, then each configuration $\mathbb{I}$ in the region $E$ will contribute $1$ to the free energy. The difference in the bottom boundary condition induces the emergent stage where the information shows linear decay instead of the sudden drop observed and predicted in previous cases. This new stage is specifically tied to the second column in Fig. \ref{fig:ruc_finiterate_dw}(a) when $(L_A-L/2)/(2v_E)<t<(L-L_A)/(2v_E)$. For $S_A$ in this case, the free energy is given as $2\mathbb{E}(v)t+L_A-2 vt$, the minimization of this term gives $\mathbb{E}'(v)=1$, which determines the velocity of the optimal entanglement membrane as $v_b$ and thus $S_A=L_A$ independent of the time $t$. However, $S_{AR}=2v_Et$ in this case relies on $t$. Therefore, the target SIC, as the differences between the two terms, has linear $t$ dependence as shown from numerical simulation. For the cases with mid-circuit measurements, we again observe the two stages of information drop: the first stage is universal as information leakage induced by measurements and the second stage resembles the behaviors for $p_m=0$ limit which is induced by information propagation.

\subsubsection{Random Floquet circuit with many-to-many encoding}

\begin{figure}[!htb]
\includegraphics[width=0.4\textwidth]{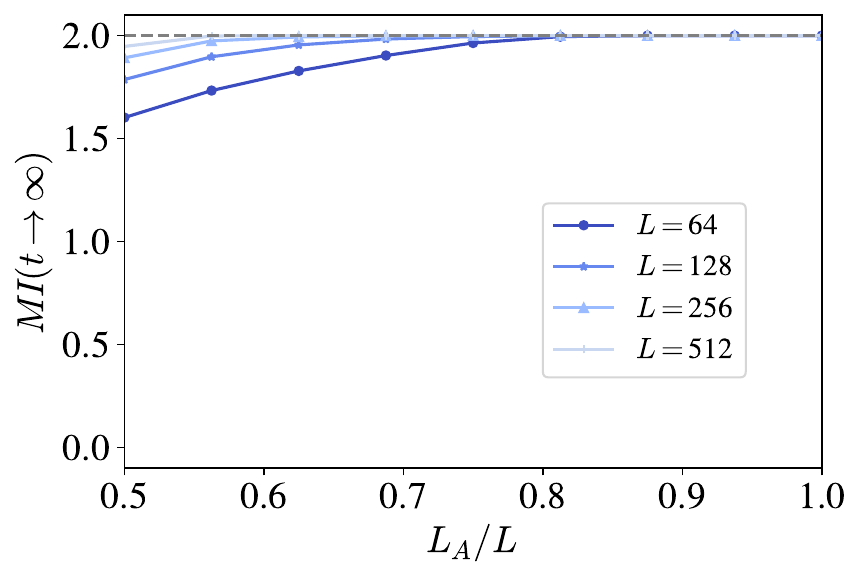}
\caption{Steady-state SIC in 1D random Clifford Floquet circuit with different subsystem sizes $L_A$ under finite rate encoding scheme that encodes $L/2$ Bell pairs with $L/2$ ancilla qubits at the beginning. The results are averaged over 400 configurations. 
} \label{fig:floquet_finiterate}   
\end{figure}

In one-dimensional random Clifford Floquet circuits of Anderson localization phases, the mutual information is observed to be localized around the $L/2$ segment $E$ as shown in Fig. \ref{fig:floquet_finiterate}. Again, consistent with Anderson localization results with one-to-one encoding, in the thermodynamic limit, the information will be fully localized in the $L_A=L/2$ subsystem, and the steady state SIC gives $2$ for any $L_A/L>1/2$. Namely, the minimal subsystem size containing full information $MI=2$ is nearly fixed as $L_A = L/2+\xi$ and this critical size approaches $L_A/L=1/2$ in the thermodynamic limit.

\section{Results for Hamiltonian dynamics} \label{sec:fermion}

In this section, we shift our focus to Hamiltonian quench dynamics described by one-dimensional fermionic systems. We investigate information dynamics in non-equilibrium phases utilizing AA model and SSH model, respectively.

One representative example we investigated is the Aubry-Andr\'e model without or with many-body interactions \cite{Harper1955, Aubry1980, Iyer2013, Khemani2017, Zhang2018, Zhang2019a}. This model is well-known for its Anderson localization (many-body localization) phase and extended (thermal) phase in non-interacting (interacting) cases. The Hamiltonian of the AA model reads as:
\begin{align}
H=&J\sum_{j=1}^{L}(c_{j+1}^\dagger c_j+h.c.)+U\sum_{j=1}^Ln_jn_{j+1}\nonumber \\ 
+& \sum_{j=1}^L V_jc_j^\dagger c_j,
\label{eq:aahamiltonian}
\end{align}
where $c_j$ represents the spinless fermionic annihilation operator on site $j$, and $n_j = c^\dagger_j c_j$ is the fermion density operator on site $j$. For AA model, $V_j=2 w\cos(2\pi \alpha j+\theta)$, where $\theta$ is the random phase to be averaged, $\alpha$ is an incommensurate wavevector for quasiperiodicity and we take $\alpha=\frac{\sqrt{5}-1}{2}$ throughout the work. In the non-interacting case $U=0$, the Anderson localization transition is known analytically via a dual mapping from real space to momentum space at $w_c/J=1$. With interaction $U>0$, we can obtain $w_c$ numerically for a many-body localization (MBL) transition. 

Besides, we also investigate information dynamics for the case with potential $V_j = C$ as a site-independent constant. The model in Eq. \eqref{eq:aahamiltonian} remains interacting but becomes integrable as it is equivalent to XXZ+Z model. 

We also investigate the interplay between information and topological properties \cite{Qi2011}, where SSH model is utilized \cite{Su1980}:
\begin{align}
    H = -\sum_{j=1}^{L} (1-(-1)^j\Delta)c_j^\dagger c_{j+1} +h.c.
    \label{eq:ssh}
\end{align}
For $\Delta<0$ ($\Delta>0$), the system is topological (trivial) exhibiting (no) edge states under open boundary conditions. Through these Hamiltonian models, we aim to unravel the intricate relationships between information dynamics, localization phenomena, and topological features.

Unlike the random circuit case, where the applied local unitaries are generic without any symmetry or conservation laws, the Hamiltonians investigated in this section are U(1) symmetric and thus follow charge conservation. U(1)-symmetric evolution often gives rise to a spectrum of novel phenomena that are distinct from those in generic evolution without any conservation laws. Examples include OTOC \cite{Khemani2018,Rakovszky2018}, R\'enyi entanglement growth  \cite{Rakovszky2019, Huang2020d, Zhou2020c_renyi, Znidaric2020}, symmetry restoration \cite{Liu2024a}, and measurement-induced charge sharpening transitions \cite{PhysRevX.12.041002}. 

In this section, we focus on the one-to-one encoding scheme and von Neumann entropy based mutual information. Notably, fermionic GHZ states generated by one-to-all encoding are beyond Gaussian states and thus cannot be simulated by Gaussian state simulators for large-size systems even the evolution Hamiltonian itself is quadratic (see Appendix \ref{sec:fghz}).  It is worth noting that in U(1) symmetric systems, R\'enyi entropy can show different dynamical behaviors (diffusive growth) than the von Neumann entropy (ballistic growth). We leave the information dynamics described by R\'enyi mutual information and Holevo information as an interesting future direction to explore.

\subsection{Quasiparticle picture for SIC dynamics in non-interacting Hamiltonians}\label{subsec:quasiparticle}

The entanglement growth in noninteracting and integrable systems following a quench can be successfully explained by the intuitive quasiparticle picture \cite{Calabrese2005, Alba2017a, Calabrese2020, Orito2023_z}. The quasiparticles of the quenched Hamiltonian are formed in pairs or multiplets and uniformly distributed in space. These quasiparticles move ballistically with the speed determined by the dispersion relations and there is no interaction between different momentum species of quasiparticles. Specifically, only quasiparticle pairs of momentum $\pm k$ with one residing on intervals $A$ and the other residing on its complementary $\bar{A}$ contribute to the entanglement entropy $S_A$ one unit $s_k$. Here $s_k$ is the entropy unit $s_k=-n_k\log n_k -(1-n_k)\log (1-n_k)$ where $n_k$ represents the quasiparticle density for momentum $k$. For N\'eel initial states and the XX Hamiltonian which is equivalent to free fermion hopping model, $n_k=1/2$ for all quasiparticle momentum $k$  with the velocity for quasiparticle as $v_k = \partial \varepsilon_k/\partial k = -2\sin k$ and entropy unit $s_k=1$ in the $\log_2$ base.
The introduction of onsite disorder renders fermion operators of different momentum coupled, which is evident in Fourier transformed Hamiltonian in momentum space as $\sum_{k, k'}\mu(k-k')c^\dagger_kc_{k'}$. Therefore, the onsite disorder leads to strong scattering between quasiparticles of different momentum, slowing down and invalidating the ballistic transport of quasiparticles.

In terms of the SIC investigated in this work, we can still apply the quasiparticle picture framework to understand the information dynamics in the non-interacting system in the clean limit. We denote $Q_{AB}=\int_{x\in A, x'\in B}\frac{dk}{2\pi} s_k$ as the effective number of quasiparticle pairs with one quasiparticle in region A and the other in region B. In the context of XX quenched Hamiltonian with N\'eel initial state, each pair of quasiparticles carry the entanglement in one unit $s_k=1$. Consequently, the entanglement of some region $A$ is reduced to the number of effective quasiparticle pairs separated between $A$ and $\bar{A}$, i.e. $S_A= Q_{A\bar{A}}$. For the one-to-one encoding scheme employed in this section, we have:
\begin{align}
    S_A &= Q_{AR}+Q_{A\bar{A}}\\
    S_{AR} &= Q_{\bar{A}R} + Q_{A\bar{A}}\\
    S_{R} &= Q_{AR}+ Q_{\bar{A}R}.
\end{align}
Given that $S_R=1$ holds for unitary dynamics considered here, we have:
\begin{align}
    I(A:R) = S_A + S_R - S_{AR} = 2Q_{AR},
\end{align}
which relates the subsystem information capacity with the number of effective quasiparticle pairs separated in $A$ and $R$.

At time $t=0$, we have $Q_{A\bar{A}}(t=0)=Q_{\bar{A}R}(t=0)=0$ implying $Q_{AR}(t=0)=1$ based on the initial entropies in the initial product state with one-to-one encoding. Consequently, the SIC at any subsequent time $t$ is determined by $Q_{AR}(t)$, i.e. the only relevant part is the propagation to region $\bar{A}$ of initial quasiparticles starting from the entangled qubit $E$, which is paired with the counterpart in $R$ as counted in $Q_{AR}(t=0)=1$. This can be quantitatively calculated as 
\begin{align}
Q_{AR}(t) = 1-\int_{\abs{(L/2+2\sin(k)t) \;\text{mod} \;L-L/2}>L_A/2}\frac{dk}{2\pi}
\end{align}
for our specific Hamiltonian and initial states. Essentially, we simply count the number of quasiparticles of different velocities that go outside the interval $A$ starting from $E$. The quasiparticle picture for SIC in our settings is summarized in Fig.~\ref{fig:free_quasi_picture}(a).

\subsection{Results for non-interacting Hamiltonians}\label{sec:freefermion}

In this subsection, we focus on the results of the non-interacting AA model ($U=0$). The model in Eq. \eqref{eq:aahamiltonian} has an analytical exact critical point separating the Anderson localization and extended phases at $w_c=J$ \cite{Harper1955, Aubry1980}.

\begin{figure}[!htb]
\includegraphics[width=0.5\textwidth]{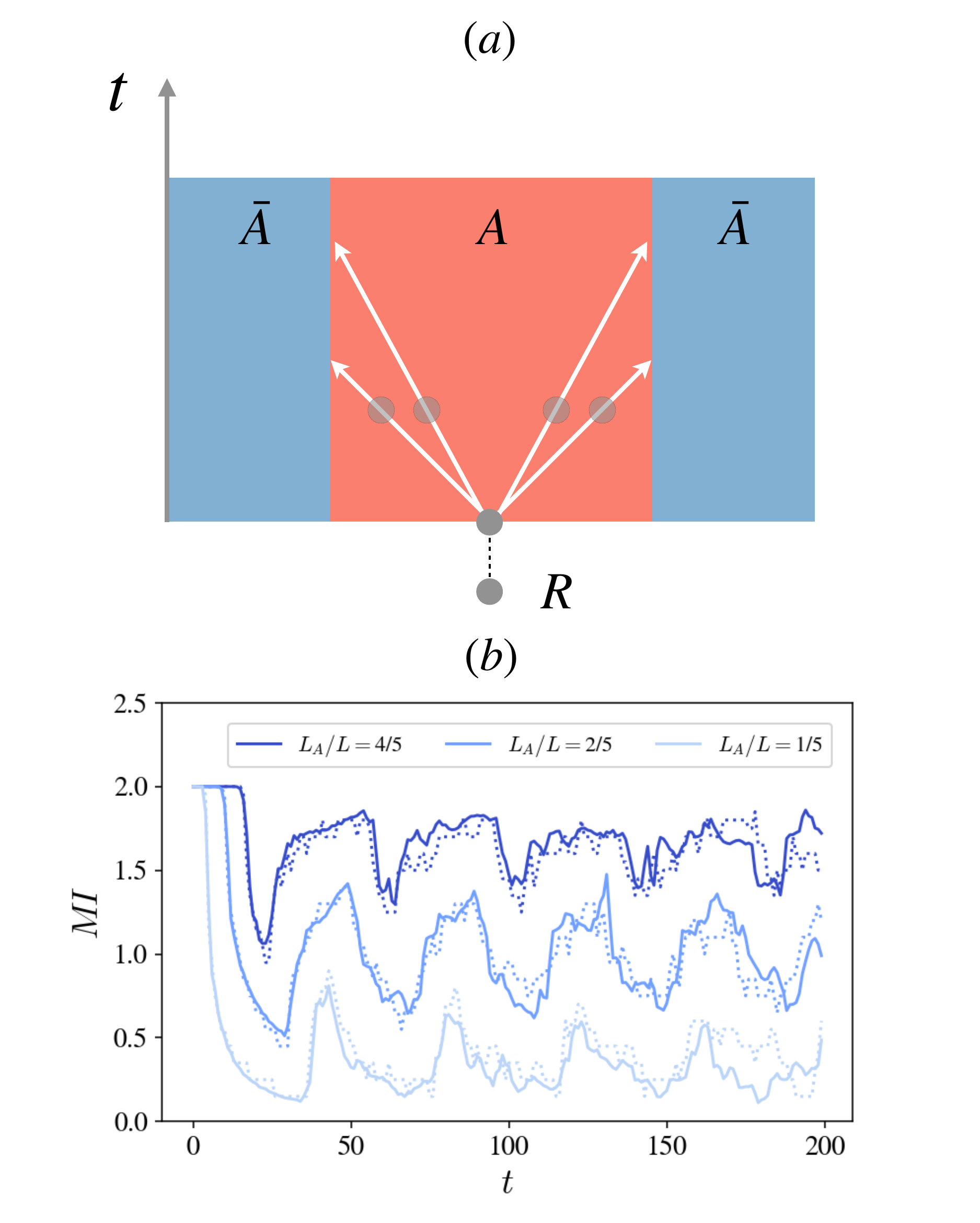}
\caption{(a) Schematic representation of the quasiparticle picture for mutual information $I(A:R)$. (b) SIC dynamics for $I(A:R)$ with $J=1, U=0, L=80, w=0$ for the model defined in Eq. \eqref{eq:aahamiltonian} with periodic boundary conditions. Initial deviating time $t^*=L_A/(2*v_{max})$. And oscillation frequency is given by $v_{max}/L$.  Dotted lines represent predictions from the quasiparticle picture, while solid lines show results from free fermion dynamics simulation.
} \label{fig:free_quasi_picture}   
\end{figure}

In the absence of quasiperiodic potential, i.e. $w=0$, the results can be interpreted by the quasiparticle picture as described above. As shown in Fig.~\ref{fig:free_quasi_picture}(b), the theoretical prediction based on the quasiparticle picture is qualitatively in agreement with the numerical results. The deviating time from the $MI=2$ plateau happens at $t^*=L_A/(2v_{max})$ where $v_{max}$ is the maximal speed of quasiparticles as $\text{max}_k \;2\sin k = 2$. After $t^*$, the fastest quasiparticle from $E$ begins to leave the interval $A$, leading to the decline in SIC. Additionally, the finite size system with periodic boundary conditions renders the oscillation of SIC dynamics in the clean limit $w=0$ when the quasiparticle picture is valid without scattering between different species. The oscillation behavior is similar as reported in \cite{Modak2020_z} where the quasiparticles can go around the system and get back to the interval $A$ in late times. It is important to note that the presence and absence of oscillation behaviors and the exact periodicity for the oscillation both depend on the quasiparticle dispersion relations $\epsilon_k$. Our results here are determined by the free fermion dispersion $\epsilon_k\propto \cos k$. Investigating the (non)oscillation behaviors for other dispersion relations presents a promising avenue for future research.

\begin{figure}[!htb]
\includegraphics[width=0.4\textwidth]{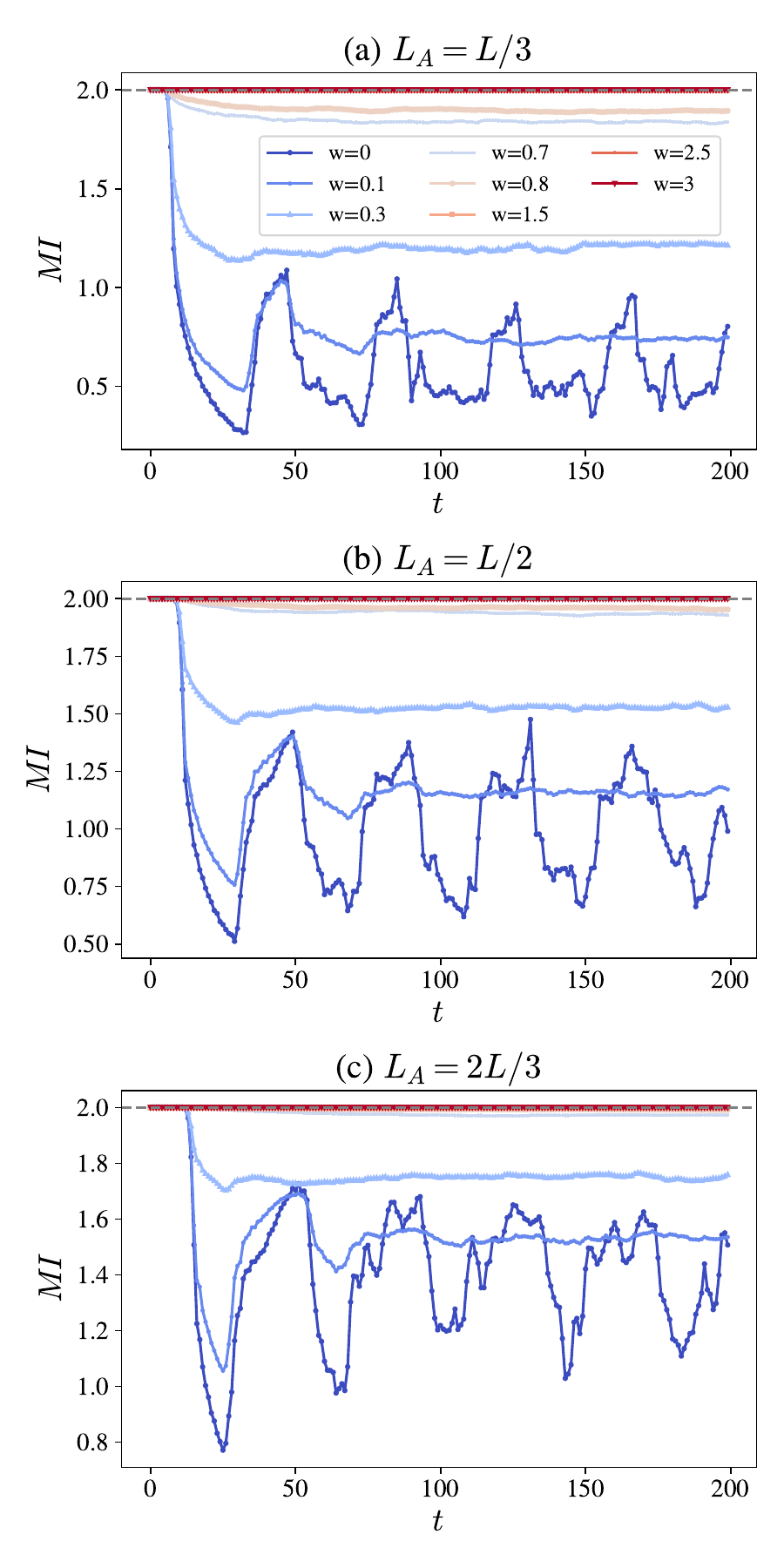}
\caption{SIC dynamics of different quasiperiodic potential strength $w$ with $J=1,U=0,L=80$ for model in Eq. \eqref{eq:aahamiltonian} with periodic boundary conditions. (a) Subsystem size $L_A=L/3$.  (b) Subsystem size $L_A=L/2$.  (c) Subsystem size $L_A=2L/3$.
} \label{fig:free_dynamics}  
\end{figure}

When the quasiperiodic potential is turned on, quasiparticles scatter with each other of different momentum, which will slow down the information dynamics -- SIC decay is slower at early times and the steady state has higher SIC compared to the clean limit with the same subsystem size. Furthermore, the dynamics are smooth with no obvious oscillation behaviors observed in $w=0$ case. For large quasiperiodic potential $w>w_c$, the information spreading is fully frozen in the thermodynamic limit due to the strong scattering of quasiparticles, namely, quasiparticles and the information they carry gets localized due to strong scattering induced by the quasiperiodic potential. The dynamics with varying potential strengths $w$ is shown in Fig. \ref{fig:free_dynamics}. For different $L_A$, the deviating time from $MI=2$ plateau, as described by $t^*=L_A/(2v_{max})$, differs accordingly, akin to the clean limit. This phenomenon can be attributed to the presence of a subset of the fastest quasiparticles that fortuitously evade scattering as they propagate through the subsystem $A$, thus leaving the same deviation time $t^*$ for different $w$ with the same $L_A$.

\begin{figure}[!htb]
\includegraphics[width=0.45\textwidth]{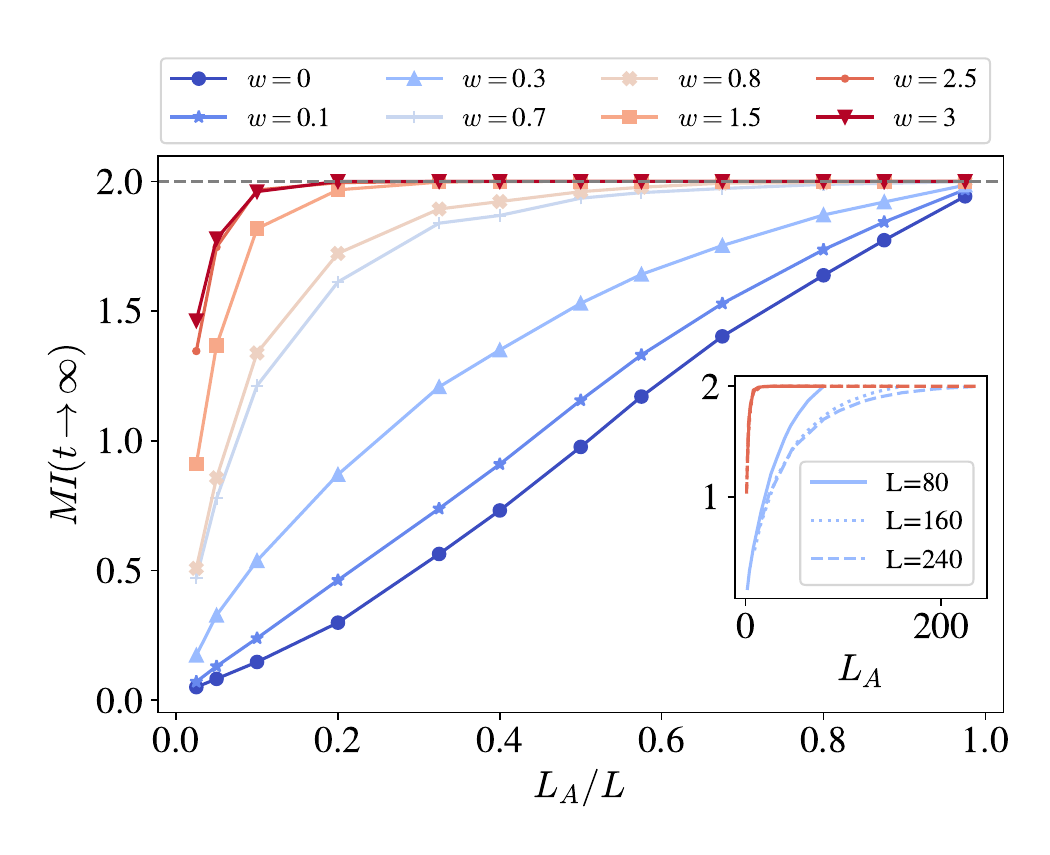}
\caption{Steady-state SIC for the model defined in Eq. \eqref{eq:aahamiltonian} with $J=1,U=0,L=80$ and periodic boundary conditions for different $w$.  Inset: Blue lines show results of $w=0.3$ (extended phase), while red lines show results of $w=2.5$ (localized phase). In Anderson localization phase, different curves of different sizes coincide with $L_A$ as the x-axis.
} \label{fig:free_steady}   
\end{figure}

\begin{figure}[!htb]
\includegraphics[width=0.4\textwidth]{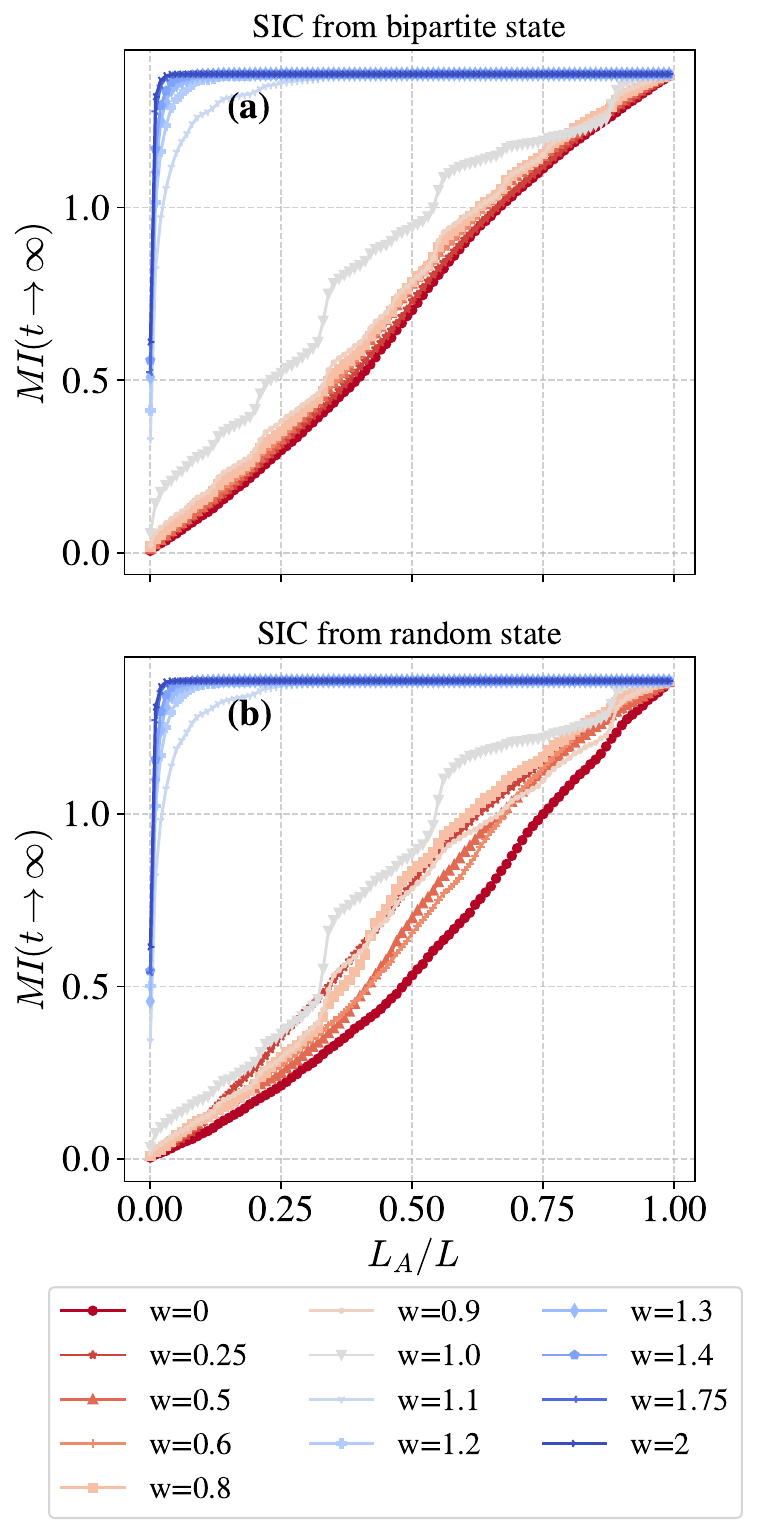}
\caption{Steady-state SIC for the model defined in Eq. \eqref{eq:aahamiltonian} with $J=1,U=0,L=200$ and periodic boundary conditions for different $w$. (a) Initial states for the dynamics is bipartite product state $\vert 000...111..\rangle$, (b) SIC is averaged over different random product state at half-filling sector.
} \label{fig:free_initial_states}   
\end{figure}

The late-time steady-state behavior for the SIC $I(A:R)$ is given in Fig. \ref{fig:free_steady}. For the clean limit $w=0$, the late-time SIC forms a straight diagonal line, indicating a uniform spatial distribution of the initial information at late times. A slightly larger $w$ will make the line bend towards the upper left. After reaching Anderson localization phases $w>w_c$, the SIC maintains the value $2$ for any $L_A/L>0$ in the thermodynamic limit. The curves for different system sizes converge in Anderson localization phases as shown in the inset of Fig. \ref{fig:free_steady}, reflecting the system size-independent nature of the localization length. This is similar to the Anderson localization phase in the random Clifford Floquet circuit. We further investigate the initial state dependence of SIC by simulation on dynamics from different initial states such as bipartite state and random product states as shown in Fig. \ref{fig:free_initial_states}. These results compare the late-time averaged SIC for different initial product states under the same Hamiltonian dynamics. Our findings indicate that while the numerical values of SIC can show some variation, the qualitative features, such as the spatial patterns of information distribution and the distinction between different dynamical phases, remain robustly consistent across different initial states. This supports the utility of SIC as a probe of the system's fundamental dynamical characteristics.

The results for the free Hamiltonian in the extended phase can be further understood via a specific type of random quantum circuits, known as swap circuits. The circuit is composed of brickwall two-qubit gates with each gate being randomly selected to be either an identity or a swap operation. In this setup, the local information encoded initially always remains inherently localized, and only the position of the information can be changed via the swap gate. When considering the average over different configurations of swap circuits, the steady-state SIC will be a straight line that is the same as $w=0$ line in Fig. \ref{fig:free_steady} and the speed of configuration-averaged information propagation is determined by the probability of swap gates. In sum, under the one-to-one encoding scheme, random circuits composed of swap and identity gates generate steady states with information distribution as given by pattern (a) in Fig.~\ref{fig:resultsummary}. The classification regarding subsystem information capacity for random quantum circuits formed by a discrete set of random gates deserves further investigation.

\subsection{Results for interacting Hamiltonians}

In this subsection, we focus on the dynamics driven by the interacting AA model. The exponential dimension of the quantum many-body Hilbert space limits our numerical investigation to systems of small sizes. Moreover, the lack of an analytical framework for information dynamics in generic many-body systems presents a challenge. Despite these limitations, we present numerical results for these small systems and offer some intuitive insights.

\subsubsection{Thermal and many-body localization systems}\label{subsec:mbl}

Many-body localization is a novel non-equilibrium phase that challenges the eigenstate thermalization hypothesis \cite{Srednicki1994}. Building upon Anderson localization \cite{Anderson1958}, the system can evade thermalization with many-body interactions \cite{Basko2006}. A variety of mechanisms are proposed to stabilize MBL phases, including strong random disorder \cite{Oganesyan2007, Pal2010a, Altman2014, Nandkishore2015, Abanin2018}, quasiperiodic potentials \cite{Iyer2013, Khemani2017, Zhang2018, Zhang2019a, Kohlert} and linear potentials \cite{Schulz2019a, VanNieuwenburg2019, Khemani2020, Doggen2021a, Liu2022}. As a fundamental basis for other emergent dynamical phases such as discrete time crystal \cite{Else2019, Zaletel2023} and Hilbert space fragmentation \cite{Khemani2020, Doggen2021a, Yang2020b}, MBL has been extensively explored on various experimental platforms of programmable quantum simulators and quantum computers \cite{Schreiber2015a, Smith2016, Liu2021c, Gong2021}. In this work, we explore MBL phases through the lens of information protection and information dynamics. 

Numerous studies have explored information scrambling in MBL systems, mainly using the probe of OTOC \cite{Fan2017, Swingle2017, Huang2017b, He2017, Chen2017} and TMI \cite{Bolter2022, MacCormack2021, Orito2022}. However, these studies have not quantitatively determined the late-time redistribution of information from local perturbations. In this work, we use the newly introduced subsystem information capacity as the probe to investigate the information dynamics in MBL systems. The results are anticipated to be very different from Anderson localization cases since entanglement growth in the two phases is very different (no growth vs. logarithmic increase) \cite{Bardarson2012, Abanin2013}.

\begin{figure}[!htb]
\includegraphics[width=0.4\textwidth]{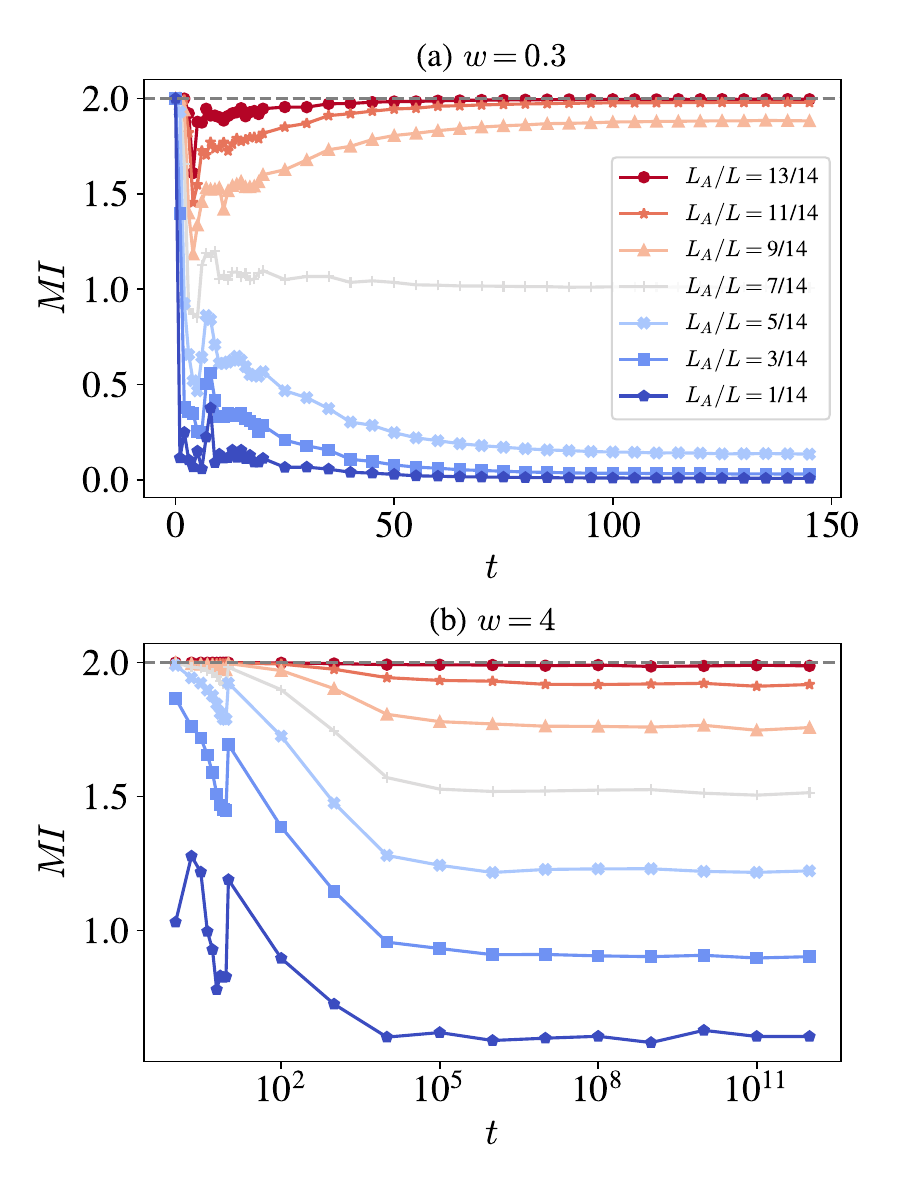}
\caption{SIC dynamics for model in Eq. \eqref{eq:aahamiltonian} with $J=1,U=0.2,L=14$ with PBC (a) $w=0.3$ (thermal phase),(b) $w=4$ (MBL phase). The dynamics at the initial stage is similar to the non-interacting case owing to the  N\'eel configuration of the initial state. To capture the rapid variations at early time, we use a constant time step of $dt=1$ for $t<10$. For the later stages, to efficiently demonstrate the exponentially slow convergence behavior, we employ an exponentially increasing time step for $t>10$.}
 \label{fig:mbl_dynamics}   
\end{figure}

\begin{figure}[!htb]
\includegraphics[width=0.42\textwidth]{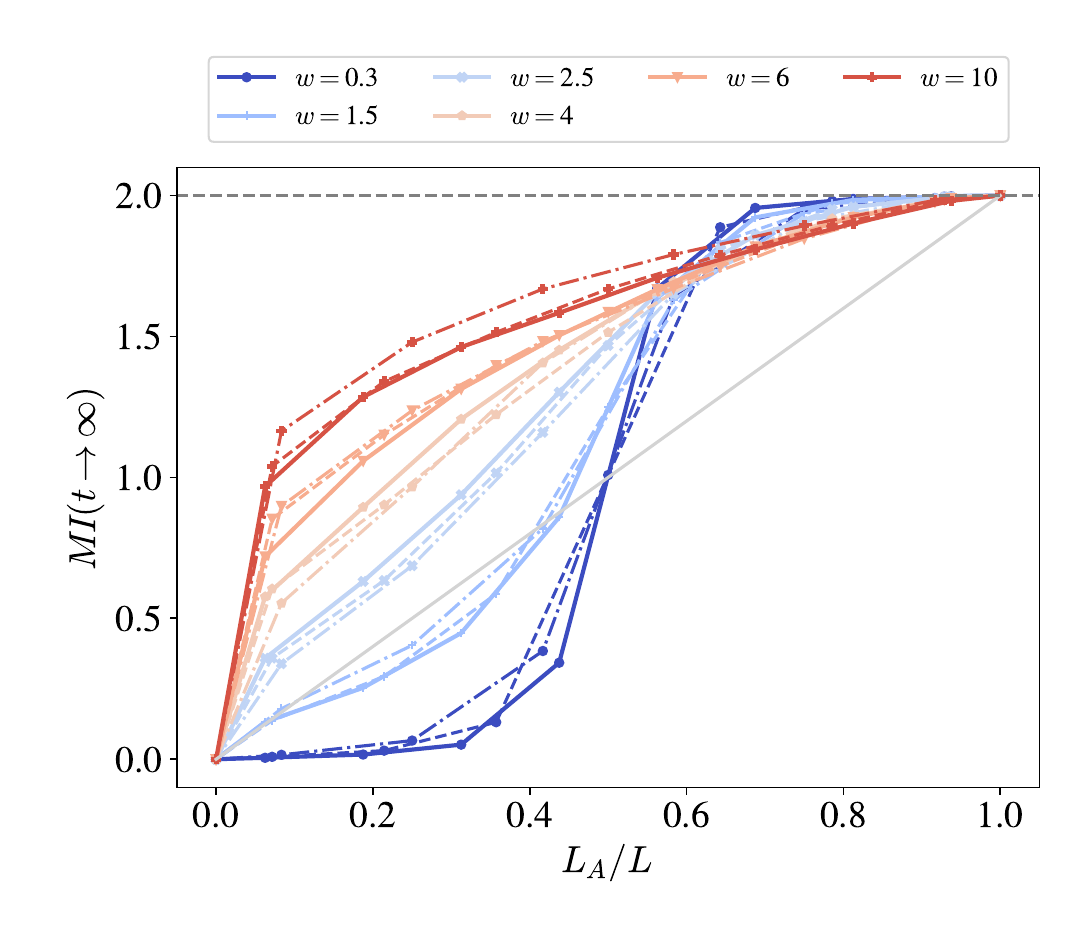}
\caption{Steady-state SIC for model in Eq. \eqref{eq:aahamiltonian} with $J=1,U=0.2$ with PBC. The results showcase the late-time SIC behaviors for thermal and MBL systems. Solid lines represent results of $L=16$, dashed lines represent results of $L=14$, while dashdot lines represent results of $L=12$.
} \label{fig:mbl_steady}   
\end{figure}

The phase diagram of the interacting AA model is determined in Appendix \ref{sec:mblr} with $w_c\approx 1.4\pm 0.2$ for MBL transitions in our parameter regimes. The dynamical and steady-state SIC for different subsystem sizes $L_A$ and different potential strengths $w$ are depicted in Fig. \ref{fig:mbl_dynamics} and \ref{fig:mbl_steady}, respectively. Regarding the dynamics, we observe a logarithmic decay in SIC in MBL phases, rendering a very slow saturation for the subsystem SIC as shown in Fig. \ref{fig:mbl_dynamics}(b). This logarithmic dynamical scaling is consistent to the logarithmic lightcone in MBL phases. Besides, the early time dynamics for both thermal and MBL phases are very similar to the non-interacting dynamics we reported before. This fact is particularly evident in the thermal region, where the SIC of a large subsystem initially drops and then returns to values close to $2$ given by information scrambling.
The early-time resemblance to the non-interacting case is because the density-density interaction plays no role at the beginning stage due to the N\'eel configuration initial states.
Therefore, despite the steady-state chaotic behavior for thermal phases being predicted and shown to be identical to the random quantum circuit case in Fig. \ref{fig:ruc_1to1}, the early-time information dynamics show very different characteristics. In other words, in the thermal phase, the information within a given subsystem $L_A>L$  will first drop and then backflows to the region with late-time convergence toward full information access $MI=2$. Such information backflow doesn't occur in chaotic random circuits with one-to-one encoding and is the unique early-time feature for chaotic Hamiltonian evolution.

In thermal phases, the steady-state SIC distribution is step function like, which is the same as reported in random quantum circuit cases, reflecting the universal aspects of quantum chaos and information scrambling. In MBL phases, the steady-state SIC is different from those in Anderson localization phases as anticipated. Albeit exponentially slow, information can eventually propagate to the entire system in MBL phases, resulting in $MI(L_A/L>0)<2$ in late times in the thermodynamic limit. On the contrary, $MI(L_A/L>0)$ is strictly 2 for Anderson localization phases in the thermodynamic limit. The steady-state SIC curve for large ratio $L_A/L$ appears to be approximately linear although the precise form remains unclear due to the limitation of small-size simulations.

\subsubsection{Integrable interacting systems}\label{subsec:integrable}

\begin{figure}[!htb]
\includegraphics[width=0.4\textwidth]{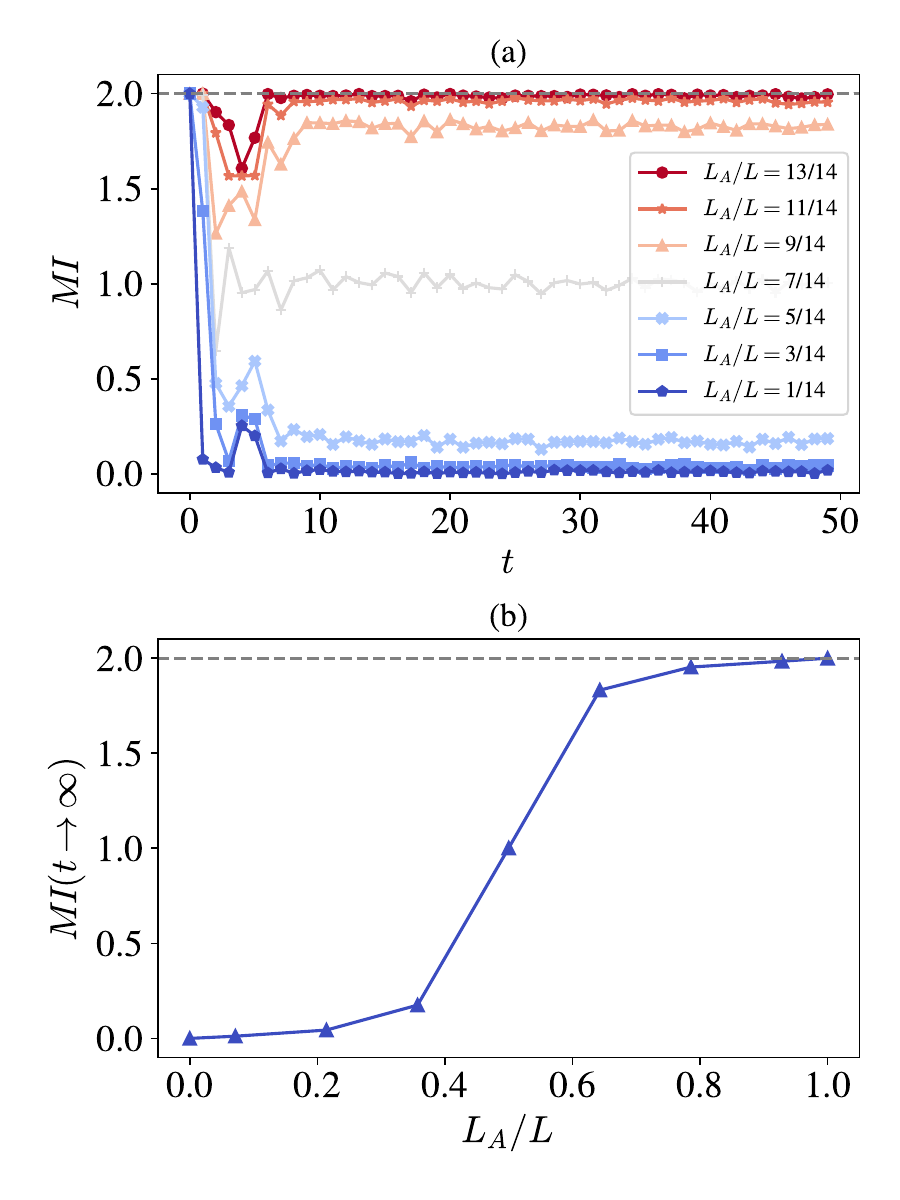}
\caption{Model defined in Eq. \eqref{eq:aahamiltonian} with $J=1,U=0.2, L=14$ and PBC is utilized. The magnetic fields correspond to uniform potentials, i.e. XXZ+Z model. (a) SIC dynamics of different subsystem sizes $L_A$. (b) Steady-state SIC for different $L_A/L$, the curve is similar to the thermalization case in finite sizes.
} \label{fig:integrable}   
\end{figure}

The nature of information scrambling in generic interacting integrable systems remains mostly elusive \cite{Iyoda2018, Schnaack2019, Caceffo2023}. Some of these works provide hints of information scrambling within integrable systems. We carry out the information dynamics simulation on XXZ+Z model, i.e. interacting AA model but with a uniform magnetic field (Z terms in spin language). This model is known as an interacting yet integrable system. The information dynamics and the steady values of SIC are shown in Fig. \ref{fig:integrable}. Notably, the late-time behavior of the integrable system exhibits striking similarities to that of the chaotic systems, as observed in the thermal phase or random quantum circuits. 
Moreover, the time scale required for the integrable system to reach the steady state, which is $O(10)$, is significantly shorter than that of the thermal case, which is $O(100)$. One possible explanation for the different equilibrium time scales is that the small accessible Hilbert space in integrable systems results in faster convergence. The dimension of the Hilbert space for integrable systems is very limited due to an extensive number of effective conservation laws while thermal systems have exponentially large charge sectors for $L/2$ occupation.

Based on our findings, we observe that while integrable systems cannot thermalize, they can still support the chaotic-like information dynamics in real space as we show here. This finding calls for a reassessment of the connection between complete information scrambling and thermalization. Further studies are required to understand the difference between thermal and integrable systems concerning information dynamics. Additionally, given that non-interacting AA model is also integrable but doesn't show chaotic behavior in steady-state SIC, it is also imperative to categorize different types of integrable systems regarding their late-time information distribution. 

\subsection{Results for topological systems}\label{sec:topological_edge}

\begin{figure}[!htb]
\includegraphics[width=0.42\textwidth]{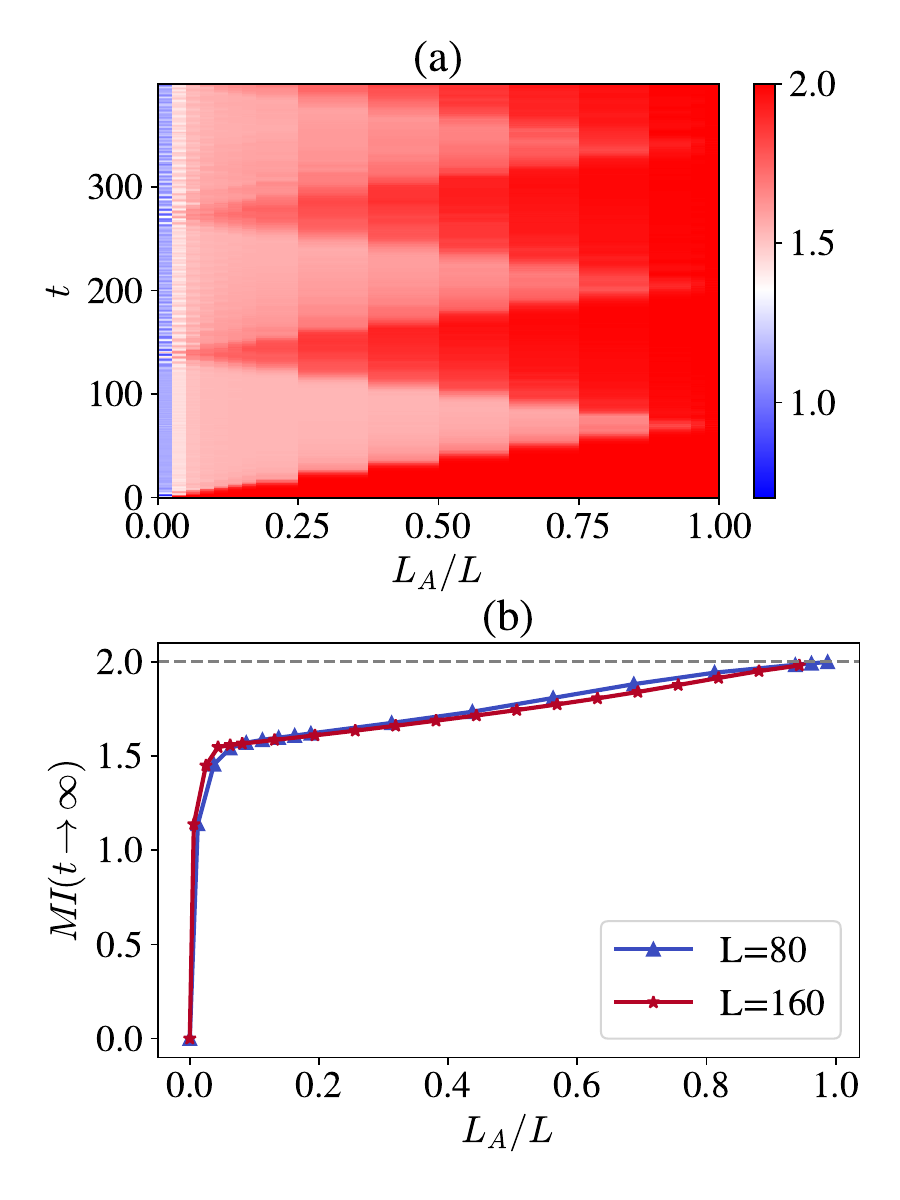}
\caption{Model defined in Eq. \eqref{eq:ssh} with $\Delta=-0.4$ (topological phase) with open boundary condition is utilized. The entanglement qubit $E$ is chosen at one of the spatial boundaries. (a) SIC $I(A:R)$ dynamics of different subsystem sizes $L_A$ and different time $t$, $L=80$. (b) Late-time averaged SIC for different $L_A/L$ with $L=80,160$.
} \label{fig:tp}   
\end{figure}

\begin{figure}[!htb]
\includegraphics[width=0.4\textwidth]{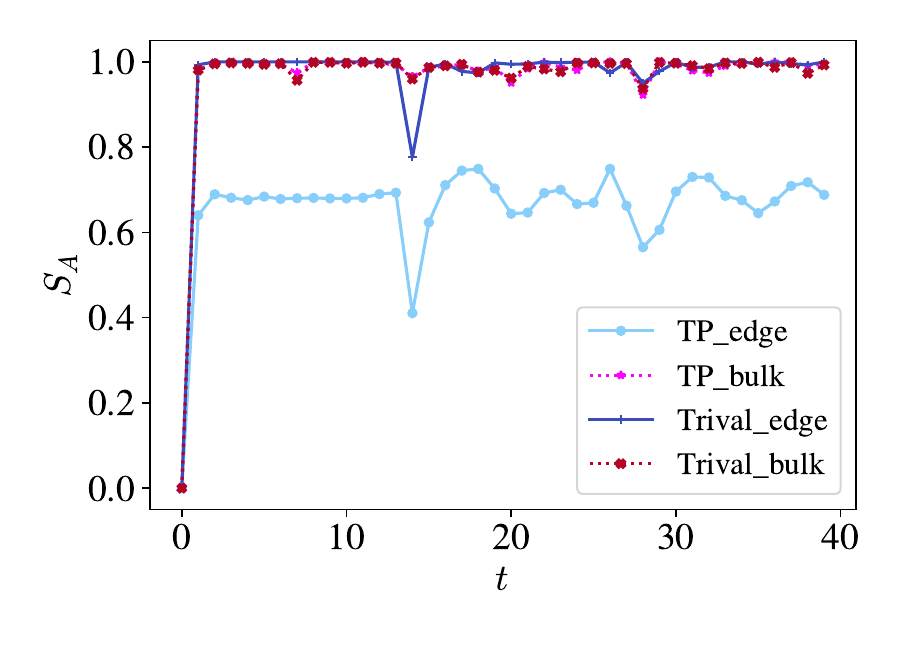}
\caption{Entanglement growth with quench Hamiltonian defined in Eq. \eqref{eq:ssh} with open boundary condition. The subregion $A$ is selected as one qubit on the boundary or in the bulk. The initial state is N\'eel state. The results $S_A(t)$ for $A$ as one qubit at the spatial boundary in topological phase ($\Delta=-0.4$, TP$\_$edge), one qubit in the bulk in topological phase (TP$\_$bulk), one qubit at the spatial boundary in trivial phase ($\Delta=0.4$, Trivial$\_$edge), and one qubit in the bulk in trivial phase (Trivial$\_$bulk) are presented.} \label{fig:eessh}   
\end{figure}

\begin{figure}[!htb]
\includegraphics[width=0.42\textwidth]{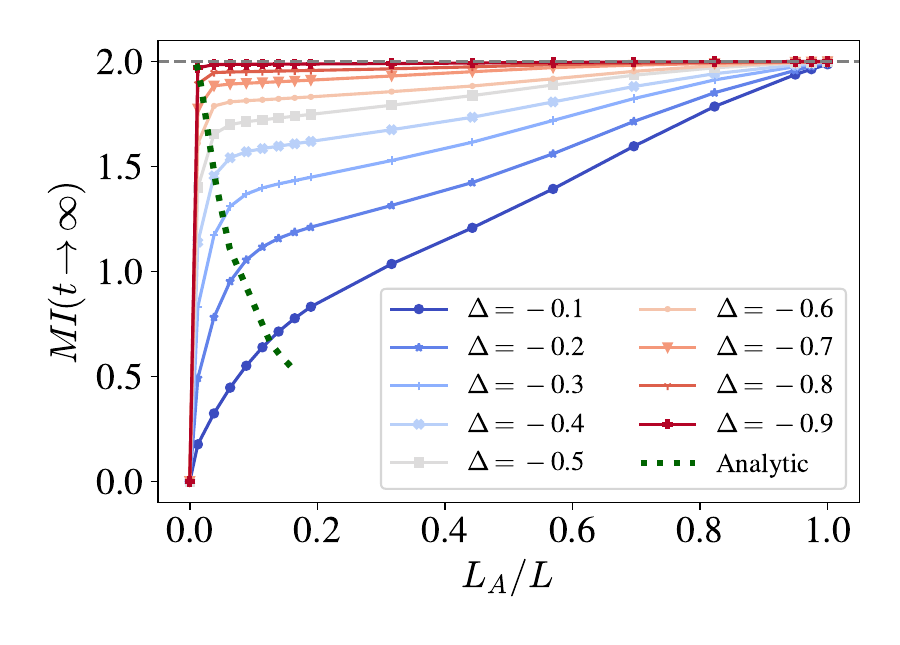}
\caption{Late-time averaged SIC $I(A:R)$  for different $L_A/L ~(L=80)$ with $\Delta$ range from $-0.1$ to $-0.9$ for model in Eq. \eqref{eq:ssh} (topological phase) with open boundary conditions. We utilize one-to-one encoding with the entangled qubit $E$ at the boundary of the one-dimensional system. The green dotted line is from theoretical predictions where the height is given by $\langle \psi\vert n_0\vert \psi\rangle_{\text{edge}}$ with respect to the relative localization length $\xi_{loc}/L=2/\ln{\frac{1-\Delta}{1+\Delta}}/L$ for different $\Delta$s. The theoretical green line successfully captures the turning points of the information distribution near the information-trapping edge.
} \label{fig:tpdelta}   
\end{figure}

\begin{figure}[!htb]
\includegraphics[width=0.45\textwidth]{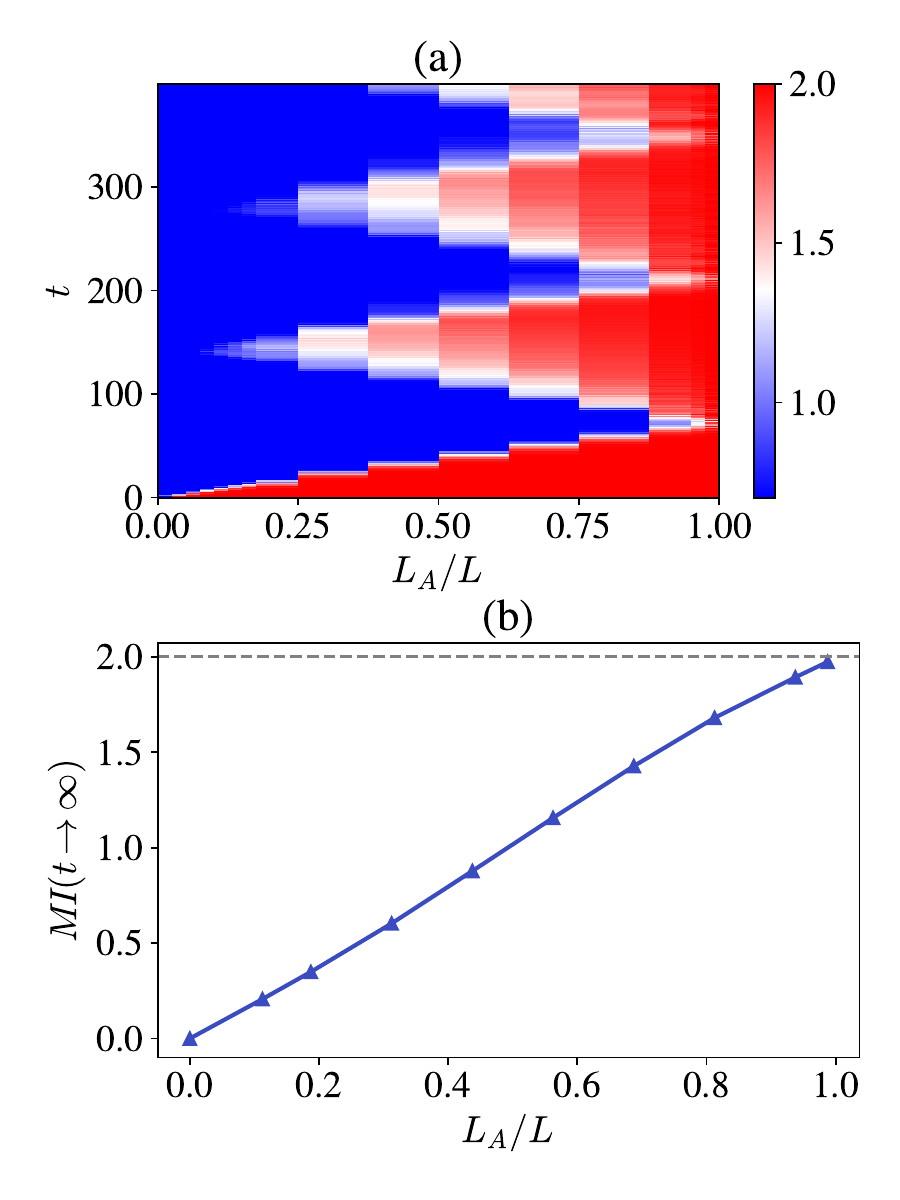}
\caption{Model defined in Eq. \eqref{eq:ssh} with $\Delta=0.4$ (trivial phase) with open boundary condition is utilized. $L=80$. The entanglement qubit is chosen at one of the spatial boundaries. (a) SIC $I(A:R)$ dynamics of different subsystem sizes $L_A$ and different time $t$. (b) Late-time averaged SIC for different $L_A/L$, the result is the same as fully extended phases given in Fig. \ref{fig:free_steady}.
} \label{fig:trivial}   
\end{figure}

The investigation on information scrambling in topological systems is limited and often restricted to the calculation of OTOC \cite{Dag2020, Sedlmayr2023, Bin2023, Sur2024}. In contrast, we investigate the interplay between information dynamics and topology characteristics through the proxy of SIC developed in this work, which provides fresh insights into information perspectives of topological systems. To explore the effect of topological edge states, we utilize models with open boundary conditions in this subsection.

The model we employed is one-dimensional SSH model as introduced in Eq. \eqref{eq:ssh}.
For $\Delta<0$, the system is in a topological phase with two edge modes localized at two ends of the one-dimensional system. Under the one-to-one encoding scheme, we investigate cases where the entangled qubit $E$ is situated either at the boundary site or within the bulk. The latter case gives similar behavior as the non-interacting model in the extended phase. However, the former case yields highly non-trivial results due to the information confinement nature of the topological edge modes.

Information dynamics results in topological phases $\Delta=-0.4$ and trivial phases $\Delta=0.4$ are shown in Fig. \ref{fig:tp} and \ref{fig:trivial}, respectively. Interestingly, in the topological phase, a substantial portion of the information remains trapped near the edge forever, at least for the non-interacting case here. In interacting topological systems, such information trapping behavior might be accompanied by a finite lifetime \cite{Dag2020}.

The information trapping phenomena near the boundary of topological systems can be attributed to the localized eigenmodes, i.e. topological edge states. Localized modes can both suppress the entanglement growth and trap the information spreading. The former is validated in Fig. \ref{fig:eessh} where only steady-state entanglement defined near the boundary in topological phases is greatly suppressed compared to the saturating value $1$. In the language of quasiparticle picture, there are a portion of quasiparticles trapped within the localized mode which leads to both small steady-state entanglement and information confinement.

With $L_A$ as the x-axis, late-time SIC curve collapses well near the edge for different system sizes, indicating that the information trapping is fully controlled by the wavefunction of the edge states, irrespective of the total system size as long as the tunneling can be safely ignored. The fraction of information trapped near the edge is thus independent of the system size $L$ and can be approximately captured by the amplitude of edge modes on the boundary site, equivalent to the density expectation $\langle \psi\vert n_0\vert \psi\rangle_{\text{edge}}$. In summary, the spatial content and magnitude of information trapping are captured by the localization length $\xi_{loc} = 2/\ln \frac{1-\Delta}{1+\Delta}$ and the edge density  $\langle \psi\vert n_0\vert \psi\rangle_{\text{edge}}$, respectively. This theoretical understanding indicated as green dotted line, together with late-time information distribution from different Hamiltonian parameter $\Delta$s, is shown in Fig. \ref{fig:tpdelta}. The green dotted line indeed successfully captures the turning point of the information curves with different $\Delta$s due to the edge modes.  Away from the edge, the late-time information distribution is uniform, which is the same as the extended case, resulting in a range of small-slope straight lines as shown in Fig. \ref{fig:tp}(b).

On the contrary, in topological trivial phases, there is no information trapping on the edge, and the steady-state SIC curve is the same as trivial non-interacting extended states (pattern (a) in Fig. \ref{fig:resultsummary}). It is also interesting to note that the information now propagates and reflects at both edges due to open boundary conditions. The velocity for the information spreading is consistent with the maximal speed given by the SSH Hamiltonian spectrum $\text{max}_k\frac{d\varepsilon_k}{dk}=2(1-\abs{\Delta})$ ($v_{max} = 1.2$ for $\Delta=\pm 0.4$). For topological trivial systems without edge modes, the quasiparticle picture can still successfully apply to the information dynamics by considering the reflection of quasiparticles at both open boundaries \cite{Khetrapal2024}, the numerical results comparing the dynamics with quasiparticle picture predictions in topological trivial systems are given in Appendix \ref{sec:app_topo_quasi}.

The results presented in this subsection not only extend the scope of information dynamics to the context of topological systems but also demonstrate that the position of entangled qubit $E$ can strongly affect the resulting information dynamics in open boundary or spatial non-uniform systems. Leveraging this insight, the subsystem information capacity can be employed as a tool to investigate spatial singularities within the system by varying the position of the entangled qubits.

\section{Discussions} \label{sec:discuss}

In this work, we propose the use of subsystem information capacity $I(A:R)$ between the output subsystem $A$ and the reference system $R$ as a metric to characterize information dynamics and information capacity for the system, or specifically for the effective quantum channel formed by the subsystem of evolution. The dynamical and steady properties of SIC successfully reveal the intrinsic nature of different dynamical phases. This metric is also closely related to quantum information processing and quantum communication, owing to its relationship with quantum coherent information.

Our settings are different from the conventional TMI settings where all the system qubits are entangled with corresponding reference qubits. Conversely, all qubits in $\bar{E}$ are initialized with pure states in our case.  On the one hand, our settings save a great number of qubits ($L+1$ in one-to-one encoding case) and are thus much more friendly for noisy intermediate-scale quantum experiments \cite{Preskill2018}. On the other hand, the results in our settings rely on the choice of the initial states in some cases. For random quantum circuit evolution, the choice of different product initial states is irrelevant as the difference is smeared out by the first layer of random gates. Instead, the results can be different for different initial states in Hamiltonian quench dynamics. In charge conserved evolution, different choices of the initial states correspond to different charge sectors of different Hilbert space dimensions, which can result in very different behaviors \cite{Liu2024a}. For different initial states within the same charge sector, the difference is less known. Therefore, it deserves further investigations into the difference in information dynamics from different initial states, particularly in the case of symmetric evolution.

For U(1) symmetric dynamics, the dynamics of mutual information characterized by von Neumann entropy and R\'enyi entropy could be drastically different.  This is because R\'enyi entropy growth shows subdiffusive behaviors for systems with conservation laws while von Neumann entropy always exhibits ballistic growth \cite{Rakovszky2019, Huang2020d, Zhou2020c, Znidaric2020}. In our work, we focus on von Neumann entropy based mutual information for Hamiltonian dynamics. The dynamical behaviors for R\'enyi mutual information in these systems as well as the interplay between other symmetries with information dynamics require further investigation.

As previously noted, the framework for information dynamics established within this work is amenable to noisy intermediate-scale quantum \cite{Preskill2018} technology and is thus well-suited for implementation on near-term quantum computing platforms.
Specifically, our settings are more qubit efficient than measuring TMI and are free from Pauli operator average and reversal evolution than measuring OTOC.
Given that $A$ and $R$ both being small regions comprising a few qubits, the von Neumann or R\'enyi entropy for these subsystems can be obtained either via full state tomography or random basis measurement \cite{VanEnk2011, Brydges2019, Huang2020b}. Apart from experiments on quantum hardware, an interesting theoretical future direction is to analyze the impact of different types of quantum noise on these settings utilizing statistical model mapping. This could lead to the identification of noise-resilient experimental protocols for the implementation of information dynamics. It is also worth noting that random circuit structures utilized in this work are helpful in alleviating coherent quantum noises and demonstrating quantum advantages \cite{Arute2019}.

This work concentrates on one-dimensional systems, and we expect that higher-dimensional systems have the potential to exhibit a broader range of behaviors in information dynamics. There are also more possible geometry configurations to arrange $E$ and $A$ regions which could lead to a richer understanding of information propagation and scrambling. Another limit is to apply information metrics developed in this work to zero-dimensional systems with all-to-all interactions such as SYK models \cite{Sachdev1993, Maldacena2016a, Chowdhury2022}, which yield faster information scrambling and are analytically tractable in the large $N$ limit.

Apart from its implications for non-equilibrium physics, SIC and a deeper understanding of information dynamics can also be helpful and insightful for designing and benchmarking quantum algorithms, particularly in the context of variational quantum algorithms \cite{Cerezo2020b, Bharti2022}. Promising examples along this direction include understanding the training dynamics and learning capability for variational quantum machine learning algorithms from information perspectives \cite{Shen2020, Garcia2022, Sajjan2023}. Besides, it is inspiring to develop better circuit ansatz design \cite{Zhang2020b, Du2020a, Lu2020, Zhang2021, Wu2021b}, parameter initialization \cite{Park2024, Cao2024} and training strategies \cite{Liu2023a} with enhanced expressive power and alleviated barren plateau issues \cite{McClean2018} by leveraging the principles of non-equilibrium dynamics and information spreading.

\section*{Acknowledgements}
We thank helpful discussions with Zhou-Quan Wan and Shuai Yin. YQC acknowledges the support by NSAF No. U2330401. SXZ acknowledges the support from Innovation Program for Quantum Science and Technology (2024ZD0301700) and the start-up grant at IOP-CAS.

\appendix

\section{Relation between subsystem information capacity and out-of-time correlators}\label{sec:tnproof}

\begin{figure}[!htb]
\includegraphics[width=0.5\textwidth]{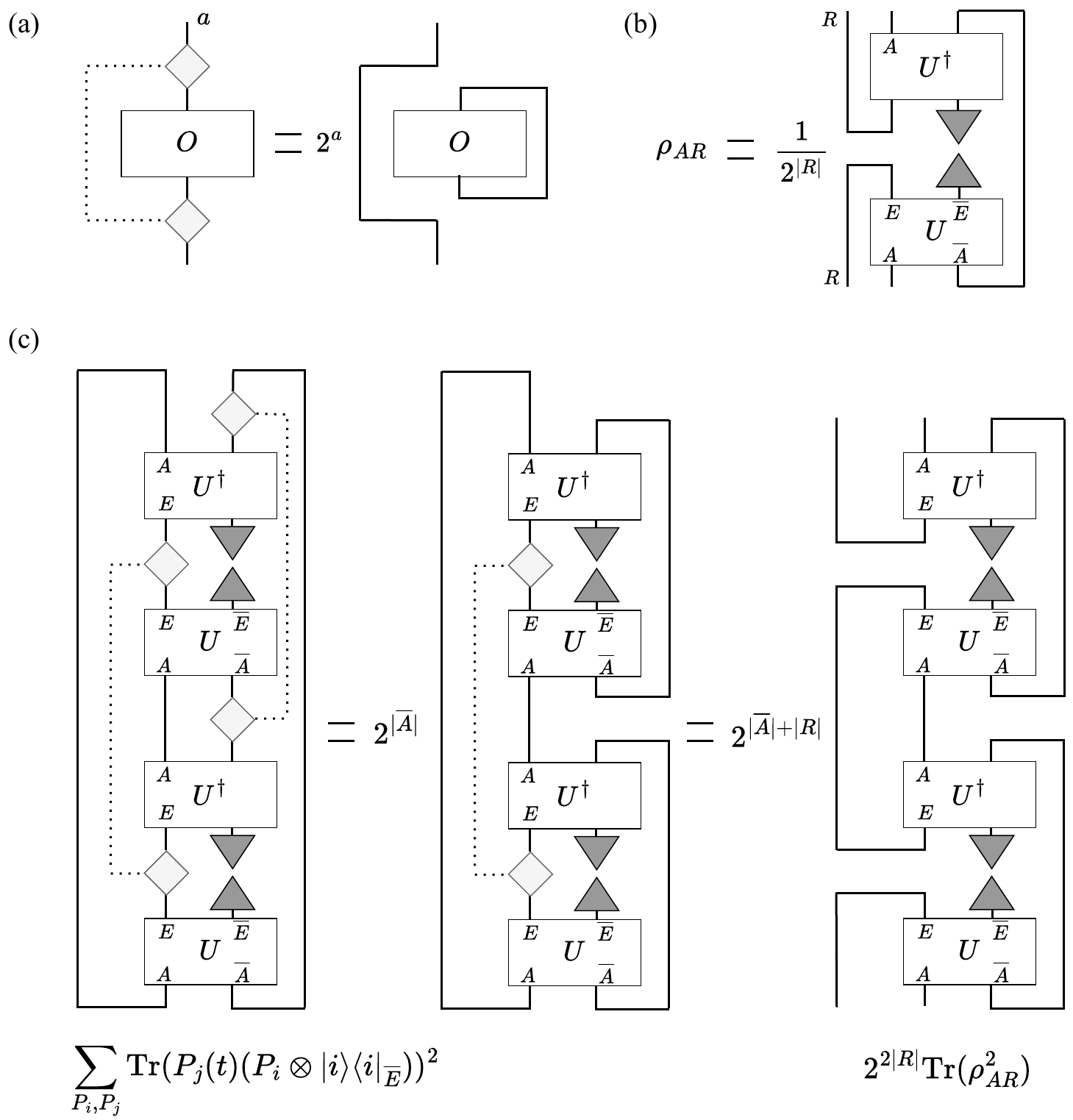}
\caption{(a) Operator identity: the diamond pairs indicate twirling the operator $O$ with all possible Pauli operators, i.e. $\sum_i P_i OP_i$.  (b) The reduced density matrix $\rho_{AR}$ by partial tracing freedom in $\bar{A}$. (c) The tensor network schematic proof for Eq. \eqref{eq:tnproof}. The grey triangle is for the pure input state in the subsystem $\bar{E}$: $\vert i\rangle_{\bar{E}}$.
} \label{fig:otocproof}   
\end{figure}
The established relation between mutual information and average OTOC over local Pauli operators \cite{Hosur2016} can be similarly extended to our case:
\begin{align}
2^{-2\abs{\bar{A}}-2\abs{R}}&\sum_{P_i\in E, P_j\in \bar{A}}\text{Tr}\left((P_j(t)(P_i\otimes \vert i\rangle\langle i\vert_{\bar{E}}))^2\right)\nonumber\\&=
2^{\abs{R}-\abs{\bar{A}}-S^{(2)}_{AR}},
\label{eq:tnproof}
\end{align}
where $P_i\in E$ implies that the sum is over all possible Pauli operators that are nontrivial on qubits in the interval $E$. The input of the system is divided into regions $E$ and $\bar{E}$, where qubits in $E$ form Bell pairs with qubits in the reference system $R$. The output of the system is divided into regions $A$ and $\bar{A}$, and in general $\abs{A}\neq \abs{E}$.

The left-hand side of Eq. \eqref{eq:tnproof} is similar to the notation of averaged OTOC over local Pauli operators, but now the zero time operator is replaced with local Pauli operator product with the initial state projector on the remaining system $\bar{E}$. The R\'enyi-2 entropy for $\rho_{AR}$ on the right-hand side of Eq. \eqref{eq:tnproof} is closed related to the R\'enyi-2 mutual information $I^{(2)}(A:R)$ focused in this work.

The schematic proof based on tensor network representation is shown in Fig.~\ref{fig:otocproof}. For the tripartite mutual information setting in Ref. \cite{Hosur2016}, the initial state projectors (grey triangles in Fig.~\ref{fig:otocproof}) are replaced with direct links in pairs since the input state is a fully mixed state for each qubit. In that case, we reproduce the equivalence between R\'enyi-2 entanglement entropy with the conventional OTOC for local Pauli operators, unlike the introduction of projectors for the pure initial state in our settings.

\section{Relation between quantum mutual information and Holevo information}\label{sec:holevo}

The two information quantities are equal under the condition $\text{Tr}_{\bar{A}}(\vert \psi_0^{(1)}\rangle \langle \psi_0^{(2)}\vert)=0$. In this Appendix, we demonstrate the equivalence. With the introduction of one single ancilla qubit as $\frac{1}{\sqrt{2}}(\vert 0\rangle_R\vert \psi_0^{(1)}\rangle +\vert 1\rangle_R\vert \psi_0^{(2)}\rangle )$, the reduced density matrix on $AR$ is given by 
\begin{align}
\rho_{AR}=\frac{1}{2}\vert 0\rangle\langle 0\vert\text{Tr}_{\bar{A}}(\rho_1)+\frac{1}{2}\vert 1\rangle\langle 1\vert\text{Tr}_{\bar{A}}(\rho_2),
\end{align}
where $\rho_i = \vert \psi_0^{(i)}\rangle \langle \psi_0^{(i)}\vert$. Note that the crossing term like $\vert 0\rangle \langle 1\vert \text{Tr}_{\bar{A}}(\vert \psi_0^{(1)}\rangle \langle \psi_0^{(2)}\vert) $ is zero only under the given condition. We thus have 
\begin{align}
S_{AR}=-\text{Tr}_{AR}(\rho_{AR}\log \rho_{AR})= \frac{1}{2}(S_A(\rho_1)+S_A(\rho_2))+1.
\end{align}
Since we have already shown that the first term of Holevo information is equivalent to $S_A$ with the reference qubit, we arrive at:
\begin{align}
\chi_A &= S_A -  \frac{1}{2}(S_A(\rho_1)+S_A(\rho_2)) \nonumber \\&= S_A + 1 - S_{AR} = I(A:R).
\end{align}
For the one-to-all encoding scheme, the condition $\text{Tr}_{\bar{A}}(\vert 0^L\rangle \langle 1^L\vert)=0$ is satisfied, resulting in the same initial value $1$ for both mutual information and Holevo information. On the contrary, the condition is generally not satisfied when states $\vert \psi_0^{(i)}\rangle$ differ only in the local region $E$, as is the case for one-to-one encoding.

\section{Mapping between random quantum circuits and effective statistical models}\label{sec:mapping}

The random quantum circuits under investigation are constructed from a sequence of random two-qudit unitary gates, each sampled from the Haar ensemble and arranged in a brickwall layout, as depicted in Fig. \ref{fig:setup}(a). In this study, we adhere to the convention of defining the unit time $\Delta t$ = 1 for an even-odd layer of two-qudit gates. 
This convention implies a lightcone speed of 
$v=2$, which differs from some literature where 
$\Delta t=1$ is associated with a single even or odd brickwall layer. 
After $T$ layers of gates, the output density matrix $\rho$ is given by
\begin{eqnarray}
    \rho = \prod_{t=1}^{T} \tilde{U}_{t} \rho_{0} \tilde{U}_{t}^{\dagger},
\end{eqnarray}
where $\rho_{0}$ is the density matrix of the initial state, and
\begin{eqnarray}
    \tilde{U}_{t}= \prod_{i=0}^{\frac{L-2}{2}} U_{t, (2i+2, 2i+3)}  \prod_{i=0}^{\frac{L-2}{2}} U_{t, (2i+1, 2i+2)}, 
\end{eqnarray}
is the unitary evolution at time step $t$. To obtain the von Neumann entropy of the subsystem for the output state, we can first express the density matrix in an $r$-fold replicated Hilbert space:
\begin{eqnarray}
    \vert \rho \rangle^{\otimes r} &=& \prod_{t=1}^{T} \left[ \tilde{U}_{t} \otimes \tilde{U}_{t}^{*}  \right]^{\otimes r} \vert \rho_{0} \rangle^{\otimes r}.
\end{eqnarray}

The mapping to the effective statistical model is given by the average over Haar random two-qudit unitary gates $U$:
\begin{eqnarray}
    \mathbb{E}_{\mathcal{U}}(U \otimes U^{*})^{\otimes r} = \sum_{\sigma, \tau \in S_{r}} \text{Wg}_{d^2}^{(r)}(\sigma \tau^{-1}) \vert \tau \tau \rangle \langle \sigma \sigma \vert,
\end{eqnarray}
where $S_{r}$ is the permutation group of degree $r$, $d$ is the local Hilbert space dimension of qudits, and $\text{Wg}_{d^{2}}^{(r)}$ is the Weingarten function with an asymptotic expansion for large $d$ as~\cite{Zhou2019, collinsIntegrationRespectHaar2006}:
\begin{eqnarray}
    \text{Wg}_{d^{2}}^{(r)} (\sigma)  = \frac{1}{d^{2r}} \left[ \frac{\text{Moeb}(\sigma)}{d^{2\vert\sigma\vert}} + \mathcal{O}(d^{-2\vert \sigma \vert -4})\right].
\end{eqnarray}
Here $\vert \sigma \vert$ is the number of transpositions required to transform from the identity permutation spin $\mathbb{I}$ to $\sigma$. By averaging all two-qudit gates independently, the quantum circuit is effectively transformed into a classical spin model, where the degrees of freedom are formed by permutation-valued spins $\sigma$ and $\tau$ on a honeycomb lattice.

The partition function $Z$ for this statistical model composed of permutation spins is obtained by summing the total weights of all possible spin configurations. For a given spin configuration, the weight is the product of the bond weights on the diagonal and vertical bonds on the honeycomb lattice.  The weight of the diagonal bond is given by the inner product between two diagonally adjacent permutation spins (between two two-qudit gates in neighboring layers and neighboring positions in the original circuit language)
\begin{eqnarray}
    w_{d}(\sigma, \tau) = \langle \sigma \vert \tau \rangle = d^{r-\vert \sigma^{-1} \tau \vert},
\end{eqnarray}
and the weight for the vertical bond (within one two-qudit gate in the original circuit language) is given by the Weingarten function. Besides, mid-circuit measurements can be treated as quenched disorder \cite{PhysRevLett.129.080501, PhysRevB.107.L201113}, and the diagonal bond with the measurement gives a weight contribution factor $d^0$. However, the partition function is ill-defined due to the negative weighted configurations since $\text{Moeb}(\sigma)$, the Moebius number of $\sigma$, can be negative~\cite{collinsIntegrationRespectHaar2006}. To obtain positive definite weights and properly defined partition function for the statistical model, we integrate out half of the freedoms (all $\tau$ spins) and obtain positive three-body weights of downward triangles as
\begin{align}
    &W^{0}(\sigma_{1}, \sigma_{2}; \sigma_{3}) =\nonumber\\ &\sum_{\tau \in S_{r}} \text{Wg}_{d^{2}}^{(r)} (\sigma_{3}\tau^{-1}) d^{2r-\vert \sigma_{1}^{-1} \tau \vert - \vert \sigma_{2}^{-1} \tau \vert}.
    \label{eq:3body}
\end{align}
After tracing out $\tau$, the total weight of a given spin configuration is the product of the weights of all the downward triangles on the space-time lattice of the circuit.
In the large $d$ limit, we focus on the most dominant spin configuration of the largest weight, namely, the partition function $Z$ is determined by the weight of the dominant spin configuration. A finite $d$ such as $d=2$ in qubit case introduces interactions between different replicas and a more general entanglement membrane picture is required as discussed in Sec. \ref{subsec:mapping}. 

The statistical model is ferromagnetic since the weights of the downward triangles with specific spin configurations as
$W^{0}(\sigma,\sigma; \sigma) 
    \approx d^{0}
    $ and  
    $W^{0}(\sigma^{\prime},\sigma; \sigma) 
    \approx d^{-\vert (\sigma^{\prime})^{-1}) \sigma \vert}<1$.
Therefore, the spin-spin interaction in the effective model is ferromagnetic, and all the spins tend to align in the same direction. However, as discussed below, due to the particular boundary conditions, domain walls may appear, with the unit free energy of $\log\, W^{0}(\sigma^{\prime},\sigma; \sigma)$ for unit length. It is also worth noting that the weight $W^{0}(\sigma, \sigma;\sigma^{\prime})=0$ due to unitary constraint~\cite{Uaverage_Qi, PRXQuantum.4.010331}. 
Consequently, the domain wall in the random unitary circuit is precluded from exhibiting horizontal configurations or upward triangular turning points. However, the presence of mid-circuit measurements and quantum noises \cite{Liu2024} can relax the unitary constraint, thereby enabling the emergence of such domain walls. These non-unitary factors introduce a new layer of complexity and versatility to the quantum circuit, potentially giving rise to a richer variety of dynamical behaviors.

Now, we utilize the fact that the von Neumann entropy $S_{\alpha}$ is $n=1$ limit for R\'enyi-$n$ entropy
\begin{eqnarray}
    S_{\alpha} = \underset{n \rightarrow 1}{\lim} S^{(n)}_{\alpha} = \underset{n \rightarrow 1}{\lim} \frac{1}{1-n} \mathbb{E}_{\mathcal{U}} \log\frac{\tr \rho_{\alpha}^{n}}{(\tr \rho)^{n}},
\end{eqnarray}
where $\rho_{\alpha}$ is the reduced density matrix of region $\alpha$ and $S^{(n)}_{\alpha}$ is the $n$-th order R\'enyi entropy. Via the statistical model mapping introduced above, we express $S^{(n)}_{\alpha}$ in $n$-fold replicated Hilbert space as
\begin{align}
    S_{\alpha}^{(n)} &= \frac{1}{1-n} \mathbb{E}_{\mathcal{U}} \log \frac{\tr \rho_{\alpha}^{n}}{(\tr \rho)^{n}} \nonumber\\&= \frac{1}{1-n} \mathbb{E}_{\mathcal{U}} \log \frac{\Tr((C_{\alpha} \otimes I_{\bar{\alpha}}) \rho^{\otimes n})}{\Tr ( (I_{\alpha} \otimes I_{\bar{\alpha}}) \rho^{\otimes n})} \nonumber\\&= \frac{1}{1-n} \mathbb{E}_{\mathcal{U}} \log \frac{Z^{(n)}_{\alpha}}{Z^{(n)}_{0}},
\end{align}
where $C = \begin{pmatrix} 1 & 2 & ... & n  \\ 2 & 3 & ... & 1\end{pmatrix} $ and $I = \begin{pmatrix} 1 & 2 & ... & n  \\ 1 & 2 & ... & n\end{pmatrix} $ for corresponding qudits, as the cyclic and identity permutations in $S_n$ group. With the help of the replica trick~\cite{nishimori2001statistical, kardar2007statistical}, we can overcome the difficulty of the average outside the logarithmic function
\begin{eqnarray}
    \mathbb{E}_{\mathcal{U}} \log Z_{\alpha}^{(n)} &=&  \underset{k \rightarrow 0}{\lim} \frac{1}{k} \log Z_{\alpha}^{(n,k)},\\ \nonumber 
    \mathbb{E}_{\mathcal{U}} \log Z_{0}^{(n)} &=&  \underset{k \rightarrow 0 }{\lim} \frac{1}{k} \log Z_{0}^{(n,k)},
\end{eqnarray}
where 
\begin{eqnarray}
    Z_{\alpha}^{(n,k)} &=& \Tr  \left\{ \mathbb{C}_{\alpha} \otimes \mathbb{I}_{\bar{\alpha}}  \left[ \mathbb{E}_{\mathcal{U}} \rho^{\otimes nk} \right] \right\}, \\ \nonumber
    Z_{0}^{(n,k)} &=& \Tr \left\{ \mathbb{I}_{\alpha} \otimes \mathbb{I}_{\bar{\alpha}}  \left[ \mathbb{E}_{\mathcal{U}} \rho^{\otimes nk} \right] \right\},
\end{eqnarray}
with $\mathbb{C} = \begin{pmatrix} 1 & 2 & ... & n  \\ 2 & 3 & ... & 1\end{pmatrix}^{\otimes k} $ and $\mathbb{I} = \begin{pmatrix} 1 & 2 & ... & n  \\ 1 & 2 & ... & n\end{pmatrix}^{\otimes k} $ are permutations in the $r$-fold replicated Hilbert space with $r=nk$. Therefore,
\begin{eqnarray}
    S_{\alpha} = \underset{n \rightarrow 1}{\underset{k \rightarrow 0 }{\lim}} \frac{1}{k(1-n)} \log \left\{ \frac{Z_{\alpha}^{(n,k)}}{Z_{0}^{(n,k)}} \right\},
\end{eqnarray}
where $Z$ is the partition function for the classical spin model via the mapping, and it corresponds to the weight of the dominant spin configuration of the ferromagnetic spin model with particular top boundary conditions in the large $d$ limit:
$\mathbb{C}_{\alpha} \otimes \mathbb{I}_{\bar{\alpha}}$ for $Z_{\alpha}$ and $\mathbb{I}_{\alpha} \otimes \mathbb{I}_{\bar{\alpha}}$ for $Z_{0}$. 

In the language of the effective statistical model, $S_{\alpha}$ is represented as the free energy difference:
\begin{eqnarray}
    S_{\alpha}^{(n,k)} = \frac{1}{k(n-1)} \left[ F_{\alpha}^{(n,k)} - F_{0}^{(n,k)}\right].
\end{eqnarray}
We note that the free energy $F^{(n,k)}$ is proportional to the length of the domain wall with unit energy $k(n-1)$, and thus $\frac{1}{k(n-1)} F^{(n,k)}$ is independent of the index $(n,k)$ in the large $d$ limit.  Last but not least, the above discussion has assumed that the initial state is a product state, which imposes a free bottom boundary condition in the language of the effective statistical model. For different input states beyond product states, we have to figure out the bottom boundary conditions by calculating the overlap between different spin configurations and the input state $\rho_0$, such as computing the boundary weight contribution from $\text{Tr}(\mathbb{C}_\alpha\rho_0)$.

\section{Experiment protocol advantage for measuring SIC}\label{sec:exp}

\begin{figure}[htbp] 
    \centering
    \includegraphics[width=0.46\textwidth]{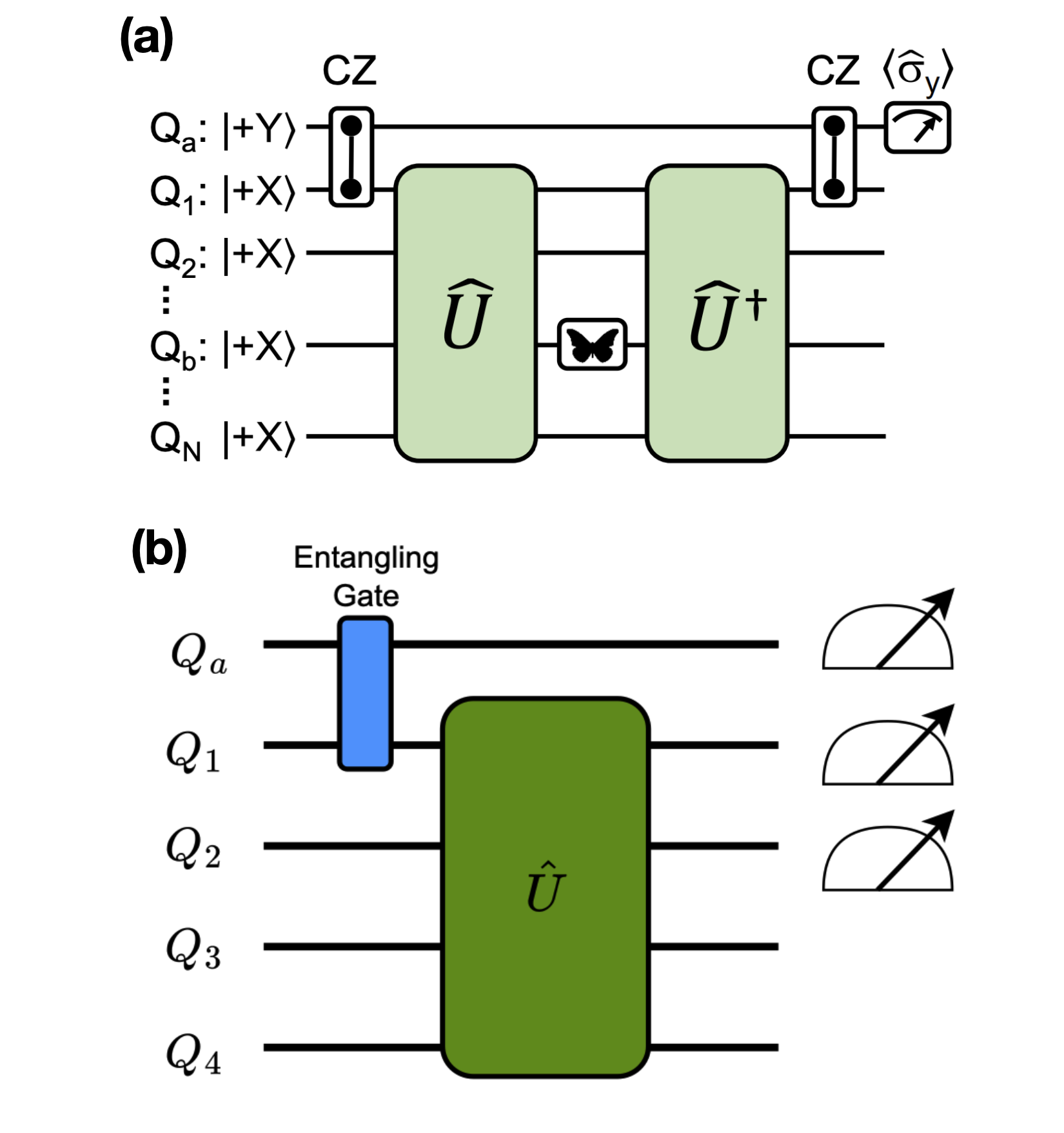} 
    \caption{Schematic comparison of experimental protocols. (a) A typical measurement protocol for OTOCs, requiring both forward ($U$) and backward ($U^\dagger$) time evolution. The figure is taken from Ref. \cite{Mi2021} (b) Protocol for measuring SIC (specifically $MI(Q_a, \{Q_1, Q_2\})$ here), requiring only a single forward time evolution ($U$) followed by subsystem entropy measurements based on random measurement scheme \cite{Brydges2019}.}
    \label{fig:exp_protocols} 
\end{figure}

Beyond its theoretical utility in characterizing quantum dynamics, SIC framework often presents significant practical advantages in terms of experimental implementation compared to other common information scrambling probes like OTOC and TMI. Here, we elaborate on these advantages, with the schematic protocols illustrated in Fig.~\ref{fig:exp_protocols}.

A primary advantage of SIC measurement lies in its reduced circuit complexity regarding time evolution.
 As depicted schematically in Fig.~\ref{fig:exp_protocols}(a), standard methods for measuring OTOCs typically require executing both the forward time evolution $U$ and the backward time evolution $U^\dagger$. Implementing $U^\dagger$ accurately can be experimentally demanding, often requiring precise knowledge and decomposition of $U$. The implementation doubles the circuit depth for gate-based evolution, or necessitates fine-tuned pulse reversal in analog systems. This increased complexity increases susceptibility to quantum noise and decoherence.

 Measuring TMI, on the other hand, often involves preparing fully mixed initial states where the entire system L is entangled with an equally sized reference system (L). Such thermofield double states require preparing and controlling $2L$ qubits in a highly entangled state.
    
    In stark contrast, the protocol for measuring SIC, particularly the one-to-one encoding scheme predominantly used in this work, is shown in Fig.~\ref{fig:exp_protocols}(b), which fundamentally requires only a single application of the forward evolution $U$ and one ancilla qubit. After the system evolves, the task reduces to measuring the necessary subsystem entropies. This elimination of the backward evolution step ($U^\dagger$) significantly simplifies the quantum circuit, reduces the overall gate count or evolution time, and consequently minimizes the accumulation of errors.

Finally, the measurement step for SIC involves estimating subsystem von Neumann entropies ($S_A$, $S_R$, $S_{AR}$). And measuring entanglement entropy requires randomized measurements, where we perform measurements in multiple random basis on copies of the evolved state. For a larger subsystem $|A|>L/2$, we can measure the entanglement entropy of $S_{R\bar{A}}$ instead, to further reduce the measurement overhead, since the whole system is closed.

In conclusion,  the requirement of only single forward evolution and the significantly reduced demand on initial entanglement resources contribute to making SIC a potentially more feasible, resource-efficient, and robust probe of quantum information dynamics on current quantum hardware platforms.

\section{Non-Gaussianity of fermionic GHZ states}\label{sec:fghz}

In this Appendix, we demonstrate that quadratic Hamiltonian quench dynamics with the one-to-all encoding scheme cannot be simulated with the large-scale fermion Gaussian state simulator. In other words, the fermionic GHZ state employed in the one-to-all encoding scheme as given by $\vert GHZ\rangle = \frac{1}{\sqrt{2}}(\vert 0101\cdots\rangle + \vert 1010\cdots\rangle)$ in Fock space is beyond Gaussian states. Multi-point correlations under Gaussian states are known to follow Wick expansion. By considering the multi-point correlator operator as $\hat{O}=c_0^\dagger c_1 c_2^\dagger c_3 \cdots$, we have the expectation value $\langle GHZ \vert\hat{O}\vert GHZ\rangle = 1/2$. However, any term resulting from the Wick expansion of the operator, such as $c_0^\dagger c_1$, gives a zero expectation under GHZ state, i.e. $\langle GHZ\vert c_0^\dagger c_1\vert GHZ \rangle=0$. Therefore, Wick expansion does not hold for the expectation of $\hat{O}$ under fermionic GHZ states, thereby establishing that these cat states are not Gaussian states.

\section{MBL phase transitions in interacting AA model}\label{sec:mblr}

\begin{figure}[!htb]
\includegraphics[width=0.4\textwidth]{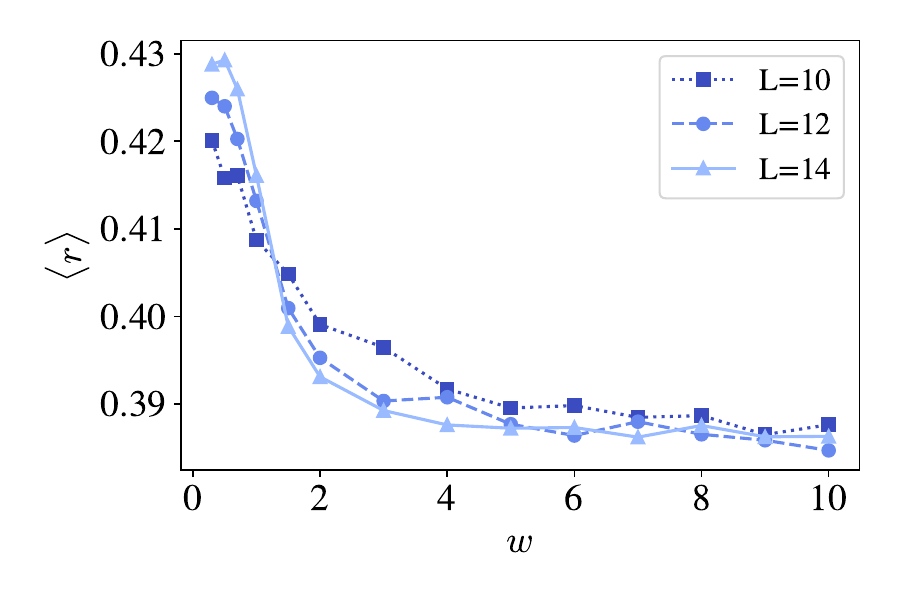}
\caption{Averaged level spacing ratio $\langle r\rangle$ for model defined in Eq. \eqref{eq:aahamiltonian} with $J=1,U=0.2$ and PBC, $w_c\approx1.4\pm  0.2$ is estimated for the corresponding many-body localization transition.
} \label{fig:mbl_r}   
\end{figure}

We determine the MBL transition for interacting AA model with $U=0.2$ as shown in Fig. \ref{fig:mbl_r}. The critical value of quasiperiodic potential strength $w_c\approx 1.4 \pm 0.2$ is estimated from the finite size crossing of the disorder averaged (phase $\theta$ average in quasiperiodic potential case) level spacing ratio $r$, defined as:
\begin{align}
    r = \frac{1}{D-2}\sum_i^{D-2}\frac{\text{min}(\delta_{i}, \delta_{i+1})}{\text{max}(\delta_{i}, \delta_{i+1})},
\end{align}
where $D$ is the Hilbert space dimension, $\delta_i = \epsilon_{i+1}-\epsilon_{i}$ is the energy level spacing between neighboring eigen-energies $\epsilon_i$ for the interacting AA Hamiltonian defined in Eq. \eqref{eq:aahamiltonian}.

\begin{figure}[!htb]
\includegraphics[width=0.45\textwidth]{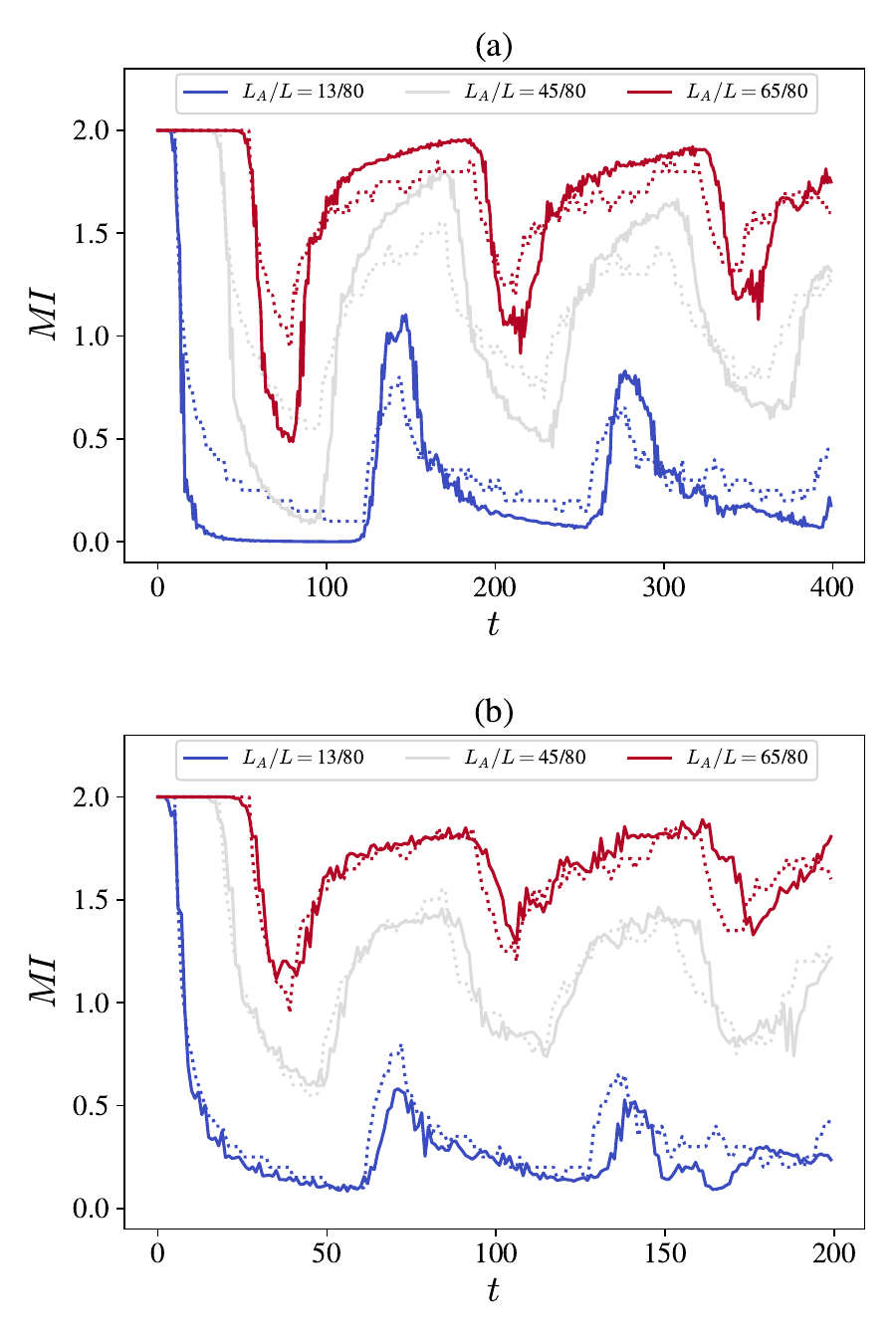}
\caption{SIC dynamics for $I(A:R)$ with $\Delta=0.4$ (trivial phase), $L=80$ for the model defined in Eq. \eqref{eq:ssh}  with open boundary conditions. (a) One-to-one encoding with the entangled qubit E at the boundary of the one-dimensional system. Initial deviating time $t^*=L_A/v_{max}$. And oscillation frequency is given by $v_{max}/(2L)$.  (b) One-to-one encoding with the entangled qubit E, positioned centrally within the one-dimensional system. Initial deviating time $t^*=L_A/(2v_{max})$. And oscillation frequency is given by $v_{max}/L$.  
Dotted lines represent predictions from quasiparticle pictures, while solid lines show results from free fermion dynamics simulation based on SSH model. Despite the quasiparticle picture firstly being formulated to describe the bulk state behavior under periodic boundary conditions, it maintains a good prediction power for systems with open boundary conditions,  especially when entangled qubits are in the bulk as shown in (b).
} \label{fig:ssh_quasi_picture}   
\end{figure}

\section{Quasiparticle picture of SSH model}\label{sec:app_topo_quasi}

In this Appendix, we provide further additional data on the information dynamics in SSH model trivial phases ($\Delta=0.4$) with open boundary conditions as defined in Eq. \eqref{eq:ssh}. Due to the lack of edge modes, SSH model in trivial phases can still be qualitatively predicted by the quasiparticle picture. The differences in the quasiparticle picture understanding from the AA model are the quasiparticle energy spectrum now is given by SSH model and the open boundary conditions lead to the reflection of quasiparticles. The information dynamics results are shown in Fig. \ref{fig:ssh_quasi_picture} with the dotted line being predictions from the quasiparticle picture. The two cases correspond to the entangled qubit $E$ on the boundary or in the bulk of the one-dimensional SSH chain. The prediction and the simulation match better for $E$ in the bulk due to less boundary effect. We find that the periodicity for the entangled qubit in the bulk is half as the entangled qubit on the boundary as a result of different distances from $E$ to the opposite boundary. Overall, the quasiparticle picture works well for both positions of the entangled qubit. However, the same quasiparticle picture totally breaks down with topological phases ($\Delta<0$), where localized edge modes are of great importance which are beyond the information given by the bulk energy spectrum.

%


\begin{thebibliography}{160}%
\makeatletter
\providecommand \@ifxundefined [1]{%
 \@ifx{#1\undefined}
}%
\providecommand \@ifnum [1]{%
 \ifnum #1\expandafter \@firstoftwo
 \else \expandafter \@secondoftwo
 \fi
}%
\providecommand \@ifx [1]{%
 \ifx #1\expandafter \@firstoftwo
 \else \expandafter \@secondoftwo
 \fi
}%
\providecommand \natexlab [1]{#1}%
\providecommand \bibnamefont  [1]{#1}%
\providecommand \bibfnamefont [1]{#1}%
\providecommand \citenamefont [1]{#1}%
\providecommand \href@noop [0]{\@secondoftwo}%
\providecommand \href [0]{\begingroup \@sanitize@url \@href}%
\providecommand \@href[1]{\@@startlink{#1}\@@href}%
\providecommand \@@href[1]{\endgroup#1\@@endlink}%
\providecommand \@sanitize@url [0]{\catcode `\\12\catcode `\$12\catcode `\&12\catcode `\#12\catcode `\^12\catcode `\_12\catcode `\%12\relax}%
\providecommand \@@startlink[1]{}%
\providecommand \@@endlink[0]{}%
\providecommand \url  [0]{\begingroup\@sanitize@url \@url }%
\providecommand \@url [1]{\endgroup\@href {#1}{\urlprefix }}%
\providecommand \urlprefix  [0]{URL }%
\providecommand \Eprint [0]{\href }%
\providecommand \doibase [0]{http://dx.doi.org/}%
\providecommand \selectlanguage [0]{\@gobble}%
\providecommand \bibinfo  [0]{\@secondoftwo}%
\providecommand \bibfield  [0]{\@secondoftwo}%
\providecommand \translation [1]{[#1]}%
\providecommand \BibitemOpen [0]{}%
\providecommand \bibitemStop [0]{}%
\providecommand \bibitemNoStop [0]{.\EOS\space}%
\providecommand \EOS [0]{\spacefactor3000\relax}%
\providecommand \BibitemShut  [1]{\csname bibitem#1\endcsname}%
\let\auto@bib@innerbib\@empty
\bibitem [{\citenamefont {Deutsch}(1991)}]{Deutsch1991}%
  \BibitemOpen
  \bibfield  {author} {\bibinfo {author} {\bibfnamefont {J.~M.}\ \bibnamefont {Deutsch}},\ }\bibfield  {title} {{Quantum statistical mechanics in a closed system},\ }\href {\doibase 10.1103/PhysRevA.43.2046} {\bibfield  {journal} {\bibinfo  {journal} {Physical Review A}\ }\textbf {\bibinfo {volume} {43}},\ \bibinfo {pages} {2046} (\bibinfo {year} {1991})}\BibitemShut {NoStop}%
\bibitem [{\citenamefont {Srednicki}(1994)}]{Srednicki1994}%
  \BibitemOpen
  \bibfield  {author} {\bibinfo {author} {\bibfnamefont {M.}~\bibnamefont {Srednicki}},\ }\bibfield  {title} {{Chaos and quantum thermalization},\ }\href {\doibase 10.1103/PhysRevE.50.888} {\bibfield  {journal} {\bibinfo  {journal} {Physical Review E}\ }\textbf {\bibinfo {volume} {50}},\ \bibinfo {pages} {888} (\bibinfo {year} {1994})}\BibitemShut {NoStop}%
\bibitem [{\citenamefont {Rigol}\ \emph {et~al.}(2008)\citenamefont {Rigol}, \citenamefont {Dunjko},\ and\ \citenamefont {Olshanii}}]{Rigol2008}%
  \BibitemOpen
  \bibfield  {author} {\bibinfo {author} {\bibfnamefont {M.}~\bibnamefont {Rigol}}, \bibinfo {author} {\bibfnamefont {V.}~\bibnamefont {Dunjko}}, \ and\ \bibinfo {author} {\bibfnamefont {M.}~\bibnamefont {Olshanii}},\ }\bibfield  {title} {{Thermalization and its mechanism for generic isolated quantum systems},\ }\href {\doibase 10.1038/nature06838} {\bibfield  {journal} {\bibinfo  {journal} {Nature}\ }\textbf {\bibinfo {volume} {452}},\ \bibinfo {pages} {854} (\bibinfo {year} {2008})}\BibitemShut {NoStop}%
\bibitem [{\citenamefont {Maldacena}\ \emph {et~al.}(2016)\citenamefont {Maldacena}, \citenamefont {Shenker},\ and\ \citenamefont {Stanford}}]{Maldacena2016}%
  \BibitemOpen
  \bibfield  {author} {\bibinfo {author} {\bibfnamefont {J.}~\bibnamefont {Maldacena}}, \bibinfo {author} {\bibfnamefont {S.~H.}\ \bibnamefont {Shenker}}, \ and\ \bibinfo {author} {\bibfnamefont {D.}~\bibnamefont {Stanford}},\ }\bibfield  {title} {{A bound on chaos},\ }\href {\doibase 10.1007/JHEP08(2016)106} {\bibfield  {journal} {\bibinfo  {journal} {Journal of High Energy Physics}\ }\textbf {\bibinfo {volume} {2016}},\ \bibinfo {pages} {106} (\bibinfo {year} {2016})}\BibitemShut {NoStop}%
\bibitem [{\citenamefont {Xu}\ \emph {et~al.}(2020)\citenamefont {Xu}, \citenamefont {Scaffidi},\ and\ \citenamefont {Cao}}]{Xu2020}%
  \BibitemOpen
  \bibfield  {author} {\bibinfo {author} {\bibfnamefont {T.}~\bibnamefont {Xu}}, \bibinfo {author} {\bibfnamefont {T.}~\bibnamefont {Scaffidi}}, \ and\ \bibinfo {author} {\bibfnamefont {X.}~\bibnamefont {Cao}},\ }\bibfield  {title} {{Does Scrambling Equal Chaos?},\ }\href {\doibase 10.1103/PhysRevLett.124.140602} {\bibfield  {journal} {\bibinfo  {journal} {Phys. Rev. Lett.}\ }\textbf {\bibinfo {volume} {124}},\ \bibinfo {pages} {140602} (\bibinfo {year} {2020})}\BibitemShut {NoStop}%
\bibitem [{\citenamefont {Dowling}\ \emph {et~al.}(2023)\citenamefont {Dowling}, \citenamefont {Kos},\ and\ \citenamefont {Modi}}]{Dowling2023}%
  \BibitemOpen
  \bibfield  {author} {\bibinfo {author} {\bibfnamefont {N.}~\bibnamefont {Dowling}}, \bibinfo {author} {\bibfnamefont {P.}~\bibnamefont {Kos}}, \ and\ \bibinfo {author} {\bibfnamefont {K.}~\bibnamefont {Modi}},\ }\bibfield  {title} {{Scrambling Is Necessary but Not Sufficient for Chaos},\ }\href {\doibase 10.1103/PhysRevLett.131.180403} {\bibfield  {journal} {\bibinfo  {journal} {Physical Review Letters}\ }\textbf {\bibinfo {volume} {131}},\ \bibinfo {pages} {180403} (\bibinfo {year} {2023})}\BibitemShut {NoStop}%
\bibitem [{\citenamefont {Hayden}\ and\ \citenamefont {Preskill}(2007)}]{Hayden2007}%
  \BibitemOpen
  \bibfield  {author} {\bibinfo {author} {\bibfnamefont {P.}~\bibnamefont {Hayden}}\ and\ \bibinfo {author} {\bibfnamefont {J.}~\bibnamefont {Preskill}},\ }\bibfield  {title} {{Black holes as mirrors: quantum information in random subsystems},\ }\href {\doibase 10.1088/1126-6708/2007/09/120} {\bibfield  {journal} {\bibinfo  {journal} {Journal of High Energy Physics}\ }\textbf {\bibinfo {volume} {2007}},\ \bibinfo {pages} {120} (\bibinfo {year} {2007})}\BibitemShut {NoStop}%
\bibitem [{\citenamefont {Brown}\ \emph {et~al.}(2016)\citenamefont {Brown}, \citenamefont {Roberts}, \citenamefont {Susskind}, \citenamefont {Swingle},\ and\ \citenamefont {Zhao}}]{Brown2016}%
  \BibitemOpen
  \bibfield  {author} {\bibinfo {author} {\bibfnamefont {A.~R.}\ \bibnamefont {Brown}}, \bibinfo {author} {\bibfnamefont {D.~A.}\ \bibnamefont {Roberts}}, \bibinfo {author} {\bibfnamefont {L.}~\bibnamefont {Susskind}}, \bibinfo {author} {\bibfnamefont {B.}~\bibnamefont {Swingle}}, \ and\ \bibinfo {author} {\bibfnamefont {Y.}~\bibnamefont {Zhao}},\ }\bibfield  {title} {{Holographic Complexity Equals Bulk Action?},\ }\href {\doibase 10.1103/PhysRevLett.116.191301} {\bibfield  {journal} {\bibinfo  {journal} {Physical Review Letters}\ }\textbf {\bibinfo {volume} {116}},\ \bibinfo {pages} {191301} (\bibinfo {year} {2016})}\BibitemShut {NoStop}%
\bibitem [{\citenamefont {Lloyd}(2000)}]{Lloyd2000}%
  \BibitemOpen
  \bibfield  {author} {\bibinfo {author} {\bibfnamefont {S.}~\bibnamefont {Lloyd}},\ }\bibfield  {title} {{Ultimate physical limits to computation},\ }\href {\doibase 10.1038/35023282} {\bibfield  {journal} {\bibinfo  {journal} {Nature}\ }\textbf {\bibinfo {volume} {406}},\ \bibinfo {pages} {1047} (\bibinfo {year} {2000})}\BibitemShut {NoStop}%
\bibitem [{\citenamefont {Choi}\ \emph {et~al.}(2020)\citenamefont {Choi}, \citenamefont {Bao}, \citenamefont {Qi},\ and\ \citenamefont {Altman}}]{PhysRevLett.125.030505}%
  \BibitemOpen
  \bibfield  {author} {\bibinfo {author} {\bibfnamefont {S.}~\bibnamefont {Choi}}, \bibinfo {author} {\bibfnamefont {Y.}~\bibnamefont {Bao}}, \bibinfo {author} {\bibfnamefont {X.-L.}\ \bibnamefont {Qi}}, \ and\ \bibinfo {author} {\bibfnamefont {E.}~\bibnamefont {Altman}},\ }\bibfield  {title} {Quantum error correction in scrambling dynamics and measurement-induced phase transition,\ }\href {\doibase 10.1103/PhysRevLett.125.030505} {\bibfield  {journal} {\bibinfo  {journal} {Phys. Rev. Lett.}\ }\textbf {\bibinfo {volume} {125}},\ \bibinfo {pages} {030505} (\bibinfo {year} {2020})}\BibitemShut {NoStop}%
\bibitem [{\citenamefont {Fan}\ \emph {et~al.}(2021)\citenamefont {Fan}, \citenamefont {Vijay}, \citenamefont {Vishwanath},\ and\ \citenamefont {You}}]{PhysRevB.103.174309}%
  \BibitemOpen
  \bibfield  {author} {\bibinfo {author} {\bibfnamefont {R.}~\bibnamefont {Fan}}, \bibinfo {author} {\bibfnamefont {S.}~\bibnamefont {Vijay}}, \bibinfo {author} {\bibfnamefont {A.}~\bibnamefont {Vishwanath}}, \ and\ \bibinfo {author} {\bibfnamefont {Y.-Z.}\ \bibnamefont {You}},\ }\bibfield  {title} {Self-organized error correction in random unitary circuits with measurement,\ }\href {\doibase 10.1103/PhysRevB.103.174309} {\bibfield  {journal} {\bibinfo  {journal} {Phys. Rev. B}\ }\textbf {\bibinfo {volume} {103}},\ \bibinfo {pages} {174309} (\bibinfo {year} {2021})}\BibitemShut {NoStop}%
\bibitem [{\citenamefont {Li}\ and\ \citenamefont {Fisher}(2021)}]{PhysRevB.103.104306}%
  \BibitemOpen
  \bibfield  {author} {\bibinfo {author} {\bibfnamefont {Y.}~\bibnamefont {Li}}\ and\ \bibinfo {author} {\bibfnamefont {M.~P.~A.}\ \bibnamefont {Fisher}},\ }\bibfield  {title} {Statistical mechanics of quantum error correcting codes,\ }\href {\doibase 10.1103/PhysRevB.103.104306} {\bibfield  {journal} {\bibinfo  {journal} {Phys. Rev. B}\ }\textbf {\bibinfo {volume} {103}},\ \bibinfo {pages} {104306} (\bibinfo {year} {2021})}\BibitemShut {NoStop}%
\bibitem [{\citenamefont {Amico}\ \emph {et~al.}(2008)\citenamefont {Amico}, \citenamefont {Fazio}, \citenamefont {Osterloh},\ and\ \citenamefont {Vedral}}]{Amico2008}%
  \BibitemOpen
  \bibfield  {author} {\bibinfo {author} {\bibfnamefont {L.}~\bibnamefont {Amico}}, \bibinfo {author} {\bibfnamefont {R.}~\bibnamefont {Fazio}}, \bibinfo {author} {\bibfnamefont {A.}~\bibnamefont {Osterloh}}, \ and\ \bibinfo {author} {\bibfnamefont {V.}~\bibnamefont {Vedral}},\ }\bibfield  {title} {{Entanglement in many-body systems},\ }\href {\doibase 10.1103/RevModPhys.80.517} {\bibfield  {journal} {\bibinfo  {journal} {Reviews of Modern Physics}\ }\textbf {\bibinfo {volume} {80}},\ \bibinfo {pages} {517} (\bibinfo {year} {2008})}\BibitemShut {NoStop}%
\bibitem [{\citenamefont {Calabrese}\ and\ \citenamefont {Cardy}(2005)}]{Calabrese2005}%
  \BibitemOpen
  \bibfield  {author} {\bibinfo {author} {\bibfnamefont {P.}~\bibnamefont {Calabrese}}\ and\ \bibinfo {author} {\bibfnamefont {J.}~\bibnamefont {Cardy}},\ }\bibfield  {title} {{Evolution of entanglement entropy in one-dimensional systems},\ }\href {\doibase 10.1088/1742-5468/2005/04/P04010} {\bibfield  {journal} {\bibinfo  {journal} {Journal of Statistical Mechanics: Theory and Experiment}\ }\textbf {\bibinfo {volume} {2005}},\ \bibinfo {pages} {P04010} (\bibinfo {year} {2005})}\BibitemShut {NoStop}%
\bibitem [{\citenamefont {Kim}\ and\ \citenamefont {Huse}(2013)}]{Kim2013}%
  \BibitemOpen
  \bibfield  {author} {\bibinfo {author} {\bibfnamefont {H.}~\bibnamefont {Kim}}\ and\ \bibinfo {author} {\bibfnamefont {D.~A.}\ \bibnamefont {Huse}},\ }\bibfield  {title} {{Ballistic Spreading of Entanglement in a Diffusive Nonintegrable System},\ }\href {\doibase 10.1103/PhysRevLett.111.127205} {\bibfield  {journal} {\bibinfo  {journal} {Physical Review Letters}\ }\textbf {\bibinfo {volume} {111}},\ \bibinfo {pages} {127205} (\bibinfo {year} {2013})}\BibitemShut {NoStop}%
\bibitem [{\citenamefont {Liu}\ and\ \citenamefont {Suh}(2014)}]{Liu2014a}%
  \BibitemOpen
  \bibfield  {author} {\bibinfo {author} {\bibfnamefont {H.}~\bibnamefont {Liu}}\ and\ \bibinfo {author} {\bibfnamefont {S.~J.}\ \bibnamefont {Suh}},\ }\bibfield  {title} {{Entanglement Tsunami: Universal Scaling in Holographic Thermalization},\ }\href {\doibase 10.1103/PhysRevLett.112.011601} {\bibfield  {journal} {\bibinfo  {journal} {Physical Review Letters}\ }\textbf {\bibinfo {volume} {112}},\ \bibinfo {pages} {011601} (\bibinfo {year} {2014})}\BibitemShut {NoStop}%
\bibitem [{\citenamefont {Alba}\ and\ \citenamefont {Calabrese}(2017)}]{Alba2017a}%
  \BibitemOpen
  \bibfield  {author} {\bibinfo {author} {\bibfnamefont {V.}~\bibnamefont {Alba}}\ and\ \bibinfo {author} {\bibfnamefont {P.}~\bibnamefont {Calabrese}},\ }\bibfield  {title} {{Entanglement and thermodynamics after a quantum quench in integrable systems},\ }\href {\doibase 10.1073/pnas.1703516114} {\bibfield  {journal} {\bibinfo  {journal} {Proceedings of the National Academy of Sciences}\ }\textbf {\bibinfo {volume} {114}},\ \bibinfo {pages} {7947} (\bibinfo {year} {2017})}\BibitemShut {NoStop}%
\bibitem [{\citenamefont {Ho}\ and\ \citenamefont {Abanin}(2017)}]{Ho2017}%
  \BibitemOpen
  \bibfield  {author} {\bibinfo {author} {\bibfnamefont {W.~W.}\ \bibnamefont {Ho}}\ and\ \bibinfo {author} {\bibfnamefont {D.~A.}\ \bibnamefont {Abanin}},\ }\bibfield  {title} {{Entanglement dynamics in quantum many-body systems},\ }\href {\doibase 10.1103/PhysRevB.95.094302} {\bibfield  {journal} {\bibinfo  {journal} {Phys. Rev. B}\ }\textbf {\bibinfo {volume} {95}},\ \bibinfo {pages} {094302} (\bibinfo {year} {2017})}\BibitemShut {NoStop}%
\bibitem [{\citenamefont {Mezei}\ and\ \citenamefont {Stanford}(2017)}]{Mezei2017}%
  \BibitemOpen
  \bibfield  {author} {\bibinfo {author} {\bibfnamefont {M.}~\bibnamefont {Mezei}}\ and\ \bibinfo {author} {\bibfnamefont {D.}~\bibnamefont {Stanford}},\ }\bibfield  {title} {{On entanglement spreading in chaotic systems},\ }\href {\doibase 10.1007/JHEP05(2017)065} {\bibfield  {journal} {\bibinfo  {journal} {J. High Energy Phys.}\ }\textbf {\bibinfo {volume} {2017}},\ \bibinfo {pages} {65} (\bibinfo {year} {2017})}\BibitemShut {NoStop}%
\bibitem [{\citenamefont {Nahum}\ \emph {et~al.}(2017)\citenamefont {Nahum}, \citenamefont {Ruhman}, \citenamefont {Vijay},\ and\ \citenamefont {Haah}}]{PhysRevX.7.031016}%
  \BibitemOpen
  \bibfield  {author} {\bibinfo {author} {\bibfnamefont {A.}~\bibnamefont {Nahum}}, \bibinfo {author} {\bibfnamefont {J.}~\bibnamefont {Ruhman}}, \bibinfo {author} {\bibfnamefont {S.}~\bibnamefont {Vijay}}, \ and\ \bibinfo {author} {\bibfnamefont {J.}~\bibnamefont {Haah}},\ }\bibfield  {title} {Quantum entanglement growth under random unitary dynamics,\ }\href {\doibase 10.1103/PhysRevX.7.031016} {\bibfield  {journal} {\bibinfo  {journal} {Phys. Rev. X}\ }\textbf {\bibinfo {volume} {7}},\ \bibinfo {pages} {031016} (\bibinfo {year} {2017})}\BibitemShut {NoStop}%
\bibitem [{\citenamefont {Bertini}\ \emph {et~al.}(2022)\citenamefont {Bertini}, \citenamefont {Klobas}, \citenamefont {Alba}, \citenamefont {Lagnese},\ and\ \citenamefont {Calabrese}}]{Bertini2022}%
  \BibitemOpen
  \bibfield  {author} {\bibinfo {author} {\bibfnamefont {B.}~\bibnamefont {Bertini}}, \bibinfo {author} {\bibfnamefont {K.}~\bibnamefont {Klobas}}, \bibinfo {author} {\bibfnamefont {V.}~\bibnamefont {Alba}}, \bibinfo {author} {\bibfnamefont {G.}~\bibnamefont {Lagnese}}, \ and\ \bibinfo {author} {\bibfnamefont {P.}~\bibnamefont {Calabrese}},\ }\bibfield  {title} {{Growth of R{\'{e}}nyi Entropies in Interacting Integrable Models and the Breakdown of the Quasiparticle Picture},\ }\href {\doibase 10.1103/PhysRevX.12.031016} {\bibfield  {journal} {\bibinfo  {journal} {Physical Review X}\ }\textbf {\bibinfo {volume} {12}},\ \bibinfo {pages} {031016} (\bibinfo {year} {2022})}\BibitemShut {NoStop}%
\bibitem [{\citenamefont {Lewis-Swan}\ \emph {et~al.}(2019)\citenamefont {Lewis-Swan}, \citenamefont {Safavi-Naini}, \citenamefont {Kaufman},\ and\ \citenamefont {Rey}}]{Lewis-Swan2019}%
  \BibitemOpen
  \bibfield  {author} {\bibinfo {author} {\bibfnamefont {R.~J.}\ \bibnamefont {Lewis-Swan}}, \bibinfo {author} {\bibfnamefont {A.}~\bibnamefont {Safavi-Naini}}, \bibinfo {author} {\bibfnamefont {A.~M.}\ \bibnamefont {Kaufman}}, \ and\ \bibinfo {author} {\bibfnamefont {A.~M.}\ \bibnamefont {Rey}},\ }\bibfield  {title} {{Dynamics of quantum information},\ }\href {\doibase 10.1038/s42254-019-0090-y} {\bibfield  {journal} {\bibinfo  {journal} {Nat. Rev. Phys.}\ }\textbf {\bibinfo {volume} {1}},\ \bibinfo {pages} {627} (\bibinfo {year} {2019})}\BibitemShut {NoStop}%
\bibitem [{\citenamefont {Xu}\ and\ \citenamefont {Swingle}(2024)}]{Xu2024}%
  \BibitemOpen
  \bibfield  {author} {\bibinfo {author} {\bibfnamefont {S.}~\bibnamefont {Xu}}\ and\ \bibinfo {author} {\bibfnamefont {B.}~\bibnamefont {Swingle}},\ }\bibfield  {title} {{Scrambling Dynamics and Out-of-Time-Ordered Correlators in Quantum Many-Body Systems},\ }\href {\doibase 10.1103/PRXQuantum.5.010201} {\bibfield  {journal} {\bibinfo  {journal} {PRX Quantum}\ }\textbf {\bibinfo {volume} {5}},\ \bibinfo {pages} {010201} (\bibinfo {year} {2024})}\BibitemShut {NoStop}%
\bibitem [{\citenamefont {Sekino}\ and\ \citenamefont {Susskind}(2008)}]{Sekino2008}%
  \BibitemOpen
  \bibfield  {author} {\bibinfo {author} {\bibfnamefont {Y.}~\bibnamefont {Sekino}}\ and\ \bibinfo {author} {\bibfnamefont {L.}~\bibnamefont {Susskind}},\ }\bibfield  {title} {{Fast scramblers},\ }\href {\doibase 10.1088/1126-6708/2008/10/065} {\bibfield  {journal} {\bibinfo  {journal} {Journal of High Energy Physics}\ }\textbf {\bibinfo {volume} {2008}},\ \bibinfo {pages} {065} (\bibinfo {year} {2008})}\BibitemShut {NoStop}%
\bibitem [{\citenamefont {Lashkari}\ \emph {et~al.}(2013)\citenamefont {Lashkari}, \citenamefont {Stanford}, \citenamefont {Hastings}, \citenamefont {Osborne},\ and\ \citenamefont {Hayden}}]{Lashkari2013a}%
  \BibitemOpen
  \bibfield  {author} {\bibinfo {author} {\bibfnamefont {N.}~\bibnamefont {Lashkari}}, \bibinfo {author} {\bibfnamefont {D.}~\bibnamefont {Stanford}}, \bibinfo {author} {\bibfnamefont {M.}~\bibnamefont {Hastings}}, \bibinfo {author} {\bibfnamefont {T.}~\bibnamefont {Osborne}}, \ and\ \bibinfo {author} {\bibfnamefont {P.}~\bibnamefont {Hayden}},\ }\bibfield  {title} {{Towards the fast scrambling conjecture},\ }\href {\doibase 10.1007/JHEP04(2013)022} {\bibfield  {journal} {\bibinfo  {journal} {Journal of High Energy Physics}\ }\textbf {\bibinfo {volume} {2013}},\ \bibinfo {pages} {22} (\bibinfo {year} {2013})}\BibitemShut {NoStop}%
\bibitem [{\citenamefont {Hosur}\ \emph {et~al.}(2016)\citenamefont {Hosur}, \citenamefont {Qi}, \citenamefont {Roberts},\ and\ \citenamefont {Yoshida}}]{Hosur2016}%
  \BibitemOpen
  \bibfield  {author} {\bibinfo {author} {\bibfnamefont {P.}~\bibnamefont {Hosur}}, \bibinfo {author} {\bibfnamefont {X.-L.}\ \bibnamefont {Qi}}, \bibinfo {author} {\bibfnamefont {D.~A.}\ \bibnamefont {Roberts}}, \ and\ \bibinfo {author} {\bibfnamefont {B.}~\bibnamefont {Yoshida}},\ }\bibfield  {title} {{Chaos in quantum channels},\ }\href {\doibase 10.1007/JHEP02(2016)004} {\bibfield  {journal} {\bibinfo  {journal} {Journal of High Energy Physics}\ }\textbf {\bibinfo {volume} {2016}},\ \bibinfo {pages} {4} (\bibinfo {year} {2016})}\BibitemShut {NoStop}%
\bibitem [{\citenamefont {Schumacher}\ and\ \citenamefont {Nielsen}(1996)}]{Schumacher1996}%
  \BibitemOpen
  \bibfield  {author} {\bibinfo {author} {\bibfnamefont {B.}~\bibnamefont {Schumacher}}\ and\ \bibinfo {author} {\bibfnamefont {M.~A.}\ \bibnamefont {Nielsen}},\ }\bibfield  {title} {{Quantum data processing and error correction},\ }\href {\doibase 10.1103/PhysRevA.54.2629} {\bibfield  {journal} {\bibinfo  {journal} {Phys. Rev. A}\ }\textbf {\bibinfo {volume} {54}},\ \bibinfo {pages} {2629} (\bibinfo {year} {1996})}\BibitemShut {NoStop}%
\bibitem [{\citenamefont {Lloyd}(1997)}]{Lloyd1997}%
  \BibitemOpen
  \bibfield  {author} {\bibinfo {author} {\bibfnamefont {S.}~\bibnamefont {Lloyd}},\ }\bibfield  {title} {{Capacity of the noisy quantum channel},\ }\href {\doibase 10.1103/PhysRevA.55.1613} {\bibfield  {journal} {\bibinfo  {journal} {Phys. Rev. A}\ }\textbf {\bibinfo {volume} {55}},\ \bibinfo {pages} {1613} (\bibinfo {year} {1997})}\BibitemShut {NoStop}%
\bibitem [{\citenamefont {DiVincenzo}\ \emph {et~al.}(1998)\citenamefont {DiVincenzo}, \citenamefont {Shor},\ and\ \citenamefont {Smolin}}]{DiVincenzo1998}%
  \BibitemOpen
  \bibfield  {author} {\bibinfo {author} {\bibfnamefont {D.~P.}\ \bibnamefont {DiVincenzo}}, \bibinfo {author} {\bibfnamefont {P.~W.}\ \bibnamefont {Shor}}, \ and\ \bibinfo {author} {\bibfnamefont {J.~A.}\ \bibnamefont {Smolin}},\ }\bibfield  {title} {{Quantum-channel capacity of very noisy channels},\ }\href {\doibase 10.1103/PhysRevA.57.830} {\bibfield  {journal} {\bibinfo  {journal} {Physical Review A}\ }\textbf {\bibinfo {volume} {57}},\ \bibinfo {pages} {830} (\bibinfo {year} {1998})}\BibitemShut {NoStop}%
\bibitem [{\citenamefont {Barnum}\ \emph {et~al.}(1998)\citenamefont {Barnum}, \citenamefont {Nielsen},\ and\ \citenamefont {Schumacher}}]{Barnum1998}%
  \BibitemOpen
  \bibfield  {author} {\bibinfo {author} {\bibfnamefont {H.}~\bibnamefont {Barnum}}, \bibinfo {author} {\bibfnamefont {M.~A.}\ \bibnamefont {Nielsen}}, \ and\ \bibinfo {author} {\bibfnamefont {B.}~\bibnamefont {Schumacher}},\ }\bibfield  {title} {{Information transmission through a noisy quantum channel},\ }\href {\doibase 10.1103/PhysRevA.57.4153} {\bibfield  {journal} {\bibinfo  {journal} {Physical Review A}\ }\textbf {\bibinfo {volume} {57}},\ \bibinfo {pages} {4153} (\bibinfo {year} {1998})}\BibitemShut {NoStop}%
\bibitem [{\citenamefont {Shor}(2003)}]{Shor2003}%
  \BibitemOpen
  \bibfield  {author} {\bibinfo {author} {\bibfnamefont {P.~W.}\ \bibnamefont {Shor}},\ }\bibfield  {title} {{Capacities of quantum channels and how to find them},\ }\href {\doibase 10.1007/s10107-003-0446-y} {\bibfield  {journal} {\bibinfo  {journal} {Math. Program.}\ }\textbf {\bibinfo {volume} {97}},\ \bibinfo {pages} {311} (\bibinfo {year} {2003})}\BibitemShut {NoStop}%
\bibitem [{\citenamefont {Devetak}\ and\ \citenamefont {Winter}(2003)}]{Devetak2003}%
  \BibitemOpen
  \bibfield  {author} {\bibinfo {author} {\bibfnamefont {I.}~\bibnamefont {Devetak}}\ and\ \bibinfo {author} {\bibfnamefont {A.}~\bibnamefont {Winter}},\ }\bibfield  {title} {{Classical data compression with quantum side information},\ }\href {\doibase 10.1103/PhysRevA.68.042301} {\bibfield  {journal} {\bibinfo  {journal} {Phys. Rev. A}\ }\textbf {\bibinfo {volume} {68}},\ \bibinfo {pages} {042301} (\bibinfo {year} {2003})}\BibitemShut {NoStop}%
\bibitem [{\citenamefont {Gyongyosi}\ \emph {et~al.}(2018)\citenamefont {Gyongyosi}, \citenamefont {Imre},\ and\ \citenamefont {Nguyen}}]{Gyongyosi2018}%
  \BibitemOpen
  \bibfield  {author} {\bibinfo {author} {\bibfnamefont {L.}~\bibnamefont {Gyongyosi}}, \bibinfo {author} {\bibfnamefont {S.}~\bibnamefont {Imre}}, \ and\ \bibinfo {author} {\bibfnamefont {H.~V.}\ \bibnamefont {Nguyen}},\ }\bibfield  {title} {{A Survey on Quantum Channel Capacities},\ }\href {\doibase 10.1109/COMST.2017.2786748} {\bibfield  {journal} {\bibinfo  {journal} {IEEE Commun. Surv. Tutorials}\ }\textbf {\bibinfo {volume} {20}},\ \bibinfo {pages} {1149} (\bibinfo {year} {2018})}\BibitemShut {NoStop}%
\bibitem [{\citenamefont {Colmenarez}\ \emph {et~al.}(2024)\citenamefont {Colmenarez}, \citenamefont {Huang}, \citenamefont {Diehl},\ and\ \citenamefont {Müller}}]{Colmenarez2023}%
  \BibitemOpen
  \bibfield  {author} {\bibinfo {author} {\bibfnamefont {L.}~\bibnamefont {Colmenarez}}, \bibinfo {author} {\bibfnamefont {Z.-M.}\ \bibnamefont {Huang}}, \bibinfo {author} {\bibfnamefont {S.}~\bibnamefont {Diehl}}, \ and\ \bibinfo {author} {\bibfnamefont {M.}~\bibnamefont {Müller}},\ }\bibfield  {title} {Accurate optimal quantum error correction thresholds from coherent information,\ }\href {\doibase 10.1103/PhysRevResearch.6.L042014} {\bibfield  {journal} {\bibinfo  {journal} {Physical Review Research}\ }\textbf {\bibinfo {volume} {6}},\ \bibinfo {pages} {L042014} (\bibinfo {year} {2024})}\BibitemShut {NoStop}%
\bibitem [{\citenamefont {Fisher}\ \emph {et~al.}(2023)\citenamefont {Fisher}, \citenamefont {Khemani}, \citenamefont {Nahum},\ and\ \citenamefont {Vijay}}]{Fisher2023}%
  \BibitemOpen
  \bibfield  {author} {\bibinfo {author} {\bibfnamefont {M.~P.}\ \bibnamefont {Fisher}}, \bibinfo {author} {\bibfnamefont {V.}~\bibnamefont {Khemani}}, \bibinfo {author} {\bibfnamefont {A.}~\bibnamefont {Nahum}}, \ and\ \bibinfo {author} {\bibfnamefont {S.}~\bibnamefont {Vijay}},\ }\bibfield  {title} {{Random Quantum Circuits},\ }\href {\doibase 10.1146/annurev-conmatphys-031720-030658} {\bibfield  {journal} {\bibinfo  {journal} {Annu. Rev. Condens. Matter Phys.}\ }\textbf {\bibinfo {volume} {14}},\ \bibinfo {pages} {335} (\bibinfo {year} {2023})}\BibitemShut {NoStop}%
\bibitem [{\citenamefont {Li}\ \emph {et~al.}(2018)\citenamefont {Li}, \citenamefont {Chen},\ and\ \citenamefont {Fisher}}]{PhysRevB.98.205136}%
  \BibitemOpen
  \bibfield  {author} {\bibinfo {author} {\bibfnamefont {Y.}~\bibnamefont {Li}}, \bibinfo {author} {\bibfnamefont {X.}~\bibnamefont {Chen}}, \ and\ \bibinfo {author} {\bibfnamefont {M.~P.~A.}\ \bibnamefont {Fisher}},\ }\bibfield  {title} {Quantum zeno effect and the many-body entanglement transition,\ }\href {\doibase 10.1103/PhysRevB.98.205136} {\bibfield  {journal} {\bibinfo  {journal} {Phys. Rev. B}\ }\textbf {\bibinfo {volume} {98}},\ \bibinfo {pages} {205136} (\bibinfo {year} {2018})}\BibitemShut {NoStop}%
\bibitem [{\citenamefont {Li}\ \emph {et~al.}(2019)\citenamefont {Li}, \citenamefont {Chen},\ and\ \citenamefont {Fisher}}]{PhysRevB.100.134306}%
  \BibitemOpen
  \bibfield  {author} {\bibinfo {author} {\bibfnamefont {Y.}~\bibnamefont {Li}}, \bibinfo {author} {\bibfnamefont {X.}~\bibnamefont {Chen}}, \ and\ \bibinfo {author} {\bibfnamefont {M.~P.~A.}\ \bibnamefont {Fisher}},\ }\bibfield  {title} {Measurement-driven entanglement transition in hybrid quantum circuits,\ }\href {\doibase 10.1103/PhysRevB.100.134306} {\bibfield  {journal} {\bibinfo  {journal} {Phys. Rev. B}\ }\textbf {\bibinfo {volume} {100}},\ \bibinfo {pages} {134306} (\bibinfo {year} {2019})}\BibitemShut {NoStop}%
\bibitem [{\citenamefont {Skinner}\ \emph {et~al.}(2019{\natexlab{a}})\citenamefont {Skinner}, \citenamefont {Ruhman},\ and\ \citenamefont {Nahum}}]{PhysRevX.9.031009}%
  \BibitemOpen
  \bibfield  {author} {\bibinfo {author} {\bibfnamefont {B.}~\bibnamefont {Skinner}}, \bibinfo {author} {\bibfnamefont {J.}~\bibnamefont {Ruhman}}, \ and\ \bibinfo {author} {\bibfnamefont {A.}~\bibnamefont {Nahum}},\ }\bibfield  {title} {Measurement-induced phase transitions in the dynamics of entanglement,\ }\href {\doibase 10.1103/PhysRevX.9.031009} {\bibfield  {journal} {\bibinfo  {journal} {Phys. Rev. X}\ }\textbf {\bibinfo {volume} {9}},\ \bibinfo {pages} {031009} (\bibinfo {year} {2019}{\natexlab{a}})}\BibitemShut {NoStop}%
\bibitem [{\citenamefont {Chan}\ \emph {et~al.}(2019)\citenamefont {Chan}, \citenamefont {Nandkishore}, \citenamefont {Pretko},\ and\ \citenamefont {Smith}}]{PhysRevB.99.224307}%
  \BibitemOpen
  \bibfield  {author} {\bibinfo {author} {\bibfnamefont {A.}~\bibnamefont {Chan}}, \bibinfo {author} {\bibfnamefont {R.~M.}\ \bibnamefont {Nandkishore}}, \bibinfo {author} {\bibfnamefont {M.}~\bibnamefont {Pretko}}, \ and\ \bibinfo {author} {\bibfnamefont {G.}~\bibnamefont {Smith}},\ }\bibfield  {title} {Unitary-projective entanglement dynamics,\ }\href {\doibase 10.1103/PhysRevB.99.224307} {\bibfield  {journal} {\bibinfo  {journal} {Phys. Rev. B}\ }\textbf {\bibinfo {volume} {99}},\ \bibinfo {pages} {224307} (\bibinfo {year} {2019})}\BibitemShut {NoStop}%
\bibitem [{\citenamefont {Farshi}\ \emph {et~al.}(2022)\citenamefont {Farshi}, \citenamefont {Toniolo}, \citenamefont {Gonz{\'{a}}lez-Guill{\'{e}}n}, \citenamefont {Alhambra},\ and\ \citenamefont {Masanes}}]{Farshi2022}%
  \BibitemOpen
  \bibfield  {author} {\bibinfo {author} {\bibfnamefont {T.}~\bibnamefont {Farshi}}, \bibinfo {author} {\bibfnamefont {D.}~\bibnamefont {Toniolo}}, \bibinfo {author} {\bibfnamefont {C.~E.}\ \bibnamefont {Gonz{\'{a}}lez-Guill{\'{e}}n}}, \bibinfo {author} {\bibfnamefont {{\'{A}}.~M.}\ \bibnamefont {Alhambra}}, \ and\ \bibinfo {author} {\bibfnamefont {L.}~\bibnamefont {Masanes}},\ }\bibfield  {title} {{Mixing and localization in random time-periodic quantum circuits of Clifford unitaries},\ }\href {\doibase 10.1063/5.0054863} {\bibfield  {journal} {\bibinfo  {journal} {Journal of Mathematical Physics}\ }\textbf {\bibinfo {volume} {63}},\ \bibinfo {pages} {032201} (\bibinfo {year} {2022})}\BibitemShut {NoStop}%
\bibitem [{\citenamefont {Farshi}\ \emph {et~al.}(2023)\citenamefont {Farshi}, \citenamefont {Richter}, \citenamefont {Toniolo}, \citenamefont {Pal},\ and\ \citenamefont {Masanes}}]{Farshi2023}%
  \BibitemOpen
  \bibfield  {author} {\bibinfo {author} {\bibfnamefont {T.}~\bibnamefont {Farshi}}, \bibinfo {author} {\bibfnamefont {J.}~\bibnamefont {Richter}}, \bibinfo {author} {\bibfnamefont {D.}~\bibnamefont {Toniolo}}, \bibinfo {author} {\bibfnamefont {A.}~\bibnamefont {Pal}}, \ and\ \bibinfo {author} {\bibfnamefont {L.}~\bibnamefont {Masanes}},\ }\bibfield  {title} {{Absence of Localization in Two-Dimensional Clifford Circuits},\ }\href {\doibase 10.1103/PRXQuantum.4.030302} {\bibfield  {journal} {\bibinfo  {journal} {PRX Quantum}\ }\textbf {\bibinfo {volume} {4}},\ \bibinfo {pages} {030302} (\bibinfo {year} {2023})}\BibitemShut {NoStop}%
\bibitem [{\citenamefont {S{\"{u}}nderhauf}\ \emph {et~al.}(2018)\citenamefont {S{\"{u}}nderhauf}, \citenamefont {P{\'{e}}rez-Garc{\'{i}}a}, \citenamefont {Huse}, \citenamefont {Schuch},\ and\ \citenamefont {Cirac}}]{Sunderhauf2018}%
  \BibitemOpen
  \bibfield  {author} {\bibinfo {author} {\bibfnamefont {C.}~\bibnamefont {S{\"{u}}nderhauf}}, \bibinfo {author} {\bibfnamefont {D.}~\bibnamefont {P{\'{e}}rez-Garc{\'{i}}a}}, \bibinfo {author} {\bibfnamefont {D.~A.}\ \bibnamefont {Huse}}, \bibinfo {author} {\bibfnamefont {N.}~\bibnamefont {Schuch}}, \ and\ \bibinfo {author} {\bibfnamefont {J.~I.}\ \bibnamefont {Cirac}},\ }\bibfield  {title} {{Localization with random time-periodic quantum circuits},\ }\href {\doibase 10.1103/PhysRevB.98.134204} {\bibfield  {journal} {\bibinfo  {journal} {Physical Review B}\ }\textbf {\bibinfo {volume} {98}},\ \bibinfo {pages} {134204} (\bibinfo {year} {2018})}\BibitemShut {NoStop}%
\bibitem [{\citenamefont {Mi}\ \emph {et~al.}(2021)\citenamefont {Mi}, \citenamefont {Roushan}, \citenamefont {Quintana}, \citenamefont {Mandrà}, \citenamefont {Marshall}, \citenamefont {Neill}, \citenamefont {Arute}, \citenamefont {Arya}, \citenamefont {Atalaya}, \citenamefont {Babbush}, \citenamefont {Bardin}, \citenamefont {Barends}, \citenamefont {Basso}, \citenamefont {Bengtsson}, \citenamefont {Boixo}, \citenamefont {Bourassa}, \citenamefont {Broughton}, \citenamefont {Buckley}, \citenamefont {Buell}, \citenamefont {Burkett}, \citenamefont {Bushnell}, \citenamefont {Chen}, \citenamefont {Chiaro}, \citenamefont {Collins}, \citenamefont {Courtney}, \citenamefont {Demura}, \citenamefont {Derk}, \citenamefont {Dunsworth}, \citenamefont {Eppens}, \citenamefont {Erickson}, \citenamefont {Farhi}, \citenamefont {Fowler}, \citenamefont {Foxen}, \citenamefont {Gidney}, \citenamefont {Giustina}, \citenamefont {Gross}, \citenamefont {Harrigan}, \citenamefont {Harrington}, \citenamefont {Hilton}, \citenamefont
  {Ho}, \citenamefont {Hong}, \citenamefont {Huang}, \citenamefont {Huggins}, \citenamefont {Ioffe}, \citenamefont {Isakov}, \citenamefont {Jeffrey}, \citenamefont {Jiang}, \citenamefont {Jones}, \citenamefont {Kafri}, \citenamefont {Kelly}, \citenamefont {Kim}, \citenamefont {Kitaev}, \citenamefont {Klimov}, \citenamefont {Korotkov}, \citenamefont {Kostritsa}, \citenamefont {Landhuis}, \citenamefont {Laptev}, \citenamefont {Lucero}, \citenamefont {Martin}, \citenamefont {McClean}, \citenamefont {McCourt}, \citenamefont {McEwen}, \citenamefont {Megrant}, \citenamefont {Miao}, \citenamefont {Mohseni}, \citenamefont {Montazeri}, \citenamefont {Mruczkiewicz}, \citenamefont {Mutus}, \citenamefont {Naaman}, \citenamefont {Neeley}, \citenamefont {Newman}, \citenamefont {Niu}, \citenamefont {O’Brien}, \citenamefont {Opremcak}, \citenamefont {Ostby}, \citenamefont {Pato}, \citenamefont {Petukhov}, \citenamefont {Redd}, \citenamefont {Rubin}, \citenamefont {Sank}, \citenamefont {Satzinger}, \citenamefont {Shvarts},
  \citenamefont {Strain}, \citenamefont {Szalay}, \citenamefont {Trevithick}, \citenamefont {Villalonga}, \citenamefont {White}, \citenamefont {Yao}, \citenamefont {Yeh}, \citenamefont {Zalcman}, \citenamefont {Neven}, \citenamefont {Aleiner}, \citenamefont {Kechedzhi}, \citenamefont {Smelyanskiy},\ and\ \citenamefont {Chen}}]{Mi2021}%
  \BibitemOpen
  \bibfield  {author} {\bibinfo {author} {\bibfnamefont {X.}~\bibnamefont {Mi}}, \bibinfo {author} {\bibfnamefont {P.}~\bibnamefont {Roushan}}, \bibinfo {author} {\bibfnamefont {C.}~\bibnamefont {Quintana}}, \bibinfo {author} {\bibfnamefont {S.}~\bibnamefont {Mandrà}}, \bibinfo {author} {\bibfnamefont {J.}~\bibnamefont {Marshall}}, \bibinfo {author} {\bibfnamefont {C.}~\bibnamefont {Neill}}, \bibinfo {author} {\bibfnamefont {F.}~\bibnamefont {Arute}}, \bibinfo {author} {\bibfnamefont {K.}~\bibnamefont {Arya}}, \bibinfo {author} {\bibfnamefont {J.}~\bibnamefont {Atalaya}}, \bibinfo {author} {\bibfnamefont {R.}~\bibnamefont {Babbush}}, \bibinfo {author} {\bibfnamefont {J.~C.}\ \bibnamefont {Bardin}}, \bibinfo {author} {\bibfnamefont {R.}~\bibnamefont {Barends}}, \bibinfo {author} {\bibfnamefont {J.}~\bibnamefont {Basso}}, \bibinfo {author} {\bibfnamefont {A.}~\bibnamefont {Bengtsson}}, \bibinfo {author} {\bibfnamefont {S.}~\bibnamefont {Boixo}}, \bibinfo {author} {\bibfnamefont {A.}~\bibnamefont {Bourassa}},
  \bibinfo {author} {\bibfnamefont {M.}~\bibnamefont {Broughton}}, \bibinfo {author} {\bibfnamefont {B.~B.}\ \bibnamefont {Buckley}}, \bibinfo {author} {\bibfnamefont {D.~A.}\ \bibnamefont {Buell}}, \bibinfo {author} {\bibfnamefont {B.}~\bibnamefont {Burkett}}, \bibinfo {author} {\bibfnamefont {N.}~\bibnamefont {Bushnell}}, \bibinfo {author} {\bibfnamefont {Z.}~\bibnamefont {Chen}}, \bibinfo {author} {\bibfnamefont {B.}~\bibnamefont {Chiaro}}, \bibinfo {author} {\bibfnamefont {R.}~\bibnamefont {Collins}}, \bibinfo {author} {\bibfnamefont {W.}~\bibnamefont {Courtney}}, \bibinfo {author} {\bibfnamefont {S.}~\bibnamefont {Demura}}, \bibinfo {author} {\bibfnamefont {A.~R.}\ \bibnamefont {Derk}}, \bibinfo {author} {\bibfnamefont {A.}~\bibnamefont {Dunsworth}}, \bibinfo {author} {\bibfnamefont {D.}~\bibnamefont {Eppens}}, \bibinfo {author} {\bibfnamefont {C.}~\bibnamefont {Erickson}}, \bibinfo {author} {\bibfnamefont {E.}~\bibnamefont {Farhi}}, \bibinfo {author} {\bibfnamefont {A.~G.}\ \bibnamefont {Fowler}},
  \bibinfo {author} {\bibfnamefont {B.}~\bibnamefont {Foxen}}, \bibinfo {author} {\bibfnamefont {C.}~\bibnamefont {Gidney}}, \bibinfo {author} {\bibfnamefont {M.}~\bibnamefont {Giustina}}, \bibinfo {author} {\bibfnamefont {J.~A.}\ \bibnamefont {Gross}}, \bibinfo {author} {\bibfnamefont {M.~P.}\ \bibnamefont {Harrigan}}, \bibinfo {author} {\bibfnamefont {S.~D.}\ \bibnamefont {Harrington}}, \bibinfo {author} {\bibfnamefont {J.}~\bibnamefont {Hilton}}, \bibinfo {author} {\bibfnamefont {A.}~\bibnamefont {Ho}}, \bibinfo {author} {\bibfnamefont {S.}~\bibnamefont {Hong}}, \bibinfo {author} {\bibfnamefont {T.}~\bibnamefont {Huang}}, \bibinfo {author} {\bibfnamefont {W.~J.}\ \bibnamefont {Huggins}}, \bibinfo {author} {\bibfnamefont {L.~B.}\ \bibnamefont {Ioffe}}, \bibinfo {author} {\bibfnamefont {S.~V.}\ \bibnamefont {Isakov}}, \bibinfo {author} {\bibfnamefont {E.}~\bibnamefont {Jeffrey}}, \bibinfo {author} {\bibfnamefont {Z.}~\bibnamefont {Jiang}}, \bibinfo {author} {\bibfnamefont {C.}~\bibnamefont {Jones}}, \bibinfo
  {author} {\bibfnamefont {D.}~\bibnamefont {Kafri}}, \bibinfo {author} {\bibfnamefont {J.}~\bibnamefont {Kelly}}, \bibinfo {author} {\bibfnamefont {S.}~\bibnamefont {Kim}}, \bibinfo {author} {\bibfnamefont {A.}~\bibnamefont {Kitaev}}, \bibinfo {author} {\bibfnamefont {P.~V.}\ \bibnamefont {Klimov}}, \bibinfo {author} {\bibfnamefont {A.~N.}\ \bibnamefont {Korotkov}}, \bibinfo {author} {\bibfnamefont {F.}~\bibnamefont {Kostritsa}}, \bibinfo {author} {\bibfnamefont {D.}~\bibnamefont {Landhuis}}, \bibinfo {author} {\bibfnamefont {P.}~\bibnamefont {Laptev}}, \bibinfo {author} {\bibfnamefont {E.}~\bibnamefont {Lucero}}, \bibinfo {author} {\bibfnamefont {O.}~\bibnamefont {Martin}}, \bibinfo {author} {\bibfnamefont {J.~R.}\ \bibnamefont {McClean}}, \bibinfo {author} {\bibfnamefont {T.}~\bibnamefont {McCourt}}, \bibinfo {author} {\bibfnamefont {M.}~\bibnamefont {McEwen}}, \bibinfo {author} {\bibfnamefont {A.}~\bibnamefont {Megrant}}, \bibinfo {author} {\bibfnamefont {K.~C.}\ \bibnamefont {Miao}}, \bibinfo {author}
  {\bibfnamefont {M.}~\bibnamefont {Mohseni}}, \bibinfo {author} {\bibfnamefont {S.}~\bibnamefont {Montazeri}}, \bibinfo {author} {\bibfnamefont {W.}~\bibnamefont {Mruczkiewicz}}, \bibinfo {author} {\bibfnamefont {J.}~\bibnamefont {Mutus}}, \bibinfo {author} {\bibfnamefont {O.}~\bibnamefont {Naaman}}, \bibinfo {author} {\bibfnamefont {M.}~\bibnamefont {Neeley}}, \bibinfo {author} {\bibfnamefont {M.}~\bibnamefont {Newman}}, \bibinfo {author} {\bibfnamefont {M.~Y.}\ \bibnamefont {Niu}}, \bibinfo {author} {\bibfnamefont {T.~E.}\ \bibnamefont {O’Brien}}, \bibinfo {author} {\bibfnamefont {A.}~\bibnamefont {Opremcak}}, \bibinfo {author} {\bibfnamefont {E.}~\bibnamefont {Ostby}}, \bibinfo {author} {\bibfnamefont {B.}~\bibnamefont {Pato}}, \bibinfo {author} {\bibfnamefont {A.}~\bibnamefont {Petukhov}}, \bibinfo {author} {\bibfnamefont {N.}~\bibnamefont {Redd}}, \bibinfo {author} {\bibfnamefont {N.~C.}\ \bibnamefont {Rubin}}, \bibinfo {author} {\bibfnamefont {D.}~\bibnamefont {Sank}}, \bibinfo {author}
  {\bibfnamefont {K.~J.}\ \bibnamefont {Satzinger}}, \bibinfo {author} {\bibfnamefont {V.}~\bibnamefont {Shvarts}}, \bibinfo {author} {\bibfnamefont {D.}~\bibnamefont {Strain}}, \bibinfo {author} {\bibfnamefont {M.}~\bibnamefont {Szalay}}, \bibinfo {author} {\bibfnamefont {M.~D.}\ \bibnamefont {Trevithick}}, \bibinfo {author} {\bibfnamefont {B.}~\bibnamefont {Villalonga}}, \bibinfo {author} {\bibfnamefont {T.}~\bibnamefont {White}}, \bibinfo {author} {\bibfnamefont {Z.~J.}\ \bibnamefont {Yao}}, \bibinfo {author} {\bibfnamefont {P.}~\bibnamefont {Yeh}}, \bibinfo {author} {\bibfnamefont {A.}~\bibnamefont {Zalcman}}, \bibinfo {author} {\bibfnamefont {H.}~\bibnamefont {Neven}}, \bibinfo {author} {\bibfnamefont {I.}~\bibnamefont {Aleiner}}, \bibinfo {author} {\bibfnamefont {K.}~\bibnamefont {Kechedzhi}}, \bibinfo {author} {\bibfnamefont {V.}~\bibnamefont {Smelyanskiy}}, \ and\ \bibinfo {author} {\bibfnamefont {Y.}~\bibnamefont {Chen}},\ }\bibfield  {title} {Information scrambling in quantum circuits,\ }\href
  {\doibase 10.1126/science.abg5029} {\bibfield  {journal} {\bibinfo  {journal} {Science}\ }\textbf {\bibinfo {volume} {374}},\ \bibinfo {pages} {1479} (\bibinfo {year} {2021})}\BibitemShut {NoStop}%
\bibitem [{\citenamefont {Yuan}\ \emph {et~al.}(2022)\citenamefont {Yuan}, \citenamefont {Zhang}, \citenamefont {Wang}, \citenamefont {Duan},\ and\ \citenamefont {Deng}}]{Yuan2022}%
  \BibitemOpen
  \bibfield  {author} {\bibinfo {author} {\bibfnamefont {D.}~\bibnamefont {Yuan}}, \bibinfo {author} {\bibfnamefont {S.-Y.}\ \bibnamefont {Zhang}}, \bibinfo {author} {\bibfnamefont {Y.}~\bibnamefont {Wang}}, \bibinfo {author} {\bibfnamefont {L.-M.}\ \bibnamefont {Duan}}, \ and\ \bibinfo {author} {\bibfnamefont {D.-L.}\ \bibnamefont {Deng}},\ }\bibfield  {title} {{Quantum information scrambling in quantum many-body scarred systems},\ }\href {\doibase 10.1103/PhysRevResearch.4.023095} {\bibfield  {journal} {\bibinfo  {journal} {Physical Review Research}\ }\textbf {\bibinfo {volume} {4}},\ \bibinfo {pages} {023095} (\bibinfo {year} {2022})}\BibitemShut {NoStop}%
\bibitem [{\citenamefont {Zhang}\ \emph {et~al.}(2023)\citenamefont {Zhang}, \citenamefont {Allcock}, \citenamefont {Wan}, \citenamefont {Liu}, \citenamefont {Sun}, \citenamefont {Yu}, \citenamefont {Yang}, \citenamefont {Qiu}, \citenamefont {Ye}, \citenamefont {Chen}, \citenamefont {Lee}, \citenamefont {Zheng}, \citenamefont {Jian}, \citenamefont {Yao}, \citenamefont {Hsieh},\ and\ \citenamefont {Zhang}}]{Zhang2022}%
  \BibitemOpen
  \bibfield  {author} {\bibinfo {author} {\bibfnamefont {S.-X.}\ \bibnamefont {Zhang}}, \bibinfo {author} {\bibfnamefont {J.}~\bibnamefont {Allcock}}, \bibinfo {author} {\bibfnamefont {Z.-Q.}\ \bibnamefont {Wan}}, \bibinfo {author} {\bibfnamefont {S.}~\bibnamefont {Liu}}, \bibinfo {author} {\bibfnamefont {J.}~\bibnamefont {Sun}}, \bibinfo {author} {\bibfnamefont {H.}~\bibnamefont {Yu}}, \bibinfo {author} {\bibfnamefont {X.-H.}\ \bibnamefont {Yang}}, \bibinfo {author} {\bibfnamefont {J.}~\bibnamefont {Qiu}}, \bibinfo {author} {\bibfnamefont {Z.}~\bibnamefont {Ye}}, \bibinfo {author} {\bibfnamefont {Y.-Q.}\ \bibnamefont {Chen}}, \bibinfo {author} {\bibfnamefont {C.-K.}\ \bibnamefont {Lee}}, \bibinfo {author} {\bibfnamefont {Y.-C.}\ \bibnamefont {Zheng}}, \bibinfo {author} {\bibfnamefont {S.-K.}\ \bibnamefont {Jian}}, \bibinfo {author} {\bibfnamefont {H.}~\bibnamefont {Yao}}, \bibinfo {author} {\bibfnamefont {C.-Y.}\ \bibnamefont {Hsieh}}, \ and\ \bibinfo {author} {\bibfnamefont {S.}~\bibnamefont {Zhang}},\
  }\bibfield  {title} {{TensorCircuit: a Quantum Software Framework for the NISQ Era},\ }\href {\doibase 10.22331/q-2023-02-02-912} {\bibfield  {journal} {\bibinfo  {journal} {Quantum}\ }\textbf {\bibinfo {volume} {7}},\ \bibinfo {pages} {912} (\bibinfo {year} {2023})}\BibitemShut {NoStop}%
\bibitem [{\citenamefont {Weinberg}\ and\ \citenamefont {Bukov}(2017)}]{Weinberg2017}%
  \BibitemOpen
  \bibfield  {author} {\bibinfo {author} {\bibfnamefont {P.}~\bibnamefont {Weinberg}}\ and\ \bibinfo {author} {\bibfnamefont {M.}~\bibnamefont {Bukov}},\ }\bibfield  {title} {{QuSpin: a Python package for dynamics and exact diagonalisation of quantum many body systems part I: spin chains},\ }\href {\doibase 10.21468/SciPostPhys.2.1.003} {\bibfield  {journal} {\bibinfo  {journal} {SciPost Physics}\ }\textbf {\bibinfo {volume} {2}},\ \bibinfo {pages} {003} (\bibinfo {year} {2017})}\BibitemShut {NoStop}%
\bibitem [{\citenamefont {Weinberg}\ and\ \citenamefont {Bukov}(2019)}]{Weinberg2019}%
  \BibitemOpen
  \bibfield  {author} {\bibinfo {author} {\bibfnamefont {P.}~\bibnamefont {Weinberg}}\ and\ \bibinfo {author} {\bibfnamefont {M.}~\bibnamefont {Bukov}},\ }\bibfield  {title} {{QuSpin: a Python package for dynamics and exact diagonalisation of quantum many body systems. Part II: bosons, fermions and higher spins},\ }\href {\doibase 10.21468/SciPostPhys.7.2.020} {\bibfield  {journal} {\bibinfo  {journal} {SciPost Physics}\ }\textbf {\bibinfo {volume} {7}},\ \bibinfo {pages} {020} (\bibinfo {year} {2019})}\BibitemShut {NoStop}%
\bibitem [{cod(2025)}]{code}%
  \BibitemOpen
  \href@noop {} {Code implementation},\ \bibinfo {howpublished} {\url{https://github.com/sxzgroup/subsystem_information_capacity}} (\bibinfo {year} {2025})\BibitemShut {NoStop}%
\bibitem [{\citenamefont {Nahum}\ \emph {et~al.}(2018{\natexlab{a}})\citenamefont {Nahum}, \citenamefont {Ruhman},\ and\ \citenamefont {Huse}}]{Nahum2018c}%
  \BibitemOpen
  \bibfield  {author} {\bibinfo {author} {\bibfnamefont {A.}~\bibnamefont {Nahum}}, \bibinfo {author} {\bibfnamefont {J.}~\bibnamefont {Ruhman}}, \ and\ \bibinfo {author} {\bibfnamefont {D.~A.}\ \bibnamefont {Huse}},\ }\bibfield  {title} {{Dynamics of entanglement and transport in one-dimensional systems with quenched randomness},\ }\href {\doibase 10.1103/PhysRevB.98.035118} {\bibfield  {journal} {\bibinfo  {journal} {Physical Review B}\ }\textbf {\bibinfo {volume} {98}},\ \bibinfo {pages} {035118} (\bibinfo {year} {2018}{\natexlab{a}})}\BibitemShut {NoStop}%
\bibitem [{\citenamefont {Nahum}\ \emph {et~al.}(2018{\natexlab{b}})\citenamefont {Nahum}, \citenamefont {Vijay},\ and\ \citenamefont {Haah}}]{Nahum2018b}%
  \BibitemOpen
  \bibfield  {author} {\bibinfo {author} {\bibfnamefont {A.}~\bibnamefont {Nahum}}, \bibinfo {author} {\bibfnamefont {S.}~\bibnamefont {Vijay}}, \ and\ \bibinfo {author} {\bibfnamefont {J.}~\bibnamefont {Haah}},\ }\bibfield  {title} {{Operator Spreading in Random Unitary Circuits},\ }\href {\doibase 10.1103/PhysRevX.8.021014} {\bibfield  {journal} {\bibinfo  {journal} {Physical Review X}\ }\textbf {\bibinfo {volume} {8}},\ \bibinfo {pages} {021014} (\bibinfo {year} {2018}{\natexlab{b}})}\BibitemShut {NoStop}%
\bibitem [{\citenamefont {von Keyserlingk}\ \emph {et~al.}(2018)\citenamefont {von Keyserlingk}, \citenamefont {Rakovszky}, \citenamefont {Pollmann},\ and\ \citenamefont {Sondhi}}]{VonKeyserlingk2018b}%
  \BibitemOpen
  \bibfield  {author} {\bibinfo {author} {\bibfnamefont {C.~W.}\ \bibnamefont {von Keyserlingk}}, \bibinfo {author} {\bibfnamefont {T.}~\bibnamefont {Rakovszky}}, \bibinfo {author} {\bibfnamefont {F.}~\bibnamefont {Pollmann}}, \ and\ \bibinfo {author} {\bibfnamefont {S.~L.}\ \bibnamefont {Sondhi}},\ }\bibfield  {title} {{Operator Hydrodynamics, OTOCs, and Entanglement Growth in Systems without Conservation Laws},\ }\href {\doibase 10.1103/PhysRevX.8.021013} {\bibfield  {journal} {\bibinfo  {journal} {Physical Review X}\ }\textbf {\bibinfo {volume} {8}},\ \bibinfo {pages} {021013} (\bibinfo {year} {2018})}\BibitemShut {NoStop}%
\bibitem [{\citenamefont {Rakovszky}\ \emph {et~al.}(2018)\citenamefont {Rakovszky}, \citenamefont {Pollmann},\ and\ \citenamefont {von Keyserlingk}}]{Rakovszky2018}%
  \BibitemOpen
  \bibfield  {author} {\bibinfo {author} {\bibfnamefont {T.}~\bibnamefont {Rakovszky}}, \bibinfo {author} {\bibfnamefont {F.}~\bibnamefont {Pollmann}}, \ and\ \bibinfo {author} {\bibfnamefont {C.~W.}\ \bibnamefont {von Keyserlingk}},\ }\bibfield  {title} {{Diffusive Hydrodynamics of Out-of-Time-Ordered Correlators with Charge Conservation},\ }\href {\doibase 10.1103/PhysRevX.8.031058} {\bibfield  {journal} {\bibinfo  {journal} {Physical Review X}\ }\textbf {\bibinfo {volume} {8}},\ \bibinfo {pages} {031058} (\bibinfo {year} {2018})}\BibitemShut {NoStop}%
\bibitem [{\citenamefont {Khemani}\ \emph {et~al.}(2018)\citenamefont {Khemani}, \citenamefont {Vishwanath},\ and\ \citenamefont {Huse}}]{Khemani2018}%
  \BibitemOpen
  \bibfield  {author} {\bibinfo {author} {\bibfnamefont {V.}~\bibnamefont {Khemani}}, \bibinfo {author} {\bibfnamefont {A.}~\bibnamefont {Vishwanath}}, \ and\ \bibinfo {author} {\bibfnamefont {D.~A.}\ \bibnamefont {Huse}},\ }\bibfield  {title} {{Operator Spreading and the Emergence of Dissipative Hydrodynamics under Unitary Evolution with Conservation Laws},\ }\href {\doibase 10.1103/PhysRevX.8.031057} {\bibfield  {journal} {\bibinfo  {journal} {Physical Review X}\ }\textbf {\bibinfo {volume} {8}},\ \bibinfo {pages} {31057} (\bibinfo {year} {2018})}\BibitemShut {NoStop}%
\bibitem [{\citenamefont {Zhuang}\ \emph {et~al.}(2023)\citenamefont {Zhuang}, \citenamefont {Wu},\ and\ \citenamefont {Duan}}]{Zhuang2023}%
  \BibitemOpen
  \bibfield  {author} {\bibinfo {author} {\bibfnamefont {J.-Z.}\ \bibnamefont {Zhuang}}, \bibinfo {author} {\bibfnamefont {Y.-K.}\ \bibnamefont {Wu}}, \ and\ \bibinfo {author} {\bibfnamefont {L.-M.}\ \bibnamefont {Duan}},\ }\bibfield  {title} {{Dynamical phase transitions of information flow in random quantum circuits},\ }\href {\doibase 10.1103/PhysRevResearch.5.L042043} {\bibfield  {journal} {\bibinfo  {journal} {Physical Review Research}\ }\textbf {\bibinfo {volume} {5}},\ \bibinfo {pages} {L042043} (\bibinfo {year} {2023})}\BibitemShut {NoStop}%
\bibitem [{\citenamefont {Ippoliti}\ and\ \citenamefont {Khemani}(2021)}]{PhysRevLett.126.060501}%
  \BibitemOpen
  \bibfield  {author} {\bibinfo {author} {\bibfnamefont {M.}~\bibnamefont {Ippoliti}}\ and\ \bibinfo {author} {\bibfnamefont {V.}~\bibnamefont {Khemani}},\ }\bibfield  {title} {Postselection-free entanglement dynamics via spacetime duality,\ }\href {\doibase 10.1103/PhysRevLett.126.060501} {\bibfield  {journal} {\bibinfo  {journal} {Phys. Rev. Lett.}\ }\textbf {\bibinfo {volume} {126}},\ \bibinfo {pages} {060501} (\bibinfo {year} {2021})}\BibitemShut {NoStop}%
\bibitem [{\citenamefont {Lu}\ and\ \citenamefont {Grover}(2021)}]{PRXQuantum.2.040319}%
  \BibitemOpen
  \bibfield  {author} {\bibinfo {author} {\bibfnamefont {T.-C.}\ \bibnamefont {Lu}}\ and\ \bibinfo {author} {\bibfnamefont {T.}~\bibnamefont {Grover}},\ }\bibfield  {title} {Spacetime duality between localization transitions and measurement-induced transitions,\ }\href {\doibase 10.1103/PRXQuantum.2.040319} {\bibfield  {journal} {\bibinfo  {journal} {PRX Quantum}\ }\textbf {\bibinfo {volume} {2}},\ \bibinfo {pages} {040319} (\bibinfo {year} {2021})}\BibitemShut {NoStop}%
\bibitem [{\citenamefont {Ippoliti}\ \emph {et~al.}(2022)\citenamefont {Ippoliti}, \citenamefont {Rakovszky},\ and\ \citenamefont {Khemani}}]{PhysRevX.12.011045}%
  \BibitemOpen
  \bibfield  {author} {\bibinfo {author} {\bibfnamefont {M.}~\bibnamefont {Ippoliti}}, \bibinfo {author} {\bibfnamefont {T.}~\bibnamefont {Rakovszky}}, \ and\ \bibinfo {author} {\bibfnamefont {V.}~\bibnamefont {Khemani}},\ }\bibfield  {title} {Fractal, logarithmic, and volume-law entangled nonthermal steady states via spacetime duality,\ }\href {\doibase 10.1103/PhysRevX.12.011045} {\bibfield  {journal} {\bibinfo  {journal} {Phys. Rev. X}\ }\textbf {\bibinfo {volume} {12}},\ \bibinfo {pages} {011045} (\bibinfo {year} {2022})}\BibitemShut {NoStop}%
\bibitem [{\citenamefont {Gullans}\ and\ \citenamefont {Huse}(2020{\natexlab{a}})}]{PhysRevX.10.041020}%
  \BibitemOpen
  \bibfield  {author} {\bibinfo {author} {\bibfnamefont {M.~J.}\ \bibnamefont {Gullans}}\ and\ \bibinfo {author} {\bibfnamefont {D.~A.}\ \bibnamefont {Huse}},\ }\bibfield  {title} {Dynamical purification phase transition induced by quantum measurements,\ }\href {\doibase 10.1103/PhysRevX.10.041020} {\bibfield  {journal} {\bibinfo  {journal} {Phys. Rev. X}\ }\textbf {\bibinfo {volume} {10}},\ \bibinfo {pages} {041020} (\bibinfo {year} {2020}{\natexlab{a}})}\BibitemShut {NoStop}%
\bibitem [{\citenamefont {Szyniszewski}\ \emph {et~al.}(2019)\citenamefont {Szyniszewski}, \citenamefont {Romito},\ and\ \citenamefont {Schomerus}}]{PhysRevB.100.064204}%
  \BibitemOpen
  \bibfield  {author} {\bibinfo {author} {\bibfnamefont {M.}~\bibnamefont {Szyniszewski}}, \bibinfo {author} {\bibfnamefont {A.}~\bibnamefont {Romito}}, \ and\ \bibinfo {author} {\bibfnamefont {H.}~\bibnamefont {Schomerus}},\ }\bibfield  {title} {Entanglement transition from variable-strength weak measurements,\ }\href {\doibase 10.1103/PhysRevB.100.064204} {\bibfield  {journal} {\bibinfo  {journal} {Phys. Rev. B}\ }\textbf {\bibinfo {volume} {100}},\ \bibinfo {pages} {064204} (\bibinfo {year} {2019})}\BibitemShut {NoStop}%
\bibitem [{\citenamefont {Hoke}\ \emph {et~al.}(2023)\citenamefont {Hoke}, \citenamefont {Ippoliti}, \citenamefont {Rosenberg}, \citenamefont {Abanin}, \citenamefont {Acharya}, \citenamefont {Andersen}, \citenamefont {Ansmann}, \citenamefont {Arute}, \citenamefont {Arya}, \citenamefont {Asfaw}, \citenamefont {Atalaya}, \citenamefont {Bardin}, \citenamefont {Bengtsson}, \citenamefont {Bortoli}, \citenamefont {Bourassa}, \citenamefont {Bovaird}, \citenamefont {Brill}, \citenamefont {Broughton}, \citenamefont {Buckley}, \citenamefont {Buell}, \citenamefont {Burger}, \citenamefont {Burkett}, \citenamefont {Bushnell}, \citenamefont {Chen}, \citenamefont {Chiaro}, \citenamefont {Chik}, \citenamefont {Cogan}, \citenamefont {Collins}, \citenamefont {Conner}, \citenamefont {Courtney}, \citenamefont {Crook}, \citenamefont {Curtin}, \citenamefont {Dau}, \citenamefont {Debroy}, \citenamefont {Del Toro~Barba}, \citenamefont {Demura}, \citenamefont {Di~Paolo}, \citenamefont {Drozdov}, \citenamefont {Dunsworth},
  \citenamefont {Eppens}, \citenamefont {Erickson}, \citenamefont {Farhi}, \citenamefont {Fatemi}, \citenamefont {Ferreira}, \citenamefont {Burgos}, \citenamefont {Forati}, \citenamefont {Fowler}, \citenamefont {Foxen}, \citenamefont {Giang}, \citenamefont {Gidney}, \citenamefont {Gilboa}, \citenamefont {Giustina}, \citenamefont {Gosula}, \citenamefont {Gross}, \citenamefont {Habegger}, \citenamefont {Hamilton}, \citenamefont {Hansen}, \citenamefont {Harrigan}, \citenamefont {Harrington}, \citenamefont {Heu}, \citenamefont {Hoffmann}, \citenamefont {Hong}, \citenamefont {Huang}, \citenamefont {Huff}, \citenamefont {Huggins}, \citenamefont {Isakov}, \citenamefont {Iveland}, \citenamefont {Jeffrey}, \citenamefont {Jiang}, \citenamefont {Jones}, \citenamefont {Juhas}, \citenamefont {Kafri}, \citenamefont {Kechedzhi}, \citenamefont {Khattar}, \citenamefont {Khezri}, \citenamefont {Kieferová}, \citenamefont {Kim}, \citenamefont {Kitaev}, \citenamefont {Klimov}, \citenamefont {Klots}, \citenamefont {Korotkov},
  \citenamefont {Kostritsa}, \citenamefont {Kreikebaum}, \citenamefont {Landhuis}, \citenamefont {Laptev}, \citenamefont {Lau}, \citenamefont {Laws}, \citenamefont {Lee}, \citenamefont {Lee}, \citenamefont {Lensky}, \citenamefont {Lester}, \citenamefont {Lill}, \citenamefont {Liu}, \citenamefont {Locharla}, \citenamefont {Martin}, \citenamefont {McClean}, \citenamefont {McEwen}, \citenamefont {Miao}, \citenamefont {Mieszala}, \citenamefont {Montazeri}, \citenamefont {Morvan}, \citenamefont {Movassagh}, \citenamefont {Mruczkiewicz}, \citenamefont {Neeley}, \citenamefont {Neill}, \citenamefont {Nersisyan}, \citenamefont {Newman}, \citenamefont {Ng}, \citenamefont {Nguyen}, \citenamefont {Nguyen}, \citenamefont {Niu}, \citenamefont {O’Brien}, \citenamefont {Omonije}, \citenamefont {Opremcak}, \citenamefont {Petukhov}, \citenamefont {Potter}, \citenamefont {Pryadko}, \citenamefont {Quintana}, \citenamefont {Rocque}, \citenamefont {Rubin}, \citenamefont {Saei}, \citenamefont {Sank}, \citenamefont
  {Sankaragomathi}, \citenamefont {Satzinger}, \citenamefont {Schurkus}, \citenamefont {Schuster}, \citenamefont {Shearn}, \citenamefont {Shorter}, \citenamefont {Shutty}, \citenamefont {Shvarts}, \citenamefont {Skruzny}, \citenamefont {Smith}, \citenamefont {Somma}, \citenamefont {Sterling}, \citenamefont {Strain}, \citenamefont {Szalay}, \citenamefont {Torres}, \citenamefont {Vidal}, \citenamefont {Villalonga}, \citenamefont {Heidweiller}, \citenamefont {White}, \citenamefont {Woo}, \citenamefont {Xing}, \citenamefont {Yao}, \citenamefont {Yeh}, \citenamefont {Yoo}, \citenamefont {Young}, \citenamefont {Zalcman}, \citenamefont {Zhang}, \citenamefont {Zhu}, \citenamefont {Zobrist}, \citenamefont {Neven}, \citenamefont {Babbush}, \citenamefont {Bacon}, \citenamefont {Boixo}, \citenamefont {Hilton}, \citenamefont {Lucero}, \citenamefont {Megrant}, \citenamefont {Kelly}, \citenamefont {Chen}, \citenamefont {Smelyanskiy}, \citenamefont {Mi}, \citenamefont {Khemani}, \citenamefont {Roushan},\ and\ \citenamefont
  {{Google Quantum AI and Collaborators}}}]{hokeMeasurementinducedEntanglementTeleportation2023}%
  \BibitemOpen
  \bibfield  {author} {\bibinfo {author} {\bibfnamefont {J.~C.}\ \bibnamefont {Hoke}}, \bibinfo {author} {\bibfnamefont {M.}~\bibnamefont {Ippoliti}}, \bibinfo {author} {\bibfnamefont {E.}~\bibnamefont {Rosenberg}}, \bibinfo {author} {\bibfnamefont {D.}~\bibnamefont {Abanin}}, \bibinfo {author} {\bibfnamefont {R.}~\bibnamefont {Acharya}}, \bibinfo {author} {\bibfnamefont {T.~I.}\ \bibnamefont {Andersen}}, \bibinfo {author} {\bibfnamefont {M.}~\bibnamefont {Ansmann}}, \bibinfo {author} {\bibfnamefont {F.}~\bibnamefont {Arute}}, \bibinfo {author} {\bibfnamefont {K.}~\bibnamefont {Arya}}, \bibinfo {author} {\bibfnamefont {A.}~\bibnamefont {Asfaw}}, \bibinfo {author} {\bibfnamefont {J.}~\bibnamefont {Atalaya}}, \bibinfo {author} {\bibfnamefont {J.~C.}\ \bibnamefont {Bardin}}, \bibinfo {author} {\bibfnamefont {A.}~\bibnamefont {Bengtsson}}, \bibinfo {author} {\bibfnamefont {G.}~\bibnamefont {Bortoli}}, \bibinfo {author} {\bibfnamefont {A.}~\bibnamefont {Bourassa}}, \bibinfo {author} {\bibfnamefont
  {J.}~\bibnamefont {Bovaird}}, \bibinfo {author} {\bibfnamefont {L.}~\bibnamefont {Brill}}, \bibinfo {author} {\bibfnamefont {M.}~\bibnamefont {Broughton}}, \bibinfo {author} {\bibfnamefont {B.~B.}\ \bibnamefont {Buckley}}, \bibinfo {author} {\bibfnamefont {D.~A.}\ \bibnamefont {Buell}}, \bibinfo {author} {\bibfnamefont {T.}~\bibnamefont {Burger}}, \bibinfo {author} {\bibfnamefont {B.}~\bibnamefont {Burkett}}, \bibinfo {author} {\bibfnamefont {N.}~\bibnamefont {Bushnell}}, \bibinfo {author} {\bibfnamefont {Z.}~\bibnamefont {Chen}}, \bibinfo {author} {\bibfnamefont {B.}~\bibnamefont {Chiaro}}, \bibinfo {author} {\bibfnamefont {D.}~\bibnamefont {Chik}}, \bibinfo {author} {\bibfnamefont {J.}~\bibnamefont {Cogan}}, \bibinfo {author} {\bibfnamefont {R.}~\bibnamefont {Collins}}, \bibinfo {author} {\bibfnamefont {P.}~\bibnamefont {Conner}}, \bibinfo {author} {\bibfnamefont {W.}~\bibnamefont {Courtney}}, \bibinfo {author} {\bibfnamefont {A.~L.}\ \bibnamefont {Crook}}, \bibinfo {author} {\bibfnamefont
  {B.}~\bibnamefont {Curtin}}, \bibinfo {author} {\bibfnamefont {A.~G.}\ \bibnamefont {Dau}}, \bibinfo {author} {\bibfnamefont {D.~M.}\ \bibnamefont {Debroy}}, \bibinfo {author} {\bibfnamefont {A.}~\bibnamefont {Del Toro~Barba}}, \bibinfo {author} {\bibfnamefont {S.}~\bibnamefont {Demura}}, \bibinfo {author} {\bibfnamefont {A.}~\bibnamefont {Di~Paolo}}, \bibinfo {author} {\bibfnamefont {I.~K.}\ \bibnamefont {Drozdov}}, \bibinfo {author} {\bibfnamefont {A.}~\bibnamefont {Dunsworth}}, \bibinfo {author} {\bibfnamefont {D.}~\bibnamefont {Eppens}}, \bibinfo {author} {\bibfnamefont {C.}~\bibnamefont {Erickson}}, \bibinfo {author} {\bibfnamefont {E.}~\bibnamefont {Farhi}}, \bibinfo {author} {\bibfnamefont {R.}~\bibnamefont {Fatemi}}, \bibinfo {author} {\bibfnamefont {V.~S.}\ \bibnamefont {Ferreira}}, \bibinfo {author} {\bibfnamefont {L.~F.}\ \bibnamefont {Burgos}}, \bibinfo {author} {\bibfnamefont {E.}~\bibnamefont {Forati}}, \bibinfo {author} {\bibfnamefont {A.~G.}\ \bibnamefont {Fowler}}, \bibinfo {author}
  {\bibfnamefont {B.}~\bibnamefont {Foxen}}, \bibinfo {author} {\bibfnamefont {W.}~\bibnamefont {Giang}}, \bibinfo {author} {\bibfnamefont {C.}~\bibnamefont {Gidney}}, \bibinfo {author} {\bibfnamefont {D.}~\bibnamefont {Gilboa}}, \bibinfo {author} {\bibfnamefont {M.}~\bibnamefont {Giustina}}, \bibinfo {author} {\bibfnamefont {R.}~\bibnamefont {Gosula}}, \bibinfo {author} {\bibfnamefont {J.~A.}\ \bibnamefont {Gross}}, \bibinfo {author} {\bibfnamefont {S.}~\bibnamefont {Habegger}}, \bibinfo {author} {\bibfnamefont {M.~C.}\ \bibnamefont {Hamilton}}, \bibinfo {author} {\bibfnamefont {M.}~\bibnamefont {Hansen}}, \bibinfo {author} {\bibfnamefont {M.~P.}\ \bibnamefont {Harrigan}}, \bibinfo {author} {\bibfnamefont {S.~D.}\ \bibnamefont {Harrington}}, \bibinfo {author} {\bibfnamefont {P.}~\bibnamefont {Heu}}, \bibinfo {author} {\bibfnamefont {M.~R.}\ \bibnamefont {Hoffmann}}, \bibinfo {author} {\bibfnamefont {S.}~\bibnamefont {Hong}}, \bibinfo {author} {\bibfnamefont {T.}~\bibnamefont {Huang}}, \bibinfo {author}
  {\bibfnamefont {A.}~\bibnamefont {Huff}}, \bibinfo {author} {\bibfnamefont {W.~J.}\ \bibnamefont {Huggins}}, \bibinfo {author} {\bibfnamefont {S.~V.}\ \bibnamefont {Isakov}}, \bibinfo {author} {\bibfnamefont {J.}~\bibnamefont {Iveland}}, \bibinfo {author} {\bibfnamefont {E.}~\bibnamefont {Jeffrey}}, \bibinfo {author} {\bibfnamefont {Z.}~\bibnamefont {Jiang}}, \bibinfo {author} {\bibfnamefont {C.}~\bibnamefont {Jones}}, \bibinfo {author} {\bibfnamefont {P.}~\bibnamefont {Juhas}}, \bibinfo {author} {\bibfnamefont {D.}~\bibnamefont {Kafri}}, \bibinfo {author} {\bibfnamefont {K.}~\bibnamefont {Kechedzhi}}, \bibinfo {author} {\bibfnamefont {T.}~\bibnamefont {Khattar}}, \bibinfo {author} {\bibfnamefont {M.}~\bibnamefont {Khezri}}, \bibinfo {author} {\bibfnamefont {M.}~\bibnamefont {Kieferová}}, \bibinfo {author} {\bibfnamefont {S.}~\bibnamefont {Kim}}, \bibinfo {author} {\bibfnamefont {A.}~\bibnamefont {Kitaev}}, \bibinfo {author} {\bibfnamefont {P.~V.}\ \bibnamefont {Klimov}}, \bibinfo {author} {\bibfnamefont
  {A.~R.}\ \bibnamefont {Klots}}, \bibinfo {author} {\bibfnamefont {A.~N.}\ \bibnamefont {Korotkov}}, \bibinfo {author} {\bibfnamefont {F.}~\bibnamefont {Kostritsa}}, \bibinfo {author} {\bibfnamefont {J.~M.}\ \bibnamefont {Kreikebaum}}, \bibinfo {author} {\bibfnamefont {D.}~\bibnamefont {Landhuis}}, \bibinfo {author} {\bibfnamefont {P.}~\bibnamefont {Laptev}}, \bibinfo {author} {\bibfnamefont {K.-M.}\ \bibnamefont {Lau}}, \bibinfo {author} {\bibfnamefont {L.}~\bibnamefont {Laws}}, \bibinfo {author} {\bibfnamefont {J.}~\bibnamefont {Lee}}, \bibinfo {author} {\bibfnamefont {K.~W.}\ \bibnamefont {Lee}}, \bibinfo {author} {\bibfnamefont {Y.~D.}\ \bibnamefont {Lensky}}, \bibinfo {author} {\bibfnamefont {B.~J.}\ \bibnamefont {Lester}}, \bibinfo {author} {\bibfnamefont {A.~T.}\ \bibnamefont {Lill}}, \bibinfo {author} {\bibfnamefont {W.}~\bibnamefont {Liu}}, \bibinfo {author} {\bibfnamefont {A.}~\bibnamefont {Locharla}}, \bibinfo {author} {\bibfnamefont {O.}~\bibnamefont {Martin}}, \bibinfo {author} {\bibfnamefont
  {J.~R.}\ \bibnamefont {McClean}}, \bibinfo {author} {\bibfnamefont {M.}~\bibnamefont {McEwen}}, \bibinfo {author} {\bibfnamefont {K.~C.}\ \bibnamefont {Miao}}, \bibinfo {author} {\bibfnamefont {A.}~\bibnamefont {Mieszala}}, \bibinfo {author} {\bibfnamefont {S.}~\bibnamefont {Montazeri}}, \bibinfo {author} {\bibfnamefont {A.}~\bibnamefont {Morvan}}, \bibinfo {author} {\bibfnamefont {R.}~\bibnamefont {Movassagh}}, \bibinfo {author} {\bibfnamefont {W.}~\bibnamefont {Mruczkiewicz}}, \bibinfo {author} {\bibfnamefont {M.}~\bibnamefont {Neeley}}, \bibinfo {author} {\bibfnamefont {C.}~\bibnamefont {Neill}}, \bibinfo {author} {\bibfnamefont {A.}~\bibnamefont {Nersisyan}}, \bibinfo {author} {\bibfnamefont {M.}~\bibnamefont {Newman}}, \bibinfo {author} {\bibfnamefont {J.~H.}\ \bibnamefont {Ng}}, \bibinfo {author} {\bibfnamefont {A.}~\bibnamefont {Nguyen}}, \bibinfo {author} {\bibfnamefont {M.}~\bibnamefont {Nguyen}}, \bibinfo {author} {\bibfnamefont {M.~Y.}\ \bibnamefont {Niu}}, \bibinfo {author} {\bibfnamefont
  {T.~E.}\ \bibnamefont {O’Brien}}, \bibinfo {author} {\bibfnamefont {S.}~\bibnamefont {Omonije}}, \bibinfo {author} {\bibfnamefont {A.}~\bibnamefont {Opremcak}}, \bibinfo {author} {\bibfnamefont {A.}~\bibnamefont {Petukhov}}, \bibinfo {author} {\bibfnamefont {R.}~\bibnamefont {Potter}}, \bibinfo {author} {\bibfnamefont {L.~P.}\ \bibnamefont {Pryadko}}, \bibinfo {author} {\bibfnamefont {C.}~\bibnamefont {Quintana}}, \bibinfo {author} {\bibfnamefont {C.}~\bibnamefont {Rocque}}, \bibinfo {author} {\bibfnamefont {N.~C.}\ \bibnamefont {Rubin}}, \bibinfo {author} {\bibfnamefont {N.}~\bibnamefont {Saei}}, \bibinfo {author} {\bibfnamefont {D.}~\bibnamefont {Sank}}, \bibinfo {author} {\bibfnamefont {K.}~\bibnamefont {Sankaragomathi}}, \bibinfo {author} {\bibfnamefont {K.~J.}\ \bibnamefont {Satzinger}}, \bibinfo {author} {\bibfnamefont {H.~F.}\ \bibnamefont {Schurkus}}, \bibinfo {author} {\bibfnamefont {C.}~\bibnamefont {Schuster}}, \bibinfo {author} {\bibfnamefont {M.~J.}\ \bibnamefont {Shearn}}, \bibinfo {author}
  {\bibfnamefont {A.}~\bibnamefont {Shorter}}, \bibinfo {author} {\bibfnamefont {N.}~\bibnamefont {Shutty}}, \bibinfo {author} {\bibfnamefont {V.}~\bibnamefont {Shvarts}}, \bibinfo {author} {\bibfnamefont {J.}~\bibnamefont {Skruzny}}, \bibinfo {author} {\bibfnamefont {W.~C.}\ \bibnamefont {Smith}}, \bibinfo {author} {\bibfnamefont {R.}~\bibnamefont {Somma}}, \bibinfo {author} {\bibfnamefont {G.}~\bibnamefont {Sterling}}, \bibinfo {author} {\bibfnamefont {D.}~\bibnamefont {Strain}}, \bibinfo {author} {\bibfnamefont {M.}~\bibnamefont {Szalay}}, \bibinfo {author} {\bibfnamefont {A.}~\bibnamefont {Torres}}, \bibinfo {author} {\bibfnamefont {G.}~\bibnamefont {Vidal}}, \bibinfo {author} {\bibfnamefont {B.}~\bibnamefont {Villalonga}}, \bibinfo {author} {\bibfnamefont {C.~V.}\ \bibnamefont {Heidweiller}}, \bibinfo {author} {\bibfnamefont {T.}~\bibnamefont {White}}, \bibinfo {author} {\bibfnamefont {B.~W.~K.}\ \bibnamefont {Woo}}, \bibinfo {author} {\bibfnamefont {C.}~\bibnamefont {Xing}}, \bibinfo {author}
  {\bibfnamefont {Z.~J.}\ \bibnamefont {Yao}}, \bibinfo {author} {\bibfnamefont {P.}~\bibnamefont {Yeh}}, \bibinfo {author} {\bibfnamefont {J.}~\bibnamefont {Yoo}}, \bibinfo {author} {\bibfnamefont {G.}~\bibnamefont {Young}}, \bibinfo {author} {\bibfnamefont {A.}~\bibnamefont {Zalcman}}, \bibinfo {author} {\bibfnamefont {Y.}~\bibnamefont {Zhang}}, \bibinfo {author} {\bibfnamefont {N.}~\bibnamefont {Zhu}}, \bibinfo {author} {\bibfnamefont {N.}~\bibnamefont {Zobrist}}, \bibinfo {author} {\bibfnamefont {H.}~\bibnamefont {Neven}}, \bibinfo {author} {\bibfnamefont {R.}~\bibnamefont {Babbush}}, \bibinfo {author} {\bibfnamefont {D.}~\bibnamefont {Bacon}}, \bibinfo {author} {\bibfnamefont {S.}~\bibnamefont {Boixo}}, \bibinfo {author} {\bibfnamefont {J.}~\bibnamefont {Hilton}}, \bibinfo {author} {\bibfnamefont {E.}~\bibnamefont {Lucero}}, \bibinfo {author} {\bibfnamefont {A.}~\bibnamefont {Megrant}}, \bibinfo {author} {\bibfnamefont {J.}~\bibnamefont {Kelly}}, \bibinfo {author} {\bibfnamefont {Y.}~\bibnamefont
  {Chen}}, \bibinfo {author} {\bibfnamefont {V.}~\bibnamefont {Smelyanskiy}}, \bibinfo {author} {\bibfnamefont {X.}~\bibnamefont {Mi}}, \bibinfo {author} {\bibfnamefont {V.}~\bibnamefont {Khemani}}, \bibinfo {author} {\bibfnamefont {P.}~\bibnamefont {Roushan}}, \ and\ \bibinfo {author} {\bibnamefont {{Google Quantum AI and Collaborators}}},\ }\bibfield  {title} {Measurement-induced entanglement and teleportation on a noisy quantum processor,\ }\href {\doibase 10.1038/s41586-023-06505-7} {\bibfield  {journal} {\bibinfo  {journal} {Nature}\ }\textbf {\bibinfo {volume} {622}},\ \bibinfo {pages} {481} (\bibinfo {year} {2023})}\BibitemShut {NoStop}%
\bibitem [{\citenamefont {Turkeshi}\ \emph {et~al.}(2020)\citenamefont {Turkeshi}, \citenamefont {Fazio},\ and\ \citenamefont {Dalmonte}}]{PhysRevB.102.014315}%
  \BibitemOpen
  \bibfield  {author} {\bibinfo {author} {\bibfnamefont {X.}~\bibnamefont {Turkeshi}}, \bibinfo {author} {\bibfnamefont {R.}~\bibnamefont {Fazio}}, \ and\ \bibinfo {author} {\bibfnamefont {M.}~\bibnamefont {Dalmonte}},\ }\bibfield  {title} {Measurement-induced criticality in $(2+1)$-dimensional hybrid quantum circuits,\ }\href {\doibase 10.1103/PhysRevB.102.014315} {\bibfield  {journal} {\bibinfo  {journal} {Phys. Rev. B}\ }\textbf {\bibinfo {volume} {102}},\ \bibinfo {pages} {014315} (\bibinfo {year} {2020})}\BibitemShut {NoStop}%
\bibitem [{\citenamefont {Sierant}\ \emph {et~al.}(2022)\citenamefont {Sierant}, \citenamefont {Schir\`o}, \citenamefont {Lewenstein},\ and\ \citenamefont {Turkeshi}}]{PhysRevB.106.214316}%
  \BibitemOpen
  \bibfield  {author} {\bibinfo {author} {\bibfnamefont {P.}~\bibnamefont {Sierant}}, \bibinfo {author} {\bibfnamefont {M.}~\bibnamefont {Schir\`o}}, \bibinfo {author} {\bibfnamefont {M.}~\bibnamefont {Lewenstein}}, \ and\ \bibinfo {author} {\bibfnamefont {X.}~\bibnamefont {Turkeshi}},\ }\bibfield  {title} {Measurement-induced phase transitions in $(d+1)$-dimensional stabilizer circuits,\ }\href {\doibase 10.1103/PhysRevB.106.214316} {\bibfield  {journal} {\bibinfo  {journal} {Phys. Rev. B}\ }\textbf {\bibinfo {volume} {106}},\ \bibinfo {pages} {214316} (\bibinfo {year} {2022})}\BibitemShut {NoStop}%
\bibitem [{\citenamefont {Gullans}\ and\ \citenamefont {Huse}(2020{\natexlab{b}})}]{Gullans2020}%
  \BibitemOpen
  \bibfield  {author} {\bibinfo {author} {\bibfnamefont {M.~J.}\ \bibnamefont {Gullans}}\ and\ \bibinfo {author} {\bibfnamefont {D.~A.}\ \bibnamefont {Huse}},\ }\bibfield  {title} {Scalable probes of measurement-induced criticality,\ }\href {\doibase 10.1103/PhysRevLett.125.070606} {\bibfield  {journal} {\bibinfo  {journal} {Physical Review Letters}\ }\textbf {\bibinfo {volume} {125}},\ \bibinfo {pages} {070606} (\bibinfo {year} {2020}{\natexlab{b}})}\BibitemShut {NoStop}%
\bibitem [{\citenamefont {Weinstein}\ \emph {et~al.}(2022)\citenamefont {Weinstein}, \citenamefont {Bao},\ and\ \citenamefont {Altman}}]{PhysRevLett.129.080501}%
  \BibitemOpen
  \bibfield  {author} {\bibinfo {author} {\bibfnamefont {Z.}~\bibnamefont {Weinstein}}, \bibinfo {author} {\bibfnamefont {Y.}~\bibnamefont {Bao}}, \ and\ \bibinfo {author} {\bibfnamefont {E.}~\bibnamefont {Altman}},\ }\bibfield  {title} {Measurement-induced power-law negativity in an open monitored quantum circuit,\ }\href {\doibase 10.1103/PhysRevLett.129.080501} {\bibfield  {journal} {\bibinfo  {journal} {Phys. Rev. Lett.}\ }\textbf {\bibinfo {volume} {129}},\ \bibinfo {pages} {080501} (\bibinfo {year} {2022})}\BibitemShut {NoStop}%
\bibitem [{\citenamefont {Liu}\ \emph {et~al.}(2023{\natexlab{a}})\citenamefont {Liu}, \citenamefont {Li}, \citenamefont {Zhang}, \citenamefont {Jian},\ and\ \citenamefont {Yao}}]{PhysRevB.107.L201113}%
  \BibitemOpen
  \bibfield  {author} {\bibinfo {author} {\bibfnamefont {S.}~\bibnamefont {Liu}}, \bibinfo {author} {\bibfnamefont {M.-R.}\ \bibnamefont {Li}}, \bibinfo {author} {\bibfnamefont {S.-X.}\ \bibnamefont {Zhang}}, \bibinfo {author} {\bibfnamefont {S.-K.}\ \bibnamefont {Jian}}, \ and\ \bibinfo {author} {\bibfnamefont {H.}~\bibnamefont {Yao}},\ }\bibfield  {title} {Universal kardar-parisi-zhang scaling in noisy hybrid quantum circuits,\ }\href {\doibase 10.1103/PhysRevB.107.L201113} {\bibfield  {journal} {\bibinfo  {journal} {Phys. Rev. B}\ }\textbf {\bibinfo {volume} {107}},\ \bibinfo {pages} {L201113} (\bibinfo {year} {2023}{\natexlab{a}})}\BibitemShut {NoStop}%
\bibitem [{\citenamefont {Liu}\ \emph {et~al.}(2024{\natexlab{a}})\citenamefont {Liu}, \citenamefont {Li}, \citenamefont {Zhang},\ and\ \citenamefont {Jian}}]{Liu2024}%
  \BibitemOpen
  \bibfield  {author} {\bibinfo {author} {\bibfnamefont {S.}~\bibnamefont {Liu}}, \bibinfo {author} {\bibfnamefont {M.-R.}\ \bibnamefont {Li}}, \bibinfo {author} {\bibfnamefont {S.-X.}\ \bibnamefont {Zhang}}, \ and\ \bibinfo {author} {\bibfnamefont {S.-K.}\ \bibnamefont {Jian}},\ }\bibfield  {title} {Entanglement structure and information protection in noisy hybrid quantum circuits,\ }\href {\doibase 10.1103/PhysRevLett.132.240402} {\bibfield  {journal} {\bibinfo  {journal} {Physical Review Letters}\ }\textbf {\bibinfo {volume} {132}},\ \bibinfo {pages} {240402} (\bibinfo {year} {2024}{\natexlab{a}})}\BibitemShut {NoStop}%
\bibitem [{\citenamefont {Liu}\ \emph {et~al.}(2024{\natexlab{b}})\citenamefont {Liu}, \citenamefont {Li}, \citenamefont {Zhang}, \citenamefont {Jian},\ and\ \citenamefont {Yao}}]{liu2024noise}%
  \BibitemOpen
  \bibfield  {author} {\bibinfo {author} {\bibfnamefont {S.}~\bibnamefont {Liu}}, \bibinfo {author} {\bibfnamefont {M.-R.}\ \bibnamefont {Li}}, \bibinfo {author} {\bibfnamefont {S.-X.}\ \bibnamefont {Zhang}}, \bibinfo {author} {\bibfnamefont {S.-K.}\ \bibnamefont {Jian}}, \ and\ \bibinfo {author} {\bibfnamefont {H.}~\bibnamefont {Yao}},\ }\bibfield  {title} {Noise-induced phase transitions in hybrid quantum circuits,\ }\href {\doibase 10.1103/PhysRevB.110.064323} {\bibfield  {journal} {\bibinfo  {journal} {Physical Review B}\ }\textbf {\bibinfo {volume} {110}},\ \bibinfo {pages} {064323} (\bibinfo {year} {2024}{\natexlab{b}})}\BibitemShut {NoStop}%
\bibitem [{\citenamefont {Lovas}\ \emph {et~al.}(2024)\citenamefont {Lovas}, \citenamefont {Agrawal},\ and\ \citenamefont {Vijay}}]{Coding_Vijay}%
  \BibitemOpen
  \bibfield  {author} {\bibinfo {author} {\bibfnamefont {I.}~\bibnamefont {Lovas}}, \bibinfo {author} {\bibfnamefont {U.}~\bibnamefont {Agrawal}}, \ and\ \bibinfo {author} {\bibfnamefont {S.}~\bibnamefont {Vijay}},\ }\bibfield  {title} {Quantum coding transitions in the presence of boundary dissipation,\ }\href {\doibase 10.1103/PRXQuantum.5.030327} {\bibfield  {journal} {\bibinfo  {journal} {PRX Quantum}\ }\textbf {\bibinfo {volume} {5}},\ \bibinfo {pages} {030327} (\bibinfo {year} {2024})}\BibitemShut {NoStop}%
\bibitem [{\citenamefont {Turkeshi}\ and\ \citenamefont {Sierant}(2024)}]{Turkeshi2024}%
  \BibitemOpen
  \bibfield  {author} {\bibinfo {author} {\bibfnamefont {X.}~\bibnamefont {Turkeshi}}\ and\ \bibinfo {author} {\bibfnamefont {P.}~\bibnamefont {Sierant}},\ }\bibfield  {title} {{Error-Resilience Phase Transitions in Encoding-Decoding Quantum Circuits},\ }\href {\doibase 10.1103/PhysRevLett.132.140401} {\bibfield  {journal} {\bibinfo  {journal} {Phys. Rev. Lett.}\ }\textbf {\bibinfo {volume} {132}},\ \bibinfo {pages} {140401} (\bibinfo {year} {2024})}\BibitemShut {NoStop}%
\bibitem [{\citenamefont {Weinstein}\ \emph {et~al.}(2023)\citenamefont {Weinstein}, \citenamefont {Kelly}, \citenamefont {Marino},\ and\ \citenamefont {Altman}}]{Weinstein2023}%
  \BibitemOpen
  \bibfield  {author} {\bibinfo {author} {\bibfnamefont {Z.}~\bibnamefont {Weinstein}}, \bibinfo {author} {\bibfnamefont {S.~P.}\ \bibnamefont {Kelly}}, \bibinfo {author} {\bibfnamefont {J.}~\bibnamefont {Marino}}, \ and\ \bibinfo {author} {\bibfnamefont {E.}~\bibnamefont {Altman}},\ }\bibfield  {title} {{Scrambling Transition in a Radiative Random Unitary Circuit},\ }\href {\doibase 10.1103/PhysRevLett.131.220404} {\bibfield  {journal} {\bibinfo  {journal} {Physical Review Letters}\ }\textbf {\bibinfo {volume} {131}},\ \bibinfo {pages} {220404} (\bibinfo {year} {2023})}\BibitemShut {NoStop}%
\bibitem [{\citenamefont {Jonay}\ \emph {et~al.}(2018)\citenamefont {Jonay}, \citenamefont {Huse},\ and\ \citenamefont {Nahum}}]{Jonay2018}%
  \BibitemOpen
  \bibfield  {author} {\bibinfo {author} {\bibfnamefont {C.}~\bibnamefont {Jonay}}, \bibinfo {author} {\bibfnamefont {D.~A.}\ \bibnamefont {Huse}}, \ and\ \bibinfo {author} {\bibfnamefont {A.}~\bibnamefont {Nahum}},\ }\bibfield  {title} {{Coarse-grained dynamics of operator and state entanglement},\ }\href {http://arxiv.org/abs/1803.00089} {\bibfield  {journal} {\bibinfo  {journal} {arXiv:1803.00089}\ } (\bibinfo {year} {2018})}\BibitemShut {NoStop}%
\bibitem [{\citenamefont {Zhou}\ and\ \citenamefont {Nahum}(2020)}]{Zhou2020c}%
  \BibitemOpen
  \bibfield  {author} {\bibinfo {author} {\bibfnamefont {T.}~\bibnamefont {Zhou}}\ and\ \bibinfo {author} {\bibfnamefont {A.}~\bibnamefont {Nahum}},\ }\bibfield  {title} {{Entanglement Membrane in Chaotic Many-Body Systems},\ }\href {\doibase 10.1103/PhysRevX.10.031066} {\bibfield  {journal} {\bibinfo  {journal} {Phys. Rev. X}\ }\textbf {\bibinfo {volume} {10}},\ \bibinfo {pages} {031066} (\bibinfo {year} {2020})}\BibitemShut {NoStop}%
\bibitem [{\citenamefont {Sierant}\ \emph {et~al.}(2023)\citenamefont {Sierant}, \citenamefont {Schir{\`{o}}}, \citenamefont {Lewenstein},\ and\ \citenamefont {Turkeshi}}]{Sierant2023}%
  \BibitemOpen
  \bibfield  {author} {\bibinfo {author} {\bibfnamefont {P.}~\bibnamefont {Sierant}}, \bibinfo {author} {\bibfnamefont {M.}~\bibnamefont {Schir{\`{o}}}}, \bibinfo {author} {\bibfnamefont {M.}~\bibnamefont {Lewenstein}}, \ and\ \bibinfo {author} {\bibfnamefont {X.}~\bibnamefont {Turkeshi}},\ }\bibfield  {title} {{Entanglement Growth and Minimal Membranes in (d+1) Random Unitary Circuits},\ }\href {\doibase 10.1103/PhysRevLett.131.230403} {\bibfield  {journal} {\bibinfo  {journal} {Phys. Rev. Lett.}\ }\textbf {\bibinfo {volume} {131}},\ \bibinfo {pages} {230403} (\bibinfo {year} {2023})}\BibitemShut {NoStop}%
\bibitem [{\citenamefont {Skinner}\ \emph {et~al.}(2019{\natexlab{b}})\citenamefont {Skinner}, \citenamefont {Ruhman},\ and\ \citenamefont {Nahum}}]{Skinner2019a}%
  \BibitemOpen
  \bibfield  {author} {\bibinfo {author} {\bibfnamefont {B.}~\bibnamefont {Skinner}}, \bibinfo {author} {\bibfnamefont {J.}~\bibnamefont {Ruhman}}, \ and\ \bibinfo {author} {\bibfnamefont {A.}~\bibnamefont {Nahum}},\ }\bibfield  {title} {{Measurement-Induced Phase Transitions in the Dynamics of Entanglement},\ }\href {\doibase 10.1103/PhysRevX.9.031009} {\bibfield  {journal} {\bibinfo  {journal} {Physical Review X}\ }\textbf {\bibinfo {volume} {9}},\ \bibinfo {pages} {031009} (\bibinfo {year} {2019}{\natexlab{b}})}\BibitemShut {NoStop}%
\bibitem [{\citenamefont {Zhou}\ and\ \citenamefont {Nahum}(2019)}]{Zhou2019}%
  \BibitemOpen
  \bibfield  {author} {\bibinfo {author} {\bibfnamefont {T.}~\bibnamefont {Zhou}}\ and\ \bibinfo {author} {\bibfnamefont {A.}~\bibnamefont {Nahum}},\ }\bibfield  {title} {{Emergent statistical mechanics of entanglement in random unitary circuits},\ }\href {\doibase 10.1103/PhysRevB.99.174205} {\bibfield  {journal} {\bibinfo  {journal} {Physical Review B}\ }\textbf {\bibinfo {volume} {99}},\ \bibinfo {pages} {174205} (\bibinfo {year} {2019})}\BibitemShut {NoStop}%
\bibitem [{\citenamefont {Bao}\ \emph {et~al.}(2020)\citenamefont {Bao}, \citenamefont {Choi},\ and\ \citenamefont {Altman}}]{Bao2020}%
  \BibitemOpen
  \bibfield  {author} {\bibinfo {author} {\bibfnamefont {Y.}~\bibnamefont {Bao}}, \bibinfo {author} {\bibfnamefont {S.}~\bibnamefont {Choi}}, \ and\ \bibinfo {author} {\bibfnamefont {E.}~\bibnamefont {Altman}},\ }\bibfield  {title} {{Theory of the phase transition in random unitary circuits with measurements},\ }\href {\doibase 10.1103/PhysRevB.101.104301} {\bibfield  {journal} {\bibinfo  {journal} {Physical Review B}\ }\textbf {\bibinfo {volume} {101}},\ \bibinfo {pages} {104301} (\bibinfo {year} {2020})}\BibitemShut {NoStop}%
\bibitem [{\citenamefont {Jian}\ \emph {et~al.}(2020)\citenamefont {Jian}, \citenamefont {You}, \citenamefont {Vasseur},\ and\ \citenamefont {Ludwig}}]{Jian2020a}%
  \BibitemOpen
  \bibfield  {author} {\bibinfo {author} {\bibfnamefont {C.-M.}\ \bibnamefont {Jian}}, \bibinfo {author} {\bibfnamefont {Y.-Z.}\ \bibnamefont {You}}, \bibinfo {author} {\bibfnamefont {R.}~\bibnamefont {Vasseur}}, \ and\ \bibinfo {author} {\bibfnamefont {A.~W.~W.}\ \bibnamefont {Ludwig}},\ }\bibfield  {title} {{Measurement-induced criticality in random quantum circuits},\ }\href {\doibase 10.1103/PhysRevB.101.104302} {\bibfield  {journal} {\bibinfo  {journal} {Physical Review B}\ }\textbf {\bibinfo {volume} {101}},\ \bibinfo {pages} {104302} (\bibinfo {year} {2020})}\BibitemShut {NoStop}%
\bibitem [{\citenamefont {Collins}\ and\ \citenamefont {Śniady}(2006)}]{collinsIntegrationRespectHaar2006}%
  \BibitemOpen
  \bibfield  {author} {\bibinfo {author} {\bibfnamefont {B.}~\bibnamefont {Collins}}\ and\ \bibinfo {author} {\bibfnamefont {P.}~\bibnamefont {Śniady}},\ }\bibfield  {title} {Integration with {Respect} to the {Haar} {Measure} on {Unitary}, {Orthogonal} and {Symplectic} {Group},\ }\href {\doibase 10.1007/s00220-006-1554-3} {\bibfield  {journal} {\bibinfo  {journal} {Communications in Mathematical Physics}\ }\textbf {\bibinfo {volume} {264}},\ \bibinfo {pages} {773} (\bibinfo {year} {2006})}\BibitemShut {NoStop}%
\bibitem [{\citenamefont {Dong}\ \emph {et~al.}(2021)\citenamefont {Dong}, \citenamefont {Qi},\ and\ \citenamefont {Walter}}]{Uaverage_Qi}%
  \BibitemOpen
  \bibfield  {author} {\bibinfo {author} {\bibfnamefont {X.}~\bibnamefont {Dong}}, \bibinfo {author} {\bibfnamefont {X.-L.}\ \bibnamefont {Qi}}, \ and\ \bibinfo {author} {\bibfnamefont {M.}~\bibnamefont {Walter}},\ }\bibfield  {title} {Holographic entanglement negativity and replica symmetry breaking,\ }\href {\doibase 10.1007/JHEP06(2021)024} {\bibfield  {journal} {\bibinfo  {journal} {Journal of High Energy Physics}\ }\textbf {\bibinfo {volume} {2021}},\ \bibinfo {pages} {24} (\bibinfo {year} {2021})}\BibitemShut {NoStop}%
\bibitem [{\citenamefont {Li}\ \emph {et~al.}(2023)\citenamefont {Li}, \citenamefont {Vijay},\ and\ \citenamefont {Fisher}}]{PRXQuantum.4.010331}%
  \BibitemOpen
  \bibfield  {author} {\bibinfo {author} {\bibfnamefont {Y.}~\bibnamefont {Li}}, \bibinfo {author} {\bibfnamefont {S.}~\bibnamefont {Vijay}}, \ and\ \bibinfo {author} {\bibfnamefont {M.~P.}\ \bibnamefont {Fisher}},\ }\bibfield  {title} {Entanglement domain walls in monitored quantum circuits and the directed polymer in a random environment,\ }\href {\doibase 10.1103/PRXQuantum.4.010331} {\bibfield  {journal} {\bibinfo  {journal} {PRX Quantum}\ }\textbf {\bibinfo {volume} {4}},\ \bibinfo {pages} {010331} (\bibinfo {year} {2023})}\BibitemShut {NoStop}%
\bibitem [{\citenamefont {Nishimori}(2001)}]{nishimori2001statistical}%
  \BibitemOpen
  \bibfield  {author} {\bibinfo {author} {\bibfnamefont {H.}~\bibnamefont {Nishimori}},\ }Statistical physics of spin glasses and information processing: an introduction,\ \bibinfo {number} {111}\ (\bibinfo  {publisher} {Clarendon Press},\ \bibinfo {year} {2001})\BibitemShut {NoStop}%
\bibitem [{\citenamefont {Kardar}(2007)}]{kardar2007statistical}%
  \BibitemOpen
  \bibfield  {author} {\bibinfo {author} {\bibfnamefont {M.}~\bibnamefont {Kardar}},\ }Statistical physics of fields\ (\bibinfo  {publisher} {Cambridge University Press},\ \bibinfo {year} {2007})\BibitemShut {NoStop}%
\bibitem [{\citenamefont {Aaronson}\ and\ \citenamefont {Gottesman}(2004)}]{Aaronson2004a}%
  \BibitemOpen
  \bibfield  {author} {\bibinfo {author} {\bibfnamefont {S.}~\bibnamefont {Aaronson}}\ and\ \bibinfo {author} {\bibfnamefont {D.}~\bibnamefont {Gottesman}},\ }\bibfield  {title} {{Improved simulation of stabilizer circuits},\ }\href {\doibase 10.1103/PhysRevA.70.052328} {\bibfield  {journal} {\bibinfo  {journal} {Physical Review A}\ }\textbf {\bibinfo {volume} {70}},\ \bibinfo {pages} {052328} (\bibinfo {year} {2004})}\BibitemShut {NoStop}%
\bibitem [{\citenamefont {Harper}(1955)}]{Harper1955}%
  \BibitemOpen
  \bibfield  {author} {\bibinfo {author} {\bibfnamefont {P.~G.}\ \bibnamefont {Harper}},\ }\bibfield  {title} {{Single Band Motion of Conduction Electrons in a Uniform Magnetic Field},\ }\href {\doibase 10.1088/0370-1298/68/10/304} {\bibfield  {journal} {\bibinfo  {journal} {Proc. Phys. Soc. Sect. A}\ }\textbf {\bibinfo {volume} {68}},\ \bibinfo {pages} {874} (\bibinfo {year} {1955})}\BibitemShut {NoStop}%
\bibitem [{\citenamefont {Aubry}\ and\ \citenamefont {Andr{\'{e}}}(1980)}]{Aubry1980}%
  \BibitemOpen
  \bibfield  {author} {\bibinfo {author} {\bibfnamefont {S.}~\bibnamefont {Aubry}}\ and\ \bibinfo {author} {\bibfnamefont {G.}~\bibnamefont {Andr{\'{e}}}},\ }\bibfield  {title} {{Analyticity breaking and Anderson localization in incommensurate lattices},\ }\href@noop {} {\bibfield  {journal} {\bibinfo  {journal} {Ann. Isr. Phys. Soc}\ }\textbf {\bibinfo {volume} {3}},\ \bibinfo {pages} {133} (\bibinfo {year} {1980})}\BibitemShut {NoStop}%
\bibitem [{\citenamefont {Iyer}\ \emph {et~al.}(2013)\citenamefont {Iyer}, \citenamefont {Oganesyan}, \citenamefont {Refael},\ and\ \citenamefont {Huse}}]{Iyer2013}%
  \BibitemOpen
  \bibfield  {author} {\bibinfo {author} {\bibfnamefont {S.}~\bibnamefont {Iyer}}, \bibinfo {author} {\bibfnamefont {V.}~\bibnamefont {Oganesyan}}, \bibinfo {author} {\bibfnamefont {G.}~\bibnamefont {Refael}}, \ and\ \bibinfo {author} {\bibfnamefont {D.~A.}\ \bibnamefont {Huse}},\ }\bibfield  {title} {{Many-body localization in a quasiperiodic system},\ }\href {\doibase 10.1103/PhysRevB.87.134202} {\bibfield  {journal} {\bibinfo  {journal} {Physical Review B}\ }\textbf {\bibinfo {volume} {87}},\ \bibinfo {pages} {134202} (\bibinfo {year} {2013})}\BibitemShut {NoStop}%
\bibitem [{\citenamefont {Khemani}\ \emph {et~al.}(2017)\citenamefont {Khemani}, \citenamefont {Sheng},\ and\ \citenamefont {Huse}}]{Khemani2017}%
  \BibitemOpen
  \bibfield  {author} {\bibinfo {author} {\bibfnamefont {V.}~\bibnamefont {Khemani}}, \bibinfo {author} {\bibfnamefont {D.~N.}\ \bibnamefont {Sheng}}, \ and\ \bibinfo {author} {\bibfnamefont {D.~A.}\ \bibnamefont {Huse}},\ }\bibfield  {title} {{Two Universality Classes for the Many-Body Localization Transition},\ }\href {\doibase 10.1103/PhysRevLett.119.075702} {\bibfield  {journal} {\bibinfo  {journal} {Physical Review Letters}\ }\textbf {\bibinfo {volume} {119}},\ \bibinfo {pages} {075702} (\bibinfo {year} {2017})}\BibitemShut {NoStop}%
\bibitem [{\citenamefont {Zhang}\ and\ \citenamefont {Yao}(2018)}]{Zhang2018}%
  \BibitemOpen
  \bibfield  {author} {\bibinfo {author} {\bibfnamefont {S.~X.}\ \bibnamefont {Zhang}}\ and\ \bibinfo {author} {\bibfnamefont {H.}~\bibnamefont {Yao}},\ }\bibfield  {title} {{Universal Properties of Many-Body Localization Transitions in Quasiperiodic Systems},\ }\href {\doibase 10.1103/PhysRevLett.121.206601} {\bibfield  {journal} {\bibinfo  {journal} {Physical Review Letters}\ }\textbf {\bibinfo {volume} {121}},\ \bibinfo {pages} {206601} (\bibinfo {year} {2018})}\BibitemShut {NoStop}%
\bibitem [{\citenamefont {Zhang}\ and\ \citenamefont {Yao}(2019)}]{Zhang2019a}%
  \BibitemOpen
  \bibfield  {author} {\bibinfo {author} {\bibfnamefont {S.-X.}\ \bibnamefont {Zhang}}\ and\ \bibinfo {author} {\bibfnamefont {H.}~\bibnamefont {Yao}},\ }\bibfield  {title} {{Strong and Weak Many-Body Localizations},\ }\href {http://arxiv.org/abs/1906.00971} {\bibfield  {journal} {\bibinfo  {journal} {arXiv:1906.00971}\ } (\bibinfo {year} {2019})}\BibitemShut {NoStop}%
\bibitem [{\citenamefont {Qi}\ and\ \citenamefont {Zhang}(2011)}]{Qi2011}%
  \BibitemOpen
  \bibfield  {author} {\bibinfo {author} {\bibfnamefont {X.~L.}\ \bibnamefont {Qi}}\ and\ \bibinfo {author} {\bibfnamefont {S.~C.}\ \bibnamefont {Zhang}},\ }\bibfield  {title} {{Topological insulators and superconductors},\ }\href@noop {} {\bibfield  {journal} {\bibinfo  {journal} {Reviews of Modern Physics}\ }\textbf {\bibinfo {volume} {83}},\ \bibinfo {pages} {1057} (\bibinfo {year} {2011})}\BibitemShut {NoStop}%
\bibitem [{\citenamefont {Su}\ \emph {et~al.}(1980)\citenamefont {Su}, \citenamefont {Schrieffer},\ and\ \citenamefont {Heeger}}]{Su1980}%
  \BibitemOpen
  \bibfield  {author} {\bibinfo {author} {\bibfnamefont {W.~P.}\ \bibnamefont {Su}}, \bibinfo {author} {\bibfnamefont {J.~R.}\ \bibnamefont {Schrieffer}}, \ and\ \bibinfo {author} {\bibfnamefont {A.~J.}\ \bibnamefont {Heeger}},\ }\bibfield  {title} {{Soliton excitations in polyacetylene},\ }\href {\doibase 10.1103/PhysRevB.22.2099} {\bibfield  {journal} {\bibinfo  {journal} {Physical Review B}\ }\textbf {\bibinfo {volume} {22}},\ \bibinfo {pages} {2099} (\bibinfo {year} {1980})}\BibitemShut {NoStop}%
\bibitem [{\citenamefont {Rakovszky}\ \emph {et~al.}(2019)\citenamefont {Rakovszky}, \citenamefont {Pollmann},\ and\ \citenamefont {von Keyserlingk}}]{Rakovszky2019}%
  \BibitemOpen
  \bibfield  {author} {\bibinfo {author} {\bibfnamefont {T.}~\bibnamefont {Rakovszky}}, \bibinfo {author} {\bibfnamefont {F.}~\bibnamefont {Pollmann}}, \ and\ \bibinfo {author} {\bibfnamefont {C.~W.}\ \bibnamefont {von Keyserlingk}},\ }\bibfield  {title} {{Sub-ballistic Growth of R{\'{e}}nyi Entropies due to Diffusion},\ }\href {\doibase 10.1103/PhysRevLett.122.250602} {\bibfield  {journal} {\bibinfo  {journal} {Physical Review Letters}\ }\textbf {\bibinfo {volume} {122}},\ \bibinfo {pages} {250602} (\bibinfo {year} {2019})}\BibitemShut {NoStop}%
\bibitem [{\citenamefont {Huang}(2020)}]{Huang2020d}%
  \BibitemOpen
  \bibfield  {author} {\bibinfo {author} {\bibfnamefont {Y.}~\bibnamefont {Huang}},\ }\bibfield  {title} {{Dynamics of R{\'{e}}nyi entanglement entropy in diffusive qudit systems},\ }\href {\doibase 10.1088/2633-1357/abd1e2} {\bibfield  {journal} {\bibinfo  {journal} {IOP SciNotes}\ }\textbf {\bibinfo {volume} {1}},\ \bibinfo {pages} {035205} (\bibinfo {year} {2020})}\BibitemShut {NoStop}%
\bibitem [{\citenamefont {Zhou}\ and\ \citenamefont {Ludwig}(2020)}]{Zhou2020c_renyi}%
  \BibitemOpen
  \bibfield  {author} {\bibinfo {author} {\bibfnamefont {T.}~\bibnamefont {Zhou}}\ and\ \bibinfo {author} {\bibfnamefont {A.~W.~W.}\ \bibnamefont {Ludwig}},\ }\bibfield  {title} {{Diffusive scaling of R{\'{e}}nyi entanglement entropy},\ }\href {\doibase 10.1103/PhysRevResearch.2.033020} {\bibfield  {journal} {\bibinfo  {journal} {Physical Review Research}\ }\textbf {\bibinfo {volume} {2}},\ \bibinfo {pages} {033020} (\bibinfo {year} {2020})}\BibitemShut {NoStop}%
\bibitem [{\citenamefont {{\v{Z}}nidari{\v{c}}}(2020)}]{Znidaric2020}%
  \BibitemOpen
  \bibfield  {author} {\bibinfo {author} {\bibfnamefont {M.}~\bibnamefont {{\v{Z}}nidari{\v{c}}}},\ }\bibfield  {title} {{Entanglement growth in diffusive systems},\ }\href {\doibase 10.1038/s42005-020-0366-7} {\bibfield  {journal} {\bibinfo  {journal} {Communications Physics}\ }\textbf {\bibinfo {volume} {3}},\ \bibinfo {pages} {100} (\bibinfo {year} {2020})}\BibitemShut {NoStop}%
\bibitem [{\citenamefont {Liu}\ \emph {et~al.}(2024{\natexlab{c}})\citenamefont {Liu}, \citenamefont {Zhang}, \citenamefont {Yin},\ and\ \citenamefont {Zhang}}]{Liu2024a}%
  \BibitemOpen
  \bibfield  {author} {\bibinfo {author} {\bibfnamefont {S.}~\bibnamefont {Liu}}, \bibinfo {author} {\bibfnamefont {H.-K.}\ \bibnamefont {Zhang}}, \bibinfo {author} {\bibfnamefont {S.}~\bibnamefont {Yin}}, \ and\ \bibinfo {author} {\bibfnamefont {S.-X.}\ \bibnamefont {Zhang}},\ }\bibfield  {title} {Symmetry restoration and quantum mpemba effect in symmetric random circuits,\ }\href {\doibase 10.1103/PhysRevLett.133.140405} {\bibfield  {journal} {\bibinfo  {journal} {Physical Review Letters}\ }\textbf {\bibinfo {volume} {133}},\ \bibinfo {pages} {140405} (\bibinfo {year} {2024}{\natexlab{c}})}\BibitemShut {NoStop}%
\bibitem [{\citenamefont {Agrawal}\ \emph {et~al.}(2022)\citenamefont {Agrawal}, \citenamefont {Zabalo}, \citenamefont {Chen}, \citenamefont {Wilson}, \citenamefont {Potter}, \citenamefont {Pixley}, \citenamefont {Gopalakrishnan},\ and\ \citenamefont {Vasseur}}]{PhysRevX.12.041002}%
  \BibitemOpen
  \bibfield  {author} {\bibinfo {author} {\bibfnamefont {U.}~\bibnamefont {Agrawal}}, \bibinfo {author} {\bibfnamefont {A.}~\bibnamefont {Zabalo}}, \bibinfo {author} {\bibfnamefont {K.}~\bibnamefont {Chen}}, \bibinfo {author} {\bibfnamefont {J.~H.}\ \bibnamefont {Wilson}}, \bibinfo {author} {\bibfnamefont {A.~C.}\ \bibnamefont {Potter}}, \bibinfo {author} {\bibfnamefont {J.~H.}\ \bibnamefont {Pixley}}, \bibinfo {author} {\bibfnamefont {S.}~\bibnamefont {Gopalakrishnan}}, \ and\ \bibinfo {author} {\bibfnamefont {R.}~\bibnamefont {Vasseur}},\ }\bibfield  {title} {Entanglement and charge-sharpening transitions in u(1) symmetric monitored quantum circuits,\ }\href {\doibase 10.1103/PhysRevX.12.041002} {\bibfield  {journal} {\bibinfo  {journal} {Phys. Rev. X}\ }\textbf {\bibinfo {volume} {12}},\ \bibinfo {pages} {041002} (\bibinfo {year} {2022})}\BibitemShut {NoStop}%
\bibitem [{\citenamefont {Calabrese}(2020)}]{Calabrese2020}%
  \BibitemOpen
  \bibfield  {author} {\bibinfo {author} {\bibfnamefont {P.}~\bibnamefont {Calabrese}},\ }\bibfield  {title} {{Entanglement spreading in non-equilibrium integrable systems},\ }\href {\doibase 10.21468/SciPostPhysLectNotes.20} {\bibfield  {journal} {\bibinfo  {journal} {SciPost Physics Lecture Notes}\ ,\ \bibinfo {pages} {20}} (\bibinfo {year} {2020})}\BibitemShut {NoStop}%
\bibitem [{\citenamefont {Orito}\ and\ \citenamefont {Imura}(2023)}]{Orito2023_z}%
  \BibitemOpen
  \bibfield  {author} {\bibinfo {author} {\bibfnamefont {T.}~\bibnamefont {Orito}}\ and\ \bibinfo {author} {\bibfnamefont {K.-I.}\ \bibnamefont {Imura}},\ }\bibfield  {title} {{Entanglement dynamics in the many-body Hatano-Nelson model},\ }\href {\doibase 10.1103/PhysRevB.108.214308} {\bibfield  {journal} {\bibinfo  {journal} {Phys. Rev. B}\ }\textbf {\bibinfo {volume} {108}},\ \bibinfo {pages} {214308} (\bibinfo {year} {2023})}\BibitemShut {NoStop}%
\bibitem [{\citenamefont {Modak}\ \emph {et~al.}(2020)\citenamefont {Modak}, \citenamefont {Alba},\ and\ \citenamefont {Calabrese}}]{Modak2020_z}%
  \BibitemOpen
  \bibfield  {author} {\bibinfo {author} {\bibfnamefont {R.}~\bibnamefont {Modak}}, \bibinfo {author} {\bibfnamefont {V.}~\bibnamefont {Alba}}, \ and\ \bibinfo {author} {\bibfnamefont {P.}~\bibnamefont {Calabrese}},\ }\bibfield  {title} {{Entanglement revivals as a probe of scrambling in finite quantum systems},\ }\href {\doibase 10.1088/1742-5468/aba9d9} {\bibfield  {journal} {\bibinfo  {journal} {J. Stat. Mech. Theory Exp.}\ }\textbf {\bibinfo {volume} {2020}},\ \bibinfo {pages} {083110} (\bibinfo {year} {2020})}\BibitemShut {NoStop}%
\bibitem [{\citenamefont {Anderson}(1958)}]{Anderson1958}%
  \BibitemOpen
  \bibfield  {author} {\bibinfo {author} {\bibfnamefont {P.~W.}\ \bibnamefont {Anderson}},\ }\bibfield  {title} {{Absence of diffusion in certain random lattices},\ }\href {\doibase 10.1103/PhysRev.109.1492} {\bibfield  {journal} {\bibinfo  {journal} {Physical Review}\ }\textbf {\bibinfo {volume} {109}},\ \bibinfo {pages} {1492} (\bibinfo {year} {1958})}\BibitemShut {NoStop}%
\bibitem [{\citenamefont {Basko}\ \emph {et~al.}(2006)\citenamefont {Basko}, \citenamefont {Aleiner},\ and\ \citenamefont {Altshuler}}]{Basko2006}%
  \BibitemOpen
  \bibfield  {author} {\bibinfo {author} {\bibfnamefont {D.~M.}\ \bibnamefont {Basko}}, \bibinfo {author} {\bibfnamefont {I.~L.}\ \bibnamefont {Aleiner}}, \ and\ \bibinfo {author} {\bibfnamefont {B.~L.}\ \bibnamefont {Altshuler}},\ }\bibfield  {title} {{Metal-insulator transition in a weakly interacting many-electron system with localized single-particle states},\ }\href {\doibase 10.1016/j.aop.2005.11.014} {\bibfield  {journal} {\bibinfo  {journal} {Annals of Physics}\ }\textbf {\bibinfo {volume} {321}},\ \bibinfo {pages} {1126} (\bibinfo {year} {2006})}\BibitemShut {NoStop}%
\bibitem [{\citenamefont {Oganesyan}\ and\ \citenamefont {Huse}(2007)}]{Oganesyan2007}%
  \BibitemOpen
  \bibfield  {author} {\bibinfo {author} {\bibfnamefont {V.}~\bibnamefont {Oganesyan}}\ and\ \bibinfo {author} {\bibfnamefont {D.~A.}\ \bibnamefont {Huse}},\ }\bibfield  {title} {{Localization of interacting fermions at high temperature},\ }\href {\doibase 10.1103/PhysRevB.75.155111} {\bibfield  {journal} {\bibinfo  {journal} {Physical Review B}\ }\textbf {\bibinfo {volume} {75}},\ \bibinfo {pages} {155111} (\bibinfo {year} {2007})}\BibitemShut {NoStop}%
\bibitem [{\citenamefont {Pal}\ and\ \citenamefont {Huse}(2010)}]{Pal2010a}%
  \BibitemOpen
  \bibfield  {author} {\bibinfo {author} {\bibfnamefont {A.}~\bibnamefont {Pal}}\ and\ \bibinfo {author} {\bibfnamefont {D.~A.}\ \bibnamefont {Huse}},\ }\bibfield  {title} {{Many-body localization phase transition},\ }\href {\doibase 10.1103/PhysRevB.82.174411} {\bibfield  {journal} {\bibinfo  {journal} {Physical Review B}\ }\textbf {\bibinfo {volume} {82}},\ \bibinfo {pages} {174411} (\bibinfo {year} {2010})}\BibitemShut {NoStop}%
\bibitem [{\citenamefont {Altman}\ and\ \citenamefont {Vosk}(2015)}]{Altman2014}%
  \BibitemOpen
  \bibfield  {author} {\bibinfo {author} {\bibfnamefont {E.}~\bibnamefont {Altman}}\ and\ \bibinfo {author} {\bibfnamefont {R.}~\bibnamefont {Vosk}},\ }\bibfield  {title} {{Universal Dynamics and Renormalization in Many-Body-Localized Systems},\ }\href {\doibase 10.1146/annurev-conmatphys-031214-014701} {\bibfield  {journal} {\bibinfo  {journal} {Annual Review of Condensed Matter Physics}\ }\textbf {\bibinfo {volume} {6}},\ \bibinfo {pages} {383} (\bibinfo {year} {2015})}\BibitemShut {NoStop}%
\bibitem [{\citenamefont {Nandkishore}\ and\ \citenamefont {Huse}(2015)}]{Nandkishore2015}%
  \BibitemOpen
  \bibfield  {author} {\bibinfo {author} {\bibfnamefont {R.}~\bibnamefont {Nandkishore}}\ and\ \bibinfo {author} {\bibfnamefont {D.~A.}\ \bibnamefont {Huse}},\ }\bibfield  {title} {{Many body localization and thermalization in quantum statistical mechanics},\ }\href {\doibase 10.1146/annurev-conmatphys-031214-014726} {\bibfield  {journal} {\bibinfo  {journal} {Annual Review of Condensed Matter Physics}\ }\textbf {\bibinfo {volume} {6}},\ \bibinfo {pages} {15} (\bibinfo {year} {2015})}\BibitemShut {NoStop}%
\bibitem [{\citenamefont {Abanin}\ \emph {et~al.}(2019)\citenamefont {Abanin}, \citenamefont {Altman}, \citenamefont {Bloch},\ and\ \citenamefont {Serbyn}}]{Abanin2018}%
  \BibitemOpen
  \bibfield  {author} {\bibinfo {author} {\bibfnamefont {D.~A.}\ \bibnamefont {Abanin}}, \bibinfo {author} {\bibfnamefont {E.}~\bibnamefont {Altman}}, \bibinfo {author} {\bibfnamefont {I.}~\bibnamefont {Bloch}}, \ and\ \bibinfo {author} {\bibfnamefont {M.}~\bibnamefont {Serbyn}},\ }\bibfield  {title} {{Colloquium : Many-body localization, thermalization, and entanglement},\ }\href {\doibase 10.1103/RevModPhys.91.021001} {\bibfield  {journal} {\bibinfo  {journal} {Reviews of Modern Physics}\ }\textbf {\bibinfo {volume} {91}},\ \bibinfo {pages} {021001} (\bibinfo {year} {2019})}\BibitemShut {NoStop}%
\bibitem [{\citenamefont {Kohlert}\ \emph {et~al.}(2019)\citenamefont {Kohlert}, \citenamefont {Scherg}, \citenamefont {Li}, \citenamefont {L{\"{u}}schen}, \citenamefont {{Das Sarma}}, \citenamefont {Bloch},\ and\ \citenamefont {Aidelsburger}}]{Kohlert}%
  \BibitemOpen
  \bibfield  {author} {\bibinfo {author} {\bibfnamefont {T.}~\bibnamefont {Kohlert}}, \bibinfo {author} {\bibfnamefont {S.}~\bibnamefont {Scherg}}, \bibinfo {author} {\bibfnamefont {X.}~\bibnamefont {Li}}, \bibinfo {author} {\bibfnamefont {H.~P.}\ \bibnamefont {L{\"{u}}schen}}, \bibinfo {author} {\bibfnamefont {S.}~\bibnamefont {{Das Sarma}}}, \bibinfo {author} {\bibfnamefont {I.}~\bibnamefont {Bloch}}, \ and\ \bibinfo {author} {\bibfnamefont {M.}~\bibnamefont {Aidelsburger}},\ }\bibfield  {title} {{Observation of Many-Body Localization in a One-Dimensional System with a Single-Particle Mobility Edge},\ }\href {\doibase 10.1103/PhysRevLett.122.170403} {\bibfield  {journal} {\bibinfo  {journal} {Physical Review Letters}\ }\textbf {\bibinfo {volume} {122}},\ \bibinfo {pages} {170403} (\bibinfo {year} {2019})}\BibitemShut {NoStop}%
\bibitem [{\citenamefont {Schulz}\ \emph {et~al.}(2019)\citenamefont {Schulz}, \citenamefont {Hooley}, \citenamefont {Moessner},\ and\ \citenamefont {Pollmann}}]{Schulz2019a}%
  \BibitemOpen
  \bibfield  {author} {\bibinfo {author} {\bibfnamefont {M.}~\bibnamefont {Schulz}}, \bibinfo {author} {\bibfnamefont {C.~A.}\ \bibnamefont {Hooley}}, \bibinfo {author} {\bibfnamefont {R.}~\bibnamefont {Moessner}}, \ and\ \bibinfo {author} {\bibfnamefont {F.}~\bibnamefont {Pollmann}},\ }\bibfield  {title} {{Stark Many-Body Localization},\ }\href {\doibase 10.1103/PhysRevLett.122.040606} {\bibfield  {journal} {\bibinfo  {journal} {Physical Review Letters}\ }\textbf {\bibinfo {volume} {122}},\ \bibinfo {pages} {040606} (\bibinfo {year} {2019})}\BibitemShut {NoStop}%
\bibitem [{\citenamefont {van Nieuwenburg}\ \emph {et~al.}(2019)\citenamefont {van Nieuwenburg}, \citenamefont {Baum},\ and\ \citenamefont {Refael}}]{VanNieuwenburg2019}%
  \BibitemOpen
  \bibfield  {author} {\bibinfo {author} {\bibfnamefont {E.}~\bibnamefont {van Nieuwenburg}}, \bibinfo {author} {\bibfnamefont {Y.}~\bibnamefont {Baum}}, \ and\ \bibinfo {author} {\bibfnamefont {G.}~\bibnamefont {Refael}},\ }\bibfield  {title} {{From Bloch oscillations to many-body localization in clean interacting systems},\ }\href {\doibase 10.1073/pnas.1819316116} {\bibfield  {journal} {\bibinfo  {journal} {Proceedings of the National Academy of Sciences}\ }\textbf {\bibinfo {volume} {116}},\ \bibinfo {pages} {9269} (\bibinfo {year} {2019})}\BibitemShut {NoStop}%
\bibitem [{\citenamefont {Khemani}\ \emph {et~al.}(2020)\citenamefont {Khemani}, \citenamefont {Hermele},\ and\ \citenamefont {Nandkishore}}]{Khemani2020}%
  \BibitemOpen
  \bibfield  {author} {\bibinfo {author} {\bibfnamefont {V.}~\bibnamefont {Khemani}}, \bibinfo {author} {\bibfnamefont {M.}~\bibnamefont {Hermele}}, \ and\ \bibinfo {author} {\bibfnamefont {R.}~\bibnamefont {Nandkishore}},\ }\bibfield  {title} {{Localization from Hilbert space shattering: From theory to physical realizations},\ }\href {\doibase 10.1103/PhysRevB.101.174204} {\bibfield  {journal} {\bibinfo  {journal} {Physical Review B}\ }\textbf {\bibinfo {volume} {101}},\ \bibinfo {pages} {174204} (\bibinfo {year} {2020})}\BibitemShut {NoStop}%
\bibitem [{\citenamefont {Doggen}\ \emph {et~al.}(2021)\citenamefont {Doggen}, \citenamefont {Gornyi},\ and\ \citenamefont {Polyakov}}]{Doggen2021a}%
  \BibitemOpen
  \bibfield  {author} {\bibinfo {author} {\bibfnamefont {E.~V.~H.}\ \bibnamefont {Doggen}}, \bibinfo {author} {\bibfnamefont {I.~V.}\ \bibnamefont {Gornyi}}, \ and\ \bibinfo {author} {\bibfnamefont {D.~G.}\ \bibnamefont {Polyakov}},\ }\bibfield  {title} {{Stark many-body localization: Evidence for Hilbert-space shattering},\ }\href {\doibase 10.1103/PhysRevB.103.L100202} {\bibfield  {journal} {\bibinfo  {journal} {Physical Review B}\ }\textbf {\bibinfo {volume} {103}},\ \bibinfo {pages} {L100202} (\bibinfo {year} {2021})}\BibitemShut {NoStop}%
\bibitem [{\citenamefont {Liu}\ \emph {et~al.}(2023{\natexlab{b}})\citenamefont {Liu}, \citenamefont {Zhang}, \citenamefont {Hsieh}, \citenamefont {Zhang},\ and\ \citenamefont {Yao}}]{Liu2022}%
  \BibitemOpen
  \bibfield  {author} {\bibinfo {author} {\bibfnamefont {S.}~\bibnamefont {Liu}}, \bibinfo {author} {\bibfnamefont {S.-X.}\ \bibnamefont {Zhang}}, \bibinfo {author} {\bibfnamefont {C.-Y.}\ \bibnamefont {Hsieh}}, \bibinfo {author} {\bibfnamefont {S.}~\bibnamefont {Zhang}}, \ and\ \bibinfo {author} {\bibfnamefont {H.}~\bibnamefont {Yao}},\ }\bibfield  {title} {{Discrete Time Crystal Enabled by Stark Many-Body Localization},\ }\href {\doibase 10.1103/PhysRevLett.130.120403} {\bibfield  {journal} {\bibinfo  {journal} {Physical Review Letters}\ }\textbf {\bibinfo {volume} {130}},\ \bibinfo {pages} {120403} (\bibinfo {year} {2023}{\natexlab{b}})}\BibitemShut {NoStop}%
\bibitem [{\citenamefont {Else}\ \emph {et~al.}(2020)\citenamefont {Else}, \citenamefont {Monroe}, \citenamefont {Nayak},\ and\ \citenamefont {Yao}}]{Else2019}%
  \BibitemOpen
  \bibfield  {author} {\bibinfo {author} {\bibfnamefont {D.~V.}\ \bibnamefont {Else}}, \bibinfo {author} {\bibfnamefont {C.}~\bibnamefont {Monroe}}, \bibinfo {author} {\bibfnamefont {C.}~\bibnamefont {Nayak}}, \ and\ \bibinfo {author} {\bibfnamefont {N.~Y.}\ \bibnamefont {Yao}},\ }\bibfield  {title} {{Discrete Time Crystals},\ }\href {\doibase 10.1146/annurev-conmatphys-031119-050658} {\bibfield  {journal} {\bibinfo  {journal} {Annual Review of Condensed Matter Physics}\ }\textbf {\bibinfo {volume} {11}},\ \bibinfo {pages} {467} (\bibinfo {year} {2020})}\BibitemShut {NoStop}%
\bibitem [{\citenamefont {Zaletel}\ \emph {et~al.}(2023)\citenamefont {Zaletel}, \citenamefont {Lukin}, \citenamefont {Monroe}, \citenamefont {Nayak}, \citenamefont {Wilczek},\ and\ \citenamefont {Yao}}]{Zaletel2023}%
  \BibitemOpen
  \bibfield  {author} {\bibinfo {author} {\bibfnamefont {M.~P.}\ \bibnamefont {Zaletel}}, \bibinfo {author} {\bibfnamefont {M.}~\bibnamefont {Lukin}}, \bibinfo {author} {\bibfnamefont {C.}~\bibnamefont {Monroe}}, \bibinfo {author} {\bibfnamefont {C.}~\bibnamefont {Nayak}}, \bibinfo {author} {\bibfnamefont {F.}~\bibnamefont {Wilczek}}, \ and\ \bibinfo {author} {\bibfnamefont {N.~Y.}\ \bibnamefont {Yao}},\ }\bibfield  {title} {{Colloquium : Quantum and classical discrete time crystals},\ }\href {\doibase 10.1103/RevModPhys.95.031001} {\bibfield  {journal} {\bibinfo  {journal} {Reviews of Modern Physics}\ }\textbf {\bibinfo {volume} {95}},\ \bibinfo {pages} {031001} (\bibinfo {year} {2023})}\BibitemShut {NoStop}%
\bibitem [{\citenamefont {Yang}\ \emph {et~al.}(2020)\citenamefont {Yang}, \citenamefont {Liu}, \citenamefont {Gorshkov},\ and\ \citenamefont {Iadecola}}]{Yang2020b}%
  \BibitemOpen
  \bibfield  {author} {\bibinfo {author} {\bibfnamefont {Z.-c.}\ \bibnamefont {Yang}}, \bibinfo {author} {\bibfnamefont {F.}~\bibnamefont {Liu}}, \bibinfo {author} {\bibfnamefont {A.~V.}\ \bibnamefont {Gorshkov}}, \ and\ \bibinfo {author} {\bibfnamefont {T.}~\bibnamefont {Iadecola}},\ }\bibfield  {title} {{Hilbert-Space Fragmentation from Strict Confinement},\ }\href {\doibase 10.1103/PhysRevLett.124.207602} {\bibfield  {journal} {\bibinfo  {journal} {Physical Review Letters}\ }\textbf {\bibinfo {volume} {124}},\ \bibinfo {pages} {207602} (\bibinfo {year} {2020})}\BibitemShut {NoStop}%
\bibitem [{\citenamefont {Schreiber}\ \emph {et~al.}(2015)\citenamefont {Schreiber}, \citenamefont {Hodgman}, \citenamefont {Bordia}, \citenamefont {L{\"{u}}schen}, \citenamefont {Fischer}, \citenamefont {Vosk}, \citenamefont {Altman}, \citenamefont {Schneider},\ and\ \citenamefont {Bloch}}]{Schreiber2015a}%
  \BibitemOpen
  \bibfield  {author} {\bibinfo {author} {\bibfnamefont {M.}~\bibnamefont {Schreiber}}, \bibinfo {author} {\bibfnamefont {S.~S.}\ \bibnamefont {Hodgman}}, \bibinfo {author} {\bibfnamefont {P.}~\bibnamefont {Bordia}}, \bibinfo {author} {\bibfnamefont {H.~P.}\ \bibnamefont {L{\"{u}}schen}}, \bibinfo {author} {\bibfnamefont {M.~H.}\ \bibnamefont {Fischer}}, \bibinfo {author} {\bibfnamefont {R.}~\bibnamefont {Vosk}}, \bibinfo {author} {\bibfnamefont {E.}~\bibnamefont {Altman}}, \bibinfo {author} {\bibfnamefont {U.}~\bibnamefont {Schneider}}, \ and\ \bibinfo {author} {\bibfnamefont {I.}~\bibnamefont {Bloch}},\ }\bibfield  {title} {{Observation of many-body localization of interacting fermions in a quasi-random optical lattice},\ }\href {\doibase 10.1126/science.aaa7432} {\bibfield  {journal} {\bibinfo  {journal} {Science}\ }\textbf {\bibinfo {volume} {349}},\ \bibinfo {pages} {842} (\bibinfo {year} {2015})}\BibitemShut {NoStop}%
\bibitem [{\citenamefont {Smith}\ \emph {et~al.}(2016)\citenamefont {Smith}, \citenamefont {Lee}, \citenamefont {Richerme}, \citenamefont {Neyenhuis}, \citenamefont {Hess}, \citenamefont {Hauke}, \citenamefont {Heyl}, \citenamefont {Huse},\ and\ \citenamefont {Monroe}}]{Smith2016}%
  \BibitemOpen
  \bibfield  {author} {\bibinfo {author} {\bibfnamefont {J.}~\bibnamefont {Smith}}, \bibinfo {author} {\bibfnamefont {A.}~\bibnamefont {Lee}}, \bibinfo {author} {\bibfnamefont {P.}~\bibnamefont {Richerme}}, \bibinfo {author} {\bibfnamefont {B.}~\bibnamefont {Neyenhuis}}, \bibinfo {author} {\bibfnamefont {P.~W.}\ \bibnamefont {Hess}}, \bibinfo {author} {\bibfnamefont {P.}~\bibnamefont {Hauke}}, \bibinfo {author} {\bibfnamefont {M.}~\bibnamefont {Heyl}}, \bibinfo {author} {\bibfnamefont {D.~A.}\ \bibnamefont {Huse}}, \ and\ \bibinfo {author} {\bibfnamefont {C.}~\bibnamefont {Monroe}},\ }\bibfield  {title} {{Many-body localization in a quantum simulator with programmable random disorder},\ }\href {\doibase 10.1038/nphys3783} {\bibfield  {journal} {\bibinfo  {journal} {Nature Physics}\ }\textbf {\bibinfo {volume} {12}},\ \bibinfo {pages} {907} (\bibinfo {year} {2016})}\BibitemShut {NoStop}%
\bibitem [{\citenamefont {Liu}\ \emph {et~al.}(2023{\natexlab{c}})\citenamefont {Liu}, \citenamefont {Zhang}, \citenamefont {Hsieh}, \citenamefont {Zhang},\ and\ \citenamefont {Yao}}]{Liu2021c}%
  \BibitemOpen
  \bibfield  {author} {\bibinfo {author} {\bibfnamefont {S.}~\bibnamefont {Liu}}, \bibinfo {author} {\bibfnamefont {S.-X.}\ \bibnamefont {Zhang}}, \bibinfo {author} {\bibfnamefont {C.-Y.}\ \bibnamefont {Hsieh}}, \bibinfo {author} {\bibfnamefont {S.}~\bibnamefont {Zhang}}, \ and\ \bibinfo {author} {\bibfnamefont {H.}~\bibnamefont {Yao}},\ }\bibfield  {title} {{Probing many-body localization by excited-state variational quantum eigensolver},\ }\href {\doibase 10.1103/PhysRevB.107.024204} {\bibfield  {journal} {\bibinfo  {journal} {Physical Review B}\ }\textbf {\bibinfo {volume} {107}},\ \bibinfo {pages} {024204} (\bibinfo {year} {2023}{\natexlab{c}})}\BibitemShut {NoStop}%
\bibitem [{\citenamefont {Gong}\ \emph {et~al.}(2021)\citenamefont {Gong}, \citenamefont {{de Moraes Neto}}, \citenamefont {Zha}, \citenamefont {Wu}, \citenamefont {Rong}, \citenamefont {Ye}, \citenamefont {Li}, \citenamefont {Zhu}, \citenamefont {Wang}, \citenamefont {Zhao}, \citenamefont {Liang}, \citenamefont {Lin}, \citenamefont {Xu}, \citenamefont {Peng}, \citenamefont {Deng}, \citenamefont {Bayat}, \citenamefont {Zhu},\ and\ \citenamefont {Pan}}]{Gong2021}%
  \BibitemOpen
  \bibfield  {author} {\bibinfo {author} {\bibfnamefont {M.}~\bibnamefont {Gong}}, \bibinfo {author} {\bibfnamefont {G.~D.}\ \bibnamefont {{de Moraes Neto}}}, \bibinfo {author} {\bibfnamefont {C.}~\bibnamefont {Zha}}, \bibinfo {author} {\bibfnamefont {Y.}~\bibnamefont {Wu}}, \bibinfo {author} {\bibfnamefont {H.}~\bibnamefont {Rong}}, \bibinfo {author} {\bibfnamefont {Y.}~\bibnamefont {Ye}}, \bibinfo {author} {\bibfnamefont {S.}~\bibnamefont {Li}}, \bibinfo {author} {\bibfnamefont {Q.}~\bibnamefont {Zhu}}, \bibinfo {author} {\bibfnamefont {S.}~\bibnamefont {Wang}}, \bibinfo {author} {\bibfnamefont {Y.}~\bibnamefont {Zhao}}, \bibinfo {author} {\bibfnamefont {F.}~\bibnamefont {Liang}}, \bibinfo {author} {\bibfnamefont {J.}~\bibnamefont {Lin}}, \bibinfo {author} {\bibfnamefont {Y.}~\bibnamefont {Xu}}, \bibinfo {author} {\bibfnamefont {C.-z.}\ \bibnamefont {Peng}}, \bibinfo {author} {\bibfnamefont {H.}~\bibnamefont {Deng}}, \bibinfo {author} {\bibfnamefont {A.}~\bibnamefont {Bayat}}, \bibinfo {author}
  {\bibfnamefont {X.}~\bibnamefont {Zhu}}, \ and\ \bibinfo {author} {\bibfnamefont {J.-W.}\ \bibnamefont {Pan}},\ }\bibfield  {title} {{Experimental characterization of the quantum many-body localization transition},\ }\href {\doibase 10.1103/PhysRevResearch.3.033043} {\bibfield  {journal} {\bibinfo  {journal} {Physical Review Research}\ }\textbf {\bibinfo {volume} {3}},\ \bibinfo {pages} {033043} (\bibinfo {year} {2021})}\BibitemShut {NoStop}%
\bibitem [{\citenamefont {Fan}\ \emph {et~al.}(2017)\citenamefont {Fan}, \citenamefont {Zhang}, \citenamefont {Shen},\ and\ \citenamefont {Zhai}}]{Fan2017}%
  \BibitemOpen
  \bibfield  {author} {\bibinfo {author} {\bibfnamefont {R.}~\bibnamefont {Fan}}, \bibinfo {author} {\bibfnamefont {P.}~\bibnamefont {Zhang}}, \bibinfo {author} {\bibfnamefont {H.}~\bibnamefont {Shen}}, \ and\ \bibinfo {author} {\bibfnamefont {H.}~\bibnamefont {Zhai}},\ }\bibfield  {title} {{Out-of-time-order correlation for many-body localization},\ }\href {\doibase 10.1016/j.scib.2017.04.011} {\bibfield  {journal} {\bibinfo  {journal} {Science Bulletin}\ }\textbf {\bibinfo {volume} {62}},\ \bibinfo {pages} {707} (\bibinfo {year} {2017})}\BibitemShut {NoStop}%
\bibitem [{\citenamefont {Swingle}\ and\ \citenamefont {Chowdhury}(2017)}]{Swingle2017}%
  \BibitemOpen
  \bibfield  {author} {\bibinfo {author} {\bibfnamefont {B.}~\bibnamefont {Swingle}}\ and\ \bibinfo {author} {\bibfnamefont {D.}~\bibnamefont {Chowdhury}},\ }\bibfield  {title} {{Slow scrambling in disordered quantum systems},\ }\href {\doibase 10.1103/PhysRevB.95.060201} {\bibfield  {journal} {\bibinfo  {journal} {Physical Review B}\ }\textbf {\bibinfo {volume} {95}},\ \bibinfo {pages} {060201} (\bibinfo {year} {2017})}\BibitemShut {NoStop}%
\bibitem [{\citenamefont {Huang}\ \emph {et~al.}(2017)\citenamefont {Huang}, \citenamefont {Zhang},\ and\ \citenamefont {Chen}}]{Huang2017b}%
  \BibitemOpen
  \bibfield  {author} {\bibinfo {author} {\bibfnamefont {Y.}~\bibnamefont {Huang}}, \bibinfo {author} {\bibfnamefont {Y.}~\bibnamefont {Zhang}}, \ and\ \bibinfo {author} {\bibfnamefont {X.}~\bibnamefont {Chen}},\ }\bibfield  {title} {{Out‐of‐time‐ordered correlators in many‐body localized systems},\ }\href {\doibase 10.1002/andp.201600318} {\bibfield  {journal} {\bibinfo  {journal} {Annalen der Physik}\ }\textbf {\bibinfo {volume} {529}},\ \bibinfo {pages} {1600318} (\bibinfo {year} {2017})}\BibitemShut {NoStop}%
\bibitem [{\citenamefont {He}\ and\ \citenamefont {Lu}(2017)}]{He2017}%
  \BibitemOpen
  \bibfield  {author} {\bibinfo {author} {\bibfnamefont {R.-Q.}\ \bibnamefont {He}}\ and\ \bibinfo {author} {\bibfnamefont {Z.-Y.}\ \bibnamefont {Lu}},\ }\bibfield  {title} {{Characterizing many-body localization by out-of-time-ordered correlation},\ }\href {\doibase 10.1103/PhysRevB.95.054201} {\bibfield  {journal} {\bibinfo  {journal} {Physical Review B}\ }\textbf {\bibinfo {volume} {95}},\ \bibinfo {pages} {054201} (\bibinfo {year} {2017})}\BibitemShut {NoStop}%
\bibitem [{\citenamefont {Chen}\ \emph {et~al.}(2017)\citenamefont {Chen}, \citenamefont {Zhou}, \citenamefont {Huse},\ and\ \citenamefont {Fradkin}}]{Chen2017}%
  \BibitemOpen
  \bibfield  {author} {\bibinfo {author} {\bibfnamefont {X.}~\bibnamefont {Chen}}, \bibinfo {author} {\bibfnamefont {T.}~\bibnamefont {Zhou}}, \bibinfo {author} {\bibfnamefont {D.~A.}\ \bibnamefont {Huse}}, \ and\ \bibinfo {author} {\bibfnamefont {E.}~\bibnamefont {Fradkin}},\ }\bibfield  {title} {{Out‐of‐time‐order correlations in many‐body localized and thermal phases},\ }\href {\doibase 10.1002/andp.201600332} {\bibfield  {journal} {\bibinfo  {journal} {Annalen der Physik}\ }\textbf {\bibinfo {volume} {529}},\ \bibinfo {pages} {1600332} (\bibinfo {year} {2017})}\BibitemShut {NoStop}%
\bibitem [{\citenamefont {B{\"{o}}lter}\ and\ \citenamefont {Kehrein}(2022)}]{Bolter2022}%
  \BibitemOpen
  \bibfield  {author} {\bibinfo {author} {\bibfnamefont {N.}~\bibnamefont {B{\"{o}}lter}}\ and\ \bibinfo {author} {\bibfnamefont {S.}~\bibnamefont {Kehrein}},\ }\bibfield  {title} {{Scrambling and many-body localization in the XXZ chain},\ }\href {\doibase 10.1103/PhysRevB.105.104202} {\bibfield  {journal} {\bibinfo  {journal} {Phys. Rev. B}\ }\textbf {\bibinfo {volume} {105}},\ \bibinfo {pages} {104202} (\bibinfo {year} {2022})}\BibitemShut {NoStop}%
\bibitem [{\citenamefont {MacCormack}\ \emph {et~al.}(2021)\citenamefont {MacCormack}, \citenamefont {Tan}, \citenamefont {Kudler-Flam},\ and\ \citenamefont {Ryu}}]{MacCormack2021}%
  \BibitemOpen
  \bibfield  {author} {\bibinfo {author} {\bibfnamefont {I.}~\bibnamefont {MacCormack}}, \bibinfo {author} {\bibfnamefont {M.~T.}\ \bibnamefont {Tan}}, \bibinfo {author} {\bibfnamefont {J.}~\bibnamefont {Kudler-Flam}}, \ and\ \bibinfo {author} {\bibfnamefont {S.}~\bibnamefont {Ryu}},\ }\bibfield  {title} {{Operator and entanglement growth in nonthermalizing systems: Many-body localization and the random singlet phase},\ }\href {\doibase 10.1103/PhysRevB.104.214202} {\bibfield  {journal} {\bibinfo  {journal} {Phys. Rev. B}\ }\textbf {\bibinfo {volume} {104}},\ \bibinfo {pages} {214202} (\bibinfo {year} {2021})}\BibitemShut {NoStop}%
\bibitem [{\citenamefont {Orito}\ \emph {et~al.}(2022)\citenamefont {Orito}, \citenamefont {Kuno},\ and\ \citenamefont {Ichinose}}]{Orito2022}%
  \BibitemOpen
  \bibfield  {author} {\bibinfo {author} {\bibfnamefont {T.}~\bibnamefont {Orito}}, \bibinfo {author} {\bibfnamefont {Y.}~\bibnamefont {Kuno}}, \ and\ \bibinfo {author} {\bibfnamefont {I.}~\bibnamefont {Ichinose}},\ }\bibfield  {title} {{Quantum information spreading in random spin chains with topological order},\ }\href {\doibase 10.1103/PhysRevB.106.104204} {\bibfield  {journal} {\bibinfo  {journal} {Phys. Rev. B}\ }\textbf {\bibinfo {volume} {106}},\ \bibinfo {pages} {104204} (\bibinfo {year} {2022})}\BibitemShut {NoStop}%
\bibitem [{\citenamefont {Bardarson}\ \emph {et~al.}(2012)\citenamefont {Bardarson}, \citenamefont {Pollmann},\ and\ \citenamefont {Moore}}]{Bardarson2012}%
  \BibitemOpen
  \bibfield  {author} {\bibinfo {author} {\bibfnamefont {J.~H.}\ \bibnamefont {Bardarson}}, \bibinfo {author} {\bibfnamefont {F.}~\bibnamefont {Pollmann}}, \ and\ \bibinfo {author} {\bibfnamefont {J.~E.}\ \bibnamefont {Moore}},\ }\bibfield  {title} {{Unbounded growth of entanglement in models of many-body localization},\ }\href {\doibase 10.1103/PhysRevLett.109.017202} {\bibfield  {journal} {\bibinfo  {journal} {Physical Review Letters}\ }\textbf {\bibinfo {volume} {109}},\ \bibinfo {pages} {017202} (\bibinfo {year} {2012})}\BibitemShut {NoStop}%
\bibitem [{\citenamefont {Serbyn}\ \emph {et~al.}(2013)\citenamefont {Serbyn}, \citenamefont {Papi{\'{c}}},\ and\ \citenamefont {Abanin}}]{Abanin2013}%
  \BibitemOpen
  \bibfield  {author} {\bibinfo {author} {\bibfnamefont {M.}~\bibnamefont {Serbyn}}, \bibinfo {author} {\bibfnamefont {Z.}~\bibnamefont {Papi{\'{c}}}}, \ and\ \bibinfo {author} {\bibfnamefont {D.~A.}\ \bibnamefont {Abanin}},\ }\bibfield  {title} {{Universal Slow Growth of Entanglement in Interacting Strongly Disordered Systems},\ }\href {\doibase 10.1103/PhysRevLett.110.260601} {\bibfield  {journal} {\bibinfo  {journal} {Physical Review Letters}\ }\textbf {\bibinfo {volume} {110}},\ \bibinfo {pages} {260601} (\bibinfo {year} {2013})}\BibitemShut {NoStop}%
\bibitem [{\citenamefont {Iyoda}\ and\ \citenamefont {Sagawa}(2018)}]{Iyoda2018}%
  \BibitemOpen
  \bibfield  {author} {\bibinfo {author} {\bibfnamefont {E.}~\bibnamefont {Iyoda}}\ and\ \bibinfo {author} {\bibfnamefont {T.}~\bibnamefont {Sagawa}},\ }\bibfield  {title} {{Scrambling of quantum information in quantum many-body systems},\ }\href {\doibase 10.1103/PhysRevA.97.042330} {\bibfield  {journal} {\bibinfo  {journal} {Phys. Rev. A}\ }\textbf {\bibinfo {volume} {97}},\ \bibinfo {pages} {042330} (\bibinfo {year} {2018})}\BibitemShut {NoStop}%
\bibitem [{\citenamefont {Schnaack}\ \emph {et~al.}(2019)\citenamefont {Schnaack}, \citenamefont {B{\"{o}}lter}, \citenamefont {Paeckel}, \citenamefont {Manmana}, \citenamefont {Kehrein},\ and\ \citenamefont {Schmitt}}]{Schnaack2019}%
  \BibitemOpen
  \bibfield  {author} {\bibinfo {author} {\bibfnamefont {O.}~\bibnamefont {Schnaack}}, \bibinfo {author} {\bibfnamefont {N.}~\bibnamefont {B{\"{o}}lter}}, \bibinfo {author} {\bibfnamefont {S.}~\bibnamefont {Paeckel}}, \bibinfo {author} {\bibfnamefont {S.~R.}\ \bibnamefont {Manmana}}, \bibinfo {author} {\bibfnamefont {S.}~\bibnamefont {Kehrein}}, \ and\ \bibinfo {author} {\bibfnamefont {M.}~\bibnamefont {Schmitt}},\ }\bibfield  {title} {{Tripartite information, scrambling, and the role of Hilbert space partitioning in quantum lattice models},\ }\href {\doibase 10.1103/PhysRevB.100.224302} {\bibfield  {journal} {\bibinfo  {journal} {Phys. Rev. B}\ }\textbf {\bibinfo {volume} {100}},\ \bibinfo {pages} {224302} (\bibinfo {year} {2019})}\BibitemShut {NoStop}%
\bibitem [{\citenamefont {Caceffo}\ and\ \citenamefont {Alba}(2023)}]{Caceffo2023}%
  \BibitemOpen
  \bibfield  {author} {\bibinfo {author} {\bibfnamefont {F.}~\bibnamefont {Caceffo}}\ and\ \bibinfo {author} {\bibfnamefont {V.}~\bibnamefont {Alba}},\ }\bibfield  {title} {{Negative tripartite mutual information after quantum quenches in integrable systems},\ }\href {\doibase 10.1103/PhysRevB.108.134434} {\bibfield  {journal} {\bibinfo  {journal} {Phys. Rev. B}\ }\textbf {\bibinfo {volume} {108}},\ \bibinfo {pages} {134434} (\bibinfo {year} {2023})}\BibitemShut {NoStop}%
\bibitem [{\citenamefont {Dağ}\ \emph {et~al.}(2020)\citenamefont {Dağ}, \citenamefont {Duan},\ and\ \citenamefont {Sun}}]{Dag2020}%
  \BibitemOpen
  \bibfield  {author} {\bibinfo {author} {\bibfnamefont {C.~B.}\ \bibnamefont {Dağ}}, \bibinfo {author} {\bibfnamefont {L.-M.}\ \bibnamefont {Duan}}, \ and\ \bibinfo {author} {\bibfnamefont {K.}~\bibnamefont {Sun}},\ }\bibfield  {title} {{Topologically induced prescrambling and dynamical detection of topological phase transitions at infinite temperature},\ }\href {\doibase 10.1103/PhysRevB.101.104415} {\bibfield  {journal} {\bibinfo  {journal} {Phys. Rev. B}\ }\textbf {\bibinfo {volume} {101}},\ \bibinfo {pages} {104415} (\bibinfo {year} {2020})}\BibitemShut {NoStop}%
\bibitem [{\citenamefont {Sedlmayr}\ \emph {et~al.}(2023)\citenamefont {Sedlmayr}, \citenamefont {Cheraghi},\ and\ \citenamefont {Sedlmayr}}]{Sedlmayr2023}%
  \BibitemOpen
  \bibfield  {author} {\bibinfo {author} {\bibfnamefont {M.}~\bibnamefont {Sedlmayr}}, \bibinfo {author} {\bibfnamefont {H.}~\bibnamefont {Cheraghi}}, \ and\ \bibinfo {author} {\bibfnamefont {N.}~\bibnamefont {Sedlmayr}},\ }\bibfield  {title} {{Information trapping by topologically protected edge states: Scrambling and the butterfly velocity},\ }\href {\doibase 10.1103/PhysRevB.108.184303} {\bibfield  {journal} {\bibinfo  {journal} {Physical Review B}\ }\textbf {\bibinfo {volume} {108}},\ \bibinfo {pages} {184303} (\bibinfo {year} {2023})}\BibitemShut {NoStop}%
\bibitem [{\citenamefont {Bin}\ \emph {et~al.}(2023)\citenamefont {Bin}, \citenamefont {Wan}, \citenamefont {Nori}, \citenamefont {Wu},\ and\ \citenamefont {L{\"{u}}}}]{Bin2023}%
  \BibitemOpen
  \bibfield  {author} {\bibinfo {author} {\bibfnamefont {Q.}~\bibnamefont {Bin}}, \bibinfo {author} {\bibfnamefont {L.-L.}\ \bibnamefont {Wan}}, \bibinfo {author} {\bibfnamefont {F.}~\bibnamefont {Nori}}, \bibinfo {author} {\bibfnamefont {Y.}~\bibnamefont {Wu}}, \ and\ \bibinfo {author} {\bibfnamefont {X.-Y.}\ \bibnamefont {L{\"{u}}}},\ }\bibfield  {title} {{Out-of-time-order correlation as a witness for topological phase transitions},\ }\href {\doibase 10.1103/PhysRevB.107.L020202} {\bibfield  {journal} {\bibinfo  {journal} {Physical Review B}\ }\textbf {\bibinfo {volume} {107}},\ \bibinfo {pages} {L020202} (\bibinfo {year} {2023})}\BibitemShut {NoStop}%
\bibitem [{\citenamefont {Sur}\ and\ \citenamefont {Sen}(2024)}]{Sur2024}%
  \BibitemOpen
  \bibfield  {author} {\bibinfo {author} {\bibfnamefont {S.}~\bibnamefont {Sur}}\ and\ \bibinfo {author} {\bibfnamefont {D.}~\bibnamefont {Sen}},\ }\bibfield  {title} {{Effects of topological and non-topological edge states on information propagation and scrambling in a Floquet spin chain},\ }\href {\doibase 10.1088/1361-648X/ad1363} {\bibfield  {journal} {\bibinfo  {journal} {Journal of Physics: Condensed Matter}\ }\textbf {\bibinfo {volume} {36}},\ \bibinfo {pages} {125402} (\bibinfo {year} {2024})}\BibitemShut {NoStop}%
\bibitem [{\citenamefont {Khetrapal}\ and\ \citenamefont {Pedersen}(2024)}]{Khetrapal2024}%
  \BibitemOpen
  \bibfield  {author} {\bibinfo {author} {\bibfnamefont {S.}~\bibnamefont {Khetrapal}}\ and\ \bibinfo {author} {\bibfnamefont {E.~T.~M.}\ \bibnamefont {Pedersen}},\ }\bibfield  {title} {{Mutual information scrambling in Ising spin chain},\ }\href {http://arxiv.org/abs/2402.13558} {\bibfield  {journal} {\bibinfo  {journal} {arXiv:2402.13558}\ } (\bibinfo {year} {2024})}\BibitemShut {NoStop}%
\bibitem [{\citenamefont {Preskill}(2018)}]{Preskill2018}%
  \BibitemOpen
  \bibfield  {author} {\bibinfo {author} {\bibfnamefont {J.}~\bibnamefont {Preskill}},\ }\bibfield  {title} {{Quantum Computing in the NISQ era and beyond},\ }\href {\doibase 10.22331/q-2018-08-06-79} {\bibfield  {journal} {\bibinfo  {journal} {Quantum}\ }\textbf {\bibinfo {volume} {2}},\ \bibinfo {pages} {79} (\bibinfo {year} {2018})}\BibitemShut {NoStop}%
\bibitem [{\citenamefont {van Enk}\ and\ \citenamefont {Beenakker}(2012)}]{VanEnk2011}%
  \BibitemOpen
  \bibfield  {author} {\bibinfo {author} {\bibfnamefont {S.~J.}\ \bibnamefont {van Enk}}\ and\ \bibinfo {author} {\bibfnamefont {C.~W.~J.}\ \bibnamefont {Beenakker}},\ }\bibfield  {title} {{The power of random measurements: measuring Tr$\rho$n on single copies of $\rho$},\ }\href {\doibase 10.1103/PhysRevLett.108.110503} {\bibfield  {journal} {\bibinfo  {journal} {Physical Review Letters}\ }\textbf {\bibinfo {volume} {108}},\ \bibinfo {pages} {110503} (\bibinfo {year} {2012})}\BibitemShut {NoStop}%
\bibitem [{\citenamefont {Brydges}\ \emph {et~al.}(2019)\citenamefont {Brydges}, \citenamefont {Elben}, \citenamefont {Jurcevic}, \citenamefont {Vermersch}, \citenamefont {Maier}, \citenamefont {Lanyon}, \citenamefont {Zoller}, \citenamefont {Blatt},\ and\ \citenamefont {Roos}}]{Brydges2019}%
  \BibitemOpen
  \bibfield  {author} {\bibinfo {author} {\bibfnamefont {T.}~\bibnamefont {Brydges}}, \bibinfo {author} {\bibfnamefont {A.}~\bibnamefont {Elben}}, \bibinfo {author} {\bibfnamefont {P.}~\bibnamefont {Jurcevic}}, \bibinfo {author} {\bibfnamefont {B.}~\bibnamefont {Vermersch}}, \bibinfo {author} {\bibfnamefont {C.}~\bibnamefont {Maier}}, \bibinfo {author} {\bibfnamefont {B.~P.}\ \bibnamefont {Lanyon}}, \bibinfo {author} {\bibfnamefont {P.}~\bibnamefont {Zoller}}, \bibinfo {author} {\bibfnamefont {R.}~\bibnamefont {Blatt}}, \ and\ \bibinfo {author} {\bibfnamefont {C.~F.}\ \bibnamefont {Roos}},\ }\bibfield  {title} {{Probing R{\'{e}}nyi entanglement entropy via randomized measurements},\ }\href {\doibase 10.1126/science.aau4963} {\bibfield  {journal} {\bibinfo  {journal} {Science}\ }\textbf {\bibinfo {volume} {364}},\ \bibinfo {pages} {260} (\bibinfo {year} {2019})}\BibitemShut {NoStop}%
\bibitem [{\citenamefont {Huang}\ \emph {et~al.}(2020)\citenamefont {Huang}, \citenamefont {Kueng},\ and\ \citenamefont {Preskill}}]{Huang2020b}%
  \BibitemOpen
  \bibfield  {author} {\bibinfo {author} {\bibfnamefont {H.~Y.}\ \bibnamefont {Huang}}, \bibinfo {author} {\bibfnamefont {R.}~\bibnamefont {Kueng}}, \ and\ \bibinfo {author} {\bibfnamefont {J.}~\bibnamefont {Preskill}},\ }\bibfield  {title} {{Predicting many properties of a quantum system from very few measurements},\ }\href {\doibase 10.1038/s41567-020-0932-7} {\bibfield  {journal} {\bibinfo  {journal} {Nature Physics}\ }\textbf {\bibinfo {volume} {16}},\ \bibinfo {pages} {1050} (\bibinfo {year} {2020})}\BibitemShut {NoStop}%
\bibitem [{\citenamefont {Arute}\ \emph {et~al.}(2019)\citenamefont {Arute}, \citenamefont {Arya}, \citenamefont {Babbush}, \citenamefont {Bacon}, \citenamefont {Bardin}, \citenamefont {Barends}, \citenamefont {Biswas}, \citenamefont {Boixo}, \citenamefont {Brandao}, \citenamefont {Buell}, \citenamefont {Burkett}, \citenamefont {Chen}, \citenamefont {Chen}, \citenamefont {Chiaro}, \citenamefont {Collins}, \citenamefont {Courtney}, \citenamefont {Dunsworth}, \citenamefont {Farhi}, \citenamefont {Foxen}, \citenamefont {Fowler}, \citenamefont {Gidney}, \citenamefont {Giustina}, \citenamefont {Graff}, \citenamefont {Guerin}, \citenamefont {Habegger}, \citenamefont {Harrigan}, \citenamefont {Hartmann}, \citenamefont {Ho}, \citenamefont {Hoffmann}, \citenamefont {Huang}, \citenamefont {Humble}, \citenamefont {Isakov}, \citenamefont {Jeffrey}, \citenamefont {Jiang}, \citenamefont {Kafri}, \citenamefont {Kechedzhi}, \citenamefont {Kelly}, \citenamefont {Klimov}, \citenamefont {Knysh}, \citenamefont {Korotkov},
  \citenamefont {Kostritsa}, \citenamefont {Landhuis}, \citenamefont {Lindmark}, \citenamefont {Lucero}, \citenamefont {Lyakh}, \citenamefont {Mandr{\`{a}}}, \citenamefont {McClean}, \citenamefont {McEwen}, \citenamefont {Megrant}, \citenamefont {Mi}, \citenamefont {Michielsen}, \citenamefont {Mohseni}, \citenamefont {Mutus}, \citenamefont {Naaman}, \citenamefont {Neeley}, \citenamefont {Neill}, \citenamefont {Niu}, \citenamefont {Ostby}, \citenamefont {Petukhov}, \citenamefont {Platt}, \citenamefont {Quintana}, \citenamefont {Rieffel}, \citenamefont {Roushan}, \citenamefont {Rubin}, \citenamefont {Sank}, \citenamefont {Satzinger}, \citenamefont {Smelyanskiy}, \citenamefont {Sung}, \citenamefont {Trevithick}, \citenamefont {Vainsencher}, \citenamefont {Villalonga}, \citenamefont {White}, \citenamefont {Yao}, \citenamefont {Yeh}, \citenamefont {Zalcman}, \citenamefont {Neven},\ and\ \citenamefont {Martinis}}]{Arute2019}%
  \BibitemOpen
  \bibfield  {author} {\bibinfo {author} {\bibfnamefont {F.}~\bibnamefont {Arute}}, \bibinfo {author} {\bibfnamefont {K.}~\bibnamefont {Arya}}, \bibinfo {author} {\bibfnamefont {R.}~\bibnamefont {Babbush}}, \bibinfo {author} {\bibfnamefont {D.}~\bibnamefont {Bacon}}, \bibinfo {author} {\bibfnamefont {J.~C.}\ \bibnamefont {Bardin}}, \bibinfo {author} {\bibfnamefont {R.}~\bibnamefont {Barends}}, \bibinfo {author} {\bibfnamefont {R.}~\bibnamefont {Biswas}}, \bibinfo {author} {\bibfnamefont {S.}~\bibnamefont {Boixo}}, \bibinfo {author} {\bibfnamefont {F.~G. S. L. S.~L.}\ \bibnamefont {Brandao}}, \bibinfo {author} {\bibfnamefont {D.~A.}\ \bibnamefont {Buell}}, \bibinfo {author} {\bibfnamefont {B.}~\bibnamefont {Burkett}}, \bibinfo {author} {\bibfnamefont {Y.}~\bibnamefont {Chen}}, \bibinfo {author} {\bibfnamefont {Z.}~\bibnamefont {Chen}}, \bibinfo {author} {\bibfnamefont {B.}~\bibnamefont {Chiaro}}, \bibinfo {author} {\bibfnamefont {R.}~\bibnamefont {Collins}}, \bibinfo {author} {\bibfnamefont {W.}~\bibnamefont
  {Courtney}}, \bibinfo {author} {\bibfnamefont {A.}~\bibnamefont {Dunsworth}}, \bibinfo {author} {\bibfnamefont {E.}~\bibnamefont {Farhi}}, \bibinfo {author} {\bibfnamefont {B.}~\bibnamefont {Foxen}}, \bibinfo {author} {\bibfnamefont {A.}~\bibnamefont {Fowler}}, \bibinfo {author} {\bibfnamefont {C.}~\bibnamefont {Gidney}}, \bibinfo {author} {\bibfnamefont {M.}~\bibnamefont {Giustina}}, \bibinfo {author} {\bibfnamefont {R.}~\bibnamefont {Graff}}, \bibinfo {author} {\bibfnamefont {K.}~\bibnamefont {Guerin}}, \bibinfo {author} {\bibfnamefont {S.}~\bibnamefont {Habegger}}, \bibinfo {author} {\bibfnamefont {M.~P.}\ \bibnamefont {Harrigan}}, \bibinfo {author} {\bibfnamefont {M.~J.}\ \bibnamefont {Hartmann}}, \bibinfo {author} {\bibfnamefont {A.}~\bibnamefont {Ho}}, \bibinfo {author} {\bibfnamefont {M.}~\bibnamefont {Hoffmann}}, \bibinfo {author} {\bibfnamefont {T.}~\bibnamefont {Huang}}, \bibinfo {author} {\bibfnamefont {T.~S.}\ \bibnamefont {Humble}}, \bibinfo {author} {\bibfnamefont {S.~V.}\ \bibnamefont
  {Isakov}}, \bibinfo {author} {\bibfnamefont {E.}~\bibnamefont {Jeffrey}}, \bibinfo {author} {\bibfnamefont {Z.}~\bibnamefont {Jiang}}, \bibinfo {author} {\bibfnamefont {D.}~\bibnamefont {Kafri}}, \bibinfo {author} {\bibfnamefont {K.}~\bibnamefont {Kechedzhi}}, \bibinfo {author} {\bibfnamefont {J.}~\bibnamefont {Kelly}}, \bibinfo {author} {\bibfnamefont {P.~V.}\ \bibnamefont {Klimov}}, \bibinfo {author} {\bibfnamefont {S.}~\bibnamefont {Knysh}}, \bibinfo {author} {\bibfnamefont {A.}~\bibnamefont {Korotkov}}, \bibinfo {author} {\bibfnamefont {F.}~\bibnamefont {Kostritsa}}, \bibinfo {author} {\bibfnamefont {D.}~\bibnamefont {Landhuis}}, \bibinfo {author} {\bibfnamefont {M.}~\bibnamefont {Lindmark}}, \bibinfo {author} {\bibfnamefont {E.}~\bibnamefont {Lucero}}, \bibinfo {author} {\bibfnamefont {D.}~\bibnamefont {Lyakh}}, \bibinfo {author} {\bibfnamefont {S.}~\bibnamefont {Mandr{\`{a}}}}, \bibinfo {author} {\bibfnamefont {J.~R.}\ \bibnamefont {McClean}}, \bibinfo {author} {\bibfnamefont {M.}~\bibnamefont
  {McEwen}}, \bibinfo {author} {\bibfnamefont {A.}~\bibnamefont {Megrant}}, \bibinfo {author} {\bibfnamefont {X.}~\bibnamefont {Mi}}, \bibinfo {author} {\bibfnamefont {K.}~\bibnamefont {Michielsen}}, \bibinfo {author} {\bibfnamefont {M.}~\bibnamefont {Mohseni}}, \bibinfo {author} {\bibfnamefont {J.}~\bibnamefont {Mutus}}, \bibinfo {author} {\bibfnamefont {O.}~\bibnamefont {Naaman}}, \bibinfo {author} {\bibfnamefont {M.}~\bibnamefont {Neeley}}, \bibinfo {author} {\bibfnamefont {C.}~\bibnamefont {Neill}}, \bibinfo {author} {\bibfnamefont {M.~Y.}\ \bibnamefont {Niu}}, \bibinfo {author} {\bibfnamefont {E.}~\bibnamefont {Ostby}}, \bibinfo {author} {\bibfnamefont {A.}~\bibnamefont {Petukhov}}, \bibinfo {author} {\bibfnamefont {J.~C.}\ \bibnamefont {Platt}}, \bibinfo {author} {\bibfnamefont {C.}~\bibnamefont {Quintana}}, \bibinfo {author} {\bibfnamefont {E.~G.}\ \bibnamefont {Rieffel}}, \bibinfo {author} {\bibfnamefont {P.}~\bibnamefont {Roushan}}, \bibinfo {author} {\bibfnamefont {N.~C.}\ \bibnamefont {Rubin}},
  \bibinfo {author} {\bibfnamefont {D.}~\bibnamefont {Sank}}, \bibinfo {author} {\bibfnamefont {K.~J.}\ \bibnamefont {Satzinger}}, \bibinfo {author} {\bibfnamefont {V.}~\bibnamefont {Smelyanskiy}}, \bibinfo {author} {\bibfnamefont {K.~J.}\ \bibnamefont {Sung}}, \bibinfo {author} {\bibfnamefont {M.~D.}\ \bibnamefont {Trevithick}}, \bibinfo {author} {\bibfnamefont {A.}~\bibnamefont {Vainsencher}}, \bibinfo {author} {\bibfnamefont {B.}~\bibnamefont {Villalonga}}, \bibinfo {author} {\bibfnamefont {T.}~\bibnamefont {White}}, \bibinfo {author} {\bibfnamefont {Z.~J.}\ \bibnamefont {Yao}}, \bibinfo {author} {\bibfnamefont {P.}~\bibnamefont {Yeh}}, \bibinfo {author} {\bibfnamefont {A.}~\bibnamefont {Zalcman}}, \bibinfo {author} {\bibfnamefont {H.}~\bibnamefont {Neven}}, \ and\ \bibinfo {author} {\bibfnamefont {J.~M.}\ \bibnamefont {Martinis}},\ }\bibfield  {title} {{Quantum supremacy using a programmable superconducting processor},\ }\href {\doibase 10.1038/s41586-019-1666-5} {\bibfield  {journal} {\bibinfo  {journal}
  {Nature}\ }\textbf {\bibinfo {volume} {574}},\ \bibinfo {pages} {505} (\bibinfo {year} {2019})}\BibitemShut {NoStop}%
\bibitem [{\citenamefont {Sachdev}\ and\ \citenamefont {Ye}(1993)}]{Sachdev1993}%
  \BibitemOpen
  \bibfield  {author} {\bibinfo {author} {\bibfnamefont {S.}~\bibnamefont {Sachdev}}\ and\ \bibinfo {author} {\bibfnamefont {J.}~\bibnamefont {Ye}},\ }\bibfield  {title} {{Gapless spin-fluid ground state in a random quantum Heisenberg magnet},\ }\href {\doibase 10.1103/PhysRevLett.70.3339} {\bibfield  {journal} {\bibinfo  {journal} {Physical Review Letters}\ }\textbf {\bibinfo {volume} {70}},\ \bibinfo {pages} {3339} (\bibinfo {year} {1993})}\BibitemShut {NoStop}%
\bibitem [{\citenamefont {Maldacena}\ and\ \citenamefont {Stanford}(2016)}]{Maldacena2016a}%
  \BibitemOpen
  \bibfield  {author} {\bibinfo {author} {\bibfnamefont {J.}~\bibnamefont {Maldacena}}\ and\ \bibinfo {author} {\bibfnamefont {D.}~\bibnamefont {Stanford}},\ }\bibfield  {title} {{Remarks on the Sachdev-Ye-Kitaev model},\ }\href {\doibase 10.1103/PhysRevD.94.106002} {\bibfield  {journal} {\bibinfo  {journal} {Physical Review D}\ }\textbf {\bibinfo {volume} {94}},\ \bibinfo {pages} {106002} (\bibinfo {year} {2016})}\BibitemShut {NoStop}%
\bibitem [{\citenamefont {Chowdhury}\ \emph {et~al.}(2022)\citenamefont {Chowdhury}, \citenamefont {Georges}, \citenamefont {Parcollet},\ and\ \citenamefont {Sachdev}}]{Chowdhury2022}%
  \BibitemOpen
  \bibfield  {author} {\bibinfo {author} {\bibfnamefont {D.}~\bibnamefont {Chowdhury}}, \bibinfo {author} {\bibfnamefont {A.}~\bibnamefont {Georges}}, \bibinfo {author} {\bibfnamefont {O.}~\bibnamefont {Parcollet}}, \ and\ \bibinfo {author} {\bibfnamefont {S.}~\bibnamefont {Sachdev}},\ }\bibfield  {title} {{Sachdev-Ye-Kitaev models and beyond: Window into non-Fermi liquids},\ }\href {\doibase 10.1103/RevModPhys.94.035004} {\bibfield  {journal} {\bibinfo  {journal} {Reviews of Modern Physics}\ }\textbf {\bibinfo {volume} {94}},\ \bibinfo {pages} {035004} (\bibinfo {year} {2022})}\BibitemShut {NoStop}%
\bibitem [{\citenamefont {Cerezo}\ \emph {et~al.}(2021)\citenamefont {Cerezo}, \citenamefont {Arrasmith}, \citenamefont {Babbush}, \citenamefont {Benjamin}, \citenamefont {Endo}, \citenamefont {Fujii}, \citenamefont {McClean}, \citenamefont {Mitarai}, \citenamefont {Yuan}, \citenamefont {Cincio},\ and\ \citenamefont {Coles}}]{Cerezo2020b}%
  \BibitemOpen
  \bibfield  {author} {\bibinfo {author} {\bibfnamefont {M.}~\bibnamefont {Cerezo}}, \bibinfo {author} {\bibfnamefont {A.}~\bibnamefont {Arrasmith}}, \bibinfo {author} {\bibfnamefont {R.}~\bibnamefont {Babbush}}, \bibinfo {author} {\bibfnamefont {S.~C.}\ \bibnamefont {Benjamin}}, \bibinfo {author} {\bibfnamefont {S.}~\bibnamefont {Endo}}, \bibinfo {author} {\bibfnamefont {K.}~\bibnamefont {Fujii}}, \bibinfo {author} {\bibfnamefont {J.~R.}\ \bibnamefont {McClean}}, \bibinfo {author} {\bibfnamefont {K.}~\bibnamefont {Mitarai}}, \bibinfo {author} {\bibfnamefont {X.}~\bibnamefont {Yuan}}, \bibinfo {author} {\bibfnamefont {L.}~\bibnamefont {Cincio}}, \ and\ \bibinfo {author} {\bibfnamefont {P.~J.}\ \bibnamefont {Coles}},\ }\bibfield  {title} {{Variational quantum algorithms},\ }\href {\doibase 10.1038/s42254-021-00348-9} {\bibfield  {journal} {\bibinfo  {journal} {Nat. Rev. Phys.}\ }\textbf {\bibinfo {volume} {3}},\ \bibinfo {pages} {625} (\bibinfo {year} {2021})}\BibitemShut {NoStop}%
\bibitem [{\citenamefont {Bharti}\ \emph {et~al.}(2022)\citenamefont {Bharti}, \citenamefont {Cervera-Lierta}, \citenamefont {Kyaw}, \citenamefont {Haug}, \citenamefont {Alperin-Lea}, \citenamefont {Anand}, \citenamefont {Degroote}, \citenamefont {Heimonen}, \citenamefont {Kottmann}, \citenamefont {Menke}, \citenamefont {Mok}, \citenamefont {Sim}, \citenamefont {Kwek},\ and\ \citenamefont {Aspuru-Guzik}}]{Bharti2022}%
  \BibitemOpen
  \bibfield  {author} {\bibinfo {author} {\bibfnamefont {K.}~\bibnamefont {Bharti}}, \bibinfo {author} {\bibfnamefont {A.}~\bibnamefont {Cervera-Lierta}}, \bibinfo {author} {\bibfnamefont {T.~H.}\ \bibnamefont {Kyaw}}, \bibinfo {author} {\bibfnamefont {T.}~\bibnamefont {Haug}}, \bibinfo {author} {\bibfnamefont {S.}~\bibnamefont {Alperin-Lea}}, \bibinfo {author} {\bibfnamefont {A.}~\bibnamefont {Anand}}, \bibinfo {author} {\bibfnamefont {M.}~\bibnamefont {Degroote}}, \bibinfo {author} {\bibfnamefont {H.}~\bibnamefont {Heimonen}}, \bibinfo {author} {\bibfnamefont {J.~S.}\ \bibnamefont {Kottmann}}, \bibinfo {author} {\bibfnamefont {T.}~\bibnamefont {Menke}}, \bibinfo {author} {\bibfnamefont {W.-K.}\ \bibnamefont {Mok}}, \bibinfo {author} {\bibfnamefont {S.}~\bibnamefont {Sim}}, \bibinfo {author} {\bibfnamefont {L.-C.}\ \bibnamefont {Kwek}}, \ and\ \bibinfo {author} {\bibfnamefont {A.}~\bibnamefont {Aspuru-Guzik}},\ }\bibfield  {title} {{Noisy intermediate-scale quantum algorithms},\ }\href {\doibase
  10.1103/RevModPhys.94.015004} {\bibfield  {journal} {\bibinfo  {journal} {Rev. Mod. Phys.}\ }\textbf {\bibinfo {volume} {94}},\ \bibinfo {pages} {015004} (\bibinfo {year} {2022})}\BibitemShut {NoStop}%
\bibitem [{\citenamefont {Shen}\ \emph {et~al.}(2020)\citenamefont {Shen}, \citenamefont {Zhang}, \citenamefont {You},\ and\ \citenamefont {Zhai}}]{Shen2020}%
  \BibitemOpen
  \bibfield  {author} {\bibinfo {author} {\bibfnamefont {H.}~\bibnamefont {Shen}}, \bibinfo {author} {\bibfnamefont {P.}~\bibnamefont {Zhang}}, \bibinfo {author} {\bibfnamefont {Y.-Z.}\ \bibnamefont {You}}, \ and\ \bibinfo {author} {\bibfnamefont {H.}~\bibnamefont {Zhai}},\ }\bibfield  {title} {{Information Scrambling in Quantum Neural Networks},\ }\href {\doibase 10.1103/PhysRevLett.124.200504} {\bibfield  {journal} {\bibinfo  {journal} {Phys. Rev. Lett.}\ }\textbf {\bibinfo {volume} {124}},\ \bibinfo {pages} {200504} (\bibinfo {year} {2020})}\BibitemShut {NoStop}%
\bibitem [{\citenamefont {Garcia}\ \emph {et~al.}(2022)\citenamefont {Garcia}, \citenamefont {Bu},\ and\ \citenamefont {Jaffe}}]{Garcia2022}%
  \BibitemOpen
  \bibfield  {author} {\bibinfo {author} {\bibfnamefont {R.~J.}\ \bibnamefont {Garcia}}, \bibinfo {author} {\bibfnamefont {K.}~\bibnamefont {Bu}}, \ and\ \bibinfo {author} {\bibfnamefont {A.}~\bibnamefont {Jaffe}},\ }\bibfield  {title} {{Quantifying scrambling in quantum neural networks},\ }\href {\doibase 10.1007/JHEP03(2022)027} {\bibfield  {journal} {\bibinfo  {journal} {J. High Energy Phys.}\ }\textbf {\bibinfo {volume} {2022}},\ \bibinfo {pages} {27} (\bibinfo {year} {2022})}\BibitemShut {NoStop}%
\bibitem [{\citenamefont {Sajjan}\ \emph {et~al.}(2023)\citenamefont {Sajjan}, \citenamefont {Singh}, \citenamefont {Selvarajan},\ and\ \citenamefont {Kais}}]{Sajjan2023}%
  \BibitemOpen
  \bibfield  {author} {\bibinfo {author} {\bibfnamefont {M.}~\bibnamefont {Sajjan}}, \bibinfo {author} {\bibfnamefont {V.}~\bibnamefont {Singh}}, \bibinfo {author} {\bibfnamefont {R.}~\bibnamefont {Selvarajan}}, \ and\ \bibinfo {author} {\bibfnamefont {S.}~\bibnamefont {Kais}},\ }\bibfield  {title} {{Imaginary components of out-of-time-order correlator and information scrambling for navigating the learning landscape of a quantum machine learning model},\ }\href {\doibase 10.1103/PhysRevResearch.5.013146} {\bibfield  {journal} {\bibinfo  {journal} {Phys. Rev. Res.}\ }\textbf {\bibinfo {volume} {5}},\ \bibinfo {pages} {013146} (\bibinfo {year} {2023})}\BibitemShut {NoStop}%
\bibitem [{\citenamefont {Zhang}\ \emph {et~al.}(2022)\citenamefont {Zhang}, \citenamefont {Hsieh}, \citenamefont {Zhang},\ and\ \citenamefont {Yao}}]{Zhang2020b}%
  \BibitemOpen
  \bibfield  {author} {\bibinfo {author} {\bibfnamefont {S.-X.}\ \bibnamefont {Zhang}}, \bibinfo {author} {\bibfnamefont {C.-Y.}\ \bibnamefont {Hsieh}}, \bibinfo {author} {\bibfnamefont {S.}~\bibnamefont {Zhang}}, \ and\ \bibinfo {author} {\bibfnamefont {H.}~\bibnamefont {Yao}},\ }\bibfield  {title} {{Differentiable quantum architecture search},\ }\href {\doibase 10.1088/2058-9565/ac87cd} {\bibfield  {journal} {\bibinfo  {journal} {Quantum Sci. Technol.}\ }\textbf {\bibinfo {volume} {7}},\ \bibinfo {pages} {045023} (\bibinfo {year} {2022})}\BibitemShut {NoStop}%
\bibitem [{\citenamefont {Du}\ \emph {et~al.}(2022)\citenamefont {Du}, \citenamefont {Huang}, \citenamefont {You}, \citenamefont {Hsieh},\ and\ \citenamefont {Tao}}]{Du2020a}%
  \BibitemOpen
  \bibfield  {author} {\bibinfo {author} {\bibfnamefont {Y.}~\bibnamefont {Du}}, \bibinfo {author} {\bibfnamefont {T.}~\bibnamefont {Huang}}, \bibinfo {author} {\bibfnamefont {S.}~\bibnamefont {You}}, \bibinfo {author} {\bibfnamefont {M.-H.}\ \bibnamefont {Hsieh}}, \ and\ \bibinfo {author} {\bibfnamefont {D.}~\bibnamefont {Tao}},\ }\bibfield  {title} {{Quantum circuit architecture search for variational quantum algorithms},\ }\href {\doibase 10.1038/s41534-022-00570-y} {\bibfield  {journal} {\bibinfo  {journal} {npj Quantum Inf.}\ }\textbf {\bibinfo {volume} {8}},\ \bibinfo {pages} {62} (\bibinfo {year} {2022})}\BibitemShut {NoStop}%
\bibitem [{\citenamefont {Lu}\ \emph {et~al.}(2021)\citenamefont {Lu}, \citenamefont {Shen},\ and\ \citenamefont {Deng}}]{Lu2020}%
  \BibitemOpen
  \bibfield  {author} {\bibinfo {author} {\bibfnamefont {Z.}~\bibnamefont {Lu}}, \bibinfo {author} {\bibfnamefont {P.-X.}\ \bibnamefont {Shen}}, \ and\ \bibinfo {author} {\bibfnamefont {D.-L.}\ \bibnamefont {Deng}},\ }\bibfield  {title} {{Markovian Quantum Neuroevolution for Machine Learning},\ }\href {\doibase 10.1103/PhysRevApplied.16.044039} {\bibfield  {journal} {\bibinfo  {journal} {Phys. Rev. Appl.}\ }\textbf {\bibinfo {volume} {16}},\ \bibinfo {pages} {044039} (\bibinfo {year} {2021})}\BibitemShut {NoStop}%
\bibitem [{\citenamefont {Zhang}\ \emph {et~al.}(2021)\citenamefont {Zhang}, \citenamefont {Hsieh}, \citenamefont {Zhang},\ and\ \citenamefont {Yao}}]{Zhang2021}%
  \BibitemOpen
  \bibfield  {author} {\bibinfo {author} {\bibfnamefont {S.-X.}\ \bibnamefont {Zhang}}, \bibinfo {author} {\bibfnamefont {C.-Y.}\ \bibnamefont {Hsieh}}, \bibinfo {author} {\bibfnamefont {S.}~\bibnamefont {Zhang}}, \ and\ \bibinfo {author} {\bibfnamefont {H.}~\bibnamefont {Yao}},\ }\bibfield  {title} {{Neural predictor based quantum architecture search},\ }\href {\doibase 10.1088/2632-2153/ac28dd} {\bibfield  {journal} {\bibinfo  {journal} {Mach. Learn. Sci. Technol.}\ }\textbf {\bibinfo {volume} {2}},\ \bibinfo {pages} {045027} (\bibinfo {year} {2021})}\BibitemShut {NoStop}%
\bibitem [{\citenamefont {Wu}\ \emph {et~al.}(2021)\citenamefont {Wu}, \citenamefont {Zhang},\ and\ \citenamefont {Zhai}}]{Wu2021b}%
  \BibitemOpen
  \bibfield  {author} {\bibinfo {author} {\bibfnamefont {Y.}~\bibnamefont {Wu}}, \bibinfo {author} {\bibfnamefont {P.}~\bibnamefont {Zhang}}, \ and\ \bibinfo {author} {\bibfnamefont {H.}~\bibnamefont {Zhai}},\ }\bibfield  {title} {{Scrambling ability of quantum neural network architectures},\ }\href {\doibase 10.1103/PhysRevResearch.3.L032057} {\bibfield  {journal} {\bibinfo  {journal} {Phys. Rev. Res.}\ }\textbf {\bibinfo {volume} {3}},\ \bibinfo {pages} {L032057} (\bibinfo {year} {2021})}\BibitemShut {NoStop}%
\bibitem [{\citenamefont {Park}\ \emph {et~al.}(2024)\citenamefont {Park}, \citenamefont {Kang},\ and\ \citenamefont {Huh}}]{Park2024}%
  \BibitemOpen
  \bibfield  {author} {\bibinfo {author} {\bibfnamefont {C.-Y.}\ \bibnamefont {Park}}, \bibinfo {author} {\bibfnamefont {M.}~\bibnamefont {Kang}}, \ and\ \bibinfo {author} {\bibfnamefont {J.}~\bibnamefont {Huh}},\ }\bibfield  {title} {{Hardware-efficient ansatz without barren plateaus in any depth},\ }\href {http://arxiv.org/abs/2403.04844} {\bibfield  {journal} {\bibinfo  {journal} {arXiv:2403.04844}\ } (\bibinfo {year} {2024})}\BibitemShut {NoStop}%
\bibitem [{\citenamefont {Cao}\ \emph {et~al.}(2024)\citenamefont {Cao}, \citenamefont {Zhou}, \citenamefont {Tannu}, \citenamefont {Shannon},\ and\ \citenamefont {Joynt}}]{Cao2024}%
  \BibitemOpen
  \bibfield  {author} {\bibinfo {author} {\bibfnamefont {C.}~\bibnamefont {Cao}}, \bibinfo {author} {\bibfnamefont {Y.}~\bibnamefont {Zhou}}, \bibinfo {author} {\bibfnamefont {S.}~\bibnamefont {Tannu}}, \bibinfo {author} {\bibfnamefont {N.}~\bibnamefont {Shannon}}, \ and\ \bibinfo {author} {\bibfnamefont {R.}~\bibnamefont {Joynt}},\ }\bibfield  {title} {{Exploiting many-body localization for scalable variational quantum simulation},\ }\href {http://arxiv.org/abs/2404.17560} {\bibfield  {journal} {\bibinfo  {journal} {arXiv:2404.17560}\ } (\bibinfo {year} {2024})}\BibitemShut {NoStop}%
\bibitem [{\citenamefont {Liu}\ \emph {et~al.}(2023{\natexlab{d}})\citenamefont {Liu}, \citenamefont {Zhang}, \citenamefont {Jian},\ and\ \citenamefont {Yao}}]{Liu2023a}%
  \BibitemOpen
  \bibfield  {author} {\bibinfo {author} {\bibfnamefont {S.}~\bibnamefont {Liu}}, \bibinfo {author} {\bibfnamefont {S.-X.}\ \bibnamefont {Zhang}}, \bibinfo {author} {\bibfnamefont {S.-K.}\ \bibnamefont {Jian}}, \ and\ \bibinfo {author} {\bibfnamefont {H.}~\bibnamefont {Yao}},\ }\bibfield  {title} {{Training variational quantum algorithms with random gate activation},\ }\href {\doibase 10.1103/PhysRevResearch.5.L032040} {\bibfield  {journal} {\bibinfo  {journal} {Phys. Rev. Res.}\ }\textbf {\bibinfo {volume} {5}},\ \bibinfo {pages} {L032040} (\bibinfo {year} {2023}{\natexlab{d}})}\BibitemShut {NoStop}%
\bibitem [{\citenamefont {McClean}\ \emph {et~al.}(2018)\citenamefont {McClean}, \citenamefont {Boixo}, \citenamefont {Smelyanskiy}, \citenamefont {Babbush},\ and\ \citenamefont {Neven}}]{McClean2018}%
  \BibitemOpen
  \bibfield  {author} {\bibinfo {author} {\bibfnamefont {J.~R.}\ \bibnamefont {McClean}}, \bibinfo {author} {\bibfnamefont {S.}~\bibnamefont {Boixo}}, \bibinfo {author} {\bibfnamefont {V.~N.}\ \bibnamefont {Smelyanskiy}}, \bibinfo {author} {\bibfnamefont {R.}~\bibnamefont {Babbush}}, \ and\ \bibinfo {author} {\bibfnamefont {H.}~\bibnamefont {Neven}},\ }\bibfield  {title} {{Barren plateaus in quantum neural network training landscapes},\ }\href {\doibase 10.1038/s41467-018-07090-4} {\bibfield  {journal} {\bibinfo  {journal} {Nat. Commun.}\ }\textbf {\bibinfo {volume} {9}},\ \bibinfo {pages} {4812} (\bibinfo {year} {2018})}\BibitemShut {NoStop}%
\end{thebibliography}

\end{document}